\documentstyle[12pt,amssymb,amscd,righttag]{amsart}

\newtheorem{thm}{Theorem}[section]
\newtheorem{lem}{Lemma}[section]
\newtheorem{pro}{Proposition}[section]
\newtheorem{cor}{Corollary}[section]

\theoremstyle{definition}

\theoremstyle{remark}

\newtheorem{rem}{Remark}[section]

\numberwithin{equation}{section}

\renewcommand{\epsilon}{\varepsilon}
\renewcommand{\phi}{\varphi}
\renewcommand{\Re}{\operatorname{Re}}
\renewcommand{\Im}{\operatorname{Im}}

\newcommand{\wC}{\Bbb C}
\newcommand{\wR}{\Bbb R}

\newcommand{\wZ}{\Bbb Z}
\newcommand{\wQ}{\Bbb Q}

\newcommand{\Id}{\operatorname{Id}}
\newcommand{\ord}{\operatorname{ord}}
\newcommand{\ind}{\operatorname{ind}}
\newcommand{\Spec}{\operatorname{Spec}}
\newcommand{\End}{\operatorname{End}}
\newcommand{\Aut}{\operatorname{Aut}}
\newcommand{\Ad}{\operatorname{Ad}}
\newcommand{\ad}{\operatorname{ad}}
\newcommand{\Ass}{\operatorname{Ass}}
\newcommand{\Ell}{\operatorname{Ell}}
\newcommand{\SEll}{\operatorname{SEll}}
\newcommand{\DEll}{\operatorname{DEll}}
\newcommand{\const}{\operatorname{const}}
\newcommand{\Tr}{\operatorname{Tr}}
\newcommand{\tr}{\operatorname{tr}}
\newcommand{\TR}{\operatorname{TR}}
\renewcommand{\Sp}{\operatorname{Sp}}
\newcommand{\res}{\operatorname{res}}
\newcommand{\Res}{\operatorname{Res}}
\newcommand{\Ker}{\operatorname{Ker}}
\newcommand{\Coker}{\operatorname{Coker}}
\newcommand{\codim}{\operatorname{codim}}
\newcommand{\rk}{\operatorname{rk}}
\newcommand{\sign}{\operatorname{sign}}
\newcommand{\var}{\operatorname{var}}
\newcommand{\Det}{\operatorname{Det}}
\newcommand{\ch}{\operatorname{ch}}
\newcommand{\Gr}{\operatorname{Gr}}
\newcommand{\Ind}{\operatorname{Ind}}
\newcommand{\cl}{\operatorname{cl}}
\newcommand{\dd}{\operatorname{d}}
\newcommand{\Alt}{\operatorname{Alt}}
\newcommand{\diam}{\operatorname{diam}}
\newcommand{\supp}{\operatorname{supp}}

\newcommand{\df}{\partial}
\newcommand{\dfg}{\df_{\gamma}}
\newcommand{\bgo}{\big|_{\gamma=0}}
\newcommand{\eps}{\epsilon}
\newcommand{\zuo}{\wZ_+\cup 0}
\newcommand{\zat}{\zeta_{A,(\theta)}}
\newcommand{\frg}{\frak{g}}
\newcommand{\sfrg}{\widetilde{\frg}}
\newcommand{\ff}{\frak{f}}
\newcommand{\fell}{\frak{e}\frak{l}\frak{l}}

\newcommand{\sX}{\widetilde{X}}
\newcommand{\sY}{\widetilde{Y}}
\newcommand{\sXY}{\widetilde{XY}}
\newcommand{\lrs}{\widetilde{\phantom{abab}}}
\newcommand{\rs}{@>>\lrs>}
\newcommand{\rss}{@>>\approx>}
\newcommand{\tpi}{\tilde{\pi}}
\newcommand{\cL}{\cal{L}}
\newcommand{\cK}{\cal{K}}
\newcommand{\cA}{\cal{A}}
\newcommand{\ttheta}{\tilde{\theta}}

\begin{document}

\pagestyle{style}
\vspace{5mm}
\title{Determinants of elliptic pseudo-differential operators}
\author{Maxim Kontsevich}
\address{Max-Planck-Institut f\"ur Mathematik, Gottfried-Claren-Stra{\ss}e 26,
53225 Bonn, Germany \\
Department of Mathematics, University of California,
Berkeley, CA 94720}
\email{maxim@@mpim-bonn.mpg.de}
\author{Simeon Vishik}
\address{Max-Planck-Institut f\"ur Mathematik, Gottfried-Claren-Stra{\ss}e 26,
53225 Bonn, Germany \\
Institute of Atmospheric Physics Russian Academy
of Sciences, Moscow, Russia}
\email{senia@@mpim-bonn.mpg.de}
\date{March 1994}
\maketitle
\vskip .5cm
\begin{abstract}
Determinants of invertible pseudo-differential operators (PDOs)
close to positive self-adjoint ones are defined through
the zeta-function regularization.

We define a multiplicative anomaly as
the ratio $\det(AB)/(\det(A)\det(B))$ considered as a function
on pairs of elliptic PDOs.
We obtained an explicit formula for the multiplicative anomaly
in terms of symbols of operators. For a certain natural class
of PDOs
on odd-dimensional manifolds generalizing the class of elliptic
differential operators, the multiplicative anomaly is identically $1$.
For elliptic PDOs from this class a holomorphic determinant and
a determinant for zero orders PDOs are introduced.
Using various algebraic, analytic, and topological tools we study
local and global properties of the multiplicative anomaly
and of the determinant Lie group closely related with it.
The Lie algebra for the determinant Lie group has a description
in terms of symbols only.

Our main discovery is that there is a {\em quadratic non-linearity} hidden
in the definition of determinants of PDOs through zeta-functions.

The natural explanation of this non-linearity follows from
complex-analytic properties of a new trace functional TR on PDOs
of non-integer orders.
Using TR we easily reproduce known facts about noncommutative residues
of PDOs and obtain several new results. In particular, we describe
a structure of derivatives of zeta-functions at zero as of functions
on logarithms of elliptic PDOs.

We propose several definitions extending zeta-regularized
determinants to general elliptic PDOs. For elliptic PDOs of nonzero
complex orders we introduce a canonical determinant in its natural domain
of definition.

\end{abstract}

\tableofcontents
\section{Introduction}
\label{S0}
Determinants of finite-dimensional matrices $A,B\in M_n(\wC)$
possess a multiplicative property:
\begin{equation}
\det(AB)=\det(A)\det(B).
\label{A3500}
\end{equation}
An invertible linear operator in a finite-dimensional linear space has
different types of generalizations to infinite-dimensional case.
One type is pseudo-differential invertible elliptic operators
$$
A\colon\Gamma(E)\to\Gamma(E),
$$
acting in the spaces of smooth sections $\Gamma(E)$ of finite rank
smooth vector bundles $E$ over closed smooth manifolds. Another type
is invertible operators of the form $\Id+K$ where $K$ is a trace class
operator, acting in a separable Hilbert space $H$.
For operators $A$, $B$
of the form $\Id+K$ the equality (\ref{A3500}) is valid.

However, for a general elliptic PDO this equality cannot be valid.
It is not trivial even to define any determinant for such an elliptic
operator. Note that there are no difficulties in defining of the Fredholm
determinant $\det_{Fr}(\Id+K)$. One of these definitions is
\begin{equation}
{\det}_{Fr}(\Id+K):=1+\Tr K+\Tr\left(\wedge^2 K\right)+\ldots+\Tr\left(
\wedge^mK\right)+\ldots
\label{A3501}
\end{equation}
The series on the right is absolutely convergent. For a finite-dimensional
linear operator $A$ its determinant is equal to the finite sum on the right
in (\ref{A3501}) with $K:=A-\Id$. Properties of the linear operators
of the form $\Id+K$ (and of their Fredholm determinants) are analogous
to the properties of finite-dimensional linear operators (and of their
determinants).

In some cases an elliptic PDO $A$ has a well-defined zeta-regularized
determinant
$$
{\det}_\zeta(A)=\exp\left(-\df/\df_s \zeta_A(s)\big|_{s=0}\right),
$$
where $\zeta_A(s)$ is a zeta-function of $A$. Such zeta-regularized
determinants were invented by D.B.~Ray and I.M.~Singer in their papers
\cite{Ra}, \cite{RS1}. They were used in these papers to define
the analytic torsion metric on the determinant line of the cohomology
of the de Rham complex. This construction was generalized by D.B.~Ray
and I.M.~Singer in \cite{RS2} to the analytic torsion metric
on the determinant line of the $\bar{\df}$-complex on a K\"ahler manifold.

However there was no definition of a determinant for a general elliptic PDO
until now.
The zeta-function $\zeta_A(s)$ is defined in the case when the order
$d(A)$ is real and nonzero and when
the principal symbol $a_d(x,\xi)$ for all $x\in M$, $\xi\in T^*_x M$,
$\xi\ne 0$, has no eigenvalues $\lambda$ in some conical neighborhood $U$
of a ray $L$ from the origin in the spectral plane
$U\subset\wC\ni\lambda$.

But even if zeta-functions are defined
for elliptic PDOs $A$, $B$, and $AB$ (so in particular, $d(A)$, $d(B)$,
$d(A)+d(B)$ are nonzero) and if the principal symbols of these three
operators possess cuts of the spectral plane, then in general
$$
\det(AB)\ne\det(A)\det(B).
$$

It is natural to investigate algebraic properties of a function
\begin{equation}
F(A,B):=\det(AB)\big/(\det(A)\det(B)).
\label{A1500}
\end{equation}
This function is defined for some pairs $(A,B)$ of elliptic PDOs.
For instance, $F(A,B)$ is defined for PDOs $A$, $B$ of positive
orders sufficiently close to self-adjoint positive PDOs (with respect
to a smooth positive density $g$ on $M$ and to a Hermitian structure $h$
on a vector bundle $E$, $A$ and $B$ act on $\Gamma(E)$).%
\footnote{The explicit formula for $F(A,B)$ in the case
of positive definite commuting elliptic differential operators $A$ and $B$
of positive orders was obtained by M. Wodzicki \cite{Kas}.
For positive definite elliptic PDOs $A$ and $B$ of positive orders
a formula for $F(A,B)$ was obtained in \cite{Fr}.
However it was obtained in another form than it is written
and used in the present paper. The authors are very indebted
to L.~Friedlander for his information about the multiplicative anomaly
formula obtained in \cite{Fr}.}
(In this case, zeta-functions of $A$, $B$ and of $AB$ can be defined
with the help of a cut of the spectral plane close to $\wR_-$. Indeed,
for self-adjoint positive $A$ and $B$ the operator $AB$
is conjugate to $A^{1/2}BA^{1/2}$ and the latter operator is self-adjoint
and positive.)

Properties of the function $F(A,B)$, (\ref{A1500}), are connected with
the following remark (due to E.~Witten). Let $A$ be an invertible
elliptic DO of a positive order possessing some cuts of the spectral plane.
Then under two infinitesimal deformations for the coefficients of $A$
in neighborhoods $U_1$ and $U_2$ on $M$ on a positive distance one
from another (i.e., $\bar{U}_1\cap\bar{U}_2=\emptyset$) we have
\begin{equation}
\delta_1\delta_2\log{\det}_\zeta(A)=-\Tr\left(\delta_1 A\cdot A^{-1}\delta_2
A\cdot A^{-1}\right).
\label{A1501}
\end{equation}
This equality is proved in Section~\ref{SS01}.
Here, $\delta_j A$ are deformations of a DO $A$ in $U_j$ without changing
of its order. The operator
on the right is smoothing (i.e., its Schwarz kernel is $C^{\infty}$
on $M\times M$). Hence it is a trace class operator and its trace
is well-defined. Note that the expression on the right in (\ref{A1501})
is independent of a cut of the spectral plane in the definition
of the zeta-regularized determinant on the left in (\ref{A1501}).

It follows from (\ref{A1501}) that $\log{\det}_\zeta(A)$ is canonically
defined up to an additional local functional on the coefficients of $A$.
Indeeed,
for two definitions, $\log{\det}_\zeta(A)$ and $\log{\det}'_\zeta(A)$,
for a given $A$, we have
\begin{equation}
\delta_1\delta_2\left(\log{\det}_\zeta(A)-\log{\det}'_\zeta(A)\right)=0.
\label{A1550}
\end{equation}
The equality $\delta_1\delta_2 F(A)=0$ for deformations $\delta_jA$
in $U_j$, $\bar{U}_1\cap\bar{U}_2=\emptyset$, is the characteristic
property of local functionals.

It follows from (\ref{A1501}) that
\begin{equation}
f(A,B):=\log\det(AB)-\log\det(A)-\log\det(B)
\label{B3701}
\end{equation}
is a local (on the coefficients of invertible DOs $A$ and $B$) functional,
if these zeta-regularized determinants are defined. Namely, if $\delta_jA$
and $\delta_jB$ are infinitesimal variations of $A$ and of $B$ in $U_j$,
$j=1,2$, $\bar{U}_1\cap\bar{U}_2=\emptyset$, then
\begin{equation}
\delta_1\delta_2 f(A,B)=0.
\label{A1511}
\end{equation}
This equality is deduced from (\ref{A1501}) in Section~\ref{SS01}.

For some natural class of classical elliptic PDOs acting in sections
$\Gamma(E)$ of a vector bundle $E$ over an odd-dimensional closed $M$,
their determinants are multiplicative (Section~\ref{SC}), if operators
are sufficiently close to positive definite ones and if their orders are
positive even numbers.
The operators from this (odd) class generalize differential operators.

As a consequence we can define (Section~\ref{SC}) determinants
for classical elliptic PDOs of order zero from this natural (odd) class.
Such determinants cannot
be defined through zeta-functions of the corresponding operators
because the zeta-function for such an operator $A$ is defined
as the analytic continuation of the trace $\Tr\left(A^{-s}\right)$.
However the operator $A^{-s}$ for a general elliptic PDO $A$ of order zero
is not of trace class for any $s\in\wC$.%
\footnote{Such an operator is defined by the integral
$(i/2\pi)\int_\Gamma\lambda_{(\theta)}^{-s}\left(A-\lambda\right)^{-1}
d\lambda$, where $\Gamma:=\Gamma_{R,\theta}$ is a smooth closed contour
defined as in (\ref{B7}), (\ref{X2}) and surrounding once
the spectrum $\Spec(A)$ of $A$ ($\Spec(A)$ is a compact set) and oriented
opposite to the clockwise, $\lambda_{(\theta)}^{-s}$ is an appropriate
branch of this multi-valued function. Here, $R$ is such that $\Spec A$
lies inside $\{\lambda\colon|\lambda|\le R/2\}$ and $\theta$ is an admissible
cut of the spectral plane for $A$ and
$\lambda_{(\theta)}^{-s}:=\exp\left(-s\log_{(\theta)}\lambda\right)$,
$\theta-2\pi\le\Im(\log_{(\theta)}\lambda)\le\theta$.}
Also such a determinant cannot be defined by standart methods
of functional analysis because such an operator $A$ is not
of the form $\Id+K$, where $K$ is a trace class operator.
Nevertheless, canonical determinants of operators from this
natural class can be defined.
Here we use the multiplicative property for the determinants
of the PDOs of positive orders from this natural (odd) class of operators
(on an odd-dimensional closed $M$).

This determinant is also defined
for an automorphism of a vector bundle on an odd-dimensional manifold
acting on global sections of this vector bundle. (Note that
the multiplication operator by a general positive smooth function has
a continuous spectrum.) The determinant of such an operator is equal
to $1$ (Section~\ref{SB}).

A {\em natural trace} $\Tr_{(-1)}$ is introduced for odd class PDOs
on an odd-dimensional closed $M$. A {\em canonical determinant}
$\det_{(-1)}(A)$ for {\em odd class} elliptic PDOs $A$ of zero orders
with given $\sigma(\log A)$
is introduced (Section~\ref{SEOdd}) with the help of $\Tr_{(-1)}$.
The determinant $\det_{(-1)}(A)$ is defined even if $\log A$ does not
exist. This $\det_{(-1)}(A)$ coincides with the determinant of $A$
(defined by the multiplicative property), if $A$ sufficiently close
to positive definite self-adjoint PDOs (Section~\ref{SEOdd}).

Let $D_u$ be a family of the Dirac operators on an odd-dimensional
spinor manifold $M$ (corresponding to a family $\left(h_u,\nabla^u\right)$
of Hermitian metrics and unitary connections on a complex vector bundle
on $M$). As a consequence of the multiplicative property we obtain
the fact that $\det\left(D_{u_1}D_{u_2}\right)$ is a real number
for any pair $\left(u_1,u_2\right)$ of parameters and that this
determinant has a form
$$
\det\left(D_{u_1}D_{u_2}\right)=\eps\left(u_1\right)\eps\left(u_2\right)
\left(\det\left(D_{u_1}^2\right)\right)^{1/2}\left(\det\left(D_{u_2}^2\right)
\right)^{1/2}
$$
for any pair of sufficiently close parameters $\left(u_1,u_2\right)$.
The factor $\eps(u)=\pm 1$ on the right is a globally defined locally
constant function on the space of invertible Dirac operators according
to the Atiyah-Patody-Singer formula for the corresponding spectral
flows.

Absolute value positive determinants $|\det|A$ for all elliptic operators
$A$ from the odd class on an odd-dimensional manifold $M$ are defined as
$$
(|\det|A)^2=\det\left(A^*A\right).
$$
They are independent of a smooth positive density on $M$ (and of a Hermitian
structure on $E$). It is proved (in Section~\ref{SC4}) that $(|\det|A)^2$
has a form $|f(A)|^2$, where $f(A)$ is a holomorphic multi-valued
function on $A$. We call it a {\em holomorphic determinant}.
The monodromy of $f(A)$ (over a closed loop) is
multiplying by a root of $1$ of degree $2^m$, where $m$ is a non-negative
integer bounded by a constant depending on $\dim M$ only
(Section~\ref{SC4}).

The algebraic interpretation of the function $F(A,B)$, (\ref{A1500}),
in the general
case is connected with a central extension of the Lie algebra
$S_{\log}(M,E)$ consisting of symbols of logarithms
for invertible elliptic PDOs
$\Ell_0^{\times}(M,E)\subset\cup_\alpha CL^\alpha(M,E)$,
$\alpha\in\wC$. (The principal symbols of elliptic PDOs
from $\Ell_0^{\times}(M,E)$ restricted to $S^*M$ are homotopic to $\Id$.)
The algebra $S_{\log}(M,E)$ is spanned
as a linear space (over $\wC$) by its subalgebra $CS^0(M,E)$
of the zero order classical PDOs symbols and by the symbol
of $\log A$. Here, $A$ is any elliptic PDO with a real nonzero order
such that its principal symbol admits a cut of the spectral plane
along some ray from the origin.

The logarithm of the zeta-regularized determinant $\det_{(\theta)}A$
for an elliptic PDO $A$ admitting a cut
$L_{(\theta)}=\{\lambda\colon\arg\lambda=\theta\}$ of the spectral plane
$\wC$ is defined as%
\footnote{Here, $\zeta'(0):=\df_s\zeta(s)|_{s=0}$. The zeta-function
is defined as the analytic continuation of the series
${\sum}'\lambda_{(\theta)}^{-s}$.}
$\exp\left(-\zat'(0)\right)$. There is a more simple function of $A$
than $\zat'(0)$. That is the value $\zat(0)$
at the origin. In the case of an invertible linear operator $A$
in a finite-dimensional Hilbert space $H$ we have $\zat(0)=\dim H$.
So $\zat(0)$ is a regularization of the dimension of the space
where the PDO acts. It is known that $\zat(0)$ is independent
of an admissible cut $L_{(\theta)}$ (\cite{Wo1}, \cite{Wo2}).
However in general $\zat(0)$
depends not only on $(M,E)$ but also on the image
of the symbol $\sigma(A)$ in $CS^\alpha(M,E)/CS^{\alpha-n-1}(M,E)$,
$\alpha:=\ord A$,
$n:=\dim M$. If $H$ is finite-dimensional, then $\zeta_A(0)=\dim H$
is constant as a function of an invertible $A\in GL(H)$.
Let invertible PDOs $A$ and $B$ of orders $\alpha$ and $\beta$
be defined in $\Gamma(M,E)$, let $\alpha,\beta,\alpha+\beta\in\wR^{\times}$,
and let there be admissible cuts $\theta_A$, $\theta_B$, and $\theta_{AB}$
of the spectral plane for their principal
symbols $a:=\sigma_\alpha(A)$, $b:=\sigma_\beta(B)$, and
for $\sigma_{\alpha+\beta}(AB)=ab$. Then the function
\begin{equation}
Z\left(\sigma(A)\right):=-\alpha\zeta_A(0)
\label{B10}
\end{equation}
is additive,
\begin{equation}
Z\left(\sigma(AB)\right)=Z\left(\sigma(A)\right)+Z\left(\sigma(B)\right).
\label{B11}
\end{equation}
The function $Z\left(\sigma(A)\right)=-\alpha\left(\zeta_A(0)+h_0(A)\right)$,
where $h_0(A)$ is the algebraic multiplicity of $\lambda=0$ for an elliptic
PDO $A\in\Ell_0^\alpha(M,E)\subset CL^\alpha(M,E)$,
was introduced by M. Wodzicki. He proved the equality (\ref{B11}).
The function $Z(\sigma(A))$ was defined by him also for zero order
elliptic symbols $\sigma(A)\in\SEll_0^0(M,E)$ which are homotopic
to $\Id$. For such $\sigma(A)$ this function coincides with
the {\em multiplicative residue}
\begin{equation}
r^{\times}(\sigma(A))=\int_0^1\res\left(a^{-1}(t)\dot{a}(t)\right)dt,
\label{B12}
\end{equation}
where $a(t)$ is a smooth loop in $\SEll_0^0(M,E)$ from $a(0)=\Id$
to $a(1)=\sigma(A)$. The integral on the right in (\ref{B12})
is independent of such a loop. This asssertion follows from the equality
which holds for all PDO-projectors $P\in CL^0(M,E)$, $P^2=P$,
\begin{equation}
\res P=0.
\label{X1}
\end{equation}
Reverse, the equalities (\ref{X1}) are equivalent to the independence
of $\zat(0)$ of an admissible cut $L_{(\theta)}$ for $\ord A\ne 0$
(\cite{Wo1}).
The additivity (\ref{B11}) holds also on the space $\SEll_0^0(M,E)$
(\cite{Kas}). Hence, the function $\zeta_A(0)$ as a function
of $\sigma\left(\log_{(\theta)}A\right)$,
where $A$ is an invertible PDO of order one, is the restriction
to the affine hyperplane $\ord A=1$ of the linear function
$-Z\left(\sigma(A)\right)$ on the linear space
$\left\{\sigma\left(\log_{(\theta)}A\right)\right\}=:S_{\log}(M,E)$
of the logarithmic symbols (defined in Section~\ref{SA}).

It occurs that $\zat'(0)$ for $\ord A=1$ is the restriction
to the hyperplane $\ord A=1$ of a quadratic form on the space
$\log_{(\theta)}(A)$.
Hence the formula
$$
\Tr(\log A)=\log(\det(A))
$$
(true for invertible operators of the form $\Id+K$, where $K$ is
a trace class operator) cannot be valid on the space of logarithms
of elliptic PDOs. (Here, we suppose that $\Tr(\log A)$ is some
linear functional of $\log A$.)

We have an analogous statement for all the derivatives of $\zat(s)$
at $s=0$. Namely for $k\in\zuo$ there is a homogeneous polynomial
of order $(k+1)$ on the space of $\log_{(\theta)}A$ such that
$\df_s^k\zat(s)|_{s=0}$ for $\ord A=1$ is the restriction of this
polynomial to the hyperplane $\ord A=1$ (Section~\ref{SB}) in logarithmic
coordinates.

These results on the derivatives $\df_s^k\zat(s)|_{s=0}$ as on functions
of $\log_{(\theta)}A$ are obtained with the help of a new
{\em canonical trace} $\TR$
for PDOs of noninteger orders introduced in Section~\ref{SB}. For a given
PDO $A\in CL^d(M,E)$, $d\notin\wZ$, such a trace $\TR(A)$ is equal
to the integral over $M$ of a canonical density $a(x)$ corresponding to $A$.
Polynomial properties of $\df_s^k\zat(s)|_{s=0}$ follows from analytic
properties of $\TR(\exp(sl+B_0))$ in $s\in\wC$ and in $B_0\in CL^0(M,E)$
for $s$ close to zero. Here, $l$ is a logarithm of an invertible
elliptic PDO $A\in\Ell_0^1(M,E)$.

This trace functional provides us with a definition
of {\em $\TR$-zeta-functions}.
These zeta-functions $\zeta_A^{\TR}(s)$ are defined for nonzero order
elliptic PDOs $A$ with given families $A^{-s}$ of their complex powers.
However, to compute $\zeta_A^{\TR}\left(s_0\right)$
(for $s_0\ord A\notin\wZ$) we do not use any analytic continuation.

The Lie algebra of the symbols for logarithms of elliptic operators
contains as a codimension one ideal the Lie algebra of the zero order
classical PDO-symbols. (We call it a cocentral one-dimensional extension.)
This Lie algebra of logarithmic symbols
has a system of one-dimensional central extensions parametrized
by logarithmic symbols of order one. On any extension of this system
a non-degenerate quadratic form is defined. We define a canonical
associative system of isomorphisms between these extensions
(Section~\ref{SD}). Hence a canonical one-dimensional central extension
is defined for the Lie algebra of logarithmic symbols. The quadratic
forms on these extensions are identified by this system of isomorphisms.
This quadratic form is invariant under the adjoint action.

The determinant Lie group is a central $\wC^{\times}$-extension
of the connected component of $\Id$ of the  Lie group of elliptic
symbols (on a given closed manifold $M$).
This Lie group
is defined as the quotient of the group of invertible elliptic PDOs
by the normal subgroup of operators of the form $\Id+\cK$, where
$\cK$ is an operator with a $C^\infty$ Schwartz kernel on $M\times M$
(i.e., a smoothing operator) and $\det_{Fr}(\Id+\cK)=1$. (Here, $\det_{Fr}$
is the Fredholm determinant.) It is proved that there is a canonical
identification of the Lie algebra for this determinant Lie group
with a canonical one-dimensional central extension of the Lie algebra
of logarithmic symbols (Section~\ref{SE}). The determinant Lie group
has a canonical section partially defined using zeta-regularized
determinants over the space of elliptic
symbols (and depending on the symbols only). Under this identification,
this section corresponds to the exponent of the null-vectors
of the canonical quadratic form on the extended Lie algebra
of logarithmic symbols (Theorem~\ref{TB570}). The two-cocycle
of the central $\wC^{\times}$-extension
of the group of elliptic symbols defined by this canonical
section is equal to the multiplicative anomaly. So this quadratic
$\wC^{\times}$-cone is deeply connected with zeta-regularized determinants
of elliptic PDOs.

An alternative proof of Theorem~\ref{TB570} without using variation
formulas is obtained in Section~\ref{S6}. This theorem claims
the canonical isomorphism between the canonical central extension
of the Lie algebra of logarithmic symbols and the determinant Lie algebra.

The multiplicative anomaly $F(A,B)$ for a pair of invertible elliptic
PDOs of positive orders sufficiently close to self-adjoint positive
definite ones gives us a partially defined symmetric $2$-cocycle
on the group of the elliptic symbols. We define a coherent system
of determinant cocycles on this group given for larger and larger
domains in the space of pairs of elliptic symbols and show that
a canonical {\em skew-symmetric} $2$-cocycle on the Lie group
of logarithmic symbols is canonically cohomologous to the {\em symmetric}
$2$-cocycle of the multiplicative anomaly (Section~\ref{SE1}).
Note that the multiplicative anomaly cocycle is singular for elliptic
PDOs of order zero.

The global structure of the determinant Lie group is defined by its
Lie algebra and by spectral invariants of a generalized spectral
asymmetry. This asymmetry is defined for pairs of a PDO-projector
of zero order and of a logarithm of an elliptic operator of a positive
order. This invariant
depends on the symbols of the projector and of the operator but this
dependence is {\em global} (Section~\ref{S7}). The first variational
derivative of this functional is given by an explicit local formula.%
{\footnote{The same properties have Chern-Simons and analytic (holomorphic)
torsion functionals.}
This functional is a natural generalization of the Atiyah-Patodi-Singer
functional of spectral asymmetry \cite{APS1}, \cite{APS2}, \cite{APS3}.
The main unsolved problem
in algebraic definition of the determinant Lie group is obtaining
a formula for this spectral asymmetry in terms of symbols.

The determinant Lie algebra over the Lie algebra of logarithmic symbols
for odd class elliptic PDOs on an odd-dimensional closed $M$
is a canonically trivial central extension. So a flat connection
on the corresponding determinant Lie group is defined. Thus
a multi-valued determinant on odd class operators is obtained.
It coincides with the holomorphic determinant defined on odd class
elliptic PDOs (Section~\ref{SEOdd}).

The exponential map from the Lie algebra of logarithms of elliptic PDOs
to the connected component of the Lie group of elliptic PDOs is not
a map ``onto'' (i.e., there are domains in this connected Lie group
where elliptic PDOs have no logarithms at all). There are some
topological obstacles (in multi-dimensional case) to the existence
of any smooth logarithm even on the level of principal symbols
(Section~\ref{SE}).

A {\em canonical determinant} $\det(A)$ is introduced for an elliptic PDO
$A$ of a nonzero complex order with a given logarithmic symbol
$\sigma(\log A)$. For this symbol to be defined, it is enough that
a smooth field of admissible cuts $\theta(x,\xi)$, $(x,\xi)\in S^*M$,
for the principal symbol of $A$ to exist and a map
$\theta\colon S^*M\to S^1=\wR/2\pi\wZ$ to be homotopic to trivial.
This canonical determinant $\det(A)$ is defined with the help of any
logarithm $B$ (such that $\sigma(B)=\sigma(\log A)$) of some invertible
elliptic PDO. However $\det(A)$ is independent of a choice of $B$.
The canonical determinant is defined in its natural domain of definition.
The ratio
\begin{equation}
d_1(A)/\det(A)=:\tilde{d}_0(\sigma(\log A))
\label{B4051}
\end{equation}
depends on $\sigma(\log A)$ only and defines a canonical (multi-valued)
section of the determinant Lie group. This section is naturally
defined over logarithmic symbols of nonzero orders (Section~\ref{SE}).
With the help of $\tilde{d}_0(\sigma(\log A))$ we can control the behavior
of $\det(A)$ near the domain where $\sigma(\log A)$ does not exist
(Section~\ref{SS93}). The canonical determinant $\det(A)$ coincides
with the $\TR$-zeta-regularized determinant, if $\log A$ exists.
However $\det(A)$ is also defined, if $\sigma(\log A)$ exists but $\log A$
does not exist.

A determinant of an elliptic operator $A$ of a nonzero complex order
is defined (Section~\ref{S8}) for a smooth curve between $A$ and
the identity operator in the space of invertible elliptic operators.
This determinant is the limit of the products of $\TR$-zeta-regularized
determinants corresponding
to the intervals of this curve (in the space of elliptic operators)
as lengths of the intervals tend to zero. This determinant is
independent of a smooth parametrization of the curve. However,
to prove the convergence of the product of $\TR$-zeta-regularized
determinants, we have to use the non-scalar language of determinant
Lie groups and their canonical sections. This determinant of $A$
is equal to a zeta-regularized determinant, if the curve is
$A^t$, $0\le t\le 1$ up to a smooth reparametrization.
Such a curve exists in the case when any $\log A$ exists.
For a product of elliptic PDOs (of nonzero orders)
and for a natural composition of monotonic curves in the space of elliptic
PDOs (corresponding to the determinants of the factors), this determinant
is equal to the product of the determinants of the factors.

Any logarithmic PDO-symbol of order one defines a connection
on the determinant Lie group over the group of elliptic PDO-symbols.
The determinant Lie group is the quotient of the Lie group of invertible
elliptic operators. The image $d_1(A)$ of an elliptic operator $A$
in the determinant Lie group is multiplicative in $A$. For any smooth
curve $s_t$ in the space $\SEll_0^{\times}$ of elliptic symbols
from $\Id$ to the symbol $\sigma(A)=s|_{t=1}$ its canonical pull-back
$\tilde{s}_t$ is a horizontal curve in the determinant Lie group
from $d_1(\Id)$. Hence $d_1(A)/\tilde{s}_1$ defines a determinant
(Section~\ref{S81}) for a general elliptic PDO $A$ of any complex order
(in particular, of zero order). This determinant depends on a smooth curve
$s_t$ from $\Id$ to $\sigma(A)$ in the space of symbols of elliptic
PDOs without a monotonic (in order) condition. It does not change
under smooth reparametrizations of the curve.

For a given logarithmic PDO-symbol of order one (i.e., for a given
connection) this determinant for a finite product of elliptic operators
is equal to the product of their determinants. (Here, the curve
in the space of elliptic symbols in the definition of the determinant
of the product is equal to the natural composition of smooth curves
corresponding to the determinants of factors.)
There are explicit formulas for the dependence of this determinant
on a first order logarithmic symbol (defining a connection
on the determinant Lie group) and on a curve $s_t$ (from $\Id$
to $\sigma(A)$) in a given homotopic class (Section~\ref{S81}).
Its dependence of an element of the fundamental group
$\pi_1\left(\SEll_0^0\right)$ is expressed with the help of the invariant
of generalized spectral asymmetry. In the case when $s_t$ is
$\sigma\left(A^t\right)$ (up to a reparametrization), $\det_{(s_t)}(A)$
is the zeta-regularized determinant corresponding to the $\log A$
definding $A^t$.

\subsection{Second variations of zeta-regularized determinants}
\label{SS01}
Let the zeta-\!regular-ized determinant ${\det}_\zeta(A)$ of an elliptic
DO $A\in\Ell^d(M,E)$, $d\in\wZ_+$, be defined with the help of a family
$A_{(\theta)}^{-s}$ of complex powers of $A$. (Such a family is defined
with the help of an admissible cut
$L_{(\theta)}=\{\lambda\colon\arg\lambda=\theta\}$ of the spectral plane,
see Section~\ref{SA}.) Then we have
\begin{equation}
\delta_1\left(-\df_s\Tr\left(A^{-s}\right)|_{s=0}\right)=\df_s\left(s\Tr
\left(\delta_1A\cdot A^{-1}A^{-s}\right)\right)|_{s=0}.
\label{A1512}
\end{equation}
The function $\Tr\left(\delta_1A\cdot A^{-1}A^{-s}\right)$ is defined
in a neighborhood of $s=0$ by the analytic continuation of this trace
from the domain $\Re s>\dim M/d$, $d=\ord A$, where the operator
$\left(\delta_1A\cdot A^{-1}A^{-s}\right)$ is of trace class.
This analytic continuation has a simple pole at $s=0$ with its residue
equal to
$$
\Res_{s=0}\Tr\left(\delta_1A\cdot A^{-1}A^{-s}\right)=-\res\left(\delta_1A
\cdot A^{-1}\right)/d,
$$
where $\res$ is the noncommutative residue \cite{Wo1}, \cite{Wo2}.
However at $s=0$ the function
$\df_s\left(s\Tr\left(\delta_1A\cdot A^{-1}A^{-s}\right)\right)$
is holomorphic.

The second variation $\delta_2\delta_1{\det}_\zeta(A)$ can be written
(by (\ref{A1512})) in the form
\begin{equation}
\delta_2\delta_1{\det}_\zeta(A)=\df_s\left(s{1\over 2\pi i}\Tr\int
_{\Gamma_{(\theta)}}\lambda^{-(s+1)}\delta_1A\left(A-\lambda\right)^{-1}
\delta_2A\left(A-\lambda\right)^{-1}d\lambda\right).
\label{A1514}
\end{equation}
(Here, $\Gamma_{(\theta)}$ is the simple contour surrounding
an admissible cut $L_{(\theta)}$, see Section~\ref{SA}, (\ref{A1}).)
The operator
$\delta_1A\left(A-\lambda\right)^{-1}\delta_2A$ is smoothing (since
its symbol is equal to zero as $\bar{U}_1\cap\bar{U}_2=\emptyset$) and
its trace-norm is uniformly bounded for $\lambda\in\Gamma_{(\theta)}$,
$|\lambda|\to\infty$. The operator norm
$\left\|\left(A-\lambda\right)^{-1}\right\|_{(2)}$ in $L_2(M,E)$ is
$O\left(\left(1+|\lambda|\right)^{-1}\right)$
for $\lambda\in\Gamma_{(\theta)}$. Hence the trace-norm of
$\delta_1A\left(A-\lambda\right)^{-1}\delta_2A\left(A-\lambda\right)^{-1}$
is $O\left(\left(1+|\lambda|\right)^{-1}\right)$
for $\lambda\in\Gamma_{(\theta)}$, and for $s$ close to zero we have
\begin{multline}
\Tr\left(\int_{\Gamma_{(\theta)}}\lambda^{-(s+1)}\delta_1A\left(A-\lambda
\right)^{-1}\delta_2A\left(A-\lambda\right)^{-1}d\lambda\right)=\\
=\int_{\Gamma_{(\theta)}}\lambda^{-(s+1)}\Tr\left(\delta_1A\left(A-\lambda
\right)^{-1}\delta_2A\left(A-\lambda\right)^{-1}\right)d\lambda.
\label{A1515}
\end{multline}
The function $\Tr\left(\delta_1A\left(A-\lambda\right)^{-1}\delta_2A\left(A-
\lambda\right)^{-1}\right)$ is holomorphic (in $\lambda$) inside
the contour $\Gamma_{(\theta)}$. Hence we can conclude from (\ref{A1514}),
(\ref{A1515}) that
\begin{equation}
\delta_2\delta_1{\det}_\zeta(A)=-\Tr\left(\delta_1A\cdot A^{-1}\delta_2A
\cdot A^{-1}\right),
\label{B3700}
\end{equation}
and the formula (\ref{A1501}) is proved.\ \ \ $\Box$

Let us deduce from (\ref{B3700}) the equality $\delta_1\delta_2f(A,B)=0$.
Here, $f(A,B)$ (given by (\ref{B3701})) is the logarithm
of the multiplicative anomaly (\ref{A1500}).

By (\ref{A1501}) we have
\begin{multline}
\delta_1\delta_2\!\left(\log\det(AB)\!-\!\log\det(A)\!-\!\log\det(B)\!\right)
\!=\!-\!\Tr\!\left(\delta_1(AB)(AB)^{-1}\delta_2(AB)(AB)^{-1}\!\right)\!+\\
+\Tr\left(\delta_1A\cdot A^{-1}\delta_2A\cdot A^{-1}\right)+
\Tr\left(\delta_1B\cdot B^{-1}\delta_2B\cdot B^{-1}\right)=\\
=\bigl(-\Tr\left(A\delta_1B\cdot B^{-1}\delta_2B\cdot B^{-1}A^{-1}\right)+
\Tr\left(\delta_1B\cdot B^{-1}\delta_2B\cdot B^{-1}\right)\bigr)-\\
-\Tr\left(\delta_1A\delta_2B\cdot B^{-1}A^{-1}\right)-\Tr\left(A\delta_1B
\cdot B^{-1}A^{-1}\delta_2A\cdot A^{-1}\right).
\label{A1510}
\end{multline}
The operator $A\delta_1B\cdot B^{-1}\delta_2B\cdot B^{-1}$ is a smoothing
operator in $\Gamma(E)$ (since its symbol is equal to zero because
$\bar{U}_1\cap\bar{U}_2=\emptyset$). Hence it is a trace class operator and
$$
\Tr\left(A\delta_1B\cdot B^{-1}\delta_2B\cdot B^{-1}A^{-1}\right)=\Tr\left(
\delta_1B\cdot B^{-1}\delta_2B\cdot B^{-1}\right).
$$
By the analogous reason we have
$$
\Tr\left(A\delta_1B\cdot B^{-1}A^{-1}\delta_2A\cdot A^{-1}\right)=\Tr\left(
\delta_1B\cdot B^{-1}A^{-1}\delta_2A\right).
$$
Since $\delta_1B\cdot B^{-1}A^{-1}\delta_2A$ is a smoothing operator
with its Schwarz kernel equal to zero in a neighborhood of the diagonal
$M\hookrightarrow M\times M$ (because $\bar{U}_1\cap\bar{U}_2=\emptyset$),
we see that
$$
\Tr\left(A\delta_1B\cdot B^{-1}A^{-1}\delta_2A\cdot A^{-1}\right)=0.
$$
Hence the equality (\ref{A1511}) is deduced.\ \ \ $\Box$

\section{Determinants and zeta-functions for elliptic PDOs.
Multiplicative anomaly}
\label{SA}

Let a classical elliptic PDO $A\in\Ell_0^d(M,E)\subset CL^d(M,E)$ be
an elliptic operator
of a positive order $d=d(A)>0$ such that its principal symbol
$a_d(x,\xi)$ has no eigenvalues in nonempty conical neighborhood $\Lambda$
of a ray $L_{(\theta)}=\left\{\lambda\in\wC,\arg\lambda=\theta\right\}$
in the spectral plane $\wC$. Suppose that $A$ is an invertible
operator $A\colon H_{(s)}(M,E)\to H_{(s-d)}(M,E)$, where $H_{(s)}$
are the Sobolev spaces (\cite{Ho2}, Appendix B).
Then there are no more than a finite number of the eigenvalues
$\lambda$ of the spectrum%
\footnote{$A$ is an invertible elliptic PDO of a positive order.
Hence $0\notin\Spec(A)$ and $\Spec(A)$ is discrete, i.e., it consists
entirely of isolated eigenvalues with finite multiplicities (\cite{Sh},
Ch.~I, \S~8, Theorem~8.4).}
$\Spec(A)$ in $\Lambda$. Let $L_{(\theta)}$ be the ray in $\Lambda\in\wC$
such that there are no eigenvalues $\lambda\in\Spec(A)$ with
$\arg\lambda=\theta$. Then the complex powers $A^z_{(\theta)}$
of $A$ are defined for $\Re z{\ll}0$ by the integral
\begin{equation}
A^z_{(\theta)}:={i\over 2\pi}\int_{\Gamma_{(\theta)}}\lambda^z(A-\lambda)^{-1}
d\lambda,
\label{A1}
\end{equation}
where $\Gamma_{(\theta)}$ is a contour
$\Gamma_{1,\theta}(\rho)\cup\Gamma_{0,\theta}(\rho)\cup\Gamma_{2,\theta}
(\rho)$,
$\Gamma_{1,\theta}(\rho)\colon\!=\!\left\{\!\lambda\!=\!x\exp(i\theta),
\!+\!\infty\!>\!x\!\ge\!\rho\!\right\}$,
$\Gamma_{0,\theta}(\rho)\colon\!=\!\left\{\lambda\!=\!\rho\exp(i\phi),
\theta\!\ge\!\phi\!\ge\!\theta\!-\!2\pi\right\}$,
$\Gamma_{2,\theta}(\rho)\colon\!=\!\left\{\lambda\!=\!x\exp i(\theta\!-\!2\pi),
\rho\!\le\!x\!<\!+\infty\right\}$,
and $\rho$ is a positive number such that all the eigenvalues
in $\Spec(A)$ are outside of the disk
$D_\rho:=\{\lambda\colon |\lambda|\le\rho\}$.
The function $\lambda^z$ on the right of (\ref{A1}) is defined
as $\exp(z\log\lambda)$, where $\theta\ge\Im\log\lambda\ge\theta-2\pi$
(i.e., $\Im\log\lambda=\theta$ on $\Gamma_{1,\theta}$,
$\Im\log\lambda=\theta-2\pi$ on $\Gamma_{2,\theta}$).
For $\Re z{\ll}0$ the operator $A^z_{(\theta)}$ defined by the integral
on the right of (\ref{A1}) is bounded in $H_{(s)}(M,E)$ for an arbitrary
$s\in\wR$ (as the integral on the right of (\ref{A1}) converges
in the operator norm on $H_{(s)}(M,E)$).
Families of operators $A^z_{(\theta)}$ depend on (admissible) $\theta$.

For $-k\in\wZ_+$ the operator $A^{-k}_{(\theta)}$ coincides with
$(A^{-1})^k$ (\cite{Sh}, Ch.~II, Proposition~10.1). Operators
$A^z_{(\theta)}$ are defined for all $z\in\wC$ by the formula
\begin{equation}
A^z_{(\theta)}=A^k A^{z-k}_{(\theta)},
\label{A2}
\end{equation}
where $z-k$ belongs to the domain of definition for (\ref{A1})
and where $A^{z-k}_{(\theta)}$ are defined
by (\ref{A1}). It is proved in \cite{Se}, Theorem~1, and in \cite{Sh},
Ch.~II, Theorem~10.1.a, that the operator $A^z_{(\theta)}$ defined
by (\ref{A2}) is independent of the choice of $k$ and that (\ref{A2})
holds for all $k\in\wZ$ for the family $A_{(\theta)}^z$.
The operators $A^z_{(\theta)}$ for $\Re z\le k\in\wZ$ form a family
of bounded linear operators from $H_{(s)}(M,E)$ into $H_{(s-d(A)k)}(M,E)$.

The operator $A^z_{(\theta)}$ is a classical elliptic PDO of order
$zd(A)$, $A^z_{(\theta)}\in\Ell_0^{zd(A)}(M,E)$. Its symbol
\begin{equation}
b^z_{(\theta)}(x,\xi)=\sum_{j\in\zuo}b^z_{zd-j,\theta}(x,\xi)
\label{A3}
\end{equation}
is defined in any local coordinate chart $U$ on $M$
(with a smooth trivialization of $E|_U$). Here, $d:=d(A)$ and
$b^z_{zd-j,\theta}(x,t,\xi)=t^{zd-j}b^z_{zd-j,\theta}(x,\xi)$
for $t\in\wR_+$ (i.e., this term is positive homogeneous of degree
$dz-j$). This symbol is defined through the symbol
\begin{equation}
b(x,\xi,\lambda)=\sum_{j\in\zuo}b_{-d-j}(x,\xi,\lambda)
\label{A4}
\end{equation}
of the elliptic operator $\left(A-\lambda\right)^{-1}$.
The term $b_{-d-j}(x,\xi,\lambda)$ is positive homogeneous
in $\left(\xi,\lambda^{1/d}\right)$ of degree $-(d+j)$.
(The parameter $\lambda$ in (\ref{A4}) has the degree $d=d(A)$.)
The symbol $b(x,\xi,\lambda)$ is defined by the following
recurrent system of equalities ($a(x,\xi):=\sum_{\zuo}a_{-d-j}(x,\xi)$
is the symbol of $A$, $D_x:=i^{-1}\df_x$)
\begin{gather}
\begin{split}
b_{-d}(x,\xi,\lambda)   & :=\left(a_d-\lambda\right)^{-1}, \\
b_{-d-1}(x,\xi,\lambda) & :=-b_{-d}\left(a_{d-1}b_{-d}+
\sum_i\df_{\xi_i}a_d D_{x_i}b_{-d}\right), \\
.\ .\ .\ .\ .\ .\ .\ .\ .\ & .\ .\ .\ .\ .\ .\ .\ .\ .\ .\ .\ .\ .\ .\ .\ .
\ .\ .\ .\ .\ .\ .\ .\ .\ .\ .\ .\ .\\
b_{-d-j}(x,\xi,\lambda) & :=-b_{-d}\sum_{|\alpha|+i+l=j}{1\over \alpha!}
\df^\alpha_\xi a_{d-i}D^\alpha_x b_{-d-l},
\end{split}
\label{B1}
\end{gather}
i.e., $\left(a(x,\xi)-\lambda\right)\circ b(x,\xi,\lambda)=\Id$, where
the composition has as its positive homogeneous
in $\left(\xi,\lambda^{1/d}\right)$ components
$$
\sum_{|\alpha|+k+l=\const}{1\over\alpha!}\df^\alpha_\xi a_{d-k}
(x,\xi,\lambda)D^\alpha_x b_{-d-l}(x,\xi,\lambda).
$$
Here, $a_{d-k}(x,\xi,\lambda):=a_{d-k}-\delta_{k,0}\lambda\Id$.
The terms $b_{-d-j}$ are regular in $(x,\xi,\lambda)$, $\xi\ne 0$, such
that the principal symbol $(a_d-\lambda\Id)$ is invertible.%
\footnote{The operators $\left(A-\lambda\right)^{-1}$ and $A-\lambda$
in general do not belong to the classes $CL^{-d}(M,E;\Lambda)$ and
$CL^d(M,E;\Lambda)$ (\cite{Sh}, Ch.~II, \S~9) of elliptic operators
with parameter. Here, $\Lambda$ is an open conical neighborhood
of the ray $L_{(\theta)}=\{\lambda\colon\arg\lambda=\theta\}$
in the spectral plane such that all the eigenvalues of the principal
symbol $a_d(x,\xi)$ of $A$ do not belong to $\bar{\Lambda}$ for any
$(x,\xi)\in T^*M$, $\xi\ne 0$.
For a general elliptic PDO $A\in CL^d(M,E)$, $d>0$, of the type
considered above and for an arbitrary $j>d$ there are no uniform
estimates in $\xi\in T^*M$, $\lambda\in\Lambda$, $(\xi,\lambda)\ne(0,0)$
of $\left|b_{-d-j}(x,\xi,\lambda)\right|$ through
$C\left(|\xi|+\left|\lambda\right|^{1/d}\right)^{-d-j}$
and of $\left|a_{d-j}(x,\xi)\right|$ through
$C_1\left(|\xi|+\left|\lambda\right|^{1/d}\right)^{d-j}$.}

If $\Re z<0$, the formula for $b^z_{(\theta)}(x,\xi)$ is (\cite{Sh},
Ch.~II, Sect.~11.2)%
\footnote{Here,
$\lambda_{(\theta)}^z:=\exp\left(z\log_{(\theta)}\lambda\right)$,
where $\theta-2\pi\le\Im\log_{(\theta)}\lambda\le\theta$.}
\begin{equation}
b^z_{zd-j,\theta}(x,\xi)={i\over 2\pi}\int_{\Gamma_{(\theta)}}\lambda
_{(\theta)}^z b_{-d-j}(x,\xi,\lambda)d\lambda.
\label{A5}
\end{equation}

For $\Re z\!<\!k$ the symbol $b_{(\theta)}^z(x,\xi)$ is defined
as the composition of classical symbols%
\footnote{The composition $a\circ b$ of the classical symbols
$a=\sum_j a_{d-j}$ and $b=\sum_j b_{m-j}$ is
$a\circ b=\sum(a\circ b)_{m+d-j}$, where
$a\circ b:=\sum_j(\alpha!)^{-1}\df^\alpha_\xi a(x,\xi)D^\alpha_x b(x,\xi)$.}

\begin{equation}
a^k(x,\xi)\circ b^{z-k}_{(\theta)}(x,\xi)=:b^z_{(\theta)}(x,\xi),
\label{A6}
\end{equation}
where $a^k(x,\xi)=\sum_{j\in\zuo}a_{kd-j}^k(x,\xi)$ is the symbol
of PDO $A^k$. The composition on the left in (\ref{A6}) is independent
of the choice of $k>\Re z$, $k\in\wZ$ (\cite{Sh}, Ch.~II, Theorem~11.1.a).
The components $b^z_{zd-j}(x,\xi)$ of the symbol of $A^z$ are the entire
functions of $z$ coinciding with $a^k_{kd-j}(x,\xi)$ for $z=k\in\wZ$
(\cite{Sh}, Ch.~II, Theorem~11.1.b,e).

The $\log_{(\theta)}A$ is a bounded linear operator from
$H_{(s)}(M,E)$ into $H_{(s-\eps)}(M,E)$ for an arbitrary $\eps>0$,
$s\in\wR$. This operator acts on smooth global sections $f\in\Gamma(E)$
as follows
\begin{equation}
\left(\log_{(\theta)}A\right)f:=\df_z\left(A_{(\theta)}^z f\right)\big|_{z=0}.
\label{A7}
\end{equation}

For arbitrary $k\in\wZ$ and $s\in\wR$ operators $A^z$ is a holomorphic
function of $z$ from $\Re z<k$ into the Banach space
$L\left(H_{(s)}(M,E),H_{(s-kd)}(M,E)\right)$ of bounded linear operators,
$d=d(A)$ (\cite{Sh}, Ch.~II, Theorem~10.1.e). Hence, the term
on the right in (\ref{A7}) is defined.
The symbol of the operator $\log_{(\theta)}A$ is
\begin{equation}
\sigma\left(\log_{(\theta)}A\right)=\df_z b^z_{(\theta)}(x,\xi)|_{z=0}
:=\sum_{j\in\zuo}\df_z b^z_{zd-j,\theta}(x,\xi)|_{z=0}.
\label{A8}
\end{equation}
The operator $A^z_{(\theta)}|_{z=0}$ is the identity operator.
Hence its symbol $b^z_{(\theta)}(x,\xi)|_{z=0}$ has as its positive
homogeneous components
\begin{equation}
b^z_{zd-j,\theta}(x,\xi)|_{z=0}=\delta_{j,0}\Id.
\label{A9}
\end{equation}

We see from (\ref{A9}) that
\begin{gather}
\begin{split}
\df_z b^z_{zd,\theta}(x,\xi)|_{z=0}   & =d(A)\log|\xi|\Id+\df_z b^z_{zd,\theta}
(x,\xi/|\xi|)|_{z=0}, \\
\df_z b^z_{zd-j,\theta}(x,\xi)|_{z=0} & =\left|\xi\right|^{-j}\df_z
b^z_{zd-j,\theta}(x,\xi/|\xi|)|_{z=0}\text{ for }j\ge 1.
\end{split}
\label{A10}
\end{gather}
hold in local coordinate charts $(U,x)$ on $M$.
Here, $|\xi|$ is taken with respect to some Riemannian metric on $TM$
(and hence on $T^*M$ also). The term $\df_z b^z_{zd,\theta}(x,\xi/|\xi|)$
on the right in (\ref{A10}) is an entire function of $z$ positive
homogeneous in $\xi$ of degree zero. By analogy,
$\df_z b^z_{zd-j,\theta}(x,\xi/|\xi|)$ is an entire function of $z$
positive homogeneous in $\xi$ of degree zero. So the symbol
of $\log_{(\theta)}A$ locally takes the form
\begin{equation}
\df_z b^z_{(\theta)}(x,\xi)|_{z=0}:=\log|\xi|\Id+\sum_{j\in\zuo}
c_{-j,\theta}(x,\xi),
\label{A11}
\end{equation}
where $c_{-j,\theta}(x,\xi):=\left|\xi\right|^{-j}\df_z b^z_{zd-j,\theta}
(x,\xi/|\xi|)$ is a smooth function on $T^*M\setminus i(M)$ positive
homogeneous in $\xi$ of degree $(-j)$. (Here, $i(M)$ is the zero section
of $T^*M$.)

The equality (\ref{A11}) means that in local coordinates $(x,\xi)$
on $T^*M$ the symbol of $\log_{(\theta)}A$ is equal to $d(A)\log|\xi|\Id$
plus a classical PDO-symbol of order zero.
The space $S_{\log}(M,E)$ is also the space of symbols of the form
\begin{equation}
\sigma=k\log_{(\theta)}A+\sigma_0,
\label{A1212}
\end{equation}
$k\in\wC$
(for an arbitrary elliptic PDO $A\in CL^d(M,E)$ of order $d>0$ and
such that there exists an admissible cut $L_{(\theta)}$ of the spectral
plane $\wC\ni\lambda$).
%
Comparing (\ref{A11}) and (\ref{A1212}) we see that the space
$S_{\log}(M,E)$ does not depend on Riemannian metric on $TM$ and on $A$
and $L_{(\theta)}$.

The zeta-regularized determinant $\det_{(\theta)}A$ is defined with the help
of the zeta-function of $A$. This function $\zeta_{A,(\theta)}(z)$
is defined for $\Re z>\dim M/d(A)$ as the trace
$\Tr\left(A^{-z}_{(\theta)}\right)$ of a trace class operator%
\footnote{A bounded linear operator $B$ acting in a separable Hilbert
space is a {\em trace class operator}, if the series of its singular
numbers (i.e., of the arithmetic square roots of the eigenvalues
for the self-adjoint operator $B^*B$) is absolutely convergent.
The operator $A^{-z}_{(\theta)}$ is a PDO of the class
$CL^{-zd}(M,E)$ with $\Re(zd)>\dim M$. Hence it has a continuous kernel
(\cite{Sh}, Ch.~II, 12.1) and this kernel is smooth of the class $C^k$
($k\in\wZ_+$) on $M\times M$ for $\Re(zd)-k>\dim M$.}
$A^{-z}_{(\theta)}$. (Here, $d(A)>0$.) This operator has a continuous kernel
on $M\times M$ for $\Re z>\dim M/d(A)$. The Lidskii theorem \cite{Li},
\cite{Kr}, \cite{ReS}, XIII.17, (177), \cite{Si}, Chapter~3, \cite{LP},
\cite{Re}, XI,
claims that for such $z$ the series of the eigenvalues
of $A^{-z}_{(\theta)}$ is absolutely convergent and that the matrix
trace
$\Sp\left(A^{-z}_{(\theta)}\right):=\sum\left(A^{-z}_{(\theta)}e_i,
e_i\right)$ of $A^{-z}_{(\theta)}$ is equal to its spectral trace%
\footnote{Here, the sum is over the eigenvalues $\lambda_i$
of $A^{-z}_{(\theta)}$, including their algebraic multiplicities
(\cite{Ka}, Ch.~1, \S~5.4), the function $\lambda^{-z}_{(\theta)}$
is defined as $\exp\left(-z\log_{(\theta)}\lambda\right)$,
$\theta-2\pi\le\Im(\log_{(\theta)}\lambda)\le\theta$.}
\begin{equation}
\Tr\left(A^{-z}_{(\theta)}\right):=\sum_{\lambda\in\Spec(A)}\lambda^{-z}
_{i,\theta}.
\label{B2}
\end{equation}
Here, $(e_i)$ is an orthonormal basis in the Hilbert space $L_2(M,E)$
(with respect to a smooth positive density $\mu$ on $M$ and to a Hermitian
metric $h$ on $E$). A bounded linear operator in a separable Hilbert space
is a trace class operator, if the series in the definition of the matrix
trace is absolutely convergent for any orthonormal basis. In this case,
the matrix trace is independent of a choice of the basis, \cite{Kr}, p.~123.
Thus for $\Re z>\dim M/d(A)$ the matrix trace of $A^{-z}_{(\theta)}$
is independent of a choice of the orthonormal basis $(e_i)$.
The Lidskii theorem claims (in particular) that this trace is independent
also of $\mu$, and of $h$. Hence for $\Re z>\dim M/d(A)$
the zeta-function $\zat(z)$ is equal to the integral of the pointwise
trace of the matrix-valued density on the diagonal
$\Delta\colon M\hookrightarrow M\times M$ defined by the restriction
to $\Delta(M)$ of the kernel $A_{-z,\theta}(x,y)$ of $A^{-z}_{(\theta)}$.

The zeta-function $\zat(z)$ possesses a meromorphic continuation
to the whole complex plane $\wC\ni z$ and $\zat(z)$ is regular
at the origin. The determinant of $A$ is a regularization
of the product of all the eigenvalues of $A$ (including their
algebraic multiplicities). The zeta-regularized determinant of $A$
is defined with the help of the zeta-function%
\footnote{In the case $d(A)<0$ the meromorphic continuation of $\zat(z)$
is done from the half-plane $\Re z<\dim M/d(A)=-\dim M/|d(A)|$.
(In this half-plane the series on the right in (\ref{B2})
is convergent.)}
$\zat(z)$ of $A$ as follows
\begin{equation}
{\det}_{(\theta)}A:=\exp\left(-\df_z\zat(z)|_{z=0}\right).
\label{B3}
\end{equation}

\begin{rem}
Note that if an admissible cut $L_{(\theta)}$ of the spectral plane
crosses a finite number of the eigenvalues of $A$, then $\det_{(\theta)}(A)$
does not change. Let $A$ possess
two essential different cuts $\theta_1$, $\theta_2$ of the spectral
plane, i.e., in the case when there are infinite number of eigenvalues
$\lambda\in\Spec(A)$ in each of the sectors
$\Lambda_1:=\left\{\lambda\colon\theta_1<\arg\lambda<\theta_2\right\}$,
$\Lambda_2:=\left\{\lambda\colon\theta_2<\arg\lambda<\theta_1+2\pi\right\}$.
Then in general $\det_{(\theta)}(A)$ depends on spectral cuts
$L_{(\theta)}$ with $\theta=\theta_j$.
\label{RB3750}
\end{rem}

\begin{rem}
If the determinant (\ref{B3}) is defined, then the order $d(A)=:d$
of the elliptic PDO $A\in\Ell_0^d(M,E)\subset CL^d(M,E)$ is nonzero.
Also for the zeta-regularized determinant of $A$ to be defined, its
zeta-function has to be defined. So a holomorphic family of complex
powers of $A$ has to be defined. Hence the principal symbol $a_{dl}(A^l)$
of some appropriate nonzero power $A^l$ of $A$ ($l\in\wC^{\times}$,
$A^l\in\Ell_0^{dl}(M,E)\subset CL^{dl}(M,E)$) has to possess a cut
$L_{(\theta)}$ of the spectral plane $\wC\ni\lambda$. This condition
is necessary for the holomorphic family $\left(A^l\right)_{(\theta)}^z$
of PDOs to be defined by an integral analogous to (\ref{A1}).
In this case, $\log_{(\theta)}\left(A^l\right)$ is defined. (Note that
$ld=l\ord A\in\wR^{\times}$.) Hence some generator
$\log A:=\log_{(\theta)}\left(A^l\right)/l$ of a family $A^z$ is also
defined. Thus the existence of a family $A^z$ is equivalent
to the existence of $\log A$.
\label{RB3710}
\end{rem}

On the algebra $CS(M,E)$ (of classical PDO-symbols) there is a natural
bilinear form defined by the noncommutative residue $\res$ (\cite{Wo1},
\cite{Wo2} or \cite{Kas})
$$
\left(a,b\right)_{\res}=\res(a\circ b).
$$
Here, $a\circ b$ is the composition of PDO-symbols $a$, $b$.
This scalar product is non-degenerate (i.e., for any $a\ne 0$
there exists $b$ such that $(a,b)_{\res}\ne 0$) and it is invariant under
conjugation with any elliptic symbol $c\in\Ell^d(M,E)\subset CS^d(M,E)$,
i.e., $c_d(x,\xi)$ is invertible for $(x,\xi)\in S^*M$. Namely
\begin{equation}
\left(cac^{-1},cbc^{-1}\right)_{\res}:=\res\left(cabc^{-1}\right)=\res(ab)=
(a,b)_{\res}.
\label{B8}
\end{equation}

\begin{rem}
The noncommutative residue is a trace type functional on the algebra
$CS^{\wZ}(M,E)$ of classical PDO-symbols of integer orders, i.e.,
$\res([a,b])=0$ for any $a,b\in CS^{\wZ}(M,E)$.
The space of trace functionals on $CS^{\wZ}(M,E)$ is one-dimensional,
\cite{Wo3}. Namely for $L:=CS^{\wZ}(M,E)$ the algebras with the discrete
topology $L/[L,L]$ and $\wC$ are isomorphic by $\res$. (Note also that
$\res a=0$ for $a\in CL^d(M,E)$ of non-integer order $d$.
So $(a,b)_{\res}=0$ for $\ord a+\ord b\notin\wZ$.) The invariance property
(\ref{B8}) of the noncommutative residue follows from the spectral
definition of $\res$ (\cite{Wo2}, \cite{Kas}).
\label{RB3725}
\end{rem}

\begin{pro}
Let $A_t\in\Ell_0^\alpha(M,E)$, $\alpha\in\wR^{\times}$, be a smooth
family of elliptic PDOs and let $B\in\Ell_0^\beta(M,E)$,
$\beta\in\wR^{\times}$, $\beta\ne-\alpha$. Let the principal symbols
$\sigma_\alpha\left(A_t\right)$ and $\sigma_\beta(B)$ be sufficiently
close to positive definite self-adjoint ones.%
\footnote{We suppose that a smooth positive density and a Hermitian
structure are given on $M$ and on $E$.}
Set
\begin{equation}
F(A,B):={\det}_{(\tpi)}(AB)\big/{\det}_{(\tpi)}(A){\det}_{(\tpi)}(B).
\label{B3872}
\end{equation}
Then the variation formula holds
\begin{equation}
\df_t\!\log F\!\left(\!A_t,B\!\right)\!=\!-\!\left(\!\sigma\!\left(\!\dot{A}_t
A_t^{-1}\!\right),\!\sigma\!\left(\!\log_{(\tpi)}\!\left(\!A_tB\!\right)
\!\right)\!/\!(\alpha\!+\!\beta)\!-\!\sigma\!\left(\!\log_{(\tpi)}\!B\right)
\!/\!\beta\!\right)_{\res}\!.
\label{B3742}
\end{equation}
Here, a cut $L_{(\tpi)}$ of the spectral plane%
\footnote{Note that $F\left(A_t,B\right)$ is independent of $L_{(\tpi)}$
by Remark~\ref{RB3750}. In general a cut $L_{(\tpi)}$ depends on $t$.}
is admissible
for $A_t$, $B$, $A_tB$ and it is sufficiently close to $L_{(\pi)}$.
The term
$\sigma\left(\log_{(\tpi)}\left(A_tB\right)\right)/(\alpha+\beta)-
\sigma\left(\log_{(\tpi)}B\right)/\beta$
on the right in (\ref{B3742}) is a classical PDO-symbol from $CS^0(M,E)$.
It does not depend on an admissible cut $L_{(\tpi)}$ close to $L_{(\pi)}$.
Hence, the scalar product $(,)_{\res}$ on the right in (\ref{B3742})
is defined. The right side of (\ref{B3742}) is locally defined.%
\footnote{The symbols $\sigma\left(\log_{(\theta)}A\right)$,
$\sigma\left(A_{(\theta)}^t\right)$ are locally defined for a PDO $A$
of an order from $\wR_+$ with its principal symbol admitting
a cut $L_{(\theta)}$ of the spectral plane.}
\label{PB3740}
\end{pro}

\begin{rem}
The principal symbols $\sigma_{\alpha+\beta}\left(A_tB\right)$ are
adjoint to $\sigma_{\alpha+\beta}\left(B^{1/2}A_tB^{1/2}\right)$.
The latter principal symbols are sufficiently close to self-adjoint
positive definite ones. The function $F(A,B)$ is called
the {\em multiplicative anomaly}.
\label{RB3741}
\end{rem}

First we formulate a corollary of this proposition.

%

\begin{cor}
Let $A$ and $B$ be invertible elliptic PDOs
$A\!\in\!\Ell_0^\alpha(M,\!E)\!\subset\! CL^\alpha(M,\!E\!)$,
$B\in\Ell_0^\beta(M,E)\subset CL^\beta(M,E)$ such that
$\alpha,\beta,(\alpha+\beta)\in\wR^{\times}$ and such that their principal
symbols $a_\alpha(x,\xi)$ and $b_\beta(x,\xi)$ are sufficiently close
to positive definite self-adjoint ones. Then the multiplicative
anomaly is defined. Its logarithm is given by a locally defined integral
\begin{multline}
\log F(A,B)=-\int_0^1 dt\bigl(\sigma\left(\dot{A}_tA_t^{-1}\right),\sigma
\left(\log_{(\tpi)}\left(A_tB\right)\right)/(\alpha+\beta)-\\
-\sigma\left(\log_{(\tpi)}\left(A_t\right)\right)/\alpha\bigr)_{\res}.
\label{B221}
\end{multline}
Here, $A_t:=\eta_{(\tpi)}^tB_{(\tpi)}^{\alpha/\beta}$,
$\eta:=AB_{(\tpi)}^{-\alpha/\beta}\in\Ell_0^0(M,E)\subset CL^0(M,E)$.
(In particular, we have $A_0:=B_{(\tpi)}^{\alpha/\beta}$, $A_1:=A$,
$F\left(A_0,B\right)=1$.)

The expression
$\sigma\left(\log_{(\tpi)}\left(A_tB\right)\right)/(\alpha+\beta)-
\sigma\left(\log_{(\tpi)}\left(A_t\right)\right)/\alpha$
on the right in (\ref{B221}) is a classical PDO-symbol from $CS^0(M,E)$.
Thus the integral formula for the multiplicative anomaly has the form
\begin{multline}
\log F(A,B)=-\int_0^1dt\bigl(\sigma\left(\log_{(\tpi)}\eta\right),
\sigma\left(\log_{(\tpi)}\left(\eta_{(\tpi)}^tB^{(\alpha+\beta)/\beta}
\right)\right)/(\alpha+\beta)-\\
-\sigma\left(\log_{(\tpi)}\left(\eta_{(\tpi)}^tB^{\alpha/\beta}\right)\right)
/\alpha\bigr)_{\res}.
\label{B3752}
\end{multline}
Operators $\log_{(\tpi)}\eta$ and $\eta_{(\tpi)}^t$ are defined
by (\ref{B7}) and (\ref{X2}) below.

\label{CB3744}
\end{cor}

The proof of Proposition~\ref{PB3740} is based on the assertions as follows.

\begin{pro}
Let $Q$ be a PDO from $CL^0(M,E)$ and let $C,A$ be PDOs of real positive
orders sufficiently close to self-adjoint positive definite PDOs.
Then the function
$$
P(s)=\zeta_{C,(\tpi)}(Q;s)-\zeta_{A,(\tpi)}(Q;s):=\Tr\left(Q\left(C_{(\tpi)}
^{-s/\ord C}-A_{(\tpi)}^{-s/\ord A}\right)\right)
$$
has a meromorphic continuation to the whole complex plane $\wC\ni s$.
The origin is a regular point of this function. Its value at the origin
is defined by the following expression through the symbols $\sigma(A)$,
$\sigma(C)$, $\sigma(Q)$
\begin{equation}
P(0)=-\left(\sigma(Q),{\sigma\left(\log_{(\tpi)}C\right)\over \ord C}-
{\sigma\left(\log_{(\tpi)}A\right)\over \ord A}\right)_{\res}.
\label{X4}
\end{equation}
The same assertions about $P(s)$ and $P(0)$ are also valid
for $Q\in CL^m(M,E)$, $m\in\wZ$.
\label{PD3}
\end{pro}

\begin{pro}
Under the conditions of Proposition~\ref{PD3}, the family of PDOs
\begin{equation}
K_s:=-Q\left(C_{(\tpi)}^{s/\ord C}-A_{(\tpi)}^{s/\ord A}\right)/s\in CL^s
(M,E)
\label{B100}
\end{equation}
is a holomorphic%
\footnote{This family of PDOs is holomorphic in the sense of \cite{Gu},
Sect.~3, (3.17), (3.18).}
family of PDOs. In particular, it is holomorphic at $s=0$.
\label{PB17}
\end{pro}

\begin{cor}
The pointwise trace on the diagonal $\tr K_s(x,x)$ of the kernel
$K_s(x,y)$ of $K_s$ is a density on $M$ for $\Re s<-\dim M$. This
density has a meromorphic continuation from $s<-\dim M$ to the whole
complex plane $\wC\ni s$.
The residue of this density at $s=0$
is equal to
\begin{multline}
\Res_{s=0}\tr K_s(x,x)=\tr\left(-Q\left(C_{(\tpi)}^{s/\ord C}-A_{(\tpi)}^{s/
\ord A}\right)\right)\big|_{s=0}(x,x)=\\
=-\res_x\sigma\left(-{Q\over s}\left(C_{(\tpi)}^{s/\ord C}-A_{(\tpi)}
^{s/\ord A}\right)\right)\big|_{s=0}=\\
=\res_x\sigma\left(Q\left({\log_{(\tpi)}C
\over \ord C}-{\log_{(\tpi)}A\over \ord A}\right)\right),
\label{B101}
\end{multline}
where $\res_x$ is the density on $M$ corresponding to the noncommutative
residue \cite{Wo2}, \cite{Kas}. These assertions follows immediately
from Proposition~\ref{PB3755} below.%
\footnote{This proposition claims that analogous assertions are true
for any holomorphic (in a weak sense) family of classical PDOs.
The proof of this proposition is based on the notion of the canonical
trace for PDOs of noninteger orders introduced in Section~\ref{SB}
(below).}
\label{CB18}
\end{cor}

\begin{rem}
The formula (\ref{X4}) follows from (\ref{B101}) since
$$
P(0)=-\Res_{s=0}\Tr K_s.
$$
\label{RB19}
\end{rem}

\noindent{\bf Proof of Proposition~\ref{PB3740}.}
The variation $\df_t\log F\left(A_t,B\right)$ is
\begin{equation}
\df_t\log F\left(A_t,B\right)=\df_t\left(-\df_s\zeta_{A_t B,(\tpi)}(s)|_{s=0}+
\df_s\zeta_{A_t,\tpi}(s)|_{s=0}\right).
\label{B5}
\end{equation}
For $\Re s\gg 1$ the operators $\left(A_t B\right)_{(\tpi)}^{-s}$ and
$A_{t,(\tpi)}^{-s}$ are of trace class. For such $s$ these operators
form smooth in $t$ families of trace class operators. By the Lidskii
theorem we have for such $s$
\begin{equation}
\Tr\left(A_{t,(\tpi)}^{-s}\right)=\sum_i\left(A_{t,(\tpi)}^{-s}e_i,e_i\right),
\label{B3515}
\end{equation}
where $e_i$ is an orthonormal basis in $L_2(M,E)$. (Here, we suppose
that a smooth positive density on $M$ and a Hermitian structure on $E$
are given.) For $\Re s\gg 1$ we have
\begin{equation}
\df_t\Tr\left(A_{t,(\tpi)}^{-s}\right)=\Tr\left(\df_tA_{t,(\tpi)}^{-s}\right).
\label{B3516}
\end{equation}
Indeed, $A_{t,(\tpi)}^{-s}$ is a smooth (in $t$) family of trace class
operators. So $\df_tA_{t,(\tpi)}^{-s}$ is a trace class operator.
Hence the series
$$
\sum_i\df_t\left(A_{t,(\tpi)}^{-s}e_i,e_i\right)\equiv\sum_i\left(\df_t
A_{t,(\tpi)}^{-s}e_i,e_i\right)
$$
is absolutely convergent. Thus the series on the right in (\ref{B3515})
can be differentiated term by term (\cite{WW}, Chapter~4, 4.7) and
the equality (\ref{B3516}) follows from (\ref{B3515}).

For $\Re s\gg 1$ the equalities hold
\begin{gather}
\begin{split}
\df_t\Tr\left(\left(A_tB\right)_{(\tpi)}^{-s}\right) & =(-s)\Tr\left(\left(
\dot{A}_t A_t^{-1}\right)\left(A_tB\right)_{(\tpi)}^{-s}\right),\\
\df_t\Tr\left(A_{t,(\tpi)}^{-s}\right)               & =(-s)\Tr\left(\left(
\dot{A}_t A_t^{-1}\right)A_{t,(\tpi)}^{-s}\right).
\end{split}
\label{B6}
\end{gather}
(Here, $\dot{A}_t$ is defined as $\df_tA_t$.) Indeed, for such $s$ we have
\begin{multline}
\df_t\Tr\left(A_{t,(\tpi)}^{-s}\right)=\Tr\left(-{i\over 2\pi}\int_{\Gamma
_{(\tpi)}}\lambda^{-s}\left(A_t-\lambda\right)^{-1}\dot{A}_t\left(A_t-\lambda
\right)^{-1}d\lambda\right)= \\
=\Tr\left(-{i\over 2\pi}\int_{\Gamma_{(\tpi)}}\lambda^{-s}\left(\df_\lambda
\left(A_t-\lambda\right)^{-1}\right)d\lambda\dot{A}_t\right)= \\
=(-s)\Tr\left({i\over 2\pi}\int_{\Gamma_{(\tpi)}}\lambda^{-(s+1)}\left(A_t-
\lambda\right)^{-1}d\lambda\dot{A}_t\right)= \\
=(-s)\Tr\left(A_{t,(\tpi)}^{-(s+1)}\dot{A}_t\right)=(-s)\Tr\left(\dot{A}_tA_t
^{-1}A_{t,(\tpi)}^{-s}\right).
\label{B3517}
\end{multline}

The zeta-function $\left(A_{t,(\tpi)}^{-s}\right)=:\zeta_{A_t,(\tpi)}(s)$
has a meromorphic continuation to the whole complex plane $\wC\ni s$ and
$s=0$ is a regular point for this zeta-function.
So $(-s)\Tr\left(\dot{A}_tA_t^{-1}A_{t,(\tpi)}^{-s}\right)$ also has
a meromorphic continuation with a regular point $s=0$. Hence
the equality holds
\begin{equation}
\df_s\left(s\Tr\left(\dot{A}_tA_t^{-1}A_{t,(\tpi)}^{-s}\right)\right)
\big|_{s=0}=\left(\left(1+s\df_s\right)\Tr\left(\dot{A}_tA_t^{-1}A_{t,(\tpi)}
^{-s}\right)\right)\big|_{s=0}
\label{B3751}
\end{equation}
and the meromorphic function on the right is regular at $s=0$.

The formula (\ref{B3742}) is an immediate consequence of (\ref{B5}),
(\ref{B6}), (\ref{B3751}), and of (\ref{X4}). In (\ref{X4})
$\dot{A}_tA_t^{-1}$ is substituted as $Q$. Proposition~\ref{PB3740}
is proved.\ \ \ $\Box$

\noindent{\bf Proof of Corollary~\ref{CB3744}.}
1. If the principal symbols $a_\alpha$, $b_\beta$ of $A$, $B$ are
sufficiently close to positive definite self-adjoint ones, then
the principal symbol $a_\alpha\left(b_\beta\right)_{(\tpi)}^{-\alpha/\beta}$
of $\eta$ possesses a cut $L_{(\pi)}$ along $\wR_-$. If all the eigenvalues
of $b_\beta$ are in a sufficiently narrow conical neighborhood of $\wR_+$,
then the principal symbol
$\left(a_\alpha\left(b_\beta\right)_{(\pi)}^{-\alpha/\beta}\right)\!_{(\pi)}^t
\left(b_\beta\right)_{(\pi)}^{\alpha/\beta}$
of $\eta_{(\tpi)}^tB_{(\tpi)}^{\alpha/\beta}$
possesses for all $0\le t\le 1$ a cut $L_{(\pi)}$ of the spectral plane.%
\footnote{This symbol is independent of a choice of an admissible cut
$L_{(\tpi)}$ close to $L_{(\pi)}$.}

Set $A_t:=\eta_{(\tpi)}^tB_{(\tpi)}^{\alpha/\beta}$.
Then $A_0=B_{(\tpi)}^{\alpha/\beta}$, $F\left(A_0,B\right)=1$, $A_1=A$,
$F(A_1,B)=F(A,B)$. We can use the variation formula
of Proposition~\ref{PB3740} and the equalities (\ref{B5}), (\ref{B6}).
Note that the operator
$\dot{A}_t A_t^{-1}\!\equiv\!\df_t\left(\eta_{(\tpi)}^t\right)
\!\left(\eta_{(\tpi)}^t\right)^{-1}$ is equal to $\log_{(\tpi)}\eta$. Here,
$\log_{(\tpi)}\eta\!\in \!CL^0(M,E)$ is the operator
\begin{equation}
{i\over 2\pi}\int_{\Gamma_{R,\tpi}}\log_{(\tpi)}\lambda\cdot\left(\eta-\lambda
\right)^{-1}d\lambda, \quad \tpi-2\pi\le\Im\log_{(\tpi)}\lambda\le\tpi,
\label{B7}
\end{equation}
$\Gamma_{R,\tpi}$ is the contour
$\Gamma_{1,R,\tpi}(\eps)\cup\Gamma_{0,\tpi}(\eps)\cup\Gamma_{2,R,\tpi}(\eps)
\cup\Gamma_R$, where
\begin{gather*}
\Gamma_{1,R,\tpi}(\eps)\!:=\!\left\{\lambda\!+\!x\exp(i\tpi),R\!\ge\!x\!\ge
\!\eps\right\}, \quad
\Gamma_{2,R,\tpi}(\eps)\!:=\!\left\{\lambda\!=\!x\exp(i(\tpi\!-\!2\pi)),\eps
\!\le\!x\!\le\!R\right\}, \\
\Gamma_{0,\tpi}(\eps):=\left\{\lambda=\eps\exp(i\phi),\tpi\ge\phi\ge\tpi-2\pi
\right\},
\end{gather*}
and $\Gamma_R$ is the circle $|\lambda|=R$, $\lambda=R\exp(i\phi)$,
oriented opposite to the clockwise ($\tpi-2\pi\le\phi\le\tpi$)
and surrounding once the whole spectrum in $L_2(M,E)$ of the bounded
operator $\eta\in CL^0(M,E)$. The radius $\eps>0$ is small enough
such that this spectrum does not intersect the domain
$\{\lambda,|\lambda|\le\eps\}$. We have
$\log_{(\tpi)}\eta=\df_t\eta_{(\tpi)}^t|_{t=0}$, where
\begin{equation}
\eta_{(\tpi)}^t:={i\over 2\pi}\int_{\Gamma_{R,\tpi}}\lambda_{(\tpi)}^t\left(
\eta-\lambda\right)^{-1}d\lambda.
\label{X2}
\end{equation}

The spectrum of elliptic PDO $\eta$ is compact.
The operator $\log_{(\tpi)}\eta$ is a classical PDO from $CL^0(M,E)$.
The symbol $\sigma\left(\log_{(\tpi)}\eta\right)\in CS^0(M,E)$
is equal to $(i/2\pi)\int_{\Gamma_+}\log\lambda\cdot
\sigma\left(\left(\eta-\lambda\right)^{-1}\right)d\lambda$.
Here, $\sigma\left(\left(\eta-\lambda\right)^{-1}\right)$ is a classical
PDO-symbol from $CS^0(M,E)$, the principal symbol $\sigma_0(\eta)(x,\xi)$,
$\xi\ne 0$, has all its eigenvalues in the half-plane
$\wC_+:=\{\lambda\colon\Re\lambda>0\}$ and $\Gamma_+$ is a contour
in $\wC_+$ oriented opposite to the clockwise and surrounding once
the compact set $\cup_{(x,\xi)\in S^*M}\Spec\left(\sigma_0(\eta)(x,\xi)\right)
\subset\wC_+$.

Hence, if the function
$\Tr\left(Q_t\left(\left(A_tB\right)_{(\tpi)}^{-s}-A_{t,(\tpi)}^{-s}\right)
\right)$ for $Q_t:=\dot{A}_tA_t^{-1}$
has an analytic continuation to the neighborhood of the origin and
if $s=0$ is a regular point of this analytic function, then we have
from (\ref{B5}), (\ref{B6}), (\ref{B3751})
\begin{multline}
\df_t\log F\left(A_t,B\right)=-\df_t\df_s\left(\Tr\left(A_tB\right)_{(\tpi)}
^{-s}-\Tr A_{t,(\tpi)}^{-s}-\Tr B_{(\tpi)}^{-s}\right)\big|_{s=0}=\\
=\!(1\!+\!s\df_s)\!\Tr\!\left(\left(\log_{(\tpi)}\!\eta\right)\!\left(\!\left(
A_tB\right)_{(\tpi)}^{-s}\!-\!A_{t,(\tpi)}^{-s}\right)\!\right)\!\big|_{s=0}\!=
\!\Tr\!\left(\!\left(\log_{(\tpi)}\!\eta\right)\!\left(\!\left(A_tB\right)
_{(\tpi)}^{-s}\!-\!A_{t,(\tpi)}^{-s}\right)\!\right)\!\big|_{s=0}\!=\\
=(1+s\df_s)\Tr\left(\dot{A}_tA_t^{-1}\left(\left(A_tB\right)_{(\tpi)}^{-s/
{\alpha+\beta}}-\Tr B_{(\tpi)}^{-s/\beta}\right)\right)\big|_{s=0}.
\label{C1}
\end{multline}

By Proposition~\ref{PB3740} we have for $0\le t\le 1$
\begin{equation}
\df_t\!\log \!F(A_t,B)\!=\!-\!\left(\sigma\!\left(\log_{(\tpi)}\eta\right)\!,
\!\sigma \!\left(\log_{(\tpi)}(\!A_t B\!)\right)\!/\!(\alpha\!+\!\beta)\!-
\!\sigma\!\left(\log_{(\tpi)}(\!A_t\!)\right)\!/\!\alpha\right)\!_{\res}\!.
\label{X5}
\end{equation}
Thus Corollary~\ref{CB3744} is proved.\ \ \ $\Box$


\begin{rem}
Let $A_t\in\Ell_0^\alpha(M,E)$, $0\le t\le 1$, be a smooth family
of invertible elliptic PDOs of order $\alpha\in\wR^{\times}$ such that
the principal symbols $a_{t,\alpha}$ of $A_t$ are sufficiently close
to positive definite ones. Let $B\in\Ell_0^\beta(M,E)$ have a real
order $\beta\ne-\alpha$ and let the principal symbol $b_\beta$ be
sufficiently close to positive definite self-adjoint one. Let $A_0$
be a power of $B$, $A_0=B_{(\tpi)}^{\alpha/\beta}$. Set $A:=A_1$.
By Proposition~\ref{PB3740} the multiplicative anomaly of $(A,B)$
is given by the locally defined integral
\begin{multline}
\log F(A,B)=-\int_0^1 dt\bigl(\sigma(Q_t),\sigma\left(\log_{(\tpi)}(A_tB)
\right)/\left(\alpha+\beta\right)-\\
-\sigma\left(\log_{(\tpi)}(A_t)\right)/\alpha\bigr)_{\res}.
\label{B3753}
\end{multline}
Here, $Q_t:=\dot{A}_tA_t^{-1}\in CL^0(M,E)$. Its symbol
$\sigma\left(Q_t\right)$ is locally defined in terms
of $\sigma\left(A_t\right)$.
The right side of it is the integral of the locally defined density
on $M$. This is a formula for the multiplicative anomaly corresponding
to a general smooth variation between $B_{(\tpi)}^{\alpha/\beta}$ and
$A$.
\label{RB3752}
\end{rem}

%

\begin{rem}
The assertions of Proposition~\ref{PD3} that $P(s)$ is regular at $s=0$
and that there exists a local expression for $P(0)$ is the contents
of Lemma~4.6 in \cite{Fr}. Proposition~\ref{PD3} is a consequence
of Proposition~\ref{PB17}, of Corollary~\ref{CB3744}, and
of Remark~\ref{RB19}.
\label{RB3749}
\end{rem}

\noindent{\bf Proof of Proposition~\ref{PB17}.} The symbols
$\sigma\left(C_{(\tpi)}^{s/\ord C}\right)$ and
$\sigma\left(A_{(\tpi)}^{s/\ord A}\right)$ at $s=0$ are equal to $\Id$.
These symbols are entire functions of $s\in\wC$ (i.e., all the homogeneous
terms of $\sigma(K_s)$ are entire functions of $s\in\wC$ in any local
coordinates on $M$.)
The family of PDOs
$C(\mu):=C_{(\tpi)}^{\mu/\ord C}\in\Ell_0^{\times}\subset CL^\mu(M,E)$
is holomorphic in the sense of \cite{Gu}, (3.18). The latter means that
for PDOs $C_k(\mu):=C(\mu)-P_kC(\mu)\in CL^{\mu-k}(M,E)$
as $\delta\to 0$ we have
\begin{equation}
\left\|\left(C_k(\mu+\delta)-C_k(\mu)\right)/\delta-\dot{C}_k(\mu)\right\|
_{\cA}^{(s)}\to 0
\label{B230}
\end{equation}
for $s>\Re\mu-k$. (Here $P_kC(\mu)\in CL^\mu(M,E)$ are the PDOs defined
by the image of $\sigma(C(\mu))$ in $CS^\mu(M,E)/CS^{\mu-k-1}(M,E)$ and
by a fixed partition of unity on $M$ subordinate to a fixed local
coordinates cover of $M$.) In (\ref{B230}) $\left\|\cdot\right\|^{(s)}$
is the operator norm from $H_{(s)}(M,E)$ into $L_2(M,E)$ of the operator
defined on the dense subspace of global $C^{\infty}$-sections
$\Gamma(E)$ in the Sobolev space $H_{(s)}(M,E)$ and
$\dot{C}_k(\mu)\colon\Gamma(E)\to\Gamma(E)$ is a linear operator.
(In (\ref{B230}), as well as in \cite{Gu}, (3.18),
$\cA=(\cA_1,\dots,\cA_m)$ is an arbitrary collection
of ordinary differential operators of order one with the scalar principal
symbols,
acting on $\Gamma(E)$.) The subscript $\cA$ in (\ref{B230}) means
\begin{equation}
\left\|B\right\|_{\cA}^{(s)}:=\left\|[\cA_1,\dots,[\cA_m,B]\ldots]
\right\|^{(s)}.
\label{B231}
\end{equation}

The operators $C(\mu)$ and $A(\mu):=A_{(\pi)}^{\mu/\ord A}$ for $\mu=0$
are the identity operators. Hence $\sigma(C(0))=\sigma(A(0))=\Id$ and
the symbol $S:=\df_\mu(\sigma(C(\mu))-\sigma(A(\mu)))|_{\mu=0}$ is
a PDO-symbol from $CS^0(M,E)$. For any PDO $\tilde{S}\in CL^0(M,E)$
with $\sigma(\tilde{S})=S$ the operator $K_0+Q\tilde{S}$ is smoothing
in $\Gamma(E)$, i.e., it has a $C^\infty$ Schwartz kernel. (Here,
$$
K_0:=-\lim_{\mu\to 0}Q(C(\mu)-A(\mu))
$$
is the value at $s=0$ of $K_s$ from (\ref{B100}).) Indeed,
$K_0+QP_{2m+1}\tilde{S}\colon H_{(-m)}\to H_{(m)}$ is a bounded linear
operator since by (\ref{B230}) it is a bounded operator from $H_{(s)}(M,E)$
to $L_2(M,E)$ for $s>-(2m+1)$. Similarly, by (\ref{B230})
$\left[\cA_1,K_0+QP_{2m+1}\tilde{S}\right]$ is a bounded operator
from $H_{(s)}(M,E)$ to $L_2(M,E)$ for such $s$ and for an arbitrary DO
$\cA_1\colon\Gamma(E)\to\Gamma(E)$ of order one with a scalar principal
symbol. Hence $K_0+QP_{2m+1}\tilde{S}$ is a bounded linear operator
from $H_{(s)}$ to $H_{(1)}$ for $s>-2m$. Applying (\ref{B230}) with
higher commutators $\cA$, we see that $K_0+QP_{2m+1}\tilde{S}$
is a bounded linear operator from $H_{(-m)}$ to $H_{(m)}$.

Operators $C_m(\mu):=C(\mu)-P_mC(\mu)$ and $A_m(\mu)$ from $H_{(s)}(M,E)$
to $L_2(M,E)$ for $|\mu|\le r$ and for $s>r-m$ are uniformly bounded
by (\ref{B230}). The analogous assertion is true by (\ref{B230})
for higher commutators of $C_m(\mu)$ (or of $A_m(\mu)$) with DOs
$\cA_j$ of first order with scalar principal symbols. (Such type
operators are defined in (\ref{B231}) and are used in (\ref{B230}).)

For a holomorphic family of PDOs we have a Cauchy integral representation.
Namely for $\Gamma_r:=\{\mu,|\mu|=r\}$ it holds
\begin{equation}
C_m(\mu)={1\over 2\pi i}\int_{\Gamma_r}{C_m(z)\over z-\mu}dz,
\label{B1500}
\end{equation}
for a family $C_m(z)$ of linear operators on $\Gamma(E)$.
This integral is absolutely convergent in the operator norm topology
in the space of bounded linear operators
$L\left(H_{(s)}(M,E),L_2(M,E)\right)$ for $s>r-m$. This integral is
convergent also with respect to the semi-norm
$\left\|\cdot\right\|_{\cA}^{(s)}$ from (\ref{B231}) for $s>r-m$,
where $\cA=\left(\cA_1,\dots,\cA_k\right)$ is an arbitrary collection
of first order DOs with the scalar principal symbols. (Indeed, the Cauchy
integral representation (\ref{B1500}) holds also for a holomorphic family
PDOs $\left[\cA_1,\dots,\left[\cA_k,C_m(\mu)\right]\ldots\right]$.)

To prove that $K_\mu$ (from (\ref{B100})) is a holomorphic at $\mu=0$
family of PDOs, it is enough to note that for
$K_m(\mu):=-Q\left(C_m(\mu)-A_m(\mu)\right)/\mu$ we have
\begin{equation}
\left(K_m(\mu)-K_m(0)\right)/\mu=-Q\left(C_m^{(2)}(\mu)-A_m^{(2)}(\mu)\right)
\label{B1503}
\end{equation}
for $C_m^{(2)}(\mu):=\left(C_m(\mu)-C_m(0)-\mu\df_\mu C_m|_{\mu=0}\right)/
\mu^2$ because $C_m(0)=A_m(0)$ ($=0$ for $m\in\wZ_+$) and because
$K_m(0):=-Q\left(\df_\mu C_m|_{\mu=0}-\df_\mu A_m|_{\mu=0}\right)$,
where $\df_\mu C_m$ is the operator $\dot{C}_m$ from (\ref{B230}).
The operator $C_m^{(2)}(\mu)\colon\Gamma(E)\to\Gamma(E)$ converges
to the operator $\df_\mu^2 C_m(\mu)|_{\mu=0}/2\colon\Gamma(E)\to\Gamma(E)$
in the semi-norms
\begin{equation}
\left\|C_m^{(2)}(\mu)-\df_\mu^2 C_m|_{\mu=0}/2\right\|_{\cA}^{(s)}\to 0
\label{B1505}
\end{equation}
as $\mu\to 0$ for $s>r-m$ because for $|\mu|<r$, $\mu\ne 0$, we have
\begin{equation}
C_m^{(2)}(\mu)={1\over 2\pi i}\int_{\Gamma_r}{C_m(z)\over z^2(z-\mu)}dz.
\label{B1504}
\end{equation}

The operator $\df_\mu^2 C_m(\mu)|_{\mu=0}/2$ is defined as this integral
with $\mu=0$. The integral in (\ref{B1504}) is convergent and continuous
in $\mu$ (for sufficiently small $|\mu|$) with respect to the semi-norms
$\left\|\cdot\right\|_{\cA}^{(s)}$ for $s>r-m$. The assertion (\ref{B1505})
is true for all $s>-m$. (It is enough to substitute in (\ref{B1505})
$\mu$ with sufficiently small $|\mu|$.) Hence the family $K_s$ from
(\ref{B100}) is holomorphic at $s=0$.\ \ \ $\Box$

\section{Canonical trace and canonical trace density for PDOs
of noninteger orders. Derivatives of zeta-functions at zero}
\label{SB}

For any classical PDO $A\in CL^\alpha(M,E)$, where $\alpha\in\wC\setminus\wZ$,
a canonical trace and a canonical trace density of $A$ are defined.
Indeed, since the symbol of $A$ has
an asymptotic expansion as $|\xi|\to+\infty$
$$
a(x,\xi)=\sum_{k\in\zuo}a_{\alpha-k}(x,\xi),
$$
the Schwartz (distributional) kernel $A(x,y)$ of $A$ has an asymptotic
expansion for $x\to y$, $x\ne y$ in any local coordinate chart $U$ on $M$
as follows
\begin{equation}
A(x,y)=\sum_{k=0,1,\dots,N}A_{-n-\alpha+k}(x,y-x)+A_{(N)}(x,y).
\label{K1}
\end{equation}
Here, $n=\dim M$, $N\in\wZ_+$ is sufficiently large%
\footnote{It is enough to take $N\in\wZ_+$ greater than $n+\Re\alpha+1$.},
$A_{(N)}(x,y)$ is continuous and this kernel is smooth enough%
\footnote{The kernel $A_{(N)}(x,y)$ is of the class $C^l(U\times U)$
for $l+\dim M+\Re\alpha<N+1$. Here we suppose that the coordinate system
is defined in some neighborhood $V$ of $\overline U$.}
near the diagonal in $U\times U$.
The term $A_{-n-\alpha+k}(x,y-x)$ is positive homogeneous in $y-x$
of degree $(-n-\alpha+k)$ for all pairs $(x,y)$ of sufficiently close
$x$ and $y$ from $U$ and for $0<t\le 1$, i.e.,
\begin{equation}
A_{-n-\alpha+k}(x,t(y-x))=t^{-n-\alpha+k}A_{-n-\alpha+k}(x,y-x).
\label{K2}
\end{equation}

\begin{rem}
The (local) kernel $A_{-n-\alpha+k}(x,y-x)$ corresponds to the integral
\begin{equation}
\int a_{\alpha-k}(x,\xi)\exp(i(x-y,\xi))d\xi.
\label{B3804}
\end{equation}
This integral is defined as follows. The analogous truncated integral
\begin{equation}
\int\rho(|\xi|)a_{\alpha-k}(x,\xi)\exp(i(x-y,\xi))d\xi
\label{B3805}
\end{equation}
is an oscillatory integral, \cite{Ho1}, 7.8, since the estimates hold
$$
\left|D_x^\beta D_\xi^\gamma\left(\rho(|\xi|)a_{\alpha-k}(x,\xi)\right)
\right|\le C_{\beta,\gamma,K}(1+|\xi|)^{\Re\alpha-k-|\gamma|}
$$
for $x\in K\subset U$. (Here, $U$ is a coordinate chart on $M$, $K$ is
a compact, $\rho(t)$ is a $C^\infty$-function, $\rho(t)\equiv 0$
for small $t$ and $\rho(t)\equiv 1$ for $t\ge 1$.) This truncated
integral defines a distribution on $C_0^\infty(x,y)$ of order%
\footnote{A distribution $u$ on $C_0^\infty(V)$ (for a local coordinate
chart on $M$) is of order $l\in\zuo$, $u\in{\cal{D}'}^l(V)$,
if for any compact $K$ in $V$ the estimates hold
$|u(f)|\le C_k\sum_{|\beta\le l}\sup\left|\df^\beta f\right|$,
$f\in C_0^\infty(K)$.}
$\le l$ for $\Re\alpha-k-l<-\dim M$, \cite{Ho1}, Theorem~7.8.2.

The integral
\begin{equation}
\int(1-\rho(|\xi|))a_{\alpha-k}(x,\xi)\exp(i(x-y,\xi))d\xi
\label{B3806}
\end{equation}
is absolutely convergent for $\Re\alpha-k>-n$, $n:=\dim M$.
All the partial derivatives in $x$, $y$ under the sign of this integral
are also absolutely convergent for such $\alpha$. So the kernel
(\ref{B3806}) is smooth in $(x,y)$ for such $\Re\alpha$.

The integral (\ref{B3804}) is the Fourier transformation $\xi\to y-x=:u$
of the homogeneous in $\xi$ distribution $a_{\alpha-k}(x,\xi)$
(depending on $x$ as on a parameter) of the degree $\alpha-k$.
For $\alpha-k\in\{m\in\wZ,m\le-\dim M\}$ and for a fixed $x$
the distribution $a_{\alpha-k}(x,\xi)$ has a unique extension
to the distribution belonging to $\cal{D}'\left(\wR^n\right)$
($\xi\in\wR^n\setminus 0$, $n:=\dim M$), \cite{Ho1}, Theorem~3.2.3.
The Fourier transformation of $a_{\alpha-k}(x,\xi)$ is a homogeneous
in $u=y-x$ distribution of the degree $(-\alpha+k-n)$, \cite{Ho1},
Theorem~7.1.16. So this integral for $\alpha-k\notin\zuo$ (and
for a fixed $x$) has a unique extension to the distribution
belonging to $\cal{D}'\left(\wR^n\right)$.

Note that $a_{\alpha-k}(x,\xi)$ is (for a fixed $x$) a temperate
distribution, i.e., it belongs to $\cal{S}'\left(\wR^n\right)$.
Indeed, $\rho(|\xi|)a_{\alpha-k}(x,\xi)$ is (for a fixed $x$)
a temperate distribution, \cite{Ho1}, 7.1.
For $\alpha-k\notin\{m\in\wZ,m\le-n\}$, $n:=\dim M$, (and for a fixed $x$)
the distribution from $\cal{D}'\left(\wR^n\setminus 0\right)$
\begin{equation}
(1-\rho(|\xi|))a_{\alpha-k}(x,\xi)
\label{B3836}
\end{equation}
(equal to $|\xi|^{\alpha-k}(1-\rho(|\xi|))a_{\alpha-k}(x,\xi/|\xi|)$
for $|\xi|$ small enough) has a unique extension
to a distribution from $\cal{D}'\left(\wR^n\right)$ with a compact
support. (This fact follows from \cite{Ho1}, Theorem~3.2.3, because
(\ref{B3836}) is homogeneous in $|\xi|$ for sufficiently small $|\xi|$.)
Hence this extension of (\ref{B3836}) is also a temperate distribution.
Its Fourier transformation provides us with an analytic continuation
in $\alpha$ of the integral (\ref{B3806}) from the domain
$\{\alpha\colon\Re\alpha>k-n\}$ to $\alpha\in\wC\setminus\{m\in\wZ,m\le k-
n\}$. So the Fourier transformation of $a_{\alpha-k}(x,\xi)$ is defined
and belongs to $\cal{S}'\left(\wR^n\right)$.

The wave front set for the oscillatory integral (\ref{B3805}) is contained
in $\{x,y-x=0,\xi\}$, \cite{Ho1}, Theorem~8.1.9. Hence the kernel
(\ref{B3805}) is smooth outside of the diagonal $x=y$. The Fourier
transformation (\ref{B3806}) from $\xi$ to $u=y-x$ (for a fixed $x$)
has its wave set belonging to $\{u=0,\xi\}$ (\cite{Ho1}, Theorem~8.1.8)
because the wave front set for $(1-\rho(|\xi|))a_{\alpha-k}(x,\xi)$
($x$ is fixed) belongs to the cotangent space at $\xi=0$. So the kernel
(\ref{B3804}) is smooth for $x\ne y$.
\label{RB3803}
\end{rem}

The {\em canonical trace density} $a_{(N)}^U(x):=\tr A_{(N)}(x,x)$ on $U$
is defined as a pointwise trace of the kernel $A_{(N)}$. This density
does not change under a shift $N\to N+k$, $k\in\wZ_+$. This density
for a large positive $N$ is denoted further by $a_U(x)$.

\begin{pro}
The denisity $a_U(x)$ at $x\in M$ is independent of a smooth coordinate
system $U\ni x$ on $M$ near $x$.
\label{PB1}
\end{pro}

\noindent{\bf Proof.} Let $z=f(Z)$ be a smooth change of local coordinates
near $x$. Then according to Taylor's formula we have for $X$ and $Y$
sufficiently close one to another
\begin{multline*}
A_{-n-\alpha+k}\left(f(X),f(Y)-f(X)\right)=\\
=A_{-n-\alpha+k}\left(f(X),\sum_{1\le|\alpha|\le N}\df_X^\beta f(X)(Y-X)
^\beta/\beta!+r_N(X,Y)\right)=\\
=B_{-n-\alpha+k}(X,Y-X)+B_{-n-\alpha+k+1}(X,Y-X)+\ldots+
R_{-n-\alpha+k,(N)}(X,Y),
\end{multline*}
where $R_{-n-\alpha+k,(N)}(X,Y)$ is a local kernel continuous near
the diagonal and such that $R_{-n-\alpha+k,(N)}(X,X)=0$. (Here, $r_N$
is $o\left(\left|X-Y\right|^N\right)$ for close $X$ and $Y$ and $r_N$
is smooth in $X$, $Y$.)
Hence $R_{(N)}$ does not alter the demsity $a_U$. Thus a local
change of coordinates does not alter this density.\ \ \ $\Box$

Hence any PDO $A\in CL^\alpha(M,E)$ of a noninteger order
$\alpha\in\wC\setminus\wZ$ defines a canonical smooth density $a(x)$
on $M$. We call $a(x)$ the {\em canonical trace density} of $A$.
Integrals of such densities provide us with a linear functional
on $CL^\alpha(M,E)\ni A$ defined as
\begin{equation}
\TR(A):=\int_M a(x).
\label{K3}
\end{equation}
We call it the {\em canonical trace} of $A\in CL^\alpha(M,E)$,
$\alpha\notin\wZ$.


\begin{rem}
Let $\Re\alpha<-n$. Then a PDO $A\in CL^\alpha(M,E)$ has a continuous
Schwartz kernel $A(x,y)$  and the density $a(x)$ on $M$ coincides
with the pointwise trace $\tr A(x,x)$ of the restriction of the kernel
$A(x,y)$ to the diagonal.
\label{RB2}
\end{rem}

\begin{rem}
Let $\alpha\in\wR_-$, $\alpha<-n$, and let the principal symbol
$\sigma_{-\alpha}(x,\xi)$ of an elliptic PDO
$A\in\Ell_0^\alpha(M,E)\subset CL^\alpha(M,E)$
possess a cut $L_{(\theta)}$ of the spectral plane $\wC$.
Then the spectrum of $A$ is discrete.%
\footnote{$A$ is a compact operator in $L_2(M,E)$. Its spectrum
is discrete in $\wC\setminus 0$. The only accumulation point of this
spectrum is $0\in\wC$, \cite{Yo}.}
According to the Lidskii theorem
\cite{Li}, \cite{Kr}, \cite{ReS}, XIII.17, (177), \cite{Si}, Chapter~3,
\cite{LP}, \cite{Re}, XI,  the operator $A$ is of trace class and
we have
$$
\Tr A=\int\tr A(x,x).
$$
Hence in this case we have
$$
\TR(A)=\Tr A.
$$
\label{RB3}
\end{rem}

\begin{rem}
Let $A\in\Ell_0^\alpha(M,E)\subset CL^\alpha(M,E)$ be an elliptic PDO
of order $\alpha\in\wR_+$ and let
its principal symbol $\sigma_{\alpha}(A)(x,\xi)$ possess
a cut $L_{(\theta)}$ of the spectral plane. Then the holomorphic
family $A_{(\ttheta)}^{-s}$ is defined. The operator $A_{(\ttheta)}^{-s}$
is of trace class for $\Re s\cdot\alpha>n$. For $s\in\wC$ such that
$s\cdot\alpha\notin\wZ$ and that $\Re s\cdot\alpha>n$ we have%
\footnote{For such $\Re s$ an operator family $A_{(\theta)}^{-s}$
is defined by the integral (\ref{A1}) with an admissible for $A$ cut
$L_{(\ttheta)}$ close to $L_{(\theta)}$. Note that
$A_{(\ttheta)}^0=\Id-P_0(A)$, where $P_0(A)$ is the projection operator
on algebraic eigenspace of $A$ corresponding to the eigenvalue
$\lambda=0$ and $P_0(A)$ is the zero operator on algebraic eigenspaces
of $A$ for nonzero eigenvalues.}
\begin{equation}
\zeta_{A,(\ttheta)}(s):=\Tr\left(A_{(\ttheta)}^{-s}\right)=\TR\left(A
_{(\ttheta)}^{-s}\right).
\label{K4}
\end{equation}
This zeta-function has a meromorphic continuation to the whole complex
plane. (This assertion also follows from Proposition~\ref{PB3755} below.)
The kernel $A_{(\ttheta)}^{-s}(x,y)$ for $x\ne y$ also has
a meromorphic continuation. Homogeneous terms of the symbol
$\sigma\left(A_{(\ttheta)}^{-s}\right)(x,\xi)$
of $A_{(\ttheta)}^{-s}\in\Ell_0^{-\alpha s}(M,E)\subset CL^{-\alpha s}(M,E)$
in any local coordinate chart $U$ on $M$ are holomorphic in $s$.
Hence by the definition of $\TR$ and by (\ref{K4}), the equality
\begin{equation}
\zeta_{A,(\ttheta)}(s)=\TR\left(A_{(\ttheta)}^{-s}\right)
\label{K5}
\end{equation}
holds for all $s\in\wC$ such that $s\cdot\alpha\notin\wZ$.
\label{RB4}
\end{rem}

\medskip
Here we use the weakest properties of a holomorphic (local) in $z$
family of PDOs it has to possess.

{\bf Definition}. A (local) family $A(z)\in CL^{f(z)}(M,E)$
is called a {\em $w$-holomorphic} family, if in an arbitrary local
coordinate chart%
\footnote{It is enough to check these conditions for a fixed finite
cover of $M$ by coordinate charts.}
$U\ni x,y$ and for any sufficiently large $N\in\wZ_+$ the difference
of the Schwartz kernel for $A(z)$ and a kernel corresponding
to a truncated symbol of $A(z)$
\begin{equation}
A_{x,y}(z)-\sum_{j=0}^N\int\rho(|\xi|)|\xi|^{f(z)-j}a_{-j}(z,x,\xi/|\xi|)\exp
(i(x-y,\xi))d\xi
\label{B3797}
\end{equation}
is a $C^m$-smooth (local) kernel on $U\times U$, where $m=m(N)$ tends
to infinity
as $N\to\infty$ and this kernel on $U\times U$ is holomorphic in $z$
together with its partial derivatives in $(x,y)$ of orders not greater
than $m(N)$. Here, $\rho(t)$ is a cutting $C^{\infty}$-function,
$\rho(t)\equiv 0$ for $0\le t\le 1/2$, $\rho(t)\equiv 1$ for $t\ge 1$,
$f(z)$ is (locally) holomorphic in $z$, and $a_{-j}(z,x,\xi/|\xi|)$ are
holomorphic in $z$ functions on $S^*M|_U$ with the values in densities
at $x$. The kernel $A_{x,y}(z)$ has to be holomorphic in $z$ for $x$, $y$
from disjoint local charts $U\ni x$, $V\ni y$,
$\overline{U}\cap\overline{V}=\emptyset$.

\begin{pro}
For classical PDOs $A$, $B$ such that $\ord A+\ord B\notin\wZ$
the equality holds
\begin{equation}
\TR(AB)=\TR(BA),
\label{B3731}
\end{equation}
i.e.,
$$
\TR([A,B])=0.
$$
\label{PB3730}
\end{pro}

\begin{rem}
The equality $\TR([A,B])=0$ means that $\TR$ is a trace class functional.
Note that the bracket $[A,B]$ is defined for classical PDOs $A$, $B$
having arbitrary orders. However the equality $\TR([A,B])=0$ is valid
only if $\ord A+\ord B\notin\wZ$ (otherwise the $\TR$-functional
is not defined).
\label{RB3734}
\end{rem}

\noindent{\bf Proof of Proposition~\ref{PB3730}.}
0. We assume that $\ord A,\ord B\in\wR$. The general case follows
by the analytic continuation (Proposition~\ref{PB3755}).

1. It is enough to prove the equality (\ref{B3731}) in the case when
$A$ is an elliptic PDO such that $\log A$ exists and such that
$\exp(z\log A)=:A^z$ is a trace class operator for $z$ from a domain
$U\subset\wC$. Indeed, any $A\in CL^\alpha(M,E)$ is the difference
$A=A_1-A_2$ such that $\ord A_j=\ord A+N$, $N\in\wZ_+$,
$\ord A_j>0$, $A_j\in\Ell_0^{\ord A+N}(M,E)$, and auch
that $\log A_1$, $\log A_2$ exist. It is enough to set
\begin{equation}
A_1=\left(\Delta_M^E+c\Id\right)_{(\pi)}^{\alpha/2+N}+A, \quad A_2=\left(
\Delta_M^E+c\Id\right)_{(\pi)}^{\alpha/2+N}.
\label{B3732}
\end{equation}

Here, $c\in\wR_+$ and $N\in\wZ_+$ are sufficiently large constants,
$\alpha:=\ord A$, and $\Delta_M^E$ is the Laplacian for $(M,E)$
corresponding to a Riemannian metric on $M$ and a unitary connection
for an Hermitian structure on $E$. Operators defined by (\ref{B3732})
are invertible and possess complex powers (for sufficiently large
$c$ and $N$).

Let $A$ be the difference $A_1-A_2$, where $A_1$, $A_2$ possess
complex powers. Suppose we can prove that
$\TR\left(A_jB\right)=\TR\left(BA_j\right)$. Then $\TR(AB)=\TR(BA)$.

2. Let $A$ be an invertible elliptic operator with $\ord A>0$ and
such that complex powers $A^z$ are defined. Let us prove that in this case
the equality (\ref{B3731}) holds (under the condition $\ord A+\ord B\ne 0$).
Let $s_0\gg 1$ be so large that for $\Re s>s_0$ the elliptic operators
$A^{(1-s)/2}$ and $A^{(1-s)/2}B$ are of trace class.
By Remark~\ref{RB3} for such $s$ we have
\begin{multline}
\TR\left(A^{-s}AB\right)=\Tr\left(A^{-s}AB\right)=\Tr\left(A^{(1-s)/2}B
A^{(1-s)/2}\right)=\\
=\Tr\left(BAA^{-s}\right)=\TR\left(BAA^{-s}\right).
\label{B3735}
\end{multline}
(Here we use the fact that $A^{(1-s)/2}$ is of trace class and that
$A^{(1-s)/2}B$ is bounded.)

Let $q_{(s)}:=\sum_{j\in\zuo}q_{(1-s)\alpha+\beta-j}$ and
$r_{(s)}:=\sum_{j\in\zuo}r_{(1-s)\alpha+\beta-j}$, $\alpha=\ord A$,
$\beta=\ord B$, be the symbols of $A^{1-s}B$ and of $BA^{1-s}$.
Note that for $\Re s>s_0$ the canonical trace density of $A$ is also equal
to the restriction to the diagonal $M$ (in $M\times M$) of the Schwartz
kernels corresponding to $A^{1-s}B$ minus the local kernel for a finite
sum of homogeneous terms in $q_{(s)}$ corresponding to $j=0,\dots,N$
(where $N\in\wZ_+$ is large enough).
This difference of the kernels is sufficiently smooth near the diagonal.
But it is continuous on the diagonal and holomorphic in $s$ also
for $\alpha s>n+1-N+\alpha+\beta$, $(1-s)\alpha+\beta\notin\wZ$,
$n:=\dim M$. So this difference is regular near the diagonal
for $s$ close to zero, if $N\in\wZ_+$ is large enough. The analogous
assertions are true also for $BA^{1-s}$ and for $r_{(s)}$ when
we take the difference with kernels corresponding to a sufficiently much
number of the first homogeneous terms in the symbols $r_{(s)}$.
But the canonical traces $\TR\left(A^{1-s}B\right)$ and
$\TR\left(BA^{1-s}\right)$ do not change for $\Re s>s_0$ when we subtract
these (positive homogeneous in $y-x$) kernels.
(Indeed, the kernels we subtract do not change the canonical trace
densities on $M$ defining the functional $\TR$.) Hence we can set $s=0$
in the equality (\ref{B3735}) as $\alpha+\beta\notin\wZ$.
So $\TR(AB)=\TR(BA)$.\ \ \ $\Box$

The use of complex powers of PDOs in the proof above looks a bit
artificial. The direct proof using only the language of distributions
also is possible but we do not give it here.

\begin{rem}
Note that families $A^{-s}$ and $A^{-s}B$ for an elliptic PDO $A$,
$\ord A\in\wR_+$, possessing complex powers and for a classical PDO $B$,
are holomorphic in $s$ families of PDOs. So the assertion (used
in (\ref{B3735})) that $\TR\left(A^{-s}B\right)$ is holomorphic in $s$
(for $-s\ord A+\ord B\notin\wZ$) can also be deduced
from Proposition~\ref{PB3755} below.
\label{RB3795}
\end{rem}

\begin{pro}
The traces of a classical elliptic PDO $A\in CL^\alpha(M,E)$,
$\alpha\notin\wZ$, and of its transpose
$^tA\in CL^\alpha\left(M,E^{\vee}\right)$ coincide
\begin{equation}
\TR(A)=\TR\left(^tA\right).
\label{B3838}
\end{equation}
(Here, $E^{\vee}$ is the tensor product of a fiber-wise dual to $E$
vector bundle and a line bundle of densities on $M$.)
\label{PB3833}
\end{pro}

\noindent{\bf Proof.} Let $A(z)$ be a holomorphic family of classical
PDOs such that $A(\alpha)=A$, $\ord A(z)\equiv z$. Then for $\Re z<-\dim M$
we have
\begin{equation}
\TR(A(z))=\Tr(A(z))=\Tr\left(^tA(z)\right)=\TR\left(^tA(z)\right).
\label{B3834}
\end{equation}

Proposition~\ref{PB3755} below claims that $\TR(A(z))$ and
$\Tr\left(^tA(z)\right)$ are holomorphic in $z$
for $z\notin\{m\in\wZ,m\ge\dim M\}$. (Here, we use that $^tA(z)$
is a holomorphic family.) So using the analytic continuation
of (\ref{B3834}), we obtain
$$
\TR(A)=\TR\left(A\left(z_0\right)\right)=\TR\left(^tA\left(z_0\right)\right)=
\TR\left(^tA(z)\right).
$$
To produce an analytic family $A(z)$, it is enough to set
$A(z):=AC^{z-\alpha}$, where $C\in\Ell_0^1(M,E)$ is an elliptic PDO
possessing complex powers.\ \ \ $\Box$

\begin{pro}
Let $A\!(\!z\!)$ be a holomorphic in $z$ family of classical PDOs,
$\ord\! A\!(\!z\!)\!=z$, where $z$ is from an open domain $U\subset\wC$.
Then $\TR(A(z))$ is a meromorphic in $z$ function regular
for $z\in U\setminus\wZ$. This function has no more than simple poles
at the points $\wZ\cap U$. Its residue at $m\in\wZ\cap U$ is given by
\begin{equation}
\Res_{z=m}\TR(A(z))=-\res A(m).
\label{B3756}
\end{equation}
(Here, $\res$ is the noncommutative residue, \cite{Wo2}, \cite{Kas}.)
For $m<-\dim M$ this function is regular at $m\in U$ (by (\ref{B3756})).

Analogous assertions are true for holomorphic families $A(z)$ of PDOs
such that $\ord A(z)=:f(z)$ is a (locally) holomorphic function.
For $\TR(A(z))$ to be a meromorphic function with simple poles
at $f^{-1}(\{m\in\wZ,m\ge\dim M\})=:S_f$, it is necessary that
$f'\left(z_0\right)\ne 0$ for any $z_0\in S_f$. If $f(z)$ satisfies
this condition, then $\TR(A(z))$ has simple poles with the residues
\begin{equation}
\Res_{z=z_0}\TR(A(z))=-{1\over f'\left(z_0\right)}\res\sigma\left(A\left(z_0
\right)\right)
\label{B3801}
\end{equation}
for $f\left(z_0\right)\in -n+\left(\zuo\right)$, $n:=\dim M$.

The equalities analogous to (\ref{B3756}) and to (\ref{B3801}) are
valid also for the densities $a_x(z)$ and
$\res_x\sigma\left(A\left(z_0\right)\right)$ on the diagonal
$x\in M\hookrightarrow M\times M$ (corresponding to the canonical trace
$\TR(A(z))$ and to the noncommutative residue
$\res\sigma\left(A\left(z_0\right)\right)$). Namely
\begin{equation}
\Res_{z=z_0}a_x(z)=-{1\over f'\left(z_0\right)}\res_x\sigma(A(z))
\label{B3802}
\end{equation}
for $z_0\in S_f$.
\label{PB3755}
\end{pro}

\begin{rem}
1. For classical PDOs it is natural to introduce a modified
trace functional
\begin{equation}
\TR_{\cl}(A):=(\exp(2\pi i\ord A)-1)\TR(A).
\label{B3831}
\end{equation}
The additional factor in this definition does not change if $\ord A$
shifts by an integer. (Note that the order of $A$ may differ
by an integer on different components of a manifold.)

For a holomorphic family $A(z)$ of classical PDOs this trace functional
is holomorphic for all $z$. Here, we do not suppose that
$f(z):=\ord A(z)$ has nonzero derivatives $f'\left(z_0\right)$ at
$\left\{z_0\colon f\left(z_0\right)\in\wZ,f\left(z_0\right)\ge-\dim M
\right\}$. This statement follows from the proof
of Proposition~\ref{PB3755}.

2. For classical elliptic PDOs it is natural to introduce a trace
functional
\begin{equation}
\TR_{\fell}(A):=\TR(A)/\Gamma(-(\ord A+\dim M)).
\label{B3832}
\end{equation}
For a holomorphic family $A(z)$ of elliptic PDOs this trace is holomorphic
for all $z$. (The proof of Proposition~\ref{PB3755} gives us such
a statement. Here we do not suppose that $f'\left(z_0\right)\ne 0$
for $f\left(z_0\right)\in\wZ$, $f\left(z_0\right)\ge-\dim M$.)
\label{RB3830}
\end{rem}

\begin{rem}
The assertion that the noncommutative residue $\res$ is a trace linear
functional on the algebra $CS_{\wZ}(M,E)$ of integer orders classical
PDO-symbols follows immediately from Propositions~\ref{PB3730}, (\ref{B3731}),
and \ref{PB3755}, (\ref{B3756}). Indeed, by (\ref{B3756}) and
by (\ref{B3731}) we have
$$
\res([A,B])=-\Res_{z=\ord A+\ord B}\TR([A(z),B(z)])=0
$$
for $A,B\in CL^{\wZ}(M,E)$ and for any holomorphic families $A(z)$,
$B(z)$ such that $A(\ord A\!+\!\ord B)\!=\!A$, $B(\ord A\!+\!\ord B)\!=\!B$.
(For example, set $A(z)\!:=\!AC^{z\!-\!\ord A\!-\!\ord B}$,
$B(z)=BC^{z-\ord A-\ord B}$,
where $C$ is an invertible first order elliptic PDO possessing complex
powers).
\label{RB3796}
\end{rem}

\noindent{\bf Proof of Proposition~\ref{PB3755}.}
The positive homogeneous in $\xi$ terms of the symbol $\sigma(A(z))$
correspond to the positive homogeneous in $y-x$ (local) summands
of the Schwartz kernel for $A(z)$. Namely (in the notations
of (\ref{B3797})) the integral
$$
\int|\xi|^{f(z)-j}a_{-j}(z,x,\xi/|\xi|)\exp(i(x-y,\xi))d\xi
$$
defined in Remark~\ref{RB3803} is the positive homogeneous of degree
$-n-f(z)+j$ in $y-x$ (local) kernel.
These kernels do not alter $\TR(A(z))$ for $f(z)\notin\wZ$ (and also
for $f(z)\in\wZ$, if $f(z)<n-j$).
Note that the kernel (\ref{B3797}) is smooth enough near the diagonal
in $U\times U$ and is locally holomorphic in $z$. So the canonical
trace $\TR$ and the canonical trace density of this kernel are regular
in $z$.

Therefore the singularities of $\TR(A(z))$ are defined by the restriction
to the diagonal of the kernel
\begin{equation}
\sum_{j=0}^N\int(\rho(|\xi|)-1)|\xi|^{f(z)-j}a_{-j}(z,x,\xi/|\xi|)\exp(i(x-y,
\xi))d\xi.
\label{B3798}
\end{equation}

The kernel (\ref{B3798}) is smooth in $(x,y)$ and holomorphic in $z$
for $f(z)-j+n\ne 0$ (because $\rho(|\xi|)\equiv 0$ small $|\xi|$).
Namely the integral (\ref{B3798}) is absolutely convergent
for $\Re f(z)>j-n$ (and the convergence is uniform in $(x,y,z)$
for $\Re f(z)\ge j-n+\eps$, $\eps>0$). The corresponding integral over
$|\xi|=\const$ is absolutely convergent for any $z$. For $x=y$ the integral
(\ref{B3798}) has an explicit analytic continuation. It is produced
with the help of the equality
$$
\int_0^1x^\lambda dx=1/(\lambda+1)
$$
for $\Re\lambda>-1$. The right side of this equality is meromorphic
in $\lambda\in\wC$.

We suppose from now on that $f'\left(z_0\right)\ne 0$ for $z_0$ such that
$f\left(z_0\right)+n\in\zuo$. The residue of the integral (\ref{B3798})
at $z=z_0$ such that $f\left(z_0\right)=-n+j$ is
\begin{equation}
-{1\over f'\left(z_0\right)}\int a_{-j}(z,x,\xi/|\xi|)d\mu_S,
\label{B3800}
\end{equation}
where $d\mu_S$ are natural densities on the fibers of $S^*M$. Note that
$a_{-j}(z,\!x,\!\xi/|\xi|)|\xi|^{f(\!z_0)\!-\!j}$ is positive homogeneous
in $\xi$ of degree $-n$. So we have
\begin{equation}
\Res_{z=z_0}\TR(A(z))=-{1\over f'\left(z_0\right)}\res\sigma\left(A\left(z_0
\right)\right).
\label{B3799}
\end{equation}
The proposition is proved.\ \ \ $\Box$

Let $\fell(M,E)$ be the Lie algebra of logarithms of (classical) elliptic
PDOs. As a linear space, $\fell(M,E)$ is spanned by its codimension one
linear subspace $CL^0(M,E)$ of zero order PDOs and by a logarithm
$l=\log_{(\theta)}A$ of an invertible elliptic PDO
$A\in\Ell_0^1(M,E)\subset CL^1(M,E)$ such that $A$ admits a cut
$L_{(\theta)}$ of the spectral plane $\wC$. The space
$\left\{sl+B_0\right\}$, $s\in\wC$, $B_0\in CL^0(M,E)$, of logarithms
of elliptic PDOs is independent of $l$. This space has a natural structure
of a Fr\'echet linear space over $\wC$. The Lie bracket
on $\fell(M,E)$ is defined by
\begin{equation}
\left[s_1l+b_1,s_2l+b_2\right]:=\left[l,s_1b_2-s_2b_1\right]+\left[b_1,b_2
\right]\in CL^0(M,E)\subset\fell(M,E).
\label{B3820}
\end{equation}
The bracket $\left[l,s_1b_2-s_2b_1\right]$ is a classical zero order
PDO because
$$
\left[l,s_1b_2-s_2b_1\right]=\df_t\left(A_{(\theta)}^t\left(s_1b_2-s_2b_1
\right)A_{(\theta)}^{-t}\right)|_{t=0}\in CL^0(M,E).
$$
(Here, $A_{(\theta)}^t:=\exp(tl)$ is a holomorphic in $t$ family
with the generator $l$. The inclusion $[l,b]\in CL^0(M,E)$
for $b\in CL^0(M,E)$ can be also deduced from (an obvious) local
inclusion of $[\log|\xi|,\sigma(b)]$ to classical zero order PDO-symbols
and of the description the corresponding Lie algebra $S_{\log}(M,E)$
in Section~\ref{SA}.)

The exponential map from $\fell(M,E)$ to the connected component
$\Ell_0(M,E)\ni\Id$ of elliptic PDOs is
\begin{equation}
sl+B_0\to\exp\left(sl+B_0\right)\in\Ell_0^s(M,E).
\label{B3821}
\end{equation}
The PDO $A_s:=\exp(sl+B_0)\in\Ell_0^s(M,E)\subset CL^s(M,E)$ is defined
as $A_s^\tau|_{\tau=1}$, where the operator $A_s^\tau$ is the solution
of the equation
\begin{gather}
\df_\tau A_s^\tau=(sl+B_0)A_s^\tau, \label{K22}\\
A_s^0:=\Id, \qquad A_0^1:=\exp(B_0). \label{K221}
\end{gather}
(Note that $A_s:=A_s^1$ depends on an element $sl+B_0\in\fell(M,E)$
only and that $A_s$ does not depend on a choice
of $l\in\log\left(\Ell_0^1(M,E)\right)$.
The solution of (\ref{K22}), (\ref{K221}) is given by the substitution
\begin{align}
A_s^\tau        & :=A^{s\tau}F_\tau, \label{K23}\\
\df_\tau F_\tau & =\left(A^{-s\tau}B_0 A^{s\tau}\right)F_\tau, \quad
F_0:=\Id. \label{K24}
\end{align}
The operator $A^{s\tau}$ in (\ref{K23}), (\ref{K24}) is defined
for $\Re(s\tau)\ll 0$ by the integral (\ref{A1}) with $z:=s\tau$.
This family is continued to $s\tau\in\wC$ by (\ref{A2}). The operator
$A^{-s\tau}B_0 A^{s\tau}$ in (\ref{K23}) is a PDO from $CL^0(M,E)$.
The operator $\exp(B_0)$ in (\ref{K221}) is defined by the integral
\begin{equation}
\exp B_0:={i\over 2\pi}\int_{\Gamma_R}\left(B_0-\lambda\right)^{-1}\exp
\lambda d\lambda,
\label{K25}
\end{equation}
where $\Gamma_R$ is a circle $|\lambda|=R$ oriented opposite
to the clockwise and surrounding $\Spec B_0$%
\footnote{The integral (\ref{K25}) is analogous to the integrals
(\ref{B7}), (\ref{X2}). We suppose here also that the principal symbol
$\sigma_0(B_0)(x,\xi)$ has all its eigenvalues inside the circle
$|\lambda|=R/2$ for all $(x,\xi)\in S^*M$.}
(Recall that this spectrum is a compact in the spectral plane $\wC$
and that the operator $\left(B_0-\lambda\right)^{-1}$ is a classical
elliptic PDO from $\Ell_0^0(M,E)\subset CL^0(M,E)$ for $\lambda\in\Gamma_R$
since $B_0\in CL^0(M,E)$.)
We have $\exp B_0\in\Ell_0^0(M,E)\subset CL^0(M,E)$.

The existence and the uniqueness of a smooth solution for such type
equations in the space of PDO-symbols is proved in Section~\ref{S8}.
So we have the solution $\sigma\left(F_\tau\right)$ of the equation
on elliptic symbols
\begin{equation}
\df_\tau\sigma\left(F_\tau\right)=\sigma\left(A^{-s\tau}B_0A^{s\tau}\right)
\sigma\left(F_\tau\right), \quad \sigma\left(F_0\right)=\Id.
\label{B3498}
\end{equation}

Let $S_\tau\in\Ell_0^0(M,E)$, $0\le\tau\le 1$, be a smooth curve
in the space of invertible elliptic operators from $S_0=\Id$ to $S_1$
with $\sigma\left(S_\tau\right)=\sigma\left(F_\tau\right)$. Then
$$
\df_\tau S_\tau=\left(\left(A^{-s\tau}B_0A^{s\tau}\right)+r_\tau\right)S_\tau,
$$
where $r_\tau$ is a smooth curve in the space $CL^{-\infty}(M,E)$
of smoothing operators (i.e., in the space of operators with smooth
kernels on $M\times M$). Set $u_\tau:=S_\tau^{-1}F_\tau-\Id$. Then
$u_\tau\in CL^{-\infty}(M,E)$ is the solution of the equation
in $CL^{-\infty}(M,E)$
\begin{equation}
\df_\tau u_\tau=-\left(S_\tau^{-1}r_\tau S_\tau\right)\left(\Id+u_\tau\right),
\quad u_0=0.
\label{B3499}
\end{equation}
This is a linear equation in the space $CL^{-\infty}(M,E)$ of smooth
kernels on $M\times M$ with known smooth in $I\times M\times M$
coefficients $S_\tau^{-1}r_\tau S_\tau\in CL^{-\infty}(M,E)$. (This
equation can be solved by using the Picard approximations.)

\begin{pro}
The exponential map (\ref{B3821}) is $w$-holomorphic,
i.e., for any (local) holomorphic map
$\phi\colon\left(\wC^N,0\right)\ni q\to s(q)l+B_0(q)\in\fell(M,E)$,
the family $\exp(\phi(q))\in\Ell_0^{s(q)}(M,E)$ is $w$-holomorphic.
\label{PB3822}
\end{pro}


The function
$$
\TR\left(\exp(sl+B_0)\right)=:T(s,B_0)
$$
is defined for any $s\in\wC\setminus\wZ$ and for any $B_0\in CL^0(M,E)$.
Note that $T(\phi(q))$ is meromorphic in $q$ with poles
at $\{q\colon s(q)\in\wZ,s(q)\ge-\dim M\}$ by Propositions~\ref{PB3755},
\ref{PB3822}.

\begin{pro}
The function $T(s,B_0)$ is meromorphic in $\left(s,B_0\right)$%
\footnote{That means that the function is meromorphic in $\left(s,B_0\right)$
on any finite-dimensional linear (or affine) subspace in $\fell(M,E)$.
In the independent of coordinates $\left(s,B_0\right)$ form
this theorem claims that the function $T$ is meromorphic near the origin
on the space of logarithms for elliptic PDOs and that $T$ has a simple
pole along the codimension one linear submanifolds of integer orders PDOs.
The residues of $T$ are given by (\ref{K6}).}
and has simple poles at the hyperplanes $s\in\wZ$, $s\ge-\dim M$. We have
\begin{equation}
\Res_{s=m}T(s,B_0)=-\res\sigma\left(\exp\left(ml+B_0\right)\right).
\label{K6}
\end{equation}
Here, $m\in\wZ$, $m\ge-\dim M$, and $\res$ is the noncommutative residue
(\cite{Wo2}).
\label{PB5}
\end{pro}

\noindent{\bf Proof.} This assertion is an immediate consequence
of Propositions~\ref{PB3822}, \ref{PB3755}.\ \ \ $\Box$

\begin{pro}
The product $A(z\!)\!B(z\!)$ of $w$-holomorphic families is $w$-holomorphic.
\label{PB3824}
\end{pro}

\noindent{\bf Proof of Proposition~\ref{PB3822}.} We can solve the equation
for symbols $\sigma\left(A_s^t\left(B_0\right)\right)$
of $A_s^t\left(B_0\right):=\exp\left(t\left(sl+B_0\right)\right)$
\begin{equation}
\df_t\sigma\left(A_s^t\left(B_0\right)\right)=\sigma\left(sl+zB_0\right)
\sigma\left(A_s^t\left(B_0\right)\right),\;\sigma\left(A_s^0\left(B_0\right)
\right)=\Id
\label{B4040}
\end{equation}
for any $\left(s,z,B_0\right)$. These symbols are holomorphic
in $\left(s,\sigma(l),z,\sigma\left(B_0\right)\right)$. (Here we use
the substitution (\ref{K23}), (\ref{K24}) but on the level of PDO-symbols.
So we don't have to inverse elliptic PDOs in solving of (\ref{B4040}).
This equation is solved above, (\ref{B3498}).)

Let $\left\{U_i\right\}$ be a finite cover of $M$ by coordinate charts,
$\phi_i$ be a smooth partition of unity subordinate to $\left\{U_i\right\}$,
and let $\psi_i\in C_0^{\infty}\left(U_i\right)$, $\psi_i\equiv 1$
on $\supp\left(\phi_i\right)$. These data define a map $f_N$
from PDO-symbols to PDOs on $M$. The difference
$A_s^t\left(B_0\right)-f_N\left(\sigma\left(A_s^t\left(B_0\right)\right)
\right)$ has a smooth enough kernel on $M\times M$ (for $N\in\wZ_+$
large enough), and this kernel $K_N^{s,t}(x,y)$ is a solution
of a linear equation with the right side smooth enough and holomorphic
in $s,B_0$ (for $s$ close to a given $s_0\in\wC$). Here, we choose $N$
depending on $s_0$. So the kernel $K_N^{(s,B_0)(q)}(x,y):=K_N^{s,t}(x,y)$
is sufficiently smooth on $M\times M$ and holomorphic in $q$ close
to $q_0$, $s\left(q_0\right)=s_0$. The PDO
$f_N\left(\sigma\left(A_s^1\left(B_0\right)\right)\right)$ is
$w$-holomorphic in $s$, $B_0$ by its definition. So
$A_s\left(B_0\right):=\exp\left(sl+B_0\right)$ is a $w$-holomorphic
in $s$, $B_0$.\ \ \ $\Box$

An alternative proof of this proposition is as follows (it uses
Proposition~\ref{PB3824}).

1. First prove that
$A_s\left(B_0\right):=\exp\left(sl+B_0\right)$ is holomorphic in $B_0$.
To prove the analyticity in $B_0$ for any fixed
$s\in\wC\setminus\wZ$, it is enough to prove that
\begin{equation}
\left\{\df_z\left(\exp(ls+zB_0)\right)\right\}\big|_{z=0}
\label{K9}
\end{equation}
exists and that we have
\begin{equation}
\left\{\df_{\bar z}\left(\exp(ls+zB_0)\right)\right\}\big|_{z=0}=0.
\label{K10}
\end{equation}

Indeed, to prove the same assertions for $z\ne 0$, we can change
the logarithm $l$ of an elliptic operator $A$ of order one
to the logarithm
$$
l_1:=l+s^{-1}zB_0
$$
of another elliptic PDO $A_1\in\Ell_0^1(M,E)\subset CL^1(M,E)$. (Note
that the principal symbols of $A$ and of $A_1$ are the same.)

By the Duhamel principle, we have%
\footnote{Note that for any $\tau$
$$
\exp(\tau(ls+zB_0))\exp((1-\tau)ls)\in\Ell_0^s(M,E), \quad
\exp(\tau(ls+zB_0))zB_0\exp((1-\tau)ls)\in CL^s(M,E).
$$}
\begin{multline}
\exp(ls+zB_0)-\exp ls=\int_0^1 d\tau\df_\tau\left(\exp(\tau(ls+zB_0))
\exp((1-\tau)ls)\right)=\\
=\int_0^1 d\tau\exp(\tau(ls+zB_0))zB_0\exp((1-\tau)ls).
\label{K11}
\end{multline}
We conclude from (\ref{K11}) that
\begin{gather}
\df_z\left\{\left(\exp(ls+zB_0)\right)\right\}\big|_{z=0}
=\int_0^1\!d\tau\exp(\tau ls)B_0\exp((1\!-\!\tau)ls),
\label{K12}\\
\df_{\bar z}\left\{\left(\exp(ls+zB_0)\right)\right\}\big|_{z=0}
=0. \label{K15}
\end{gather}
To deduce (\!\ref{K12}),\;(\ref{K15}\!) from (\ref{K11}),
note that the equation
for $A_s^\tau\!(z)\!\!:=\!\exp\!\left(\!\tau(\!sl\!+\!zB_0\!)\!\right)$
\begin{gather}
\df_\tau A_s^\tau(z)=(sl+zB_0)A_s^\tau(z), \label{K27}\\
A_s^0(z)=\Id, \quad A_0^\tau(z)=\exp(\tau z B_0), \quad A_s^\tau(0)=A^{s\tau}
\label{K28}
\end{gather}
is solved by the substitution
\begin{gather}
A_s^\tau(z)=A^{s\tau}F_\tau(z), \label{K29}\\
\df_\tau F_\tau(z)=z\left(A^{-s\tau}B_0 A^{s\tau}\right)F_\tau(z),
\label{K30}\\
F_0(z)=\Id, \quad F_\tau(0)=\Id. \label{K31}
\end{gather}
Hence for $\df_z F_\tau(z)|_{z=0}=:Q_s(\tau)$ we have
\begin{equation}
\df_\tau Q_s(\tau)=A^{-s\tau}B_0 A^{s\tau}, \qquad Q_s(0)=0,
\label{K32}
\end{equation}
and $Q_s(\tau)\in CL^0(M,E)$ depends smoothly on $\tau$. Thus we have
\begin{gather}
\exp(sl+zB_0)-\exp(sl)          =\!\int_0^1\!d\tau A^{s\tau}\left(\Id\!+
\!zQ_s(\tau)\!+\!o(z)\right)zB_0 A^{s(1\!-\!\tau)}, \label{K33}\\
\df_z\exp(sl+zB_0)\big|_{z=0}   =\int_0^1 d\tau A^{s\tau}B_0 A^{s(1-\tau)}.
\label{K34}
\end{gather}
Here $o(z)$ is considered with respect to a Fr\'echet structure
on $CL^0(M,E)$. This structure is defined by natural semi-norms
(\ref{B3233}) on $CS^0(M,E)$ (with respect to a finite cover
$\left\{U_i\right\}$ of $M$) and by natural semi-norms on the kernels
of $A-f_N\sigma(A)\in C^k(M\times M)$ for appropriate $k\in\zuo$,
$N\in\wZ_+$ are large enough.%
\footnote{We use the notations of the first proof of this proposition.}

\medskip
The expression on the left in (\ref{K34}) is the derivative
of the function with its values in $CL^s(M,E)$ and the operator
on the right in (\ref{K34}) is also from $CL^s(M,E)$.
%
%
%

\medskip
2. The family of elliptic PDOs
\begin{equation}
A_\mu\left(B_0\right):=\exp(\mu l+B_0)\in\Ell_0^\mu(M,E)\subset CL^\mu(M,E)
\label{K66}
\end{equation}
is $w$-holomorphic in $\mu$. Indeed, set $l_\gamma:=l+B_0/\gamma$.
Then $A_{\mu,\gamma}:=\exp\left(\mu l_\gamma\right)$ is holomorphic
in $\mu$, $\gamma$ for $\gamma\ne 0$. We have also
$A_\mu=A_{\mu,\gamma}|_{\gamma=\mu}$.

Thus it is enough to prove that $A_\mu$ is $w$-holomorphic in $\mu$
at $\mu=0$.
Set $A_\mu^\tau(z):=\exp\left(\tau\left(\mu l+zB_0\right)\right)$,
$F_\tau(\mu,z):=\exp(-\tau\mu l)A_\mu^\tau(z)$. By (\ref{K28}),
(\ref{K29}) we have
\begin{equation}
\df_\tau F_\tau(\mu,z)=z\left(\Ad_{\exp(-\mu\tau l)}B_0\right)F_\tau(\mu,z),
\quad F_0(\mu,z)\equiv\Id\equiv F_\tau(\mu,0).
\label{B3510}
\end{equation}
We have to prove that the family $F_\tau(\mu,1)|_{\tau=1}$ is $w$-holomorphic
in $\mu$ at $\mu=0$. The coefficient
$z\Ad_{\exp(-\mu\tau l)}B_0=:zv(\mu\tau)\in CL^0(M,E)$ in (\ref{B3510})
is holomorphic in $\mu\tau$. Set $\df_\mu F_\tau(\mu,z)=f_\tau(\mu,z)$.
(We know that $f_\tau$ exists, if $\mu\ne 0$.) Then
\begin{equation}
\df_\tau f_\tau(\mu,z)\!=\!zv(\mu\tau)f_\tau(\mu,z)\!+\!z\df_\mu v(\mu\tau)F
_\tau(\mu,z), \quad f_\tau(\mu,0)\!=\!0\!=\!f_0(\mu,z).
\label{B3511}
\end{equation}
Let us substitute $\mu=0$ to the right side of this equation. Then
(\ref{B3511}) takes the form
\begin{equation}
\df_\tau f_\tau(0,z)=zB_0f_\tau(0,z)-z\tau\left[l,B_0\right]\exp\left(\tau z
B_0\right), \quad f_0(0,z)=0.
\label{B3512}
\end{equation}
This is a linear equation in $CL^0(M,E)$ and it has a unique solution.
(The analogous assertion is proved in Section~\ref{S8}.)
So $f_1(0,z):=\df_\mu F_1(\mu,z)|_{\mu=0}$ exists. Hence the family
$F_1(\mu,z)$ is holomorphic in $\mu$. (In particular, it is holomorphic
in $\mu$ for $z=1$.)\ \ \ $\Box$

\begin{rem}
The holomorphic in $\mu$ dependence of $A_\mu\left(B_0\right)$
in the sense of \cite{Gu} (Section~3, (3.17), (3.18)) means
that the image of $\sigma\left(A_\mu\right)$ in $CS^\mu/CS^{\mu-N}$
is holomorphic in $\mu$ (for $N\in\wZ_+$) and that for any $\mu\in\wC$
and for any $m\in\wZ_+$ there exists a linear operator
$\dot{A}_m(\mu)\colon\Gamma(E)\to\Gamma(E)$ such that the asymptotics
(analogous to (\ref{B230})) hold%
\footnote{With respect to the semi-norms
$\left\|\cdot\right\|_{\cA}^{(s)}$ from (\ref{B231}), (\ref{B230}).}
for $s>\Re\mu-m$ as $|\delta|\to 0$
\begin{equation}
\left\|\left(A_m(\mu+\delta)-A_m(\mu)\right)/\delta-\dot{A}_m(\mu)\right\|
_{\cA}^{(s)}\to 0
\label{B1550}
\end{equation}
Here, $A_m(\mu):=A_\mu-P_mA_\mu$ (where $P_mA_\mu\in CL^\mu(M,E)$ is
the PDO defined by the image of $\sigma\left(A_\mu\right)$
in $CS^\mu(M,E)/CS^{\mu-m-1}(M,E)$ and by a fixed partition of unity
subordinate to a finite cover of $M$ by local charts).

This assertion follows from the Cauchy integral representation
for $A_m(\mu)$ analogous to (\ref{B1500}). The Cauchy integral
representation for $A_m(\mu)$ (with $0<r<|\mu|$ in (\ref{B1552}),
(\ref{B1555}) below) can be deduced from Proposition~\ref{PB3822}.
This integral representation implies the expression of
the operator $\dot{A}_m(\mu)$ in (\ref{B1550})
as of the Cauchy integral of linear operators in $\Gamma(E)$
\begin{equation}
\dot{A}_m(\mu)={1\over 2\pi i}\int_{\Gamma_r(\mu)}{A_m(z)\over(z-\mu)^2}dz.
\label{B1552}
\end{equation}
Here, $\Gamma_r(\mu)$ is the contour $\{z\colon|z-\mu|=r\}$ oriented
opposite to the clockwise.
(This integral is analogous to (\ref{B1500}).)

The Cauchy integral formulas for $A_m(\mu)$ and $\dot{A}_m(\mu)$
hold and these
integrals are convergent with respect to the operator semi-norms
$\left\|\cdot\right\|_{\cA}^{(s)}$ for $s>r+\Re\mu-m$.
(The same assertion is true for any smooth simple contour $\Gamma$
surrounding once the point $\mu$ and belonging
to $D_r(\mu):=\{z\in\wC\colon|z-\mu|\le r\}$.)

Set $A_m^2(\mu,\delta):=\left(A_m(\mu+\delta)-A_m(\mu)\right)/\delta-\dot{A}
_m(\mu)$. Then for $r>|\delta|$ we have
\begin{equation}
A_m^2(\mu,\delta)
={1\over 2\pi i}\int_{\Gamma_r(\mu)}\left({1\over (z-\mu)(z-\mu-\delta)}-
{1\over (z-\mu)^2}\right)A_m(z)dz.
\label{B1555}
\end{equation}
This integral converges with respect to the operator semi-norms
$\left\|\cdot\right\|_{\cA}^{(s)}$ for $s>r+\Re\mu-m$. Its semi-norm
$\left\|\cdot\right\|_{\cA}^{(s)}$ (for any $s>r-m$) is $O(|\delta|)$
as $|\delta|\to 0$.
%
%

The convergence of these integrals in appropriate semi-norms
$\left\|\cdot\right\|_{\cal{A}}^{(s)}$
is a consequence of a holomorphic in $\mu$ dependence of $F_1(\mu,1)$
(defined by (\ref{B3510})).
\label{RB3839}
\end{rem}

\noindent{\bf Proof of Proposition~\ref{PB3824}.}
Let $A(z)\in CL^{f(z)}(M,E)$ and $B\in CL^{g(z)}(M,E)$ be $w$-holomorphic
families ($f(z)$ and $g(z)$ are holomorphic in $U\subset\wC$).
In the notations of the proof of Proposition~\ref{PB3822} the kernels
of $A(z)-f_N\sigma_N(A(z))=:r_NA(z)$%
\footnote{Here, $\sigma_N(A)$ is the image of $A$ in
$$
CL^{\ord A}(M,E)/CL^{\ord A-N-1}(M,E)=CS^{\ord A}(M,E)/CS^{\ord A-N-1}(M,E)
$$.}
and of $r_NB(z)$ are holomorphic for $z$ close to $z_0$ and are
sufficiently smooth on $M\times M$ for such $z$.

The product $f_NA(z)\cdot f_NB(z)$ is $w$-holomorphic by the standard
proof of the composition formula for classical PDOs (see for example,
\cite{Sh}, 3.6, the proof of Theorem~3.4). For $N\in\wZ_+$ large enough
(depending on $z_0\in U$) and for $z$ close to $z_0$ the kernels
of $r_NA(z)\cdot f_NB(z)$, $f_NA(z)\cdot r_NB(z)$,
$r_NA(z)\cdot r_NB(z)$ are sufficiently smooth on $M\times M$ and
holomorphic in $z$.\ \ \ $\Box$

\subsection{Derivatives of zeta-functions at zero as homogeneous
polynomials on the space of logarithms for elliptic operators}
For an element $J=\alpha l+B_0$, $\alpha\in\wC^{\times}$, of $\fell(M,E)$
the {\em $\TR$-zeta-function}
of $\exp J\in\Ell_0^\alpha(M,E)\subset CL^\alpha(M,E)$ is defined
for $s\in\wC$,
$\alpha s\notin\wZ$ by
\begin{equation}
\zeta_{\exp J}^{\TR}(s):=\TR(\exp(-sJ)).
\label{K40}
\end{equation}
%
%

Let $\alpha\in\wR^{\times}$ and let for $A\in\Ell_0^\alpha(M,E)$
its complex powers $A_{(\theta)}^s$ be defined. (Here, $\theta$ is
fixed and the cut $L_{(\theta)}$ of the spectral plane has to be
admissible for $A$.)
Let $\df_sA_{(\theta)}^s|_{s+0}=J$ (i.e., for any $C^\infty$-section
$f\in\Gamma(E)$ we have $\df_s\left(A_{(\theta)}^sf\right)|_{s=0}=Jf$).
This equality can be written as
\begin{equation}
\log_{(\theta)}A=J.
\label{B3843}
\end{equation}
By Remarks~\ref{RB3} and \ref{RB4} the $\TR$-zeta-function of $A$
coincides for $\alpha\Re s<-\dim M$ with the classical $\zeta$-function
\begin{equation}
\zeta_{\exp J}^{\TR}(s)=\zeta_{\exp J,(\theta)}(s).
\label{B3844}
\end{equation}
By Propositions~\ref{PB3833}, \ref{PB3822} the $\TR$-zeta-function
$\zeta_{\exp J}^{\TR}(s)$ is meromorphic in $s$ with no more than
simple poles at $s\in\wZ$, $s\le\dim M$.
Hence (\ref{B3844}) holds everywhere.
By (\ref{K6}) we have
$$
\Res_{s=0}\zeta_{\exp J}^{\TR}(s)=\res\Id=0.
$$
So $\zeta_{\exp J,(\theta)}^{\TR}(s)$ (and $\zeta_{\exp J}^{\TR}(s)$)
are regular at $s=0$. Hence the derivatives at $s=0$ are defined
$$
\zeta_{\exp J}^{(k)}(0):=\df_s^k\zeta_{\exp J}(s)\big|_{s=0}:=\df_s^k
\zeta_{\exp J}^{\TR}(s)\big|_{s=0}.
$$


Our definition of the $\TR$-function differs from the usual one
in two aspects. Firstly we consider it as a function depending
on a logarithm of an elliptic operator and not on an operator
with an admissible cut. Secondly, the order $\alpha$ should not
be real.

\begin{rem}
The main difference between a $\TR$-zeta-function and a classical one
is that we do not use an analytic continuation of the $\TR$-zeta-function
in its definition. This function $\zeta_A^{\TR}(s):=\TR\left(A^{-s}\right)$
is canonically defined at any point $s_0$ such that $s_0\ord A\notin\wZ$.
This definition uses a family $A^{-s}$ of complex powers of a nonzero
order elliptic PDO $A$. However, if we know a PDO $A^{-s_0}$,
$s_0\ord A\notin\wZ$, then we know $\zeta_A^{\TR}\left(s_0\right)$.
For example, in the classical definition of zeta-functions it was not
clear, if the equality holds
\begin{equation}
\zeta_A\left(s_0\right)=\zeta_B\left(s_1\right),
\label{B4015}
\end{equation}
where $A^{-s_0}=B^{-s_1}$, $s_0\ord A=s_1\ord B\notin\wZ$, for nonzero
orders elliptic PDOs $A$, $B$ with existing complex powers $A^{-s}$,
$B^{-s}$. For $\TR$-zeta-functions the equality (\ref{B4015}) follows
from their definitions. These zeta-functions coincide with the classical
ones for $\Re(s\ord A)>\dim M$, $\Re(s\ord B)>\dim M$. Hence the equality
(\ref{B4015}) holds for classical zeta-functions also.

Note that the equality (\ref{B4015}) is not valid in general,
if $s_0\ord A=s_1\ord B\in\wZ$. If $s_0\ord A$ is an integer and
if $s_0\ord A\le\dim M$, then $\zeta_A(s)$ has a pole at $s_0$
for a general elliptic PDO $A$. If such $A$ is an elliptic DO and
if $s_0\ord A\in\zuo$, then $s_0$ is a regular point.
\label{RB4016}
\end{rem}

\begin{rem}
The existence of the complex powers of an invertible elliptic
operator $A\in\Ell_0^\alpha(M,E)\subset CL^\alpha(M,E)$
(for $\alpha\in\wC^{\times}$) is equivalent to the existence of a logarithm
$\log A$.%
\footnote{Indeed, let $A^s$ be defined. Then for $s_0$ sufficiently
close to zero and such that $s_0\alpha\in\wR_+$ the principal symbol
$\sigma\left(A^{s_0}\right)$ of the operator $A^{s_0}$ possesses
a cut along $\wR_-=L_{(\pi)}$ on the spectral plane.
So in this case, $\log_{(\tpi)}\left(A^{s_0}\right)$ is defined
for a cut $L_{(\tpi)}$ close to $L_{(\pi)}$. Thus
$\log A:=s_0^{-1}\log_{(\tpi)}A^{s_0}$ is also defined.}
This condition is not equivalent to the existence
of a spectral cut $L_{(\theta)}$ for $\sigma_\alpha(A)$. For instance,
such a cut does not exist in the case $\alpha\in\wC\setminus\wR$.
However, if such a cut $L_{(\ttheta)}$ exists
for $A_1\in\Ell_0^c(M,E)\subset CL^c(M,E)$, $c\in\wR^{\times}$, and
if $A\in\Ell_0^{\alpha c}(M,E)\subset CL^{\alpha c}(M,E)$ is equal
to $A_1^\alpha$, then $\log A$ defined as $\alpha\log_{(\ttheta)}A_1$ exists.
\label{RB3840}
\end{rem}


\begin{thm}
The function $\zeta_{\exp J}^{(k)}(0)$ on the hyperplane
$\{J\!=\!l\!+\!B_0,B_0\!\in \!CL^0\!(M,E\!)\!\}$ (where $l=\log A$ and $A$
is an elliptic operator from $\Ell_0^1(M,E)\subset CL^1(M,E)$ such that
$\log A$ exists) is the restriction
to this hyperplane of a homogeneous polynomial of order $(k+1)$
on the space $\fell(M,E):=\{J=cl+B_0,c\in\wC,B_0\in CL^0(M,E)\}$
of logarithms of elliptic operators.
\label{TB10}
\end{thm}

\noindent{\bf Proof.} According to Proposition~\ref{PB5} the function
$sT(s,B_0):=s\TR(\exp(sl+B_0))$ is equal to the sum of a convergent
near $(s_0,B_0)=(0,0)$ power series%
\footnote{This power series is uniformly convergent in $B_0$ from
a neighborhood of zero in any finite-dimensional linear subspace
of $CL^0(M,E)\ni B_0$.}
\begin{equation}
sT(s,B_0)=\sum_{m\in\zuo}s^mQ_m(B_0).
\label{K52}
\end{equation}
The functions $Q_m(B_0)$ are holomorphic near $B_0=0$ (in the same
sense as in Proposition~\ref{PB5}). Hence we have
\begin{equation}
Q_m(B_0)=\sum_{q\in\zuo}Q_{m,q}(B_0),
\label{K53}
\end{equation}
where $Q_{m,q}$ is a homogeneous polynomial of order $q$
on the linear space $CL^0(M,E)\ni B_0$.

The function $\zeta_{\exp J}^{(k)}(0)$ is expressed through
$Q_{m,q}(B_0)$ as follows. For $s\in\wC\setminus\wZ$ we have
\begin{multline}
s\TR\left(\exp(s(l+B_0))\right)=\sum_{m\in\zuo}s^m\sum_{q\in\zuo}Q_{m,q}
(sB_0)=\\
=\sum_{m,q\in\zuo}s^{m+q}Q_{m,q}(B_0).
\label{K54}
\end{multline}
Hence we have
\begin{equation}
s\sum_{k\in\zuo}\zeta_{\exp(l+B_0)}^{(k)}(0)\left(-s\right)^k/k!=
\sum_{m,q\in\zuo}s^{m+q}Q_{m,q}(B_0),
\label{K60}
\end{equation}
i.e., we have for an arbitrary $k\in\zuo$ that
\begin{equation}
\zeta_{\exp(l+B_0)}^{(k)}(0)=k!\left(-1\right)^k\sum_{m,q\in\zuo,m+q=k+1}
Q_{m,q}(B_0).
\label{K61}
\end{equation}

The function
\begin{equation}
T_{k+1}(cl+B_0):=\sum_{m,q\in\zuo,m+q=k+1}c^mQ_{m,q}(B_0)
\label{B211}
\end{equation}
(where $c\in\wC$) is a homogeneous polynomial of order $(k+1)$
on $\fell(M,E)=\{cl+B_0,B_0\in CL^0(M,E)\}$.
Hence according to (\ref{K61})
$\left(-1\right)^k(k!)^{-1}\zeta_{\exp(l+B_0)}^{(k)}(0)$
is the restriction of this homogeneous polynomial of order $(k+1)$
to the hyperplane $c=1$. The theorem is proved.\ \ \ $\Box$

\begin{pro}
$\zeta_{\exp J}^{(k)}(0)$ is a homogeneous function
on $\fell(M,E)\setminus CL^0(M,E)$ of degree $k$.
\label{PB3845}
\end{pro}

\noindent{\bf Proof.}
We have
\begin{gather*}
\zeta_{\exp\lambda J}(s)=\TR(\exp\lambda sJ)=\zeta_{\exp J}(\lambda s), \\
\df_s^k\zeta_{\exp\lambda J}(s)=\lambda^k\df_s^k\zeta_{\exp J}(s)(\lambda s).
\end{gather*}
Then substitute $s=0$.\ \ \ $\Box$

\begin{rem}
The homogeneous function of degree $k$ on $\fell(M,E)\setminus CL^0(M,E)$
defined in Proposition~\ref{PB3845} has the form
$T_{k+1}/(\ord J)\equiv T_{k+1}/\alpha$, where $T_{k+1}$ is
a homogeneous polynomial of order $k+1$ on $\fell(M,E)$ defined
by (\ref{B211}). So
\begin{equation}
\zeta_{\exp(\alpha l+B_0)}^{(k)}(0)=T_{k+1}\left(\alpha,B_0\right)/\alpha=
k!\left(-1\right)^k\sum_{m+q=k+1,\;m,q\in\wZ_+}\alpha^{m-1}Q_{m,q}\left(B_0
\right).
\label{B3847}
\end{equation}
The polynomial $T_{k+1}\left(\alpha,B_0\right)$ is invariantly defined
on the linear space $\fell(M,E)$ (i.e., it does not depend
on a choice of $l$). By (\ref{B3847}) we conclude that
$\zeta_{\exp(\alpha l+B_0)}^{(k)}(0)$ has a singularity
$O\left(\alpha^{-1}\right)=O\left((\ord J)^{-1}\right)$ as $\alpha$
tends to zero.
\label{RB3846}
\end{rem}

\begin{rem}
The linear form $Q_{0,1}(B_0)$ (defined by (\ref{K53}) depends
on $\sigma(B_0)$ only. It coincides
(up to a sign) with the {\em multiplicative residue}
$-\res^{\times}\sigma(\exp(B_0))$ (defined by (\ref{B12})) for the symbol
of $\exp(B_0)$. The linear function $T_1(cl+B_0)$ depends
on $\sigma\left(\exp(cl+B_0)\right)$ only. By (\ref{B3847})
$T_1$ coincides (up to a sign) with the defined by (\ref{B10}) function
$-Z\!\left(\sigma\!\left(\!\exp\!(\!cl\!+\!B_0\!)\right)\right)$.
Hence $T_1(cl+B_0)$ possesses the property (\ref{B11}).
\label{RB205}
\end{rem}

\begin{pro}
The term $Q_{0,k+1}(B_0)$ in the formula (\ref{K61})
for $\zeta_{\exp(l+B_0)}^{(k)}(0)$ is as follows
\begin{equation}
Q_{0,k+1}(B_0)=-\res\left(B_0^{k+1}\right)\big/(k+1)!.
\label{B804}
\end{equation}
\label{PB800}
\end{pro}

\noindent{\bf Proof.} It follows from Proposition~\ref{PB5} that
\begin{equation}
\left\{s\TR\exp\left(sl+B_0\right)\right\}\big|_{s=0}=-\res\sigma\left(\exp
B_0\right).
\label{B801}
\end{equation}
Here, the expression on the left is a continuous function of $s$ at $s=0$.
($\TR$ is defined for $s\notin\wZ$ and $s\TR\exp(sl+B_0)$ is continuous
in $s$ at $s=0$.) According to (\ref{K52}), (\ref{K53}) we have power
series at $s=0$, $B_0=0$
\begin{gather}
\begin{split}
s\TR\exp\left(sl+B_0\right) & =\sum_{m\in\zuo}s^mQ_m\left(B_0\right), \\
Q_m\left(B_0\right)         & =\sum_{j\in\zuo}Q_{m,j}\left(B_0\right).
\end{split}
\label{B802}
\end{gather}

We deduce from (\ref{B801}), (\ref{B802}) that
\begin{gather}
\begin{split}
Q_0\left(B_0\right)     & =-\res\sigma\left(\exp B_0\right), \\
Q_{0,j}\left(B_0\right) & =-\res\sigma\left(B_0^j\right)/j!.
\end{split}
\label{B803}
\end{gather}
In particular,
\begin{equation}
Q_{0,2}\left(B_0\right)=-\res\sigma\left(B_0^2\right)/2=-\left(\sigma\left(
B_0\right),\sigma\left(B_0\right)\right)_{\res}/2.
\label{B807}
\end{equation}

The proposition is proved.\ \ \ $\Box$

\begin{rem}
For all $k\in\zuo$ we have%
\footnote{The homogeneous polynomial $T_{k+1}$ of order $k+1$ on $\fell(M,E)$
is defined by (\ref{B211}).}
\begin{equation}
T_{k+1}\left(\log_{(\theta)}(AB)\right)=T_{k+1}\left(\log_{(\theta)}(BA)
\right)
\label{B207}
\end{equation}
for an arbitrary pair $(A,B)$ of invertible elliptic PDOs
$A\in\Ell^\alpha(M,E)\subset CL^\alpha(M,E)$ and
$B\in\Ell^\beta(M,E)\subset CL^\beta(M,E)$ such that
$\log_{(\theta)}(AB)$ is defined for some cut $L_{(\theta)}$
of the spectral plane. (In this case
$\log_{(\theta)}(BA)$ is also defined. It is enough to suppose
the existence of $\log_{(\theta)}\sigma_{\alpha+\beta}(AB)$
for the principal symbol
of $AB\in\Ell_0^{\alpha+\beta}(M,E)\subset CL^{\alpha+\beta}(M,E)$. Then
$\log_{(\tilde{\theta})}(AB)$ is defined for a cut $L_{(\tilde{\theta})}$
close to $L_{(\theta)}$.)
\label{RB206}
\end{rem}

%

\begin{rem}
Let $A=\exp\left(\alpha l+A_0\right)\in\Ell_0^\alpha(M,E)$,
$B=\exp\left(\beta l+A_0\right)\in\Ell^\beta(M,E)$ be of nonzero orders.
Here we suppose that $A$, $B$ are sufficiently close to positive
definite self-adjoint ones. These conditions are satisfied, if $A_0$
and $B_0$ are sufficiently small. Then
\begin{equation}
\log F(A,\!B)\!=\!-\!T_2\!\left(\alpha l\!+\!A_0\right)/\alpha\!-\!T_2\!\left(
\beta l\!+\!B_0\right)/\beta\!+\!T_2\!\left((\alpha\!+\!\beta)l\!+\!D_0\right)
/(\alpha\!+\!\beta),
\label{B3851}
\end{equation}
where $(\alpha+\beta)l+D_0:=\log_{(\tpi)}\left(\exp\left(\alpha l+A_0\right)
\exp\left(\beta l+B_0\right)\right)$. Hence (by Proposition~\ref{PB3740})
the expression on the right
in (\ref{B3851}) depends on the symbols $\sigma\left(\alpha l+A_0\right)$
and $\sigma\left(\beta l+B_0\right)$ only.
\label{RB3850}
\end{rem}

\begin{rem}
Let $A_1,\dots,A_k$ be elliptic PDOs from $CL^{\alpha_i}(M,E)\ni A_i$
($1\le i\le k$) of nonzero orders $\alpha_i$ such that their
complex powers are defined.
Let $B_i\in CL^{\beta_i}(M,E)$ ($1\le i\le k$) be a set of $k$ PDOs.
In this situation a generalized $\TR$-zeta-function is defined by
\begin{equation}
f_{\{A_i\},\{B_i\}}(s_1,\dots,s_k):=\TR\left(B_1A_1^{s_1}\ldots B_kA_k^{s_k}
\right).
\label{B214}
\end{equation}
This function
is defined on the complement $U$ to the hyperplanes in $\wC^k$, namely
on $U:=\wC^k\setminus\left\{(s_1,\dots,s_k)\colon\sum_{i=1}^k(\beta_i+s_i
\alpha_i)\in\wZ\right\}$. This function is analytic and non-ramified
on $U$. Indeed, this function is defined by $\TR$ for any point
$\bold{s}\in U$ without an analytic continuation in parameters
$\bold{s}:=\left(s_1,\dots,s_k\right)$ of the holomorphic family
$B_1A_1^{s_1}\ldots B_kA_k^{s_k}$. By Proposition~\ref{PB3755}
the expression (\ref{B214}) is meromorphic in $\bold{s}$ with simple
poles on the hyperplanes $\sum_i\left(\beta_i+s_i\alpha_i\right)=m\in\wZ$,
$m\ge-\dim M$. Note that by (\ref{B3801}) we have for $m\in\wZ$,
$m\ge-\dim M$,
\begin{equation}
\Res_{\sum_i(\beta_i+s_i\alpha_i)\!=\!m}f_{\{A_i\},\{B_i\}}\left(s_1,\ldots,s_k
\right)\!=\!-\!\res\sigma\left(B_1A_1^{s_1}\ldots B_kA_k^{s_k}|_{z(\bold{s})=m}
\right)\!.
\label{B3857}
\end{equation}
(Here, $\Res$ is taken with respect to a natural parameter
$z=z(\bold{s}):=\sum_i\left(\beta_i+s_i\alpha_i\right)$ transversal
to hyperplanes $\{\bold{s},z(\bold{s})=m\}$.)
For $\sum_i\left(\beta_i+s_i\alpha_i\right)=m<-\dim M$, $m\in\wZ$,
the function $f_{\{A_i\},\{B_i\}}\left(s_1,\ldots,s_k\right)$ is regular
on the hyperplane $z(\bold{s})=m$.
\label{RB212}
\end{rem}

\begin{rem}
Let $A\in\Ell_0^\alpha(M,E)\subset CL^\alpha(M,E)$,
$\alpha\!\in\!\wR^{\times}$, be an invertible elliptic operator such that
a holomorphic family of its complex powers $A^s$ exists. (This family
depends on a choice of $\log A$.)
Let $B\in CL^\beta(M,E)$. Then a generalized zeta-function of $A$
$$
\TR\left(BA^{-s}\right)=:\zeta_{A,B}^{\TR}(s)=:\zeta
_{A,B}(s)
$$
is defined on the complement $U$ to the arithmetic progression, namely
on $U:=\wC\setminus\{s:-\alpha s+\beta\in\wZ\}$. Suppose for simplicity
that $\beta\in\wZ$. Then the function $\alpha s\cdot\zeta_{A,B,(\theta)}(s)$
is holomorphic%
\footnote{For $\beta\notin\wZ$ this function is also holomorphic
at $s=0$ but in this case, $s=0$ is not a distinguished point
for the $\TR$ of the family $BA_{(\theta)}^{-s}$.}
at $s=0$ and for $s$ close to zero we have
\begin{equation}
\alpha s\cdot\zeta_{A,B}^{\TR}(s)=\sum_{k\in\zuo}s^k F_k(A,B),
\label{B3852}
\end{equation}
where $F_k$ are homogeneous polynomial of orders $k$ in $\log A$
with their coefficients linear in $B$. Indeed, by Propositions~\ref{PB3755},
\ref{PB3822} we know that the left side of (\ref{B3852}) is a holomorphic
in $s$, $\log A$ function (for $s$ close to zero). Namely we have
$$
s\TR\left(B\exp\left(sl+A_0\right)\right)=\sum_{k\in\zuo}s^kP_k\left(B,A_0
\right),
$$
$P_k\left(B,A_0\right)$ is a regular in $A_0\in CL^0(M,E)$ analytic
function. Set
$$
P_k\left(B,A_0\right)=\sum_{m\in\zuo}P_{k,m}\left(B,A_0\right),
$$
where $P_{k,m}\left(B,A_0\right)$ is a homogeneous in $A_0$ polynomial
of order $m$ with its coefficients linear in $B$. So
\begin{gather*}
s\TR\left(B\exp\left(s\left(l+A_0\right)\right)\right)=\sum s^{k+m}P_{k,m}
\left(B,A_0\right), \\
\alpha s\TR\left(B\exp\left(-s\left(\alpha l+A_0\right)\right)\right)=\sum
\alpha^k(-1)^{m+k-1}s^{k+m}P_{k,m}\left(B,A_0\right).
\end{gather*}
Thus the coefficients $F_k(A,B)$ in (\ref{B3852}) are homogeneous
polynomials of orders $k$ in $\log A:=\alpha l+A_0$. Namely
\begin{equation}
F_k(A,B)=(-1)^{k-1}\sum_{r+m=k,\;r,m\in\zuo}\alpha^rP_{l,m}\left(B,A_0\right),
\label{B3855}
\end{equation}
Note that for $\beta:=\ord B\in\wZ$, $\beta<-\dim M$, $F_0(A,B)$ is zero.
For $\beta\ge-\dim M$ this term $F_0(A,B)$ is equal to $-\res\sigma(B)$.
It is independent of $A$ and depends on $\sigma(B)$ only.


More generallly, we can define this zeta-function for arbitrary
$\log A\in\fell(M,E)\setminus CL^0(M,E)$ and $B\in CL^{\wZ}(M,E)$.
\label{RB215}
\end{rem}

\begin{rem}
An analogous to the power series expansion (\ref{B3852}) is also valid
for a generalized $\TR$-zeta-function (\ref{B214}). Suppose for simplicity
that $\sum\beta_i=:q\in\wZ$, $q\ge-\dim M$. Then
$$
\left(\alpha_1s_1+\ldots+\alpha_ks_k\right)f_{\{A_i\},\{B_i\}}\left(s_1,
\ldots,s_k\right)=\sum_{n_j\in\zuo}s_1^{n_1}\ldots s_k^{n_k}
F_{n_1,\ldots,n_k}\left(\left\{A_i\right\},\left\{B_i\right\}\right)
$$
for $s_j\in\wC$ close to zero. Indeed, by Propositions~\ref{PB3755},
\ref{PB3822} for $C_j\in CL^0(M,E)$ we have
\begin{multline*}
\left(s_1+\ldots+s_k\right)\TR\left(B_1\exp\left(s_1l+C_1\right)\ldots B_k
\exp\left(s_kl+C_k\right)\right)=\\
=\sum_{m_j\in\zuo}s_1^{m_1}\ldots s_k^{m_k}P_{m_1,\dots,m_k}\left(C_1,\dots,
C_k\right),
\end{multline*}
where $P_{\bold{m}}\left(\left\{C_j\right\}\right)$ is a holomorphic
in $C_j$ regular function on $CL^0(M,E)^{\oplus k}$. So
\begin{multline*}
\left(\alpha_1s_1+\ldots+\alpha_ks_k\right)\TR\left(B_1\exp\left(s_1\left(
\alpha_1l+C_1\right)\right)\ldots B_k\exp\left(s_k\left(\alpha_kl+C_k\right)
\right)\right)=\\
=\sum\alpha_1^{m_1}s_1^{m_1+n_1}\ldots\alpha_k^{m_k}s_k^{m_k+n_k}
P_{m_1,\dots,m_k}^{n_1,\dots,n_k}\left(C_1,\dots,C_k\right),
\end{multline*}
where $P_{\bold{m}}^{\bold{n}}\left(\left\{C_j\right\}\right)$ is
polyhomogeneous in $C_j$ of orders $m_j$ polynomial
with its coefficients polylinear in $B_1,\dots,B_k$. Thus
$$
\left(\alpha_1s_1+\ldots+\alpha_ks_k\right)f_{\{A_i\},\{B_i\}}\left(s_1,
\ldots,s_k\right)=\sum_{n_j\in\zuo}s_1^{n_1}\ldots s_k^{n_k}
F_{n_1,\ldots,n_k}\left(\left\{A_i\right\},\left\{B_i\right\}\right),
$$
where the coefficients
\begin{equation}
F_{n_1,\ldots,n_k}\!\left(\!\left\{A_i\right\}\!,\!\left\{B_i\right\}\!
\right)\!=\!\sum_{m_j+r_j\!=\!n_j,\;m_j,r_j\in\zuo}\alpha_1^{m_1}\!\ldots
\!\alpha_k^{m_k}P_{m_1,\dots,m_k}^{r_1,\dots,r_k}\!\left(C_1,\!\dots\!,C_k
\right)
\label{B3858}
\end{equation}
are polyhomogeneous in $\log A_j=\alpha_jl+C_j$ of orders $n_j$ and polylinear
in $B_1,\dots,B_k$ polynomials. The coefficient
$F_{0,\dots,0}\left(\left\{A_i\right\},\left\{B_i\right\}\right)$
is independent of $\left\{A_i\right\}$ and it is equal
to $-\res\sigma\left(B_1,\dots,B_k\right)$. For $\sum\beta_i=q<-\dim M$
($q\in\wZ$) this coefficient is equal to zero.
\label{RB3856}
\end{rem}

\begin{rem}
Invertible elliptic operators $A_i$, $A$ in (\ref{B214}), (\ref{B3857})
can have different logarithms in $\fell(M,E)$. Let $\ord A\ne 0$.
Then by Remark~\ref{RB3830}, 2., and by Propositions~\ref{PB3755},
\ref{PB3822} we conclude that
\begin{equation}
\tilde{\zeta}_{A,B}(s):=\TR\left(BA^{-s}\right)/\Gamma(s\ord A-\ord B-\dim M)
\label{B3896}
\end{equation}
is an entire function of $s$ and of $\log A\in\fell(M,E)$ linear
in $B\in CL^m(M,E)$ and depending on a holomorphic family $A^{-s}$.%
\footnote{That is $\tilde{\zeta}_{A,B}(s)$ is defined by $\log A$, $B$,
$s$.}
Note that $\TR$ is canonically defined for $s\ord A-\ord B\notin\wZ$
(and also for $s\ord A-\ord B=m\in\wZ$, $m>\dim M$). But the proof
of Proposition~\ref{PB3755} gives us the regularity of (\ref{B3896})
for all $s\in\wC$. The value of $\tilde{\zeta}_{A,B}(s)$ for $s=m\in\wZ$
depends on $A$, $B$, $m$ but not on $\log A$. (If $q:=-m\ord A+\ord B\in\wZ$,
then the latter assertion follows from Proposition~\ref{PB3755},
(\ref{B3801}). If $q\notin\wZ$, then it follows from the definition
of $\TR$ becase this definition does not use any analytic continuations.)
So the values of $\tilde{\zeta}_{A,B}(m)$ at $m\in\wZ$ as of a function
on $\fell(M,E)\ni\log A$ are the same at all different $\log A$
(for a given $A$).

So we have a power expansion of $\tilde{\zeta}_{A,B}(s)$
\begin{equation}
\tilde{\zeta}_{A,B}(s)=\sum_{k\in\zuo}\alpha^ks^k\tilde{P}_k\left(B,sA_0
\right),
\label{B3898}
\end{equation}
where $\tilde{P}_k\left(B,A_0\right)$ is the $s^k$-coefficient
(as $s\to 0$) for
$\TR\left(B\exp\left(-\left(sl+A_0\right)\right)\right)/\Gamma(s\alpha\!-\beta-
n)$, $\beta:=\ord B$, $n:=\dim M$. Here, $\log A:=\alpha l+A_0$,
$\alpha\in\wC$, $l\in\fell(M,E)$ is a logarithm of an order one elliptic
PDO, $A_0\in CL^0(M,E)$. Note that $\tilde{P}_k$ is an entire function
in $A_0$ linear in $B\in CL(M,E)$. The series
(\ref{B3898}) is convergent for all $s\in\wC$. As well as in (\ref{B3852})
we have
$$
\tilde{P}_k\left(B,A_0\right)=\sum_{k,m\in\zuo}\tilde{P}_{k,m}\left(B,A_0
\right),
$$
where $\tilde{P}_{k,m}$ are homogeneous polynomials of order $m$
in $A_0$ linear in $B$. So
\begin{equation}
\tilde{\zeta}_{A,B}(s)=\sum_{k,m\in\zuo}s^{k+m}\alpha^k\tilde{P}_{k,m}\left(B,
A_0\right).
\label{B3899}
\end{equation}
The coefficient $\sum_{k+m=r}\alpha^k\tilde{P}_{k,m}\left(B,A_0\right)$
in (\ref{B3899}) is a homogeneous polynomial of order $r$
in $\left(\alpha,A_0\right)$ (i.e., in $\log A$) and it is linear in $B$.
The values of the convergent series (\ref{B3899}) at $s\in\wZ$ are equal
for different $\log A=\alpha l+A_0\in\fell(M,E)$ of $A$.

The analogous assertion is true for
\begin{equation}
\tilde{f}_{\{A_i\},\{B_i\}}(\bold{s}):=f_{\{A_i\},\{B_i\}}(\bold{s})/\Gamma
\left(-\dim M-\sum\left(s_i\ord A_i+\ord B_i\right)\right).
\label{B3897}
\end{equation}
Here, $f_{\{A_i\},\{B_i\}}(\bold{s})$ is defined by (\ref{B214}),
$\bold{s}:=\left(s_1,\dots,s_k\right)\in\wC^k$. The function (\ref{B3897})
is an entire function of $\bold{s}$, $\left\{B_i\right\}$,
$\left\{\log A_i\right\}$. (However $f_{\{A_i\},\{B_i\}}(\bold{s})$
depends on $\log A_i$ and not on $A_i$ only.) For $\bold{s}\in\wZ^k$
the values $\tilde{f}_{\{A_i\},\{B_i\}}(\bold{s})$ on
$\left\{\log A_i\right\}\in\fell(M,E)^k$ do not depend on $\log A_i$
and depend on $\left\{A_i\right\}$ only. So this function has the same
values at all the points $\left\{\log A_i\right\}\in\fell(M,E)^k$,
where the $i$-th component of a vector $\left\{\log A_i\right\}$
is any $\log A_i$.
\label{RB3895}
\end{rem}

\section{Multiplicative property for determinants on odd-dimensional
manifolds}
\label{SC}

The symbol $\sigma(P)(x,\xi)$ of a differential operator $P\in CL^d(M,E)$,
$d\in\zuo$, is polynomial in $\xi$. Hence $\sigma_k(P)(x,\xi)$ is not
only positive homogeneous in $\xi$ (i.e., $\sigma_k(P)(x,t\xi)=
t^k\sigma_k(P)(x,\xi)$ for $t\in\wR_+^{\times}$) but it also possesses
the property
\begin{equation}
\sigma_k(P)(x,-\xi)=\left(-1\right)^k\sigma_k(P)(x,\xi).
\label{B240}
\end{equation}
This property of a symbol $\sigma(A)$ makes sense for $A\in CL^m(M,E)$,
where $m\in\wZ$. The condition (\ref{B240}) is invariant under a change
of local coordinates on $M$. Hence it is enough to check it for a fixed
finite cover of $M$ by local charts.

Denote by $CL_{(-1)}^m(M,E)$ the class of PDOs from $CL^m(M,E)$ whose
symbols possess the property (\ref{B240}) (for $m\in\wZ$). We call
operators from $CL_{(-1)}^{\bullet}(M,E)$ the {\em odd class operators}.

\begin{rem}
Let $A\in CL_{(-1)}^m(M,E)$, $m\in\wZ$, be an invertible elliptic operator.
Let all the eigenvalues of its principal symbol $\sigma_m(A)(x,\xi)$
have positive real parts for $\xi\ne 0$. Then $m$ is an even integer,
$m=2l$, $l\in\wZ$.
\label{RB245}
\end{rem}

\begin{rem}
Let $A\in CL_{(-1)}^{m_1}(M,E)$ and $B\in CL_{(-1)}^{m_2}(M,E)$,
$m_1,m_2\in\wZ$. Then $AB\in CL_{(-1)}^{m_1+m_2}(M,E)$.
If besides $B$ is an invertible elliptic operator, then
$B^{-1}\in CL_{(-1)}^{-m_2}(M,E)$ and $AB^{-1}\in CL_{(-1)}^{m_1-m_2}(M,E)$.
\label{RB246}
\end{rem}

The following proposition defines a {\em canonical trace} for odd class
PDOs on odd-dimensional manifold $M$.

\begin{pro}
Let $A\in CL_{(-1)}^m(M,E)$, $m\in\wZ$, be an odd class PDO. Let $C$
be any odd class elliptic PDO
$C\in\Ell_0^{2q}(M,E)\cap CL_{(-1)}^{2q}(M,E)$, $q\in\wZ_+$, $2q>m$,
sufficiently close to positive definite and self-adjoint PDOs. Then
a generalized $\TR$-zeta-function $\TR\left(AC_{(\tpi)}^{-s}\right)$
is regular at $s=0$. Its value at $s=0$
\begin{equation}
\TR\left(AC_{(\tpi)}^{-s}\right)|_{s=0}=:\Tr_{(-1)}(A)
\label{B3887}
\end{equation}
is independent of $C$. We call it the {\em canonical trace} of $A$.
\label{PB3886}
\end{pro}

\noindent{\bf Proof.} 1. By Proposition~\ref{PB3755} and
by Remark~\ref{RB3795} the $\TR\left(AC_{(\tpi)}^{-s}\right)$ has
a meromorphic continuation to the whole complex plane $\wC\ni s$ and
its residue at $s=0$ is equal to $-\res\sigma(A)$ (where $\res$ is
the noncommutative residue, \cite{Wo2}, \cite{Kas}).
By Remark~\ref{RB3785} below, $\res\sigma(A)=0$ for any odd class PDO
$A$ on an odd-dimensional manifold $M$.

2. Let $B\in\Ell_0^{2r}(M,E)\cap CL_{(-1)}^{2r}(M,E)$, $r\in\wZ_+$,
be another positive definite odd class elliptic operator. Then
by Corollary~\ref{CB18}, (\ref{B101}), and by Remark~\ref{RB4} we have
\begin{gather}
\begin{split}
\TR\left(AC_{(\tpi)}^{-s}\right)-\TR\left(AB_{(\tpi)}^{-s}\right) &
=\Tr\left(AC_{(\tpi)}^{-s}\right)-\Tr\left(AB_{(\tpi)}^{-s}\right)
\text{ for }\Re s\gg 1, \\
\Tr\left(A\left(C_{(\tpi)}^{-s}-B_{(\tpi)}^{-s}\right)\right)\big|_{s=0}\!= &
-\!\left(\sigma(A),\sigma\left(\log_{(\tpi)}C\right)\!/2q\!-\!\sigma\left
(\log_{(\tpi)}B\right)\!/2r\right)_{\res}.
\end{split}
\label{B3888}
\end{gather}
(Note that (\ref{B101}) is valid for any $Q\in CL^m(M,E)$, $m\in\wZ$.)
Applying Corollary~\ref{CB3787} below, (\ref{B3889}), (\ref{B3890}),
to pairs $(CB,B)$ and $(CB,C)$, we obtain
$$
\log_{(\tpi)}C/2q-\log_{(\tpi)}B/2r\in CL_{(-1)}^0(M,E).
$$
Hence on the right in (\ref{B3888}) we have a product of odd class
symbols. By Remark~\ref{RB3785} the residue $\res$ of this product
is equal to zero (as $M$ is odd-dimensional).
Thus $\TR\left(AC_{(\tpi)}^{-s}\right)|_{s=0}=\TR\left(AB_{(\tpi)}^{-s}\right)
|_{s=0}$.\ \ \ $\Box$


\begin{rem}
Let $A\in CL_{(-1)}^0(M,E)$ be an odd class operator on an odd-dimensional
manifold $M$. Then%
\footnote{A PDO $\exp(zA)$ is defined by (\ref{K25}).}
$\exp(zA)\in CL_{(-1)}^0(M,E)\cap\Ell_0^0(M,E)$ is a holomorphic
family of odd class elliptic operators of zero orders.
By Proposition~\ref{PB3886}
\begin{equation}
\Tr_{(-1)}\exp(zA):=\TR\left(\exp(zA)C_{(\pi)}^{-s}\right)|_{s=0}
\label{B3894}
\end{equation}
is defined for $z\in\wC$. It is an entire function of $z\in\wC$ since
by (\ref{B3894}) and by the equalities analogous to (\ref{B3515}),
(\ref{B3516}) we have
\begin{align*}
\df_z\Tr_{(-1)}(\exp(zA)) & =\Tr_{(-1)}(A\exp(zA)), \\
\df_{\overline{z}}\Tr_{(-1)}(\exp(zA)) & =0.
\end{align*}

The problem is to estimate the entire function $\exp(zA)$ as $|z|\to\infty$.
Note that in general the spectrum $\Spec A$ contains a continuous
part. So the nature of the entire function $\Tr_{(-1)}\exp(zA)$
is different from the Dirichlet series.
\label{RB3893}
\end{rem}

\begin{thm}
Let $M$ be an odd-dimensional smooth closed manifold.
Let $A\in CL_{(-1)}^{m_1}(M,E)$ and $B\in CL_{(-1)}^{m_2}(M,E)$
be invertible elliptic PDOs (where $m_1,m_2,m_1+m_2\in\wZ\setminus 0$).
Let their principal symbols $\sigma_{m_1}(A)(x,\xi)$
and $\sigma_{m_2}(B)(x,\xi)$ be sufficiently close to positive definite
self-adjoint ones.
Then $\det_{(\tpi)}(A)$, $\det_{(\tpi)}(B)$ and $\det_{(\tpi)}(AB)$ are
defined (with the help of zeta-functions with the cut $L_{(\tpi)}$
of the spectral plane close to $L_{(\pi)}$). We have
\begin{equation}
{\det}_{(\tpi)}(AB)={\det}_{(\tpi)}(A){\det}_{(\tpi)}(B).
\label{B242}
\end{equation}
\label{TB241}
\end{thm}

\begin{cor}
Let $A\in CL_{(-1)}^0(M,E)\cap\Ell_0^0(M,E)=:\Ell_{(-1),0}^0(M,E)$
be an invertible zero order elliptic PDO
on a closed odd-dimensional $M$ such that its principal symbol
$\sigma_0(A)(x,\xi)$ is sufficiently close to a positive definite
self-adjoint symbol. Then such PDO $A$ of zero order has a correctly
defined determinant. Namely
\begin{equation}
{\det}_{(\tpi)}(A):={\det}_{(\tpi)}(AB)\big/{\det}_{(\tpi)}(B)
\label{B244}
\end{equation}
for an arbitrary invertible elliptic $B\in CL_{(-1)}^m(M,E)$, $m\in\wZ_+$,
such that its principal symbol is sufficiently close to a positive
definite self-adjoint symbol.

The correctness of the definition (\ref{B244}) follows
from the multiplicative property (\ref{B242}). Indeed, for two such
elliptic operators $B\in\Ell_0^{m_1}(M,E)\cap CL_{(-1)}^{m_1}(M,E)$,
$C\in\Ell_0^{m_2}(M,E)\cap CL_{(-1)}^{m_2}(M,E)$, $m_1,m_2\in\wZ_+$,
we have
$$
{\det}_{(\tpi)}(AB)\big/{\det}_{(\tpi)}(B)={\det}_{(\tpi)}(AC)\big/{\det}
_{(\tpi)}(C)\equiv{\det}_{(\tpi)}(CA)\big/{\det}_{(\tpi)}(C)
$$
as (by (\ref{B242})) we have
$$
{\det}_{(\tpi)}(AB){\det}_{(\tpi)}(C)={\det}_{(\tpi)}(CAB)={\det}_{(\tpi)}(B)
{\det}_{(\tpi)}
(CA).
$$
\label{CB243}
\end{cor}

\begin{cor}
The multiplicative property holds for odd class elliptic PDOs
$A,B\in\Ell_{(-1),0}^0(M,E)$ sufficiently close to positive definite
self-adjoint ones ($M$ is closed and odd-dimensional). Namely
\begin{equation}
{\det}_{(\tpi)}(AB)={\det}_{(\tpi)}(A){\det}_{(\tpi)}(B).
\label{B4011}
\end{equation}
\label{CB4007}
\end{cor}

Indeed, let $C=C_1C_2$ be a product of positive definite self-adjoint
odd class elliptic PDOs on $M$ of positive orders.
Then by Theorem~\ref{TB241}, (\ref{B242}), we have
\begin{multline}
{\det}_{(\tpi)}(AB)={\det}_{(\tpi)}\left(ABC_1C_2\right)\big/{\det}_{(\tpi)}
\left(C_1C_2\right)=\\
={\det}_{(\tpi)}\left(C_2A\right){\det}_{(\tpi)}\left(BC_1\right)\big/{\det}
_{(\tpi)}\left(C_1\right){\det}_{(\tpi)}\left(C_2\right)={\det}_{(\tpi)}(A)
{\det}_{(\tpi)}(B).
\label{B4008}
\end{multline}

\begin{rem}
Let $A^*$ be an adjoint operator to $A\in\Ell_{(-1),0}^{2m}(M,E)$,
$m\in\wZ$, and let $M$ be close and odd-dimensional. Let $A$ be
sufficiently close to a positive definite self-adjoint one (with
respect to some positive density and to a Hermitian structure). Then
\begin{equation}
{\det}_{(\tpi)}(A^*)=\overline{{\det}_{(\tpi)}(A)}.
\label{B4010}
\end{equation}
\label{RB4009}
\end{rem}

\begin{rem}
Let $A$ be a PDO of the odd class $CL_{(-1)}^m(M,E)$, $m\in\wZ$,
on odd-dimensional manifold. Then the noncommutative residue of $A$
is equal to zero, $\res\sigma(A)=0$.

To prove this equality, note that since $M$ is odd-dimensional,
the density $\res_x\sigma(A)$ (corresponding to the noncommutative
residue of $A$) on $M$ is the identity zero.
Indeed, the density $\res_x\sigma(A)$ at $x\in M$ is represented
in any local coordinates $U\ni x$ on $M$ by the integral over the fiber
$S_x^*M$ over $x$ (of the cospherical fiber bundle for $M$)
of the homogeneous component $\sigma_{-n}(A)$, $n:=\dim M$. Since
$n$ is odd, we have
$$
\sigma_{-n}(A)(x,-\xi)=-\sigma_{-n}(A)(x,\xi).
$$
Thus we have to integrate over $S_x^*M$ the product of an odd (with respect
to the center of the sphere $S_x^*M$) function $\sigma_{-n}(A)(x,\xi)$
and a natural density on the unit sphere $S_x^*M$. This integral
$\res_x\sigma(A)$ is equal to zero.
\label{RB3785}
\end{rem}

\begin{rem}
The equality (\ref{B242}) means that the multiplicative anomaly
for elliptic operators (nearly positive and nearly self-adjoint)
of the odd class is zero. To apply the general variation formula
(\ref{B3742}) of Proposition~\ref{PB3740} to prove that the multiplicative
anomaly is zero, we have to use deformations $A_t$ belonging to the odd
class $CL_{(-1)}^m(M,E)\cap\Ell_0^m(M,E)$. This is a rather restrictive
condition. Note that for $A$ and $B$ from the odd class a deformation
$A_t$ in the integral formula (\ref{B3752}) for the multiplicative anomaly
is not inside the odd class. For a deformation not inside the odd class
we cannot conclude from (\ref{B3742}) and (\ref{B3752}) that $F(A,B)=1$.
\label{RB3760}
\end{rem}

\begin{pro}
Let the principal symbol of $\eta\in CL_{(-1)}^0(M,E)\cap\Ell_0^0(M,E)$
be sufficiently close to a positive definite self-adjoint one. Then
the PDOs $\eta_{(\tpi)}^t$ and $\log_{(\tpi)}\eta$ defined by (\ref{X2})
and by (\ref{B7}) are from $CL^0(M,E)$ and they are {\em odd class
operators}.
\label{PB3761}
\end{pro}

\noindent{\bf Proof of Proposition~\ref{PB3761}.}
By (\ref{B1}) we see that
\begin{gather}
\begin{split}
\sigma_0\left(\left(\eta-\lambda\right)^{-1}\right)(x,\xi,\lambda) &
=\left(\sigma_0(\eta)(x,\xi)-\lambda\right)^{-1}=\sigma_0\left(\left(\eta-
\lambda\right)^{-1}\right)(x,-\xi,\lambda),\\
\sigma_{-j}\left(\left(\eta-\lambda\right)^{-1}\right)(x,\xi,\lambda) & =\\
=-\sigma_0\left(\left(\eta-\lambda\right)^{-1}\right) & \bigl(\sum_{\overset
{|\alpha|+i+l=j}{i\le j-1}}{1\over \alpha!}\df_\xi^\alpha\sigma_{-i}(\eta-
\lambda)D_x^\alpha\sigma_{-l}\left(\left(\eta-\lambda\right)^{-1}\right)
\bigr)=\\
 & =\left(-1\right)^j\sigma_{-j}\left(\left(\eta-\lambda\right)^{-1}\right)
(x,-\xi,\lambda).
\end{split}
\label{B247}
\end{gather}
(Here, $\sigma_{-j}(\eta-\lambda):=\sigma_{-j}(\eta)-\delta_{j,0}\lambda$.)
Thus we have
\begin{multline}
\sigma_{-j}\left(\eta_{(\tpi)}^t\right)(x,\xi):={i\over 2\pi}\int
_{\Gamma_R,\tpi}
\lambda_{(\tpi)}^t\sigma_{-j}\left(\left(\eta-\lambda\right)^{-1}\right)
(x,\xi,\lambda)=\\
=\left(-1\right)^j\sigma_{-j}\left(\eta_{(\tpi)}^t\right)(x,-\xi).
\label{B3998}
\end{multline}
Hence $\eta_{(\pi)}^t\in CL_{(-1)}^0(M,E)$.

By (\ref{B247}), we have $\log_{(\tpi)}\eta\in CL_{(-1)}^0(M,E)$ since
\begin{multline}
\sigma_{-j}\left(\log_{(\tpi)}\eta\right)(x,\xi):={i\over 2\pi}\int
_{\Gamma_R,\tpi}\log_{(\tpi)}\lambda\cdot\sigma_{-j}\left(\left(\eta-
\lambda\right)^{-1}\right)(x,\xi,\lambda)d\lambda=\\
=\left(-1\right)^j\sigma_{-j}\left(\log_{(\tpi)}\eta\right)(x,-\xi).
\label{B248}
\end{multline}
The proposition is proved.\ \ \ $\Box$

The proof of Theorem~\ref{TB241} is based on the assertions as follows.

\begin{pro}
Let $A_1$ and $A_2$ be invertible elliptic operators of the odd class
$\Ell_{(-1)}^m(M,E):=CL_{(-1)}^m(M,E)\cap\Ell^m(M,E)$, $m\in\wZ\setminus 0$,
such that their principal symbols are sufficiently close to positive
definite self-adjoint ones. Then
\begin{equation}
\log_{(\tpi)}A_1-\log_{(\tpi)}A_2\in CL_{(-1)}^0(M,E).
\label{B3788}
\end{equation}
\label{PB249}
\end{pro}

\begin{cor}
Let $A\in CL_{(-1)}^{m_1}(M,E)$ and $B\in CL_{(-1)}^{m_2}(M,E)$ be
invertible elliptic PDOs of the odd class and let their principal symbols
$\sigma_{m_1}(A)(x,\xi)$ and $\sigma_{m_2}(B)(x,\xi)$
be sufficiently close to positive definite self-adjoint ones.
Let $m_1,m_2,m_1+m_2\in\wZ\setminus0$. Then the following PDO
\begin{equation}
\left(m_1+m_2\right)^{-1}\log_{(\tpi)}(AB)-m_1^{-1}\log_{(\tpi)}(A)\in CL^0
(M,E)
\label{B3889}
\end{equation}
is defined and it belongs to $CL_{(-1)}^0(M,E)$. By (\ref{B3788}) and
(\ref{B3889}) we have also
\begin{equation}
\left(m_1+m_2\right)^{-1}\log_{(\tpi)}(AB)-m_2^{-1}\log_{(\tpi)}(B)\in CL
_{(-1)}^0(M,E).
\label{B3890}
\end{equation}
\label{CB3787}
\end{cor}

\begin{rem}
Proposition~\ref{PB3761} and its proof are valid for any admissible
spectral cut $\theta$ for $\sigma_0(\eta)$. Proposition~\ref{PB249}
and Corollary~\ref{CB3787} are valid for any admissible spectral cuts
for $A_1$, $A_2$, $AB$, $A$, $B$. (The logarithms in (\ref{B3788}),
(\ref{B3889}) are defined with respect to these spectral cuts.
We do not use in the proofs that the cuts for $A_1$ and for $A_2$
in (\ref{B3889}) are the same.)
\label{RB3976}
\end{rem}

\begin{rem}
The proofs of Propositions~\ref{PB3761}, \ref{PB249} and
Corollary~\ref{CB3787} are done with using symbols of PDOs (but not
the PDOs of the form $\left(A-\lambda\right)^{-1}$ themselves). Hence
a spectral cut $L_{(\theta)}$ admissible
for $\sigma\left(\left(A-\lambda\right)^{-1}\right)(x,\xi)$ can smoothly
depend on a point $(x,\xi)\in S^*M$. However it has to be the same
at the points $(x,\xi)$ and $(x,-\xi)$. Hence this spectral cut defines
a smooth map
\begin{equation}
\theta\colon P^*M:=S^*M/(\pm 1)\to S^1=\wR/2\pi\wZ.
\label{B3997}
\end{equation}
Here, $(-1)$ transforms $(x,\xi)\in S^*M$ into $(x,-\xi)$.

The map (\ref{B3997}) has to be homotopic to a trivial one for a smooth
family of branches $\lambda_{(\theta)}^{-s}$ over points $(x,\xi)\in P^*M$
in the formulas (\ref{B3998}), (\ref{B260}) to exist.

The condition of the existence of a field of admissible for the symbol
$\sigma(A)$ cuts (\ref{B3997}) homotopic to a trivial field is analogous
to the sufficient condition of the existence of a $\sigma(\log A)$
given by Remark~\ref{RB3964} below.
So Propositions~\ref{PB3761}, \ref{PB249} and Corollary~\ref{CB3787}
are valid in this more general situation of existing spectral cuts
for $\sigma\left(\left(A-\lambda\right)^{-1}\right)$ depending
on $p\in P^*M$. In this situation there exists a $\sigma(\log A)$
defined with the help of this smooth field of spectral cuts. These
fields of cuts may be different for $AB$ and for $A$ (or for $A_1$ and
$A_2$) in Proposition~\ref{PB249} and in Corollary~\ref{CB3787}.
\label{RB3994}
\end{rem}

\noindent{\bf Proof of Corollary~\ref{CB3787}.} Indeed, set
$A_1:=A_{(\tpi)}^{m_1+m_2}$, $A_2:=(AB)_{(\tpi)}^{m_1}$. Then
$A_1,A_2\in\Ell_{(-1)}^{m_1(m_1+m_2)}(M,E)$. So by (\ref{B3788}) we have
$$
\left(m_1+m_2\right)\log_{(\tpi)}A-m_1\log_{(\tpi)}(AB)=\log_{(\tpi)}A_1-
\log_{(\tpi)}A_2\in CL_{(-1)}^m(M,E).
$$
$\Box$

\noindent{\bf Proof of Proposition~\ref{PB249}.}
Because $A_1\in CL_{(-1)}^m(M,E)$ and $A_2\in CL_{(-1)}^m(M,E)$
($m$ is even),
the formulas analogous to (\ref{B247}) hold for the homogeneous
components of $\sigma\left(\left(A_1-\lambda\right)^{-1}\right)$ and
of $\sigma\left(\left(A_2-\lambda\right)^{-1}\right)$.%
\footnote{For the sake of brievity we suppose here that $m_1=2l_1$
and $m_2=2l_2$ are positive even numbers. If $l_1\in\wZ_-$, we have
to change $(A-\lambda)^{-1}$ by $(A^{-1}-\lambda)^{-1}$ and $s$ by $-s$.}
Hence for $\Re s>0$ we have
\begin{multline}
\sigma_{-2l_1s-j}\left(A_{1,(\tpi)}^{-s}\right)(x,\xi):={i\over 2\pi}\int
_{\Gamma_{(\tpi)}}
\lambda^{-s}\sigma_{-2l_1-j}\left(\left(A_1-\lambda\right)^{-1}\right)
(x,\xi,\lambda)d\lambda=\\
=\left(-1\right)^j\sigma_{-2l_1s-j}\left(A_{1,(\tpi)}^{-s}\right)(x,-\xi).
\label{B260}
\end{multline}
Here we use that since $A\in CL_{(-1)}^{2l_1}(M,E)$, the formulas
analogous to (\ref{B247})
are true for $\sigma\left(\left(A_1-\lambda\right)^{-1}\right)(x,\xi,
\lambda)$. Namely
\begin{equation}
\sigma_{-2l_1-j}\left(\left(A_1-\lambda\right)^{-1}\right)(x,-\xi,\lambda)=
\left(-1\right)^j\sigma_{-2l_1-j}\left(\left(A_1-\lambda\right)^{-1}\right)
(x,\xi,\lambda).
\label{B255}
\end{equation}

According to (\ref{A10}) for $j\in\wZ_+$ we have
$$
\sigma_{-j}\left(\log_{(\tpi)}A_1\right)=-\df_s\sigma_{-2l_1s-j}\left(A
_{1,(\tpi)}^{-s}\right)\big|_{s=0}.
$$
Hence by (\ref{B260}) we have for $j\in\wZ_+$
$$
\sigma_{-j}\left(\log_{(\tpi)}A_1\right)(x,-\xi)=\left(-1\right)^j\sigma_{-j}
\left(\log_{(\tpi)}A_1\right)(x,\xi).
$$
The same equality holds for $\sigma_{-j}\left(\log_{(\tpi)}(A_2)\right)$,
$j\in\wZ_+$.
According to (\ref{A10}) and to (\ref{B260}) we have also
\begin{multline*}
\df_s\sigma_{-ms}\left(A_{1,(\tpi)}^{-s}\right)\big|_{s=0}(x,\xi)-\df_s\sigma
_{-ms}\left(A_{2,(\tpi)}^{-s}\right)\big|_{s=0}=\\
=\df_s\sigma_{-ms}\left(A_{1,(\tpi)}^{-s}\right)\big|_{s=0}(x,\xi/|\xi|)-\\
-\df_s\sigma_{-ms}\left(A_{2,(\tpi)}^{-s}\right)\big|_{s=0}(x,
\xi/|\xi|)\in CL_{(-1)}^0(M,E)\big/CL_{(-1)}^{-1}(M,E).
\end{multline*}
Hence $\log_{(\tpi)}A_1-\log_{(\tpi)}A_2\in CL_{(-1)}^0(M,E)$.
The proposition is proved.\ \ \ $\Box$


\noindent{\bf Proof of Theorem~\ref{TB241}.}
By Remark~\ref{RB245}, the orders $m_1=2l_1$ and $m_2=2l_2$ of $A$
and $B$ are nonzero even integers.
We have $l_1,l_2,l_1+l_2\in\wZ\setminus 0$.
By Remark~\ref{RB3752}, (\ref{B3753}), we have
\begin{equation}
\log F(A,B)=
-\int_0^1 dt\left(\sigma(Q_t),{\sigma\left(\log_{(\tpi)}A_tB\right)\over
2l_1+2l_2}-{\sigma\left(\log_{(\tpi)}A_t\right)\over 2l_1}\right)_{\res}.
\label{B253}
\end{equation}
Here, $Q_t:=\dot{A}_tA_t^{-1}$ and $A_t$ is a smooth family of operators
between $B_{(\tpi)}^{l_1/l_2}$ and $A$ in the odd class elliptic operators
such that the principal symbols $\sigma_{2l_1}\left(A_t\right)$ are
sufficiently close to positive definite self-adjoint ones.

The numbers $m_1$ and $m_2$ are even. So in the odd class elliptic PDOs
we can find deformations $A_t$ and $B_t$ of $A$ and $B$ from $A=A_1$
and $B=B_1$ to $\left(\Delta_{M,E}+\Id\right)^{l_1}=A_0$ and
$\left(\Delta_{M,E}+\Id\right)^{l_2}=B_0$. Here, $\Delta_{M,E}$ is
the Laplacian on $(M,g)$ corresponding to some unitary connection
on $\left(E,h_E\right)$. The deformations can be chosen so that
the principal symbols $\sigma_{2l_1}\left(A_t\right)$,
$\sigma_{2l_2}\left(B_t\right)$ are sufficiently close to positive
definite self-adjoint ones. Hence, applying the formula (\ref{B253})
twice (namely first to $\left(A_t,B\right)$ and then
to $\left(\left(\Delta_{M,E}+\Id\right)^{l_1},B_t\right)$), we have
by Proposition~\ref{PB3761} and by Remark~\ref{RB3785})
$$
F(A,B)=F\left(\left(\Delta_{M,E}+\Id\right)^{l_1},\left(\Delta_{M,E}+
\Id\right)^{l_2}\right)=0.
$$

We can deduce from (\ref{B253}) that $F(A,B)=0$ using only one
explicit deformation of $A$.
Set $\eta:=A_{(\tpi)}^{1/l_1}B_{(\tpi)}^{-1/l_2}$,
$A_t:=\left(\eta_{(\tpi)}^t B_{(\tpi)}^{1/l_2}\right)_{(\tpi)}^{l_1}$,
where $\eta_{(\tpi)}^t$ is defined by (\ref{X2}).
Let $l_1\in\wZ_+$. Then according to (\ref{A5}) we have
\begin{equation}
\sigma_{-2-j}\left(A_{(\tpi)}^{-1/l_1}\right)(x,\xi):={i\over 2\pi}\int
_{\Gamma_{(\tpi)}}\lambda^{-1/l_1}\sigma_{-2l_1-j}\left(\left(A-\lambda\right)
^{-1}\right)(x,\xi,\lambda)d\lambda.
\label{B254}
\end{equation}
(Here, $\Gamma_{(\tpi)}$ is the contour $\Gamma_{(\theta)}$ from (\ref{A5})
with $\theta=\tpi$ close to $\pi$. The integral on the right in (\ref{B254})
is absolutely convergent for $l_1\in\wR_+$.)

Hence according to (\ref{B254}) and to (\ref{B255}) we have
\begin{equation}
\sigma_{-2-j}\left(A_{(\tpi)}^{-1/l_1}\right)(x,\xi)=\left(-1\right)^j
\sigma_{-2-j}\left(A_{(\tpi)}^{-1/l_1}\right)(x,-\xi),
\label{B256}
\end{equation}
i.e., we have $A^{-1/l_1}\in CL_{(-1)}^{-2}(M,E)\cap\Ell_0^{-2}(M,E)$
for $l_1\in\wZ_+$.
For $l_1\in\wZ_-$ we can conclude that
$A^{1/l_1}\in CL_{(-1)}^{-2}(M,E)\cap\Ell_0^{-2}(M,E)$
(changing $\lambda^{-1/l_1}$ by $\lambda^{1/l_1}$ in (\ref{B254})).
Hence according to Remark~\ref{RB246}, we have
\begin{equation}
\eta=A^{1/l_1}B^{-1/l_2}\in CL_{(-1)}^0(M,E)\cap\Ell_0^0(M,E).
\label{B257}
\end{equation}

By Proposition~\ref{PB3761} we conclude that
\begin{equation}
\begin{split}
\eta_{(\tpi)}^t\in CL_{(-1)}^0(M,E)                   & \cap\Ell_0^0(M,E), \\
\eta_{(\tpi)}^tB_{(\tpi)}^{1/l_2}\in CL_{(-1)}^2(M,E) & \cap\Ell_0^2(M,E). \\
A_t\in CL_{(-1)}^{2l_1}(M,E)\cap\Ell_0^{2l_1}(M,E),   & \ Q_t\in CL_{(-1)}^0
(M,E),
\end{split}
\label{B258}
\end{equation}

According to (\ref{B258}), to Remark~\ref{RB246}, and
to Corollary~\ref{CB3787}, the operator
$$
G_t:=Q_t\left({\sigma\left(\log_{(\tpi)}A_tB\right)\over 2l_1+2l_2}-
{\sigma\left(\log_{(\tpi)}A_t\right)\over 2l_1}\right)\in CL^0(M,E)
$$
is defined and it belongs to $CL_{(-1)}^0(M,E)$.
By the equality (\ref{B253}) and by Remark~\ref{RB3785} we have
$$
\log F(A,B)=-\int_0^1 dt\int_M\res_x\sigma\left(G_t\right)=0.
$$
Thus $\det_{(\tpi)}(AB)=\det_{(\tpi)}(A)\det_{(\tpi)}(B)$.
Theorem~\ref{TB241} is proved.\ \ \ $\Box$



%
%

\subsection{Dirac operators}
An important example of odd class elliptic operators is a family
of the Dirac operators $D=D(M,E,g,h)$ on a spinor odd-dimensional
closed manifold $M$ (with a given spinor structure). The Dirac operator
$D$ acts on the space of global smooth sections $\Gamma(S\otimes E)$
\begin{equation}
D=\sum e_i\nabla_{e_i}.
\label{B262}
\end{equation}
Here, $S$ is a spinor bundle on $M$, $(E,h)$ is a Hermitian vector
bundle on $M$, $\{e_i\}$ is a local orthonormal basis in $T_xM$
(with respect to a Riemannian metric $g$),
$\nabla=1\otimes\nabla^R+\nabla^E\otimes 1$, $\nabla^R$ is the Riemannian
connection (for $g$), $\nabla^E$ is a unitary connection on $(E,h)$,
$e_i\left(\nabla_{e_i}f\right)$ is the Clifford multiplication.
The family $\{D\}$ of Dirac operators is parametrized by $g$ and
by $\nabla^E$.

The operator $D_{g,\nabla^E}$ is a formally self-adjoint (with respect
to the natural scalar product on $\Gamma(S\otimes E)$ defined by $g$
and by $h$) elliptic differential operator of the first order.
Its spectrum $\Spec(D)$ is discrete. All the eigenvalues $\lambda$
of $D_{g,\nabla^E}$ are real. However $\Spec(D)$ has infinite number
of points from $\wR_+$ as well as points from $\wR_-$.

For the sake of simplicity suppose that $D_1$ and $D_2$ are invertible
Dirac operators corresponding to the same Riemannian metric $g$ and
to sufficiently close $\left(\nabla_1^E,h_1\right)$,
$\left(\nabla_2^E,h_2\right)$.%
\footnote{If a Riemannian metric $g$ on $M$ varies, then the spinor
bundles $S_{(g)}$ on $M$ also varies, i.e., to identify the spaces
$\Gamma(S_{(g_1)}\otimes E)$ and $\Gamma(S_{(g_2)}\otimes E)$ we have
to use a connection in the directions $\{g\}$ on the total vector
bundle $S_{\{g\}}$ over $M\times\{g\}$. Here, $\{g\}$ is the space
of smooth Riemannian metrics on $M$.}
Then the operator $D_1D_2\in\Ell_0^2(M,E)\subset CL_{(-1)}^2(M,E)$
is an invertible elliptic operator with positive real parts
of all the eigenvalues of its principal symbol $\sigma_2(D_1D_2)$.
Hence for any pairs $(D_1,D_2)$ and
$(D'_1,D'_2)$ of sufficiently close elements of the family $D_{g,\nabla^E}$
Theorem~\ref{TB241} claims that
\begin{equation}
{\det}_{(\tpi)}(D_1D_2){\det}_{(\tpi)}(D'_1D'_2)={\det}_{(\tpi)}(D_1D_2D'_1
D'_2).
\label{B263}
\end{equation}

Let all four Dirac operators $(D_1,D_2,D'_1,D'_2)$ be sufficiently
close. Then we have
\begin{equation}
{\det}_{(\tpi)}(D_1D_2){\det}_{(\tpi)}(D'_1D'_2)={\det}_{(\tpi)}(D_1D'_2)
{\det}_{(\tpi)}(D_2D'_1).
\label{B264}
\end{equation}

Indeed, according to (\ref{B263}) we have
\begin{multline}
{\det}_{(\tpi)}(D_1D_2){\det}_{(\tpi)}(D'_1D'_2)={\det}_{(\tpi)}(D_1D_2D'_1
D'_2)={\det}_{(\tpi)}(D'_2D_1D_2D'_1)=\\
={\det}_{(\tpi)}(D'_2D_1){\det}_{(\tpi)}
(D_2D'_1)={\det}_{(\tpi)}(D_1D'_2){\det}_{(\tpi)}(D_2D'_1).
\label{B265}
\end{multline}
The equality (\ref{B264}) can be written in the form
\begin{equation}
\rk\left(
\begin{split}
\det(D_1D_2)  & \quad \det(D_1D'_2)\\
\det(D_1D'_2) & \quad \det(D'_1D'_2)
\end{split}
\right)=1.
\label{B266}
\end{equation}
Hence, if all the Dirac operators parametrized by a set
$U\subset\{(\nabla^E,h)\}$ are sufficiently close one to another,
we see that the matrix
$$
A_U:=\left(A_{u_1,u_2}\right):=\left({\det}_{(\tpi)}\left(D_{u_1}D_{u_2}\right)
\right)
$$
is a rank one matrix. Hence there are scalar functions $f(u)$ on $U$
such that
\begin{equation}
{\det}_{(\tpi)}\left(D_{u_1}D_{u_2}\right)=A_{u_1,u_2}=:f(u_1)f(u_2).
\label{B267}
\end{equation}

For instance, for $u_1=u_2\in U$ we have
\begin{gather}
\begin{split}
{\det}_{(\tpi)}\left(D_{u_1}^2\right)={\det}_{(\pi)}\left(D_{u_1}^2\right)
 & =f(u_1)^2,\\
f(u)                                 & =\eps(u)\left({\det}_{(\pi)}\left(D
_{u_1}^2\right)\right)^{1/2}, \quad \eps(u)=\pm 1.
\end{split}
\label{B268}
\end{gather}
The operator $D_{u_1}^2$ is a positive definite elliptic operator
from $CL_{(-1)}^2(M,E)$. (It can have a nontrivial kernel
$\Ker D_u^2=\Ker D_u$, $\dim\Ker D_u<\infty$.)
The sign $\eps(u)$ has to be a definite number for $\Ker D_u=0$.
Hence ${\det}_{(\pi)}\left(D_{u_1}^2\right)\in\wR_+\cup 0$ (because
the spectrum of $D_{u_1}^2$ is real and discrete). This determinant
belongs to $\wR_+$ if $\Ker D_{u_1}=0$.

\begin{cor}
For any pair of Dirac operators $D_{u_1}$, $D_{u_2}$, $u_j\in U$,
we have
$$
{\det}_{(\tpi)}\left(D_{u_1}D_{u_2}\right)\in\wR.
$$
Let besides $D_{u_j}$ be invertible. Then
${\det}_{(\tpi)}\left(D_{u_1}D_{u_2}\right)\in\wR^{\times}$ and we have
\begin{equation}
{\det}_{(\tpi)}\left(D_{u_1}D_{u_2}\right)=\eps(u_1)\eps(u_2)\left({\det}
_{(\pi)}\left(D_{u_1}^2\right)\right)^{1/2}\left({\det}_{(\pi)}\left(D_{u_2}
^2\right)\right)^{1/2}.
\label{B272}
\end{equation}
The determinants on the right belongs to $\wR_+$.
\label{CB269}
\end{cor}

\begin{rem}
The function $\eps(u)=\pm 1$ is constant on the connected components
of $U\setminus\{u\in U,\Ker D_u\ne 0\}$.

If in a smooth one-parameter family $D_{u(t)}$ ($t$ is a local parameter
near $0\in\wR$) only one eigenvalue $\lambda(t)$ for $D_{u(t)}$
(of multiplicity one) crosses the origin $0\in\wR\ni\lambda$ at $t=t_0$
transversally (i.e., $\df_t\lambda_u(t)|_{t=t_0}\ne 0$), then the sign
$\eps(u(t))=\pm 1$ changes at $t=t_0$ to an opposite one.
\label{RB270}
\end{rem}

\begin{rem}
It follows from (\ref{B268}) that
\begin{equation}
\left|f(u)\right|=\left({\det}_{(\pi)}\left(D_u^2\right)\right)^{1/2}
\label{B273}
\end{equation}
is a globally defined function of $u\in\left\{g,\left(\nabla^E,h\right)
\right\}$. The sign $\eps(u)=\pm 1$ can be defined as a locally constant
continuous function on the set $u\in\left\{g,\left(\nabla^E,h\right)\right\}$
such that $\Ker D_u=0$. The last assertion follows from Remark~\ref{RB270}
and from the equality to zero of the corresponding spectral flow
\cite{APS3}.
\label{RB271}
\end{rem}

\begin{lem}
The spectral flow $SF\left(D_{\{u(\phi)\}}\right)$ is equal to zero
for a family $D_{u(\phi)}$ of the Dirac operators parametrized
by a smooth map $\phi\colon S^1\to\left\{g,\left(\nabla^E,h\right)\right\}$.
\label{LB275}
\end{lem}

\noindent{\bf Proof.} Theorem~7.4 from \cite{APS3} (p.~94) computes
the spectral flow $SF\left(D_{\{y\}}\right)$ for a family $D_{\{y\}}$
of Dirac operators parametrized by a circle. Namely for Dirac operators
$D_y=D_y\left(M_y,E|_{M_y}\right)$ on the fibers of a smooth fibration
$\pi\colon P\to S^1$ with closed spinor, odd-dimensional and oriented
fibers $M_y:=\pi^{-1}(y)$. These Dirac operators act in
$\Gamma\left(F_y\otimes E|_{M_y}\right)$, where $F_y$ is the spinor bundle
on $M_y$ and $E\to P$ is a Hermitian vector bundle with a unitary connection
$\nabla^E$. This theorem claims that $SF\left(D_{\{y\}}\right)=-\ind D_+$,
where $D_+\colon\Gamma(F_+\otimes E)\to\Gamma(F_-\otimes E)$ is the Dirac
operator on an even-dimensional spinor manifold $P$ (with orientation
$(\df_y,\bold{e})$, where $\bold{e}$ is an orientation basis in $T_xM_y$).
Hence by the Atiyah-Singer index theorem we have
\begin{equation}
SF\left(D_{\{y\}}\right)=-\left(A\left(T(P)\right)\ch(\tilde{E})\right)
[P],
\label{A1552}
\end{equation}
where $A$ is the $A$-genus, $T(P)$ is the tangent bundle.
In our case $P=M_{y_0}\times S^1$
is a canonical direct product and $\tilde{E}=\pi^*_2E_{y_0}$
(for the projection $\pi_2\colon P\to M_{y_0}$). Hence the right side
in (\ref{A1552}) is equal to zero.\ \ \ $\Box$

Thus we obtain the following result.

\begin{thm}
Let $D_{\{u\}}$ be a family of Dirac operators
$D_u=D_u\left(M,\nabla_u^E\right)$ on an odd-dimensional closed spinor
manifold $(M,g)$ corresponding to a Hermitian structure $h_u$ on a vector
bundle $E\to M$ and to unitary connections $\nabla_u^E$ on $(E,h)$.
Then there exists a function $\eps(u)=\pm 1$ defined for $u$ such that
$\Ker D_u\ne 0$, continuous and locally constant for these $u$, and such
that
\begin{equation}
{\det}_{(\pi)}\left(D_{u_1}D_{u_2}\right)=\eps(u_1)\eps(u_2)\left({\det}
_{(\pi)}\left(D_{u_1}^2\right)\right)^{1/2}\left({\det}_{(\pi)}\left(D_{u_2}^2
\right)\right)^{1/2}
\label{A1554}
\end{equation}
for all pairs $u_1$, $u_2$ of sufficiently close parameters in the family
$D_{\{u\}}$. (The square roots on the right in (\ref{A1554}) are
arithmetical.)
\label{TA1553}
\end{thm}

\begin{rem}
For a family of Dirac operators $D_u$ on an odd-dimensional closed
spinor manifold $M$ the expression on the right in (\ref{B272})
makes sense for all pairs $(u_1,u_2)$,
$u_j\in\left\{g,\left(\nabla^E,h\right)\right\}$ according
to Remarks~\ref{RB270}, \ref{RB271}, and to Lemma~\ref{LB275}.
However we don't claim that the expression on the left in (\ref{B272})
also makes sense. The expression on the right in (\ref{B272}) may be
proposed as one of possible definitions of $\det\left(D_{u_1}D_{u_2}\right)$
for a pair $\left(D_{u_j}\right)$ of Dirac operators corresponding
to $(M,E,g_j,h_j)$.
\label{RB276}
\end{rem}

\subsection{Determinants of products of odd class elliptic operators}
Note that the assertion (\ref{B272}) can be generalized as follows.

\begin{pro}
Let $D_u$, $u\in U$,
be a smooth family of elliptic differential operators
acting on the sections $\Gamma(E)$ of a smooth vector bundle $E$
over a closed odd-dimensional manifold $M$. Let $D_u$ be (formally)
self-adjoint with respect to a scalar product on $\Gamma(E)$ defined
by a smooth positive density $\rho(u)$ on $M$ and by a Hermitian structure
$h(u)$ on $E$. Let for any pair $u_1,u_2\in U$ the principal symbol
$\sigma_2\left(D_{u_1}D_{u_2}\right)(x,\xi)$ be sufficiently close
to a positive definite self-adjoint symbol. Then the equality
(\ref{B272}) with $\eps(u_j)=\pm 1$ holds
for ${\det}_{(\tpi)}\left(D_{u_1}D_{u_2}\right)$.
The elliptic operator $D_{u_1}D_{u_2}$ for general $(u_1,u_2)$
is {\em not a self-adjoint} but (according to (\ref{B272})) its determinant
is a {\em real} number for sufficiently close $u_1$, $u_2$.
Remark~\ref{RB270} is also true for the family $D_u$, $u\in U$.
\label{PB3770}
\end{pro}

\begin{pro}
The equality analogous to (\ref{B272}) is also valid for a smooth
family $D_u$, $u\in U$, of (formally)\;self-adjoint elliptic PDOs
from $\Ell_{(-1)}^m(M,E)\!:=\!\Ell^m\!(M,E)\cap CL_{(-1)}^m(M,E)$,
$m\in\wZ_+$.
Then the principal symbols $\sigma_{2m}\left(D_{u_1}D_{u_2}\right)$
are sufficiently close to positive definite ones for $\left(u_1,u_2\right)$
in a neighborhood of the diagonal in $U\times U$. So in particular
for such $\left(u_1,u_2\right)$ we have
$\det_{(\tpi)}\left(D_{u_1}D_{u_2}\right)\in\wR$. (But $D_{u_1}D_{u_2}$
is not self-adjoint in general.) However, in this case (as well
as for families $D_u$ from Proposition~\ref{PB3770}), the assertion
of Lemma~\ref{LB275} is not valid in general. So in these cases
the appropriate spectral flows are not identity zeroes. Hence in general
the factors $\eps(u)$ in (\ref{B272}) are not globally defined
for such families.
\label{PB3771}
\end{pro}

\begin{pro}
Let $D_u$, $u\in U$, be a smooth family of PDOs
from $\Ell_{(-1)}^m(\!M;\!E,\!F\!)$, $m\in\wZ_+$. (This class consists
of elliptic PDOs acting from $\Gamma(E)$ to $\Gamma(F)$ and such that
their symbols possess the property analogous to (\ref{B240}).)
Suppose that a smooth positive density on $M$ and Hermitian structures
on $E$, $F$ are given.
Then a family $V_u:=D_u^*$ (adjoint to the family $D_u$) is defined,
$V_u$ is a smooth family from $\Ell_{(-1)}^m(M;F,E)$.
The assertion analogous to (\ref{B272}) is valid in the form
\begin{equation}
{\det}_{(\tpi)}\left(V_{u_1}D_{u_2}\right)=\eps\left(u_1\right)\eps\left(u_2
\right)\left({\det}_{(\pi)}\left(V_{u_1}D_{u_1}\right)\right)^{1/2}\left(
{\det}_{(\pi)}\left(V_{u_2}D_{u_2}\right)\right)^{1/2}
\label{B3727}
\end{equation}
for any sufficiently close $u_1,u_2\in U$. The factors $\eps\left(u_j\right)$
in (\ref{B3727}) are $\pm 1$. However, they are not globally defined
on $U$ for a general family $D_u$.
\label{PB3772}
\end{pro}

\noindent{\bf Proof of Proposition~\ref{PB3772}.}
Set
$A_{u_1,u_2}:=\det_{(\tpi)}\left(V_{u_1}D_{u_2}\right)$ for $u_1$, $u_2$
from a sufficiently close neighborhood of the diagonal in $U\times U$.
(For such $\left(u_1,u_2\right)$ the principal symbol of $V_{u_1}D_{u_2}$
is sufficiently close to positive definite definite ones.) Then
the matrix $\left(A_{u_1,u_2}\right)$ (for such pairs
$\left(u_1,u_2\right)$) has the rank one. Indeed, for any four
sufficiently close $\left(u_1,u_2,u_3,u_4\right)$ such that $D_{u_j}$,
$V_{u_j}$ are invertible we have by (\ref{B263})
\begin{multline*}
A_{u_1,u_2}A_{u_3,u_4}={\det}_{(\tpi)}\left(V_{u_1}D_{u_2}V_{u_3}D_{u_4}
\right)={\det}_{(\tpi)}\left(D_{u_4}V_{u_1}D_{u_2}V_{u_3}\right)= \\
={\det}_{(\tpi)}\left(D_{u_4}V_{u_1}\right){\det}_{(\tpi)}\left(D_{u_2}
V_{u_3}\right)=A_{u_1,u_4}A_{u_3,u_2}.
\end{multline*}
Hence we have
$$
{\det}_{(\tpi)}\left(V_{u_1}D_{u_2}\right)=:k\left(u_1\right)k\left(u_2
\right), \quad {\det}_{(\tpi)}\left(V_{u_1}D_{u_1}\right)=k\left(u_1\right)^2.
$$
The proposition is proved.\ \ \ $\Box$

\begin{rem}
A geometrical origin of Dirac operators manifests itself
in the structure of the determinants of their products (Theorem~\ref{TA1553},
(\ref{A1554})). (Here, $M$ is odd-dimensional.) Namely the structure
of the expression for such determinants of products of odd class
elliptic PDOs on $M$ (Proposition~\ref{PB3772}, (\ref{B3727})).
However the factors $\eps\left(u_j\right)$ in (\ref{B3727}) cannot
in general be globally defined. Indeed, in general the spectral flow
for a family of odd class elliptic PDOs on $M$ (parametrized by a circle
$S^1$) is nonzero. So for such a family the multiple $\eps\left(u_t\right)$
cannot be defined as a locally constant function of $t\in S^1$ such
that the corresponding operators are invertible. The corresponding
spectral flow for a family of Dirac operators on $M$ (parametrized
by $S^1$) is zero (Lemma~\ref{LB275}). This fact is connected with
the geometrical origin of Dirac operators.
\label{RB3835}
\end{rem}

\subsection{Determinants of multiplication operators}
Let $M$ be an odd-dimensional closed manifold. Let $E$ be
a finite-dimensional smooth vector bundle over $M$. Let $Q\in\End E$
be a smooth fiberwise endomorphism of $E$ such that for any $x\in M$
all the eigenvalues $\lambda_i(Q_x)$ possess the property
\begin{equation}
\left|\Im\lambda_i(Q_x)\right|<\pi-\eps, \eps>0.
\label{B279}
\end{equation}
Then for all the eigenvalues $\lambda_i\left(\exp(tQ_x)\right)$
for $0\le t\le 1$ we have
$$
\left|\arg\lambda_i\left(\exp(tQ_x)\right)\right|<\pi-\eps.
$$
The determinant $\det_{(\pi)}(\exp Q)$ is defined according to (\ref{B244})
by
\begin{equation}
{\det}_{(\pi)}(\exp Q):={\det}_{(\pi)}(A\exp Q)\big/{\det}_{(\pi)}(A),
\label{B280}
\end{equation}
where $A:=\Delta+\Id$.%
\footnote{We use the fact that a principal symbol of the Laplacian
is scalar. Namely
$$
\sigma_2(\Delta)(x,\xi)=\sigma_2(\Delta_M)(x,\xi)\otimes\Id_E,
$$
where $\Delta_M$ is the Laplacian for scalar functions on $(M,g)$.
Hence $\sigma_2(\exp(tQ)\cdot\Delta)$ possesses a cut $L_{(\pi)}$
for $0\le t\le 1$.}
Here, $\Delta$ is the Laplacian
$\Delta:=\Delta_{g,\nabla^E}$ on $\Gamma(E)$ corresponding
to a Riemannian metric $g$ on $M$ and to a unitary connection $\nabla^E$
on $(E,h)$. Then we have
\begin{equation}
{\det}_{(\pi)}(\exp Q):={\det}_{(\pi)}\left(\exp Q\cdot(\Delta+\Id)\right)
\big/{\det}_{(\pi)}(\Delta+\Id).
\label{B281}
\end{equation}

For $0\le t\le 1$ we have an analogous definition
\begin{equation}
{\det}_{(\pi)}\left(\exp(tQ)\right):={\det}_{(\pi)}\left(\exp(tQ)A\right)
\big/{\det}_{(\pi)}(A)=:F(Q,t).
\label{B282}
\end{equation}

Thus we have $F(Q,0)=1$,
\begin{multline}
\df_t\log F(Q,t)=-\df_t\df_s\Tr\left(\left(\exp(tQ)\cdot(\Delta+\Id)\right)
^{-s}\right)\big|_{s=0}=\\
=(1+s\df_s)\Tr\left(Q\left(\exp(tQ)\cdot(\Delta+\Id)\right)^{-s}\right)
\big|_{s=0}=\int_M\tr\left(Q(x)K_{t,s}(x,x)\right)\big|_{s=0},
\label{B283}
\end{multline}
where $K_{t,s}(x,x)$ is an analytic continuation in $s$ from the domain
$\Re s>\dim M/2$ of the restriction to the diagonal the kernel
of $\left(\Delta+\Id\right)^{-s}$. Set $A_{1,t}:=\exp(tQ)\cdot A_1$.
Then we have (\cite{Se}, \cite{Gr})
$$
K_{t,s}(x,x)\big|_{s=0}=a_0(x,A_t),
$$
where $a_0$ is the $\tau^0$-coefficient in the asymptotic expansion
as $\tau\to+0$ for the kernel on the diagonal $P_{\tau,t}(x,x)$
of the operator $\exp(-\tau A_t)$.
Since $A_t$ is an elliptic DO of the second order and since all the real
parts of all the eigenvalues $\lambda_i\left(\sigma_2(A_t)(x,\xi)\right)$
are positive (for $\xi\ne 0$), there is (\cite{Gr}) an asymptotic
expansion as $\tau\to+0$
\begin{multline}
P_{\tau,t}(x,x)\sim a_{-n}(x,A_t)\tau^{-n/2}+a_{-(n-2)}(x,A_t)\tau^{1-n/2}+
\ldots+\\
+a_{-1}(x,A_t)\tau^{-1/2}+a_1(x,A_t)\tau^{1/2}+\ldots
\label{B286}
\end{multline}
Hence we have
\begin{gather}
a_0(x,A_t)=0, \qquad \df_t\log F(Q,t)=0,
\label{B284}\\
{\det}_{(\pi)}\exp(Q)=1.
\label{B285}
\end{gather}

Thus we obtain the following.
\begin{pro}
The determinant of a multiplication operator
on an odd-dimensio-nal
closed manifold is equal to one.
\label{PA1552}
\end{pro}

\begin{rem}
To see that the coefficients of $\tau^{-(1-n)/2}$, $\tau^{-(3-n)/2}$, \dots
in (\ref{B286}) are zero, it is enough to note that the coefficient
of $\tau^{-(j-n)/2}$ is defined by a noncommutative residue density
$\res_x$ of the symbol $\sigma\left(A_t^{-(j-n)/2}\right)$ (\cite{Sh},
Ch.~II, (12.5)).
If $n-j/2=k\in\wZ_+$, then we have
$$
\sigma_{-2k-j}\left(A_t^{-k}\right)(x,\xi)={i\over 2\pi}\int_{\Gamma_{(\pi)}}
\lambda^{-k}\sigma_{-2-j}\left(\left(A_t-\lambda\right)^{-1}\right)(x,\xi).
$$
Since $A_t\in CL_{(-1)}^2(M,E)$, we obtain (as in the proof
of Theorem~\ref{TB241})
$$
\sigma_{-2k-j}\left(A_t^{-k}\right)(x,-\xi)=\left(-1\right)^j\sigma_{-2k-j}
\left(A_t^{-k}\right)(x,\xi).
$$
Hence $\res_x\sigma\left(A_t^{-k}\right)=0$.
\label{RB287}
\end{rem}

\subsection{Absolute value determinants}
Let $A$ be an invertible elliptic differential operator
on an odd-dimensional closed manifold $M$, $A\in\Ell^d(M,E)\subset CL^d(M,E)$,
$d\in\wZ_+$. Then we can define
\begin{equation}
|\det|A:=\left(\det(A^*A)\right)^{1/2}\in\wR_+,
\label{B292}
\end{equation}
where $A^*$ is adjoint to $A$ with respect to a scalar product
on $\Gamma(E)$ defined by a smooth positive density $\rho$ on $M$
and by a Hermitian structure $h$ on $E$.

\begin{rem}
The determinant on the right in (\ref{B292}) is independent of $\rho$
and of $h$.

Indeed, let a pair $(\rho_1,h_1)$ be changed by $(\rho_2,h_2)$. Then
\begin{equation}
A_{\rho_2,h_2}^*=Q^{-1}A_{\rho_1,h_1}^*Q,
\label{B293}
\end{equation}
where $Q\in\Aut\left(E\otimes{\wedge}^nT^*M\right)$, $n=\dim M$,
is defined by $(f_1,f_2)_{\rho_2,h_2}=(f_1,Qf_2)_{\rho_1,h_1}$.
The operator $Q$ belongs to $CL_{(-1)}^0(M,E)$ and for $(\rho_2,h_2)$
close to $(\rho_1,h_1)$ this operator is close to $\Id$. Hence
for $(\rho_2,h_2)$ close to $(\rho_1,h_1)$ we have by Theorem~\ref{TB241}
and by Proposition~\ref{PA1552}
\begin{multline}
{\det}_{(\pi)}\left(A_{\rho_2,h_2}^*A\right)={\det}_{(\pi)}\left(Q^{-1}
A_{\rho_1,h_1}^*QA\right)=\\
={\det}_{(\pi)}\left(A_{\rho_1,h_1}^*QA\right)={\det}_{(\pi)}\left(QAA_{\rho
_1,h_1}^*\right)=\\
={\det}_{(\pi)}\left(AA_{\rho_1,h_1}^*\right)={\det}_{(\pi)}\left(A_{\rho_1,
h_1}^*A\right).
\label{B294}
\end{multline}

The equality (\ref{B294}) was obtained in \cite{Sch}.
\label{RB291}
\end{rem}

\begin{pro}
The functional $A\to|\det|A$ is multiplicative, i.e., for a pair
$(A,B)$ of elliptic differential operators in $\Gamma(E)$
on an odd-dimensional closed $M$ we have
\begin{equation}
|\det|(AB)=|\det|A\cdot|\det|B.
\label{B295}
\end{equation}
\label{PB292}
\end{pro}

\noindent{\bf Proof.} By Theorem~\ref{TB241} we have the following
expression for $\left(|\det|(AB)\right)^2$
\begin{multline}
{\det}_{(\tpi)}\left(B^*A^*AB\right)=
{\det}_{(\pi)}\left(BB^*A^*A\right)={\det}_{(\pi)}\left(A^*A
\right){\det}_{(\pi)}\left(BB^*\right)=\\
=\left(|\det|A\cdot|\det|B\right)^2.
\label{B3789}
\end{multline}
\ \ \ $\Box$

\begin{rem}
All the assertions about absolute value determinants given above
are true also for elliptic PDOs from $CL_{(-1)}^m(M,E)$,
$m\in\wZ$, on a closed odd-dimensional manifold $M$.

For $m=0$ the operator $A^*A$, where $A\in CL_{(-1)}^0(M,E)$,
is a self-adjoint positive definite PDO from $CL_{(-1)}^0(M,E)$.
Hence its determinant is defined by (\ref{B244}) as
\begin{equation}
\left(|\det|A\right)^2:={\det}_{(\pi)}(A^*A):={\det}_{(\pi)}\left(A^*A
\left(\Delta_E+\Id\right)\right)\big/{\det}_{(\pi)}\left(\Delta_E+\Id\right),
\label{B401}
\end{equation}
where $\Delta_E$ is the Laplacian for $(M,g,E,\nabla_E,h)$ ($\nabla_E$
is an $h$-unitary connection on $(E,h)$ and $g$ is a Riemannian metric
on $M$).

Thus absolute value determinants are defined for all elliptic PDOs
$A$ of odd class $CL_{(-1)}^{\bullet}(M,E)\cap\Ell^{\bullet}(M,E)$
on a closed odd-dimensional $M$.
All the assertions about $|\det|A$ given above are true for such PDOs $A$.
\label{RB400}
\end{rem}

Let $A$ be an invertible elliptic PDO from $\Ell_{(-1)}^m(M;E,F)$,
$m\in\wZ_+$. (This class is introduced in Proposition~\ref{PB3772}.)
Let a smooth positive density $\rho$ on $M$ and Hermitian structures
$h_E$, $h_F$ on $E$, $F$ be defined. Then the operator $A^*$ is defined,
and the absolute value determinant of $A$ is defined by
$$
|\det|A:=\left(\det\left(A^*A\right)\right)^{1/2}\in\wR_+.
$$

\begin{rem}
This absolute value determinant is independent of $\rho$, $h_E$, $h_F$.
Indeed, under small deformations of $\rho$, $h_E$, $h_F$, the operator
$A^*$ transforms to $Q_1A^*Q_2$, where $Q_j$ are the automorphism
operators of the appropriate vector bundles and $Q_j$ are sufficiently
close to $\Id$. So by Proposition~\ref{PA1552} and by Theorem~\ref{TB241}
we can produce equalities similar to (\ref{B294}). Hence
$\det\left(A^*A\right)$ is independent of $\left(\rho,h_E,h_F\right)$.
(Note that the set of $\left(\rho,h_E,h_F\right)$ is convex and so
it is a connected set.)
\label{RB3781}
\end{rem}

\begin{pro}
An absolute value determinant is multiplicative, i.e., for invertible
elliptic PDOs of the odd class $A\in\Ell_{(-1)}^{m_1}\left(M;E,F_1\right)$,
$B\in\Ell_{(-1)}^{m_2}\left(M;F_1,F_2\right)$ we have
\begin{equation}
|\det|AB=|\det|A\cdot|\det|B.
\label{B3784}
\end{equation}
\label{PB3782}
\end{pro}

\noindent{\bf Proof.} The equalities (\ref{B3789}) are applicable
in this case.\ \ \ $\Box$

\begin{rem}
The absolute value determinant $|\det|A$ is canonically defined
for $A\in\Ell_{(-1)}^m(M;E,F)$ for any $m\in\wZ$. Indeed, this
determinant is defined for $m\in\wZ\setminus 0$. For $m=0$ it is
defined by (\ref{B401}). The multiplicative property (\ref{B3784})
holds for absolute value determinants of the odd class elliptic PDOs
on an odd-dimensional closed manifold having arbitrary orders.
\label{RB3783}
\end{rem}

\subsection{A holomorphic on the space of PDOs determinant and its
monodromy}
\label{SC4}

\begin{pro}
The function $\left(|\det|A\right)^2$ on the space
$\Ell^m(M,E)\cap CL_{(-1)}^m(M,E)$, $m\in\wZ$, on an odd-dimensional
closed manifold $M$
is equal to $\left|f(A)\right|^2$, where $f$ is a multi-valued
analytic function on the space of elliptic pseudo-differential operators
from $CL_{(-1)}^m(M,E)$, i.e., on $\Ell^m(M,E)\cap CL_{(-1)}^m(M,E)$.
\label{PB1801}
\end{pro}

\begin{pro}
The assertion $(|\det|A)^2=|f(A)|^2$ of Proposition~\ref{PB1801} holds
for $A\in\Ell_{(-1)}(M;E,F)$. (This class is introduced
in Proposition~\ref{PB3772}.) Here, $f(A)$ is a holomorphic in $A$
multi-valued function on the space $\Ell_{(-1)}^m(M;E,F)$.
\label{PB3790}
\end{pro}

The proofs of these propositions are in the end of this subsection.

\begin{rem}
A natural complex structure on the space
$\Ell_{(-1)}^m\!(\!M,\!E\!)\!:=\!CL_{(-1)}^m\!(\!M,\!E\!)\cap\Ell^m(M,E)=:X$
is  defined as follows.
Note that $X$ is a fiber bundle over the space of principal
symbols $\SEll_{(-1)}^m(M,E)/CS_{(-1)}^{m-1}(M,E):=ps_{(-1)}^m(M,E)$.
Its fiber
is the space $\Ell_{\Id,(-1)}^0(M,E):=\Id+CL_{(-1)}^{-1}(M,E)$ of zero order
elliptic operators with the principal symbol $\Id$. The fiber has
a natural structure of an affine linear space over $\wC$. Let the order
$m$ be even. Then the complex structure
on $ps_{(-1)}^m(M,E)=\Aut\left(\pi^*E|_{P^*M}\right)$%
\footnote{Here, $P^*M:=\Ass\left(T^*M,RP^{n-1}\right)$, and
$\pi\colon P^*M\to M$ is a natural projection.}
is induced by complex linear structures
of fibers $\pi^*E|_{P^*M}$. Let $s\in ps_{(-1)}^m(M,E)$. Then
$T_s\Aut\left(\pi^*E|_{P^*M}\right)=\End\left(\pi^*E|_{P^*M}\right)$
has a natural structure of an infinite-dimensional space $\wC$.
(Any $v\in\End\left(\pi^*E|_{P^*M}\right)$ defines the tangent vector
$vs\in T_s\Aut\left(\pi^*E|_{P^*M}\right)$.) This complex structure
on the tangent bundle to the group $\Aut\left(\pi^*E|_{P^*M}\right)=:G$
is invariant under right multiplications $vs\to vss_1$ and under
left multiplications $vs\to\Ad_{s_1}v\cdot s_1s$ on elements $s_1\in G$.
So $\Aut\left(\pi^*E|_{P^*M}\right)$ is an infinite-dimensional complex
manifold.

Let $X^{\times}$ be the space of invertible elliptic operators
from $X$. Then the group $H^{\times}:=\Ell_{\Id,(-1)}^{0,\times}(M,E)$
of invertible operators from $\Ell_{\Id,(-1)}^{0,\times}(M,E)$ acts
on $X^{\times}$ from the right, $R_h\colon x\to xh$ for $h\in H^{\times}$,
$x\in X^{\times}$. This action defines a principal fibration
$q\colon X^{\times}\to G$ with the fiber $H^{\times}$. The group
$H^{\times}$ acts from the left on $\Ell_{\Id,(-1)}^0(M,E)$,
$L_h\colon y\to h^{-1}y$, and $X$ is canonically the total space
of the bundle associated with the principal bundle $q$. The complex
structure on $\Ell_{\Id,(-1)}^0(M,E)$ (defined by a natural
$\wC$-structure on $CL_{(-1)}^{-1}(M,E)$) is invariant under this
action of $H^{\times}$.

The natural complex structure on $T\left(X^{\times}\right)$ is defined
by the natural $\wC$-structure on $CL_{(-1)}^0(M,E)=T_{\Id}X_0^{\times}$
(where $X_0^{\times}$ corresponds to the case $m=0$)
under the identification $T_AX^{\times}\rs T_{\Id}X_0^{\times}$,
$\delta A\in T_AX^{\times}\to\delta A\cdot A^{-1}\in T_{\Id}X_0^{\times}$.
(Here, $A\in X^{\times}$.)
The complex structure on $T_{\Id}X_0^{\times}$
is invariant under the adjoint action of the $X_0^{\times}$
on $T_{\Id}X_0^{\times}$. Hence $X_0^{\times}$ is an analytic
infinite-dimensional manifold.
The complex structure on $T\left(X_0^{\times}\right)$ (induced
from $X_0^{\times}$) is invariant under the natural left and right
actions of the elements of $X_0^{\times}$ on $X^{\times}$.
So $X^{\times}$ possesses a natural structure
of an infinite-dimensional complex manifold. This complex structure
together with the complex structure on the fibers
$\Ell_{\Id,(-1)}^0(M,E)$ of the associated vector bundle defines
a natural structure of an infinite-dimensional complex manifold on $X$.
This structure is in accordance with the complex structures
on the base $G$ and on the fibers $\Ell_{\Id,(-1)}^0(M,E)$ of the natural
fibration $X\to G$.

Let $m\in\wZ$ be odd. Then any invertible elliptic operator $A\in X_m$
gives us the isomorphism $A^{-1}\colon X_m\rs X_0$, $x\to A^{-1}x$.
The complex structure on $X_0$ defines a complex structure on $X_m$.
The induced complex structure on $X_m$ is independent of an invertible
operator $A$ from $X_m$.

A natural complex structure on $\Ell_{(-1)}^m(M;E,F)$ is induced
by the identification
\begin{equation}
\Ell_{(-1)}^m(M,E)\rs\Ell_{(-1)}^m(M;E,F)
\label{B3791}
\end{equation}
given by multiplying by an invertible operator $A\in\Ell_{(-1)}^0(M;E,F)$.
\label{RB3600}
\end{rem}

\begin{pro}
Let a branch of the holomorphic determinant $f(A)$, Proposition~\ref{PB1801},
be equal to $\det_{(\tpi)}\left(\Delta_E^m+\Id\right)$ at the point
$A_0:=\Delta_E^m+\Id\in\Ell_{(-1),0}^{2m}(M,E)$. (This can be done
because the operator $\Delta_E^m+\Id$ is self-adjoint and positive
definite.) Then for any element $A$ of $\Ell_{(-1),0}^{2m}(M,E)$
sufficiently close to positive definite self-adjoint ones (with respect
to a given smooth positive density on $M$ and a Hermitian structure
on $E$) we have
\begin{equation}
{\det}_{(\tpi)}(A)=f(A).
\label{B4003}
\end{equation}
(Here, $\det_{(\tpi)}$ is defined by an admissible for $A$ cut.)
\label{PB4002}
\end{pro}

\begin{cor}
The equality (\ref{B4003}) holds for PDOs $A$ sufficiently close
to a positive definite self-adjoint PDO (with respect to any smooth
density and any Hermitian structure).
\label{CB4004}
\end{cor}

\noindent{\bf Proof.} For $\Delta_E^m$ defined by any smooth density
and any Hermitian structure we have
$$
|\det|\left(\Delta_E^m+\Id\right)={\det}_{(\tpi)}\left(\Delta_E^m+\Id\right).
$$
The set $\cal{D}$ of these operators is connected
in $\Ell_{(-1),0}^{2m}(M,E)$. A branch of $f(A)$ and
$\det_{(\tpi)}\left(\Delta_E^m+\Id\right)$ are restrictions to this set
of holomorphic functions which are equal in a neighborhood of a point
$A_0\in\cal{D}$. Hence these functions are equal on $\cal{D}$. Then
we can apply Proposition~\ref{PB4002} for any Riemannian and Hermitian
structures.\ \ \ $\Box$

The statement of Proposition~\ref{PB4002} follows immediately
from Theorem~\ref{TB241}, Remark~\ref{RB287}, (\ref{B295}),
Corollary~\ref{CB4007}, (\ref{B4011}), Remark~\ref{RB4009}, (\ref{B4010}),
or from Lemma~\ref{LB1991} (and from its proof (\ref{B1970})) below.

\begin{lem}
The monodromy of the functions $f(A)$ defined in Propositions~\ref{PB1801},
\ref{PB3790} is given by a homomorphism
$$
\phi\colon K\left(\Ass\left(T^*M,RP^{n-1}\right)\right)/\pi^*K(M)\to\wC
^{\times},
$$
where $K=K^0$ is the topological $K$-functor,
$\pi\colon\Ass\left(T^*M,RP^{n-1}\right)\to M$ is a fiber bundle with its
fiber $RP^{n-1}$ associated with $T^*M^n$ ($n:=\dim M$).
\label{LB1810}
\end{lem}

\noindent{\bf Proof of Lemma~\ref{LB1810}.} First we prove this assertion
for $f(A)$ defined on $\Ell_{(-1)}^m(\!M,\!E\!)$.
By the multiplicative property (\ref{B295})
of the absolute value determinants, it is enough to investigate monodromy
of $f(A)$ over a closed loop $A_t$ in $\Ell_{(-1)}^{2k}(M,E)$ ($k\in\wZ_+$
is fixed). Let $E_1$ be a smooth bundle over $M$ such that $E\oplus E_1$
is isomorphic to a trivial $N$-dimensional complex vector bundle $1_N$,
where $N\in\wZ_+$ is large enough. Then the monodromy of $f(A)$ over
a loop $\left(A_t\right)\in\Omega^1\Ell_{(-1)}^{2k}(M,E)$ is the same
as the monodromy of $f(A)$ for $\left(M,1_N\right)$ over a loop
$\left(A_t\oplus\left(\Delta_{E_1}+\Id\right)^k\right)\in\Omega^1\Ell_{(-1)}
^{2k}(M,1_N)$. (Indeed,
$f\left(A_t\oplus\left(\Delta_{E_1}+\Id\right)^k\right)=cf\left(A_t\right)$,
where $c\ne 0$ is independent of $t\in[0,1]$ and is defined up to a constant
complex factor of absolute value one. We can set
$c:=\det_{(\pi)}\left(\left(\Delta_{E_1}+\Id\right)^k\right)$.)

The group $K^1\left(\Ass\left(T^*M,RP^{n-1}\right)\!\right)$ is
in one-to-one correspondence with the connected components of the space
$\Ell_{(-1)}^m(M,1_N)$ of elliptic PDOs from $CL_{(-1)}^m(M,1_N)$
($m$ is fixed).
The fundamental group
$\pi_1\left(\Ell_{(-1)}^{2k}\left(M,1_N\right)\right)=\pi_1\left(\Ell_{(-1)}^0
\left(M,1_N\right)\right)$ can be interpreted as follows.
Let $P\in CL_{(-1)}^0\left(M,1_N\right)$ be a PDO-projector, i.e.,
$P^2=P\in CL^0$ and its symbol $\sigma(P)$ belong to an odd class
(\ref{B240}). (To remind, for $\sigma(P)$ to be of this odd class,
it is enough for all the homogeneous components of $\sigma(P)$ to satisfy
(\ref{B240}) in some local cover of $M$ by coordinate charts.)
The one-parametric cyclic subgroups $\exp(2\pi itP)$, $0\le t\le 1$,
($\exp(2\pi iP)=\Id$) are the generators
of $\pi_1\left(\Ell_{(-1)}^0\left(M,1_N\right)\right)$.

Indeed, it follows from the Bott periodicity that
$K^0\left(\Ass\left(T^*M,RP^{n-1}\right)\right)$ is canonically
identified with $\pi_1\left(GL_N(C(X))\right)$
for $X:=\Ass\left(T^*M,RP^{n-1}\right)$ and for $N\in\wZ_+$ large enough
(\cite{Co}, II.1). Here, $C(X)$ is an algebra of continuous functions.
Any continuous map $\phi\colon X\times S^1\to GL_N(\wC)$ such that
$\phi(X\times a)=\Id$, $a\in S^1$ is fixed, is homotopic
to a $C^\infty$-map in this class of continuous maps.
So $K^0(X)$ (for $X=\Ass\left(T^*M,RP^{n-1}\right)$) is the fundamental
group of the space of principal symbols for operators
from $\Ell_{(-1)}^0\left(M,1_N\right)$. Finite type projective
modules over $C(X)=:A$ correspond canonically to finite rank vector
bundles over $X$. For every such a module $\cal{E}$ there exists
a projector $e\in M_N(A)$, $e^2=e$, such that
$\cal{E}\approx\left\{f\in M_N(A),ef=f\right\}$ (as a right $A$-module,
$A:=C(X)$). Such a projector corresponds to a projection
$p\in\End\left(\pi^*1_N\right)$ from $\pi^*1_N$ onto a finite rank
vector bundle over $X$. Every such projection $p$ is homotopic
to a $C^\infty$-projection. The space of elliptic operators
$\Ell_{(-1)}^0\left(M,1_N\right)$ is homotopic to the space
of their principal symbols. For every smooth projection
$p\in\End\left(\pi^*1_N\right)$ there exists a zero order PDO-projector
$f\in CL_{(-1)}^0\left(M,1_N\right)$ with the principal symbol $p$.
(For projectors $P$ from $CL^0\left(M,1_N\right)$ this assertion
is proved in \cite{Wo3}.)

The principal symbol $\sigma_0(P)=:p$ defines a fiber-wise projector
$p\in\End\left(\pi^*1_N\right)$ of a trivial vector bundle $\pi^*1_N$
over $\Ass\left(T^*M,RP^{n-1}\right)$, $p^2=p$. The image $\Im(p)$ of $p$
is a smooth vector subbundle of $\pi^*1_N$ and $\Im(p)$ represents
an element of $K^0\left(\Ass\left(T^*M,RP^{n-1}\right)\right)$ and
any element of this $K$-functor can be represented as $\Im(p)$
for a projector $p\in\End\left(\pi^*1_N\right)$, $p^2=p$,
under the condition that $N\in\wZ_+$ is large enough.

For any projector $p\in\End\left(\pi^*1_N\right)$ there is a PDO-projector
$P\in CL_{(-1)}^0\left(M,1_N\right)$ with $\sigma_0(P)=p$. (An analogous
result is obtained in \cite{Wo3}.) If $\Im(p)$ and $\Im\left(p_1\right)$
represent the same element
of $K^0\left(\Ass\left(T^*M,RP^{n-1}\right)\right)$, then these projectors
are homotopic (under the condition that $N\in\wZ_+$ is large enough
with respect to $\dim M$ and to $\rk(\Im(p))$). If the principal symbols
$p$ and $p_1$ of PDO-projectors $P$ and $P_1$
(from $CL_{(-1)}^0\left(M,1_N\right)$) are homotopic, then $\exp(2\pi itP)$
and $\exp\left(2\pi itP_1\right)$ define the same element
of $\pi_1\left(\Ell_{(-1)}^0\left(M,1_N\right)\right)$. Hence
the monodromy of $f(A)$ on $\Ell_{(-1)}^{2k}\left(M,1_N\right)$ ($k\in\wZ$)
defines a homomorphism
\begin{equation}
\phi_0\colon K^0\left(\Ass\left(T^*M,RP^{n-1}\right)\right)\to\wC^{\times}.
\label{B1970}
\end{equation}

Indeed, the value of $\phi_0[\Im p]$ for an element
$[\Im p]\in K^0\left(\Ass\left(T^*M,RP^{n-1}\right)\right)=:K$,
$p=\sigma_0(P)$ (for a PDO-projector $P$ from $CL_{(-1)}^0\left(M,1_N\right)$
and for $N\in\wZ_+$ large enough), is defined as the ratio
\begin{equation}
\exp(2\pi itP)\circ f_0(A)\big/f_0(A)\big|_{t=1}=:\phi_0([\Im p])\in\wC
^{\times}.
\label{B1971}
\end{equation}
Here, $f_0(A)$ is a branch of a multi-valued function $f(A)$ near $A_0$
and $\exp(2\pi itP)\circ f_0(A)|_{t=1}$ is the analytic continuation
of $f_0(A)$ along a closed curve $S_P:=\exp(2\pi itP)\cdot A_0$,
$0\le t\le 1$. This ratio is independent of a branch $f_0(A)$ of $f(A)$
since for any two branches $f_0(A)$ and $f'_0(A)$ of $f(A)$ (defined
for $A$ close to $A_0$) their ratio $f_0(A)/f'_0(A)$ is a complex
constant (with the absolute value equals one) and so the analytic
continuation of $f_0(A)/f'_0(A)$ along $S_P$ is the same constant.

We suppose from now on that $N\in\wZ_+$ is large enough. The homomorphism
$\phi_0$ is defined since the elements $[\Im p]$ span the group $K=K^0$
and since if $\Im p_1\oplus\Im p_2=\Im p_3$, then the curve
$\exp\left(2\pi itP_3\right)$, $0\le t\le 1$, represents the sum
in the commutative group
$\pi_1\left(\Ell_{(-1)}^0\left(M,1_N\right),\Id\right)$ ($=\pi_1\left(
\SEll_{(-1)}^0\left(M,1_N\right),\Id\right)$) of the elements
represented by the curves $\exp\left(2\pi itP_j\right)$, $0\le t\le 1$.
(Here, $\sigma_0\left(P_j\right)=p_j$.)

Let $\Im(p)$ belong to a subgroup $\pi^*K^0(M)$
of $K:=K^0\left(\Ass\left(T^*M,RP^{n-1}\right)\right)$ (i.e., there is
a smooth vector bundle $V$ over $M$ such that $\pi^*V$ represents
the same element of $K$ as $\Im(p)$ does). We can suppose that
$\Im(p)=\pi^*V$.
Indeed, let $\Im p\oplus\pi^*1_{N_1}=\pi^*V\oplus\pi^*1_{N_1}$. Then
this equality holds with $N_1\in\wZ_+$ bounded by a constant depending
on $\dim M$. We suppose that $N\in\wZ$ is large enough. Then
for a projector $p_1\in\End\left(\pi^*1_{N_1}\right)$ such that
$\Im p_1=\Im p\oplus\pi^*1_{N_1}\subset\pi^*1_N$ we can conclude that
$\Im p_1$ and $\pi^*\left(V\oplus 1_{N_1}\right)$ are smoothly isotopic
as subbundles of $\pi^*1_N$. (Here, $V\oplus 1_{N_1}$ is a subbundle
of $1_N$.) Hence the monodromies coincide
$\phi_0\left(\left[\Im p_1\right]\right)=\phi_0\left(\pi^*\left(V\oplus 1
_{N_1}\right)\right)$.

The assertion of Lemma~\ref{LB1810} follows from (\ref{B1970}) and
from Lemma~\ref{LB1989} below. The identification of monodromies
$f(A)$ over $\Ell_{(-1)}^m(M,E)$ and over $\Ell_{(-1)}^m(M;E,F)$
is given by the identification (\ref{B3791}) of these spaces and
by the multiplicative property of absolute value determinants
(Proposition~\ref{PB3782}, (\ref{B3784})).\ \ \ $\Box$

\begin{lem}
The monodromy $\phi_0([\Im p])$ (defined by (\ref{B1971}))
of the multi-valued holomorphic function $f(A)$ on $\Ell_{(-1)}^m(M,E)$
is equal to $1$ for $[\Im p]\in\pi^*K^0(M)$ on an odd-dimensional
manifold $M$.
\label{LB1989}
\end{lem}


\noindent{\bf Proof of Lemma~\ref{LB1989}.} We can suppose that
$\Im p=\pi^*V$ (as it is shown above). Then there is a smooth homotopy
of $p=\sigma_0(P)$ (in the class of projectors from
$\End\left(\pi^*1_N\right)$ with the rank equal to $\rk V$) to a projector
$p_0$ constant along the fibers of $\pi$, i.e., to $p_0=\pi^*p_M$
for a projector $p_M\in\End\left(1_N\right)$ over $M$ (where $N\in\wZ_+$
is large enough). Then $\exp(tp_0)$ is the symbol of the multiplication
operator $\exp\left(tp_M\right)\in\Aut\left(1_N\right)$ over $M$
($|t|$ is small). It is shown in Proposition~\ref{PA1552} that
$\det\left(\exp\left(tp_M\right)\right)$ is defined (for such $t$) and
that this determinant is equal to one.\ \ \ $\Box$

\begin{lem}
For elliptic PDOs $A$ from $\Ell_{(-1)}^d\left(M,1_N\right)$ sufficiently
close to positive definite self-adjoint ones (where $d$ is even and
nonzero), the locally defined branch $f_0(A)$ of a holomorphic in $A$
function $f(A)$ (from Proposition~\ref{PB1801}) is
\begin{equation}
f_0(A)=c\cdot{\det}_{(\tpi)}(A),
\label{B1992}
\end{equation}
where $c\in\wC$ is a constant such that $|c|=1$. (Here, ${\det}_{(\tpi)}(A)$
is the zeta-regularized determinant of $A$ defined by an admissible
spectral cut $L_{(\theta)}$ with $\theta$ close to $\pi$.)
\label{LB1991}
\end{lem}

\noindent{\bf Proof.} By Theorem~\ref{TB241}, Corollary~\ref{CB4007},
and Remark~\ref{RB4009} we have for such $A$
\begin{gather}
\begin{split}
{\det}_{(\tpi)}(A^*A) & ={\det}_{(\tpi)}(A^*){\det}_{(\tpi)}(A), \\
{\det}_{(\tpi)}(A^*)  & =\overline{\left({\det}_{(\tpi)}(A)\right)}, \\
{\det}_{(\tpi)}(A^*A) & =\overline{f_0(A)}f_0(A),
\end{split}
\label{B1993}
\end{gather}
where ${\det}_{(\tpi)}(A)$ and $f_0(A)$ are holomorphic in $A$ (and
$f_0(A)$ is locally defined). Hence locally we have
$f_0(A)=c{\det}_{(\tpi)}(A)$ with a constant $c$ whose absolute value
is equal to one.\ \ \ $\Box$

\begin{rem}
The assertion of Lemma~\ref{LB1991} is also true
for $A\in\Ell_{(-1)}^d(M,E)$.
\label{RB19911}
\end{rem}

\begin{rem}
It follows from (\ref{B1971}) and from (\ref{B1992}) that the monodromy
$\phi_0([\Im p])$ is given by the equality
\begin{equation}
\phi_0([\Im p])={\det}_{(\tpi)}(\exp(2\pi itP)\cdot A)\big/{\det}_{(\tpi)}(A)
\big|_{t=1}.
\label{B1995}
\end{equation}
Here, $A\in\Ell_{(-1)}^d\left(M,1_N\right)$ is an invertible elliptic
PDO close to a positive definite self-adjoint one, $d$ is even and
nonzero, $P$ is a PDO-projector from $CL_{(-1)}^0\left(M,1_N\right)$
with the principal symbol $\sigma_0(P)=p$, and
${\det}_{(\tpi)}(\exp(2\pi itP)\cdot A)$ is the analytic continuation
in $t$ of the zeta-regularized determinant ${\det}_{(\tpi)}$ from small
$t\in[0,1]$ to a point $t=1$.
\label{RB1994}
\end{rem}

\begin{rem}
The analytic continuation of ${\det}_{(\tpi)}(\exp(2\pi itP)\cdot A)$
to $t=1$ for a fixed $A=A_0$ depends on the homotopy class
of $[\Im p]\subset\pi^*1_N$ only. Indeed, it is equal (up to a constant
factor $c$, $|c|=1$, locally independent of $A$) to the analytic continuation
of a holomorphic in $A$ function $f_0(A)$%
\footnote{$f_0(A)$ is a branch of a holomorphic
on $\Ell_{(-1)}^m\left(M,1_N\right)$ multi-valued function $f(A)$
locally defined near $A_0$.}
along the closed curve
$\exp(2\pi itP)\cdot A_0$ in the space $\Ell_{(-1)}^d\left(M,1_N\right)$
($d\in 2(\wZ\setminus 0)$). Hence it depends on the homotopy class
of a closed curve in this space from a fixed point $A_0$. Such homotopy
classes are defined by homotopy classes of $[\Im p]\subset\pi^*1_N$.
\label{RB1996}
\end{rem}

\begin{rem}
By Theorem~\ref{TB241} and by Corollary~\ref{CB243} we have for small
$|t|$, $t\in\wC$,
\begin{equation}
{\det}_{(\tpi)}(\exp(tP)A)={\det}_{(\tpi)}(\exp(tP)){\det}_{(\tpi)}(A).
\label{B1997}
\end{equation}
Here, $A\in\Ell_{(-1)}^d\left(M,1_N\right)$ is sufficiently close
to a positive definite self-adjoint PDO, $d\in 2(\wZ\setminus 0)$,
$P$ is a PDO-projector from $CL_{(-1)}^0\left(M,1_N\right)$.
The determinant of the zero order PDO $\exp(tP)$,
${\det}_{(\tpi)}(\exp(tP))$, is defined by (\ref{B244}).
\label{RB1998}
\end{rem}

\begin{rem}
It is shown above that for $\Im p=\pi^*V$, $V\subset 1_N$, $p=\sigma_0(P)$,
we have
\begin{equation}
{\det}_{(\tpi)}(\exp(2\pi itP)\cdot A)|_{t=1}={\det}_{(\tpi)}\left(\exp
\left(2\pi itp_M\right)\cdot A\right)|_{t=1},
\label{B2000}
\end{equation}
where $p_M\in\End\left(1_N\right)$ is a projector from $1_N$ onto $V$
(over $M$) and
where $\exp\left(tp_M\right)\in\Aut\left(1_N\right)$ is the multiplication
operator. By (\ref{B1997}) and by Proposition~\ref{PA1552} we have
for small $|t|$
\begin{equation}
{\det}_{(\tpi)}\left(\exp\left(2\pi itp_M\right)\cdot A\right)={\det}_{(\tpi)}
\left(\exp\left(2\pi itp_M\right)\right)\cdot{\det}_{(\tpi)}(A)={\det}_{(\tpi)}
(A).
\label{B2001}
\end{equation}
Here, $A\in\Ell_{(-1)}^d\left(M,1_N\right)$ is sufficiently close
to a positive definite self-adjoint PDO, $d\in 2(\wZ\setminus 0)$, and $P$
is a PDO-projector from $CL_{(-1)}^0\left(M,1_N\right)$.
\label{RB1999}
\end{rem}

\begin{cor}
Under the conditions of Remark~\ref{RB1999}, we have by (\ref{B1997})
and with using the analytic continuation in $t$ of the equality (\ref{B2001})
\begin{equation}
{\det}_{(\tpi)}(\exp(2\pi itP)\cdot A)|_{t=1}={\det}_{(\tpi)}\left(\exp\left(
2\pi itp_M\right)\cdot A\right)|_{t=1}={\det}_{(\tpi)}(A).
\label{B2003}
\end{equation}
\label{CB2002}
\end{cor}

Hence $\phi_0([\Im p])=\Id$ for $[\Im p]\in\pi^*K^0(M)$. Lemma~\ref{LB1989}
is proved.\ \ \ $\Box$

\begin{lem}
Any element $f$ of the abelian group
$K^0\left(\Ass\left(T^*M,RP^{n-1}\right)\right)/\pi^*K^0(M)$ has as its
order a power
of two, $f^{2^k}=\Id$. The number $k\in\wZ_+$ is estimated from the above
by a constant depending on $n:=\dim M$ only. (Here, $n$ is odd.)
\label{LB2010}
\end{lem}

\begin{pro}
For any closed loop in the space
$\left(\cup_{m\in\wZ}\Ell_{(-1)}^m(M,E),A_0\right)$
the monodromy of a multi-valued holomorphic in $A\in\Ell_{(-1)}^m(M,E)$
function $f(A)$ defined by Proposition~\ref{PB1801}
is multiplying by $\eps_k^q$, $\eps_k:=\exp(2\pi i/2^k)$, $q\in\wZ$.
The number $k$ is bounded by a constant depending on $n:=\dim M$ only.
(Here, the monodromy of $f(A)$ is defined by (\ref{B1971}) and $n$ is odd.)
\label{PB1802}
\end{pro}

This statement is an immediate consequence of Lemmas~\ref{LB1810},
\ref{LB2010} and of Theorem~\ref{TB241}

\noindent{\bf Proof of Lemma~\ref{LB2010}.}
%
1. For $m\in\wZ_+$ the group
$\tilde{K}^0\left(RP^m\right):=K^0\left(RP^m\right)/\pi^*K^0(pt)$
is a finite cyclic group $\wZ_{2^e}$ of order $2^e$, where $e:=[m/2]$
is the integer part of $m/2$ (\cite{A1}; \cite{Hu}, 15.12.5; \cite{Kar},
IV.6.47). The Atiyah-Hirzebruch spectral sequence (\cite{AH}, 2.1)
for $K^q\left(RP^{2m}\right)$, $m\in\wZ_+$, implies
$K^1\left(RP^{2m}\right)=0$. Indeed, the term $E_2^{p,q}$ of this spectral
sequence is $E_2^{p,q}=H^p\left(RP^{2m},K^q(pt)\right)$. If $q$ is odd,
then $E_2^{p,q}=0$, if $q$ is even we have $E_2^{p,q}=0$ for odd $p$,
$E_2^{p,q}=\wZ_2$ for even $p$, $0<p\le 2m$, and $E_2^{0,q}=0$.
The terms $E_{\infty}^{p,q}$ for $p+q=1$ are the graded groups associated
with the filtration
$F^pK^1\left(RP^{2m}\right)=\Ker \left(K^1\left(RP^{2m}\right)\to K^1\left(
RP_{p-1}^{2m}\right)\right)$, where $RP_{p-1}^{2m}$ is the $(p-1)$-skeleton
of $RP^{2m}$. So $\oplus_{p+q=1}E_2^{p,q}=0=\oplus E_{\infty}^{p,q}$
and $K^1\left(RP^{2m}\right)=0$.

2. Let $M$ be a compact closed smooth $(2m+1)$-dimensional manifold.
Then there exists a smooth tangent vector field $v(x)$ on $M$ without
zeroes. This vector field (together with the identification of $TM$
with $T^*M$ given by a Riemannian metric on $M$) defines a section
$v\colon M\hookrightarrow\Ass\left(T^*M,RP^{2m}\right)$. The composition
of maps
\begin{equation}
K^{\bullet}(M) @>\pi^*>> K^{\bullet}\left(\Ass\left(T^*M,RP^{2m}\right)\right)
@>v^*>> K^{\bullet}(M)
\label{B2610}
\end{equation}
is the identity map (since $\pi v\colon M\to M$ is the identity map).
Hence
\begin{equation}
K^{\bullet}\left(\Ass\left(T^*M,RP^{2m}\right)\right)=K^{\bullet}(M)\oplus
\Ker v^*.
\label{B2620}
\end{equation}
Here, the subgroup $\Ker v^*$
of $K^{\bullet}\left(\Ass\left(T^*M,RP^{2m}\right)\right)$ is
independent of a tangent to $M$ vector field $v$ without zeroes and
is isomorphic to
$K^{\bullet}\left(\Ass\left(T^*M,RP^{2m}\right)\right)/\pi^*K^{\bullet}(M)$.

There is a generalized Atiyah-Hirzebruch spectral sequence
for the $K$-functor of the fiber bundle of $\Ass\left(T^*M,RP^{2m}\right)$
over $M$ (\cite{AH}, 2.2; \cite{Do}, 4, Theorem~3; \cite{A2}, \S~12,
pp. 167--177; \cite{Hi}, Ch.~III; \cite{Hu}, 15.12.2).
Its $E_2^{\bullet,\bullet}$-term is
\begin{equation}
E_2^{p,q}=H^p\left(M,k_M^q\left(RP^{2m}\right)\right),
\label{B2622}
\end{equation}
where $k_M^q\left(RP^{2m}\right)$ is a local system with the fiber
$K^q\left(RP^{2m}\right)$ associated with the fibration
$\pi\colon \Ass\left(T^*M,RP^{2m}\right)\to M$. (The fiber of $\pi$
is $RP^{2m}$.) Its $E_{\infty}^{p,q}$-terms with $p+q=i$ are groups
associated with the fibration
\begin{equation}
F^p\!\left(\!K^i\!\left(\!\Ass\!\left(\!T^*\!M,\!RP^{2m}\!\right)\!\right)
\!\right)\!:=\!\Ker\!\left(\!K^i\!\left(\!\Ass\!\left(\!T^*\!M,\!RP^{2m}
\!\right)\!\right)\!\to\!K^i\!\left(\pi^{-1}\!\left(\!M_{p-1}\!\right)
\!\right)\!\right)\!,
\label{B2621}
\end{equation}
where $M_{p-1}$ is the $(p-1)$-skeleton of $M$.

The $E_2^{p,q}$-term of the Atiyah-Hirzebruch spectral sequence
for $K^{\bullet}(M)$ is equal to $H^p\left(M,K^q(pt)\right)$.
The $F^p$-filtration (\ref{B2621})
in $K^{\bullet}\left(\Ass\left(T^*M,RP^{2m}\right)\right)$ (and
in the spectral sequence (\ref{B2620})) is in accordance with
the $F^p$-filtration in $K^{\bullet}(M)$ (with respect to the direct-sum
decomposition (\ref{B2620})). The analogous direct-sum decomposition
is valid for $H^p\left(M,k_M^q\left(RP^{2m}\right)\right)$ and for further
terms $E_r^{p,q}$. Hence
\begin{gather}
\begin{split}
K^{\bullet}\left(\Ass\left(T^*M,RP^{2m}\right)\right)=K^{\bullet}(M) & \oplus
K^{\bullet}\left(\Ass\left(T^*M,RP^{2m}\right)\right)\big/\pi^*\left(
K^{\bullet}(M)\right), \\
\Gr^pK^i\left(\Ass\left(T^*M,RP^{2m}\right)\right) & =E_{\infty}^{p,i-p}(\pi)=
E_{\infty}^{p,i-p}(M)\oplus\tilde{E}_{\infty}^{p,i-p}(\pi),
\end{split}
\label{B2623}
\end{gather}
where
$\tilde{E}_2^{p,i-p}(\pi):=H^p\left(M,\tilde{k}_M^{i-p}\left(RP^{2m}\right)
\right)$
and $\tilde{E}_r^{p,i-p}(\pi)$ are the further terms in the corresponding
spectral sequence. Here, $\tilde{k}_M^q$ is the local system (analogous
to $k_M^q$) with the reduced $K$-functor $\tilde{K}^q\left(RP^{2m}\right)$
as its fiber. We have for $i=0$
$$
\tilde{E}_2^{p,-p}(\pi)=H^p\left(M,\tilde{k}_M^{-p}\left(RP^{2m}\right)
\right).
$$
So $\tilde{E}_2^{p,-p}=0$ for odd $p$ and each element
of $H^p\left(M,\tilde{k}_M^{-p}\left(RP^{2m}\right)\right)$ is of a finite
order $2^{\alpha}$, $\alpha\in\wZ$, $0\le\alpha\le m$. Hence only
the terms $\tilde{E}_{\infty}^{2l,-2l}(\pi)$, $l\in\wZ$, $0\le l\le m$,
may be unequal to zero. Thus every element
of $K^0\left(\Ass\left(T^*M,RP^{2m}\right)\right)/\pi^*K^0(M)$ has a finite
order $2^\beta$ with $\beta\in\wZ$, $0\le\beta\le m^2:=((n-1)/2)^2$,
$n=\dim M=2m+1$. The lemma is proved.\ \ \ $\Box$

\begin{rem}
The commutative diagram
\begin{gather*}
\begin{CD}
K^{\bullet}\left(\pi^{-1}M\right) @>>> K^{\bullet}\left(\pi^{-1}\left(M_{p-1}
\right)\right) @>>> K^{\bullet}\left(\pi^{-1}\left(M_{p-2}\right)\right) \\
@V{\pi^*\Big\uparrow}V v^* V @V{\pi^*\Big\uparrow}V v^* V @V{\pi^*\Big\uparrow}
V v^* V  \\
K^{\bullet}(M) @>>> K^{\bullet}\left(M_{p-1}\right) @>>> K^{\bullet}\left(
M_{p-2}\right)
\end{CD}
\end{gather*}
is used in the derivation of (\ref{B2623}).
\label{RB2624}
\end{rem}

\begin{rem}
Theorem~IV.6.45 in \cite{Kar} provides us with the exact sequence
$$
\to K^i(X)\oplus K^i(X) @>(\pi^*,-\theta^*)>> K^i(P(V))\to K^{i+1}\left(
\cal{E}^{V\oplus 1}(X)\right)\to K^{i+1}(X)\oplus K^{i+1}(X).
$$
Here, $\pi\colon P(V)\to X$ is the projective bundle of a real vector
bundle $V$ over $X$, $\theta\colon E\to\xi\otimes\pi^*E$, $\xi$ is
the canonical line bundle over $P(V)$, $\cal{E}^V(X)$ is the category
of vector bundles over $X$ with an action of the Clifford bundle $C(V)$
(\cite{Kar}, IV.4.11).
\label{RB2625}
\end{rem}

\noindent{\bf Proof of Propositions~\ref{PB1801} and \ref{PB3790}.}
First we prove that the absolute value determinant $|\det|A$
as a function on $\Ell_{(-1)}^m(M,E)$ has the form $|f(A)|$, where
$f$ is a multi-valued holomorphic function on $\Ell_{(-1)}^m(M,E)$.
Let $X$ be the infinite-dimensional analytic manifold
$$
\Ell_{(-1)}^{\bullet}(M,E):=\Ell^{\bullet}(M,E)\cap CL_{(-1)}(M,E).
$$
Let $X'$ be a manifold with the conjugate complex structure on it.
An element $A\in X$ corresponds to an operator $A^*$ as to an element
of $X'$ ($=X$). Then the function
$$
\tilde{f}(A,B):={\det}_{(\tpi)}(AB)
$$
is defined on a sufficiently close neighborhood of the diagonal
$X\hookrightarrow X\times X'$ for an admissible cut $\tpi$ (close to $\pi$)
depending on $AB$. Here we suppose that $m\in\wZ_+$. Then
$\zeta_{AB,(\tpi)}(s)$
is defined for $\Re s>\dim M/2m$ and its analytic continuation is regular
at zero. If $m\in\wZ_-$, the same is true for $\zeta_{(AB)^{-1},(\tpi)}(s)$,
$\Re s>\dim M/2|m|$. If $m=0$, the function $\det(AA^*)$ is defined
by the multiplicative property
$$
\det(AA^*):=\det\left(\left(\Delta_E+\Id\right)AA^*\left(\Delta_E+\Id\right)
\right)\big/\left(\det\left(\Delta_E+\Id\right)\right)^2.
$$
Note that ${\det}_{(\tpi)}(AB)$ is defined for $(A,B)$ with $m=0$
in a close neighborhood of the diagonal.

For pairs $(A,B)$ and $\left(A_1,B_1\right)$ of sufficiently close points
of $X\times X'$ in a close neighborhood of the diagonal $X$ we have
by Theorem~\ref{TB241}
$$
{\det}_{(\tpi)}\left(ABA_1B_1\right)={\det}_{(\tpi)}\left(B_1A\right)
{\det}_{(\tpi)}\left(BA_1\right)={\det}_{(\tpi)}\left(AB_1\right)
{\det}_{(\tpi)}\left(A_1B\right).
$$
Hence the matrix with elements $\tilde{f}(A,B)=\det_{(\tpi)}(AB)$
for $(A,B)$ from a close
neighborhood of a point $(A,A^*)\in X\hookrightarrow X\times X'$ has
the rank one. Hence there exist locally defined functions $f_1(A)$ and
$f_2(B)$ such that for sufficiently close $A$ and $B$ (belonging
to the domain of definition of $f_1$ and $f_2$) we have
$$
{\det}_{(\tpi)}(A\cdot B)\equiv\tilde{f}(A,B)=f_1(A)f_2(B).
$$
The function $f_1(A)$ is holomorphic in $A$ since for invertible
$A,B\in\Ell_{(-1)}^m(M,E)$ we have
$$
\delta_A{\det}_{(\tpi)}(A\cdot B)=\df_s\left(s\Tr\left(\delta A\cdot A^{-1}
\left(AB\right)_{(\tpi)}^{-s}\right)-\res\sigma(\delta AA^{-1})/2sm\right)
\big|_{s=0}.
$$
Here, the expression on the right has an analytic continuation in $s$
to $s=0$ and it is regular at $s=0$. The same is true for $f_2(B)$
(with respect to the holomorphic structure of $X$).

The function $f_1(A)/\overline{f_2\left(A^*\right)}$ is (locally)
analytic in $A$ and it is a real function.
(Indeed, $\det(AA^*)=f_1(A)f_2(A^*)$ and
$|f_2(A^*)|^2=f_2(A^*)\overline{f_2(A^*)}$ are real functions.)
Hence it is a real constant $c$.
It is the ratio of two positive functions, ${\det}_{(\tpi)}(AA^*)$ and
$\left|f_2(A^*)\right|^2$. Hence $c\in\wR_+$. The function $f(A)$
is defined as $c^{-1/2}f_1(A)$. Thus $f(A)$ is an analytic function
of $A$. The assertion that $|\det|A$ as a function
on $\Ell_{(-1)}^m(M;E,F)\ni A$ has the form $|f(A)|$ with a multi-valued
holomorphic $f$ is obtained from the analogous assertion for $|\det|A$
on $\Ell_{(-1)}^m(M;E,F)$ with using the identification (\ref{B3791})
of the spaces $\Ell_{(-1)}^m(M,E)\rs\Ell_{(-1)}^m(M;E,F)$ and with
using the multiplicative property (\ref{B3784}) of absolute value
determinants (Proposition~\ref{PB3782}).\ \ \ $\Box$

\section{Lie algebra of logarithmic symbols and its central extension}
\label{SD}

Symbols $\sigma\left(A_{(\theta)}^s\right)$ for complex powers
$A_{(\theta)}^s$ of elliptic PDOs $A\in\Ell_0^d(M,E)\subset CL^d(M,E)$,
$d\in\wR^{\times}$, are defined by (\ref{A5}), (\ref{A6}).
(Here we suppose that the principal symbol $\sigma_d(A)$ possesses
a cut $L_{(\theta)}$ of the spectral plane.) The symbol
of $\log_{(\theta)}A$ is defined as
\begin{equation}
\df_s\sigma\left(A_{(\theta)}^s\right)\big|_{s=0}=\sum_{j\in\zuo}\df_s
b_{sd-j,\theta}^s(x,\xi)\big|_{s=0}.
\label{B300}
\end{equation}

The equalities (\ref{A10}) hold for the components on the right
in (\ref{B300}). Hence the Lie algebra $S_{\log}(M,E)$ of symbols
$\sigma\left(\log_{(\theta)}A\right)$ is spanned as a linear space
by its subalgebra $CS^0(M,E)$ of symbols for $CL^0(M,E)$ and
by one element $\sigma\left(\log_{(\theta)}A\right)$.
Here, $A$ is an elliptic operator from $\Ell_0^d(M,E)\subset CL^d(M,E)$
admitting a cut $L_{(\theta)}$, $d\in\wR^{\times}$.
For $l:=(1/d)\sigma\left(\log_{(\theta)}A\right)$ every element
$B\in S_{\log}(M,E)$ has a form
\begin{equation}
B=ql+B_0,
\label{B301}
\end{equation}
where $q\in\wC$ and $B_0\in CS^0(M,E)$. The number $q$ in (\ref{B301})
is independent of $A$ and of $\theta$. Set $r(B):=q$.
In $S_{\log}(M,E)$ we have
\begin{equation}
\left[q_1l+B_0,q_2l+C_0\right]=\left[l,q_1C_0-q_2B_0\right]+\left[B_0,C_0
\right]\in CS^0(M,E),
\label{B302}
\end{equation}
since $[l,B_0]\in CS^0(M,E)$ according to (\ref{A10}). Note that
$CS^0(M,E)$ is a Lie ideal of codimension one in $S_{\log}(M,E)$.
We call $S_{\log}(M,E)$ a one-dimensional cocentral extension
of the Lie algebra $CS^0(M,E)$,
\begin{equation}
0\to CS^0(M,E)\to S_{\log}(M,E){\stackrel{r}{\to}}\wC\to 0.
\label{B303}
\end{equation}
The left arrow of (\ref{B303}) is the natural inclusion.
\begin{lem}
An element $l$ defines a $2$-cocycle for the Lie algebra $\frg:=S_{\log}(M,E)$
(with the coefficients in the trivial $\frg$-module)
\begin{equation}
K_l\left(q_1l+B_0,q_2l+C_0\right):=-\left(\left[l,B_0\right],C_0\right)
_{\res}.
\label{B304}
\end{equation}
The cocycles $K_l$ for different $l\in S_{\log}(M,E)$ with $r(l)=1$
are cohomologous (i.e., for logarithmic symbols $l$ of degree one;
here, $r$ is from (\ref{B303})).
\label{LB1590}
\end{lem}

\noindent{\bf Proof.} The linear form $K_l(B_0,C_0)$ is skew-symmetric
in $B_0$, $C_0$ as it follows from (\ref{B8}). (Here, we substitute
$c=\sigma\left(A_{(\theta)}^s\right)$,
$l:=\sigma\left(\log_{(\theta)}A\right)$,
$a=B_0$, $b=C_0$ into (\ref{B8}) and then take $\df_s|_{s=0}$.)

Note that $K_l$ is a cocycle because the antisymmetrization
of the $3$-linear form on $S_{\log}(M,E)$
$$
K_l\left(\left[q_0l+A_0,q_1l+B_0\right],q_2l+C_0\right)=K_l([q_0l+A_0,q_1l+
B_0],C_0)
$$
is equal to zero. Indeed, we have
\begin{multline*}
K_l([A_0,B_0],C_0)+K_l([B_0,C_0],A_0)-K_l([A_0,C_0],B_0)=\\
=\left([l,C_0],[A_0,B_0]\right)_{\res}-\left([l,A_0],[B_0,C_0]\right)_{\res}+
\left([l,B_0],[A_0,C_0]\right)_{\res}=\\
=\left(\left[[l,C_0],A_0\right],B_0\right)_{\res}+
\left(\left[[l,A_0],C_0\right],B_0\right)_{\res}-
\left(\left[l,[A_0,C_0]\right],B_0\right)_{\res}=0
\end{multline*}
by the Jacobi indentity in $S_{\log}(M,E)$. We have also
\begin{gather*}
\begin{split}
K_l\left(\left[q_0 l,B_0\right]+\left[A_0,q_1 l\right],C_0\right) &
=\left(\left[q_0 l,B_0\right]+\left[A_0,q_1 l\right],\left[l,C_0\right]\right)
_{\res}, \\
\left(\left[q_0 l,B_0\right]+\left[A_0,q_1 l\right],\left[l,C_0\right]\right)
_{\res} & -\left(\left[q_0 l,C_0\right]+\left[A_0,q_2 l\right],\left[l,B_0
\right]\right)_{\res}+ \\
 & +\left(\left[q_1 l,C_0\right]+\left[B_0,q_2 l\right],\left[l,A_0\right]
\right)_{\res}\equiv 0.
\end{split}
\end{gather*}
Hence $K_l$ is a $2$-cocycle for $\frg=S_{\log}(M,E)$ (with the coefficients
in the trivial $\frg$-module
$\wC$). For $l_1\in r^{-1}(1)$ we have $l_1=l-L_0$, $L_0\in CS^0(M,E)$,
$$
K_{l_1}(A,B)-K_l(A,B)=([L_0,A],B)_{\res}=(L_0,[A,B])_{\res},
$$
where $A,B\in CS^0(M,E)$. If $A,B\in S_{\log}(M,E)$, $A=q_0 l+A_0$,
$B=q_1 l+B_0$, $q_j\in\wC$, $A_0,B_0\in CS^0(M,E)$, then we have
\begin{multline*}
K_{l_1}(A,B)-K_l(A,B)=-\left(\left[l_1,A_0+q_0 L_0\right],B_0+q_1 L_0\right)
_{\res}+\left(\left[l,A_0\right],B_0\right)_{\res}= \\
=\left(L_0,\left[A_0+q_0 L_0,B_0+q_1 L_0\right]\right)_{\res}+\left(L_0,\left[
A_0,q_1 l\right]\right)_{\res}+\left(L_0,\left[q_0 l,B_0+q_1 L_0\right]\right)
_{\res}= \\
=\left(L_0,[A,B]\right)_{\res}.
\end{multline*}
Hence $K_{l_1}$ and $K_l$ are cohomologous $2$-cocycles.\ \ \ $\Box$

\begin{rem}
The cocycle $K_l$ defines a central extension of the Lie algebra
$\frg=S_{\log}(M,E)$
\begin{equation}
0\to\wC\to\sfrg_{(l)}\to\frg\to 0.
\label{B307}
\end{equation}
The Lie algebra structure on $\sfrg_{(l)}$ is given by
\begin{equation}
\left[q_1l+a_1+c_1\cdot 1,q_2l+a_2+c_2\cdot 1\right]=\left[q_1l+a_1,q_2l+
a_2\right]+K_l(a_1,a_2)\cdot 1.
\label{B624}
\end{equation}
Here, $q_jl+a_j\in S_{\log}(M,E)=:\frg$, $c_j\in\wC$, $1$ is the generator
of the kernel $\wC$ in the extension (\ref{B307}).
\label{RB2005}
\end{rem}

\begin{rem}
The extension of the Lie algebra of classical PDO-symbols of integer
orders analogous to (\ref{B307}), (\ref{B624}) (in the case of PDOs
acting on scalar
functions and where $l$ is the symbol of $\log S$, $S$ is an elliptic DO
with a positive principal symbol) was considered in \cite{R}, Section~3.4.
The extension of the algebra of PDO-symbols of integer orders on the circle
defined by the cocycle $K_{\log(d/dx)}$ (on this algebra) is considered
in \cite{KrKh}, \cite{KhZ1}, \cite{KhZ2}. A canonical associative
system of isomorphisms of the Lie algebras $\sfrg_{(l)}$
for $l\in S_{\log}(M,E)$, $r(l)=1$ (for $r$ as in (\ref{B303})) is defined
in Proposition~\ref{PB403} below. Thus the Lie algebra $\sfrg$,
a one-dimensional canonical central extension of $S_{\log}(M,E)$,
is defined. A determinant line bundle over the connected component
of the space of elliptic symbols $\SEll_0^{\times}(M,E)$ is defined
in Section~\ref{SE} below. The nonzero elements of the fibers of this
line bundle form a Lie group $G(M,E)$, Proposition~\ref{PB603}.
(We call it a {\em determinant Lie group}.) The Lie algebra of $G(M,E)$
is canonically isomorphic to $\sfrg$ by Theorem~\ref{TB570} below.
This connection of the extensions $\sfrg_{(l)}$, (\ref{B307}), and
the determinants of elliptic PDOs is a new fact.
\label{RB2006}
\end{rem}


\begin{rem}
The determinant group is defined in Section~\ref{SE}. It is the central
extension of the group $\SEll_0^{\times}(M,E)$ with the help
of $\wC^{\times}$.
By Theorem~\ref{TB570} its Lie algebra is canonically isomorphic
to the central extension $\sfrg_{(l)}$ (defined with the help
of the cocycle $K_l$). Lemma~\ref{LB3006} claims that (in the case
of a trivial bundle $E:=1_N$, where $N\in\wZ_+$ is large enough)
over an orientable closed manifold $M$ the determinant Lie group
is a nontrivial $\wC^{\times}$-extension of $\SEll_0^{\times}(M,E)$.
Namely for any orientable closed $M$, the assotiated line bundle
$L$ over $\SEll_0^{\times}(M,E)$ has a non-trivial
(in $H^2\left(\SEll_0^{\times}(M,E),\wQ\right)$) the first Chern class
$c_1(L)$. If the cocycle $K_l$ would be a coboudary of a continuous
one-cochain on $S_{\log}(M,E)=:\frg$, then the Lie algebra splitting
\begin{equation}
\sfrg_{(l)}=\frg\oplus\wC
\label{B4050}
\end{equation}
would give us a flat connection on the determinant Lie group over
$\SEll_0^{\times}(M,E)$. So in this case $c_1(L)$ would be zero
in $H^2\left(\SEll_0^{\times}(M,E),\wQ\right)$. Hence $K_l$ is not
a coboundary of a continuous cochain.
\label{RB3005}
\end{rem}

The cocycles $K_l$ for different $l\in r^{-1}(1)\subset\frg$
are cohomologous. We define a system of isomorphisms
of Lie algebras
\begin{equation}
W_{l_1l_2}\colon\sfrg_{(l_1)}\rs\sfrg_{(l_2)}
\label{B308}
\end{equation}
which is associative, i.e., $W_{l_2l_3}W_{l_1l_2}=W_{l_1l_3}$.
These isomorphisms $W_{l_1l_2}$ transform $ql_1+a_1$ into the same
element $ql_2+a'_1=ql_1+a_1$ of $S_{\log}(M,E)=\frg$, i.e.,
the following diagram is commutative
\begin{gather}
\begin{CD}
0 @>>> \wC @>>> \sfrg_{(l_1)}         @>>> \frg @>>> 0 \\
@.     @.       @V{\wr\wr}V W_{l_1l_2} V   \big\|@.  @.\\
0 @>>> \wC @>>> \sfrg_{(l_2)}         @>>> \frg @>>> 0
\end{CD}
\label{B402}
\end{gather}

\begin{pro}
The system of such isomorphisms $W_{l_1l_2}$, $l_j\in r^{-1}(1)$,
given by
$$
W_{l_1l_2}(ql_1+a+c\cdot 1)=(ql_2+a'+c'\cdot 1),
$$
where $ql_1+a+c\cdot 1\in\sfrg_{(l_1)}$, $ql_2+a'+c'\cdot 1\in\sfrg_{(l_2)}$,
and
\begin{gather}
\begin{split}
ql_1+a & =ql_2+a' \quad \text{ in }\frg,\\
c'     & =c+\Phi_{l_1l_2}(a)+q\Psi_{l_1l_2}, \\
\Phi_{l_1l_2}(a):=\left(l_1-l_2,a\right)_{\res}, & \quad
\Psi_{l_1l_2}:=\left(l_2-l_1,l_2-l_1\right)_{\res}/2.
\end{split}
\label{B404}
\end{gather}
is associative.
\label{PB403}
\end{pro}

\noindent{\bf Proof.} We try to construct $W_{l_1l_2}$ using conditions
of compatibility with the Lie brackets. \\
1. {\em Compatibility with the Lie brackets}.
%
Let $W_{l_1l_2}\left(q_jl_1+a_j+c_j\cdot 1\right)=q_jl_2+b_j+f_j\cdot 1$,
$j=1,2$. We want to prove that
\begin{multline}
\left[q_1l_1\!+\!a_1\!+\!c_1\cdot 1,q_2l_1\!+\!a_2\!+\!c_2\cdot 1\right]
_{\sfrg_{(l_1)}}+\left(\Phi_{l_1l_2}\left([a_1,a_2]+[l_1,q_1a_2-q_2a_1]\right)
\right)\cdot 1=\\
=\left[q_1l_2\!+\!b_1\!+\!f_1\cdot 1,q_2l_2\!+\!b_2\!+
\!f_2\cdot 1\right]_{\sfrg_{(l_2)}},
\label{B310}
\end{multline}
Using the equality $q_j(l_1-l_2)=b_j-a_j$ we can rewrite (\ref{B310}) as
$$
K_{l_2}(b_1,b_2)-K_{l_1}(a_1,a_2)=\Phi_{l_1l_2}\left([a_1,a_2]+[l_1,q_1a_2-q
_2a_1]\right).
$$
The left side of the last equality by the definitions of $K_{l_1}(a_1,a_2)$
and of $K_{l_2}(b_1,b_2)$ and according to (\ref{B8}) and
to the skew-symmetry of (\ref{B304}) is equal to
\begin{multline}
K_{l_2}(b_1,b_2)-K_{l_1}(a_1,a_2)=\\
=\left(\left[l_1-l_2,a_1\right],a_2\right)_{\res}-\left(\left[l_1,q_1(l_1-
l_2)\right],a_2\right)_{\res}-\left(\left[l_1,a_1\right],q_2(l_1-l_2)\right)
_{\res}=\\
=\left(l_1-l_2,\left[a_1,a_2\right]+\left[l_1,q_1a_2-q_2a_1\right]\right)
_{\res}.
\label{B410}
\end{multline}

We conclude comparing (\ref{B410}) and (\ref{B310}) that if we set
\begin{gather}
\begin{split}
\Phi_{l_1l_2}(a) & =\left(l_1-l_2,a\right)_{\res},\\
c'-c             & =\left(l_1-l_2,a\right)_{\res}+q\Psi_{l_1l_2}.
\end{split}
\label{B411}
\end{gather}
(for $\Psi_{l_1l_2}$ defined by (\ref{B404})), then the condition
(\ref{B310}) is satisfied.

\medskip
\noindent 2. {\em Associativity}. We want to show that
$W_{l_2l_3}W_{l_1l_2}=W_{l_1l_3}$ (for $W_{l_il_j}$ defined by (\ref{B404})).
We have
\begin{equation}
c''-c'=\left(l_2-l_3,a'\right)_{\res}+q\Psi_{l_2l_3},
\label{B412}
\end{equation}
where $W_{l_2l_3}\left(ql_2+a'+c'\cdot 1\right)=ql_3+a''+c''\cdot 1\in
\sfrg_{(l_3)}$. Thus we have to show that
\begin{equation}
q\Psi_{l_1l_3}=q\Psi_{l_1l_2}+q\Psi_{l_2l_3}+\left(l_2-l_3,a'-a\right)_{\res},
\label{B3860}
\end{equation}
where $l_j\in r^{-1}(1)$, $a'-a=q(l_1-l_2)$. We can rewrite (\ref{B3860})
in the form
\begin{equation}
\Psi_{l_1l_3}=\Psi_{l_1l_2}+\Psi_{l_2l_3}+\left(l_3-l_2,l_2-l_1\right)_{\res}.
\label{B414}
\end{equation}

%
It is clear that $\Psi_{l_1l_2}:=\left(l_2-l_1,l_2-l_1\right)_{\res}/2$
provides us with a solution of the system (\ref{B414}). The proposition
is proved.\ \ \ $\Box$

\begin{pro}
A system of quadratic forms
\begin{equation}
A_l(ql+a+c\cdot 1):=(a,a)_{\res}-2qc
\label{B418}
\end{equation}
on $\sfrg_{(l)}\ni ql+a+c\cdot 1$, $l\in r^{-1}(1)$, is invariant
under the identifications $W_{l_1l_2}$.
\label{PB417}
\end{pro}

\noindent{\bf Proof.} For $W_{l_1l_2}(ql+a+c\cdot 1)=:ql_1+a_1+c_1\cdot 1$
we have
\begin{gather}
\begin{split}
a_1-a & =q(l-l_1),\\
c_1-c &=\left(l-l_1,a\right)_{\res}+q\left(l_1-l,l_1-l\right)_{\res}/2.
\end{split}
\label{B419}
\end{gather}
Hence we have
\begin{multline}
A_{l_1}\left(ql_1+a_1+c_1\cdot 1\right):=\left(a_1,a_1\right)_{\res}-2qc_1=\\
=(a,a)_{\res}+q^2\left(l_1-l,l_1-l\right)_{\res}+2q\left(a,l-l_1\right)_{\res}-
2qc+2q\left(l_1-l,a\right)_{\res}+\\
+(-q^2)\left(l_1-l,l_1-l\right)_{\res}=A_l(ql+a+c\cdot 1).
\label{B420}
\end{multline}

The proposition is proved.\ \ \ $\Box$

\begin{cor}
The cones $C_l$ in $\sfrg_{(l)}$, $l\in r^{-1}(1)$, defined by null vectors
for $A_l$, i.e.,
$$
C_l:=\left\{ql+a+c\cdot 1\in\sfrg_{(l)},A_l(ql+a+c\cdot 1)=0\right\},
$$
are invariant under the identifications $W_{l_1l_2}$
$$
W_{l_1l_2}C_{l_1}=C_{l_2}.
$$
\label{CB421}
\end{cor}

Indeed, $W_{l_2l_1}W_{l_1l_2}=\Id$, $W_{l_1l_2}W_{l_2l_1}=\Id$ and
the quadratic forms $A_l$ are invariant under $W_{l_1l_2}$.

\begin{rem}
Let $\frg_0$ be a Lie algebra over $\wC$ with a conjugate-invariant
scalar product $(a,b)_{\frg_0}$,
\begin{equation}
([c,a],b)_{\frg_0}+(a,[c,b])_{\frg_0}=0 \quad \text{ for }a,b,c\in\frg_0.
\label{B423}
\end{equation}

Let $\frg$ be a cocentral Lie algebra one-dimensional extension of $\frg_0$,
\begin{equation}
0\to\frg_0\to\frg\stackrel{r}{\to}\wC\to 0,
\label{B424}
\end{equation}
i.e., let $\frg_0$ be a Lie ideal in $\frg$ and let $\frg_0$ be
of codimension
one in $\frg$ (the left arrow in (\ref{B424}) is the natural inclusion,
$[a,b]\in\frg_0$ for $a,b\in\frg$). Then the expression on the left
in (\ref{B423}) makes sense for $c\in\frg$.

Let the scalar product $\left(,\right)_{\frg}$ be also conjugate-invariant
under $\frg$, i.e., let (\ref{B423}) hold for $a,b\in\frg_0$ and
for $c\in\frg$. (Note that this condition is satisfied for the scalar
product $\left(,\right)_{\res}$ on the Lie algebra $CS^0(M,E)=:\frg_0$
for its central extension $S_{\log}(M,E)=:\frg$.) Then we define
a central extension
\begin{equation}
0\to\wC\to\sfrg_{(l)}@>p>>\frg\to 0
\label{B425}
\end{equation}
given by the $2$-cocycle of $\frg$ (with the coefficients in the trivial
$\frg$-module)
\begin{equation}
K_l\left(q_1l+a_1,q_2l+a_2\right):=-\left([l,a_1],a_2\right)_{\frg_0}
\label{B426}
\end{equation}
on $\frg$, where $a_j\in\frg_0$ and $l\in r^{-1}(1)\in\frg$ ($r$ is
from (\ref{B424})). These Lie algebras $\sfrg_{(l)}$ for $l\in r^{-1}(1)$
are identified by an associative system of Lie algebra isomorphisms
$W_{l_1l_2}\colon\sfrg_{(l_1)}\rss\sfrg_{(l_2)}$ defined by the same
formulas as isomorphisms (\ref{B404}) (with changing $\left(,\right)_{\res}$
by the scalar product $\left(,\right)_{\frg_0}$).
This system of isomorphisms defines the canonical central Lie algebra
extension $0\to\wC\to\sfrg\to\frg\to 0$.
The quadratic form
\begin{equation}
A_l(ql+a+c\cdot 1):=(a,a)_{\frg_0}-2qc
\label{B428}
\end{equation}
is defined on $\sfrg_{(l)}$. This system of quadratic forms $A_l$,
$l\in r^{-1}(1)$, is invariant under identifications $W_{l_1l_2}$.
The cones $C_l\subset\sfrg_{(l)}$ of zero vectors for $A_l$ are
identified under $W_{l_1l_2}$. So these quadratic forms define a canonical
quadratic form $A$ on $\sfrg$.
\label{RB422}
\end{rem}

\begin{rem}
The previous construction can be reversed. Namely, let $\sfrg'$ be
a Lie algebra over $\wC$ with an invariant scalar product, $1\in\sfrg'$
be a central element with $(1,1)=0$. We assume that the linear form
$f\colon x\to(1,x)$ on $\sfrg'$ is not zero. Denote by $\frg$
the quotient algebra $\sfrg'/\wC\cdot 1$ and by $\frg_0$ the subalgebra
of $\frg$ consisting of the kernel of $f$. Then we have a scalar
product on $\frg_0$ invariant under the adjoint action of $\frg$,
e.i., the situation at the beginning of Remark~\ref{RB422}.

We claim that $\sfrg'$ is canonically isomorphic to the central
extension $\sfrg$ constructed from $\frg$.

The idea is to use null-vectors $l$ of the quadratic form on $\sfrg'$
such that $f(l)=1$ for the system of splittings (as vector spaces)
$$
\sfrg'\rs\frg\oplus\wC\cdot 1.
$$
\label{Rmax2}
\end{rem}

\begin{rem}
The associative system of Lie algebra isomorphisms $W_{l_1l_2}$ defined
by the formulas (\ref{B404}) is the only associative system of Lie
algebra isomorphisms which is universal. This means that the system is
functorial on the category of one-dimensional cocentral extensions
of Lie algebras with invariant scalar products. This is the category
of cocentral extensions (\ref{B424}) having as its morphisms
the morphisms of the diagrams (\ref{B424}) which are equal to identity
on $\wC$ and which save invariant scalar products on the components
$\frg_0$. (This class of extensions is considered in Remark~\ref{RB422}.)
The universality of the system $W_{l_1l_2}$ (\ref{B404}) follows
immediately from the proof of Proposition~\ref{PB403}.
\label{RB3861}
\end{rem}

\begin{rem}
Let $\frg$ be a complex Lie algebra endowed with an invariant scalar
product $B\colon\frg\otimes\frg\to\wC$,
$$
B(x,y)=B(y,x), \quad B\left(x,[y,z]\right)=B\left([x,y],z\right).
$$
We will construct a map
$$
I_k\colon H^k(\frg,\frg)\to H^{k+1}(\frg,\wC)
$$
for each integer $k\ge 0$.
Here we consider $\frg$ as a $\frg$-module via the adjoint action.

First of all, we can associate with $B$ an element
$\widetilde{B}\in H^1\left(\frg,\frg^{\vee}\right)$ ($\frg^{\vee}$ is
the dual space) as the cohomology class of $1$-cochain
$$
\widetilde{B}(x)(y):=B(x,y), \quad x,y\in\frg.
$$
The cup product (with coefficients) by $\widetilde{B}$ defines a map
$$
\cup\widetilde{B}\colon H^k(\frg,\frg)\to H^{k+1}\left(\frg,\frg\otimes\frg
^{\vee}\right).
$$
The composition of this map with the map
$H^{\bullet}\left(\frg,\frg\otimes\frg^{\vee}\right)\to H^{\bullet}(\frg,\wC)$
induced by the morphism of $\frg$-modules
$$
\frg\otimes\frg^{\vee}\to\wC, \quad x\otimes\phi\to\phi(x)
$$
gives the desired map $I_k$. On the level of cochains,
$I_k$ is given by the formula
$$
I_k(\alpha)\left(x_1,\dots,x_{k+1}\right)=\Alt\left(B\left(x_1,\alpha\left(x_2,
\dots,x_{k+1}\right)\right)\right).
$$
Note that $I_0$ maps the center of $\frg$, $Z(\frg)=H^0(\frg,\frg)$,
into the ``cocenter'' $\left(\frg/[\frg,\frg]\right)^{\vee}=H^1(\frg,\wC)$.

Analogously, $I_1$ maps the space of derivations of $\frg$ modulo interior
derivations ($=H^1(\frg,\frg)$) into the space of equivalence classes
of one-dimensional central extensions ($=H^2(\frg,\wC)$).
The space $H^1(\frg,\frg)$ can also be viewed as the set of equivalence
classes of ``cocentral extensions''
$$
0\to\frg\to\sfrg\to\wC\to 0.
$$

Let us denote by $H_{skew}^1(\frg,\frg)$ the subspace of $H^1(\frg,\frg)$
represented by cocycles $\alpha\colon\frg\to\frg$ which are skew-symmetric
with respect to $B$
$$
B(\alpha x,y)+B(x,\alpha y)=0, \quad x,y\in\frg.
$$
{\bf Claim}: for a non-degenerate scalar product $B$ the maps $I_0$ and
$I_1|_{H_{skew}^1}$ are isomorphisms.

It follows almost immediately from standard formulas for differentials
in $C^*(\frg,\frg)$, $C^*(\frg,\wC)$ and from the invariance of $B$.

In our concrete situation we see that the central extension
of $S_{\log}(M,E)$ corresponds to the homomorphism of degree
$$
S_{\log}(M,E)\to\wC
$$
and the noncommutative residue
$$
\res\colon CS^0(M,E)\to\wC
$$
corresponds to the central element
$$
\Id_E\in CS^0(M,E).
$$
We can also change our Lie algebras in such a way that the scalar
products are non-degenerate. One way is to replace $CS^0(M,E)$
by the Lie algebra of integer orders PDO-symbols. Another way is
to consider the quotient algebra modulo the ideal $CS^{-\dim M-1}(M,E)$.
\label{Rmax1}
\end{rem}

\section{Determinant Lie groups and determinant bundles over spaces
of elliptic symbols. Canonical determinants}
\label{SE}

Let $\Ell_0^{\times}(M,E)$ be the connected component of $\Id$
in the group of invertible elliptic PDOs. The determinant line bundle
$\det\Ell_0^{\times}(M,E)$ is canonically defined
over the space $\SEll_0^{\times}(M,E)$ of symbols for invertible
elliptic operators with their principal symbols homotopic
to $\Id\left|\xi\right|^\alpha$ ($\alpha\in\wC$) in Section~\ref{SE52}.
Its associated $\wC^{\times}$-bundle (with a Lie group structure on it)
is defined as follows.

The associated fiber bundle $\det_*\SEll_0^{\times}(M,E)$ (with its
fiber $\wC^{\times}$) of nonzero elements in fibers of
$p\colon\det\Ell_0^{\times}(M,E)\to\SEll_0^{\times}(M,E)$
is defined as $F_0\backslash\Ell_0^{\times}(M,E)$. Here, $F_0$
is a subgroup of the group $F$ of invertible operators of the form
$\Id+\cK$, where $\cK$ is a smoothing operator (i.e.,
an operator with a $\wC^{\infty}$-kernel on $M\times M$), and $F_0$
is the set of operators from $F$ such that
\begin{equation}
{\det}_{Fr}(\Id+\cK)=1
\label{B600}
\end{equation}
($\det_{Fr}$ is the Fredholm determinant).
The operator $\cK$ is a trace class operator in $L_2(M,E)$ and
hence the Fredholm determinant in (\ref{B600}) is defined.

We have
\begin{equation}
F\backslash\Ell_0^{\times}(M,E)=\SEll_0^{\times}(M,E).
\label{B601}
\end{equation}
Hence there is a natural projection
\begin{equation}
p\colon{\det}_*\SEll_0^{\times}(M,E)\to\SEll_0^{\times}(M,E)
\label{B602}
\end{equation}
with its fiber $F_0\backslash F=\wC^{\times}$.

\begin{pro}
The bundle $\det_*\SEll_0^{\times}(M,E)$ has a natural group structure.
\label{PB603}
\end{pro}

\noindent{\bf Proof.} For an arbitrary $A\in\Ell_0^{\times}(M,E)$
we have
$$
F_0A=AF_0
$$
since for $1+K_1\in F$ there exists $K_2\in F$ such that $(1+K_1)A=A(1+K_2)$.
Indeed, $K_2:=A^{-1}K_1A$ is a smoothing operator. We have
\begin{equation}
{\det}_{Fr}(1+K_2)={\det}_{Fr}\left(A(1+K_1)A^{-1}\right)={\det}_{Fr}(1+K_1).
\label{B625}
\end{equation}
So $1+K_2\in F_0$ for $1+K_1\in F_0$.
Hence $F_0$ is a normal subgroup in $\Ell_0^{\times}(M,E)$ and
the quotient on the left in (\ref{B601}) has the group structure
induced from the group $\Ell_0^{\times}(M,E)$.\ \ \ $\Box$

We call this group $\det_*\SEll_0^{\times}(M,E)=:G(M,E)$
the {\em determinant Lie group}.

A fiber-product of the groups $\Ell_0^{\times}(M,E)$ and $G(M,E)$ over
their common quotient $\SEll_0^{\times}(M,E)$ is defined by
\begin{equation}
\DEll_0^{\times}(M,E):=\Ell_0^{\times}(M,E){\stackrel{\times}{_{\SEll_0
^{\times}(M,E)}}}F_0\backslash\Ell_0^{\times}(M,E).
\label{B610}
\end{equation}
This fiber-product consists of classes of equivalence for pairs
$$
(A,B)\in\Ell_0^{\times}(M,E)\times\Ell_0^{\times}(M,E)
$$
with equal symbols $\sigma(A)=\sigma(B)$, where the equivalence relation
is $(A_1,B_1)\sim(A_2,B_2)$ if $A_1=A_2$ and $B_1B_2^{-1}\in F_0$.
There is a natural projection $(A,B)\to A$,
\begin{equation}
p_1\colon\DEll_0^{\times}(M,E)\to\Ell_0^{\times}(M,E).
\label{B611}
\end{equation}
We have a commutative diagram
\begin{gather}
\begin{CD}
1 @>>>\wC^{\times} @>>> \DEll_0^{\times}(M,E) @>>p_1> \Ell_0^{\times}(M,E)
@>>> 1 \\
@.    @.         @VVp_2V                       @VV\sigma V               @. \\
1 @>>>\wC^{\times} @>>> F_0\backslash\Ell_0^{\times}(M,E) @>>> \SEll_0
^{\times}(M,E) @>>> 1
\end{CD}
\label{B612}
\end{gather}
where $p_1(A,B)=A$, $p_2(A,B)$ is the class of $B$ in $G(M,E)$
($=F_0\backslash\Ell_0^{\times}(M,E)=\det_*\SEll_0^{\times}(M,E)$),
$\sigma$ is the symbol map. The horizontal lines
in this diagram are group extensions.

\begin{pro}
The extension $\DEll_0^{\times}(M,E)$ of $\Ell_0^{\times}(M,E)$
is trivial, i.e., the pullback under $\sigma$ of the extension
$\det_*\SEll_0^{\times}(M,E)\to\SEll_0^{\times}(M,E)$
to $\Ell_0^{\times}(M,E)$ is isomorphic to the direct product
of groups $\wC^{\times}\times\Ell_0^{\times}(M,E)$.
\label{PB614}
\end{pro}

\noindent{\bf Proof.} The fiber $p_1^{-1}(A)$ (in the top line
of (\ref{B612})) is the set of $B\in\Ell_0^{\times}(M,E)$ with
$\sigma(B)=\sigma(A)$ up to equivalence relation $B\sim B_1$
if $B\in F_0B_1$.

There is a canonical element $F_0A$ in $p_1^{-1}(A)$ which is
the equivalence class of $A$, Thus we define a section of $p_1$.
It is obviously a group homomorphism.\ \ \ $\Box$

%

To any $A\in\Ell_0^{\times}(M,E)$ corresponds a point
$d_1(A)\in{\det}_*\SEll_0^{\times}(M,E)=G(M,E)$. Namely $d_1(A)$
is the image of $A$ in $F_0\backslash\Ell_0^{\times}(M,E)=G(M,E)$.
The group structure on $\det_*\SEll_0^{\times}(M,E)$ comes
from $\Ell_0^{\times}(M,E)$. So we have
\begin{equation}
d_1(AB)=d_1(A)d_1(B)
\label{B2022}
\end{equation}
for $A,B\in\Ell_0^{\times}(M,E)$.

Let $A_1=QA$, where $Q\in F$. Then we have
\begin{equation}
d_1(A_1)={\det}_{Fr}(Q)\cdot d_1(A),
\label{B621}
\end{equation}
where $\det_{Fr}(Q)$ is defined by the image of $Q$
in $F_0\backslash F=\wC^{\times}$.

The problem is to describe the Lie group
\begin{equation}
{\det}_*\SEll_0^{\times}(M,E)=:G(M,E)
\label{B623}
\end{equation}
without the use of Fredholm determinants.

It occurs that the Lie algebra of this group is explicitly isomorphic
to the Lie algebra $\sfrg$. (This Lie algebra is defined by the associative
system of identifications $W_{l_1l_2}\colon\sfrg_{(l_1)}\to\sfrg_{(l_2)}$
of the Lie algebras $\sfrg_{(l_j)}$. These Lie algebras are defined
by (\ref{B307}), (\ref{B624}) and are identified by $W_{l_1l_2}$ given
by Proposition~\ref{PB403}.) We call $\sfrg$ the {\em determinant Lie
algebra}.

The fiber bundle (\ref{B602}) has a partially defined canonical
section. Let a symbol $S\in\SEll_0^d(M,E)$ of an order $d\in\wR^{\times}$
elliptic operator admit a cut $L_{(\theta)}$ of the spectral plane.
Let $A\in\Ell_0^d(M,E)$ be an elliptic operator with the symbol
$S=\sigma(A)$ and such that $\Spec(A)\cap L_{(\theta)}=\emptyset$.
Then $\det_{(\theta)}(A)$ is defined by (\ref{B3}). An element $d_1(A)$
of the fiber $p^{-1}(S)$ of (\ref{B602}),
$p\colon G(M,E)\to\SEll_0^{\times}(M,E)$, is also defined. This fiber
$p^{-1}(S)$ is a principal homogeneous $\wC^{\times}$-space.
Hence the element
\begin{equation}
d_0(A):=d_1(A)\big/{\det}_{(\theta)}(A)\in p^{-1}(S)
\label{B626}
\end{equation}
is defined. We suppose from now on that $\theta=\pi$.

\begin{pro}
The element $d_0(A)$ is independent of $A\in p^{-1}(S)$.
\label{PB627}
\end{pro}

\noindent{\bf Proof.} Let $A_1,A_2\in p^{-1}(S)$. Then $A_2=QA_1$,
$Q\in F$, $d_1(A_2)=\det_{Fr}(Q)\det_{(\pi)}(A_1)$.
According to Proposition~\ref{PB630} below
we have
\begin{equation}
{\det}_{(\pi)}(QA_1)={\det}_{Fr}(Q){\det}_{(\pi)}(A_1).
\label{B628}
\end{equation}
(We suppose that $\Spec(QA_1)\cap L_{(\pi)}=\emptyset$.)\ \ \ $\Box$

\begin{rem}
To define $d_1(A)$, we don't need the order of $A$ to be real.
To defined $\det(A)$ for an elliptic PDO $A$ of a nonzero order,
we need a holomorphic family $A^{-s}$ only. (Such a family may exist
even if $A$ does not have an admissible cut of the spectral plane.)
If such a family is given, then the element
$d_1(A)/\det(A)\in p^{-1}(\sigma(A))$ is defined. (This element
depends on a family $A^{-s}$ and not on $A$ only.) We denote the element
$d_1(A)/\det(A)$ by $\tilde{d}_0(\log A)$. (Here, the family $A^{-s}$
is defined by $\log A$.)
\label{RB3870}
\end{rem}

For a zeta-regularized  determinant $\det_\zeta(A)$ of an elliptic
operator $A\in\Ell_0^{\times}(M,E)$ to be defined, its complex powers
$A^{-s}$ have to be defined.
Hence a logarithm $\log A$ of $A$ has to be defined.
However for $(M,E)$ such that $\dim M\ge 2$ and $\rk E\ge 2$ there are not
any continuous logarithms for a nonempty open set of the principal symbols
of elliptic operators from $\Ell_0^{\times}(M,E)$. Hence for operators
$A$ with such principal symbols their $\log A$ and $\det_\zeta(A)$
are not defined.

\begin{rem}
The principal symbol $a_\alpha$ of an elliptic operator
$A\in\Ell_0^\alpha(M,E)$ defines the element
$a_\alpha|_{S^*M}\in\Aut(\pi^*E)$, where $\pi\colon S^*M\to M$ is
the natural projection. For $\rk E\ge 3$ there is an open nonempty set
of the automorphisms as follows. There is a point $q\in S^*M$ such that
$a_\alpha(q)$ has a form
\begin{equation*}
a_\alpha(q)=\left(
\begin{split}
\lambda & \ 1 \\
0       & \ \lambda
\end{split}
\right)\oplus a_\alpha^1(q),
\end{equation*}
where $a_\alpha^1(q)$ acts on an invariant (with respect to $a_\alpha(q)$)
complement to the two-dimensional
$\lambda$-eigenspace of $a_\alpha(q)$ in $\left(\pi^*E\right)_q$.
(In general, multiple eigenvalues of $\Aut(\pi^*E)$ appear over a subset
of codimension two in $S^*M$, and $\dim S^*M\ge 3$ for $\dim M\ge 2$.
In general, multiple eigenvalues appear in Jordan blocks.) Then there is
a smooth curve $f\colon\left(S^1,pt\right)\to(S^*M,q)$, $t\to f(t)$,
such that two eigenvalues $\lambda$ over $q=f(t_0)$ vary as $\lambda_1(t)$
and $\lambda_2(t)$, where $\lambda_1(t)\ne\lambda_2(t)$
at $t\in S^1\setminus t_0$ and $\lambda_1(t)/|\lambda_1(t)|$ and
$\lambda_2(t)/|\lambda_2(t)|$ are the maps
$f_i\colon\left(S^1,pt\right)\to S^1$, $i=1,2$, of different degrees.
Hence there is no continuous logarithm
$\log\left(f^*a_\alpha\right)\in\End(f^*\pi^*E)$ of $a_\alpha(f(t))$
over the circle of parameters $t\in S^1$. (Here, we suppose that
$f^*\pi^*E$ is a trivial bundle over $S^1$.) Such a curve $f(t)$ can
appear in a coordinate neighborhood of a point $q\in S^*M$ (and $\pi^*E$
is a trivial bundle over this curve). To see this, it is enough to take
$\lambda_3(t)$ sufficiently close
to $\left(\lambda_1(t)\lambda_2(t)\right)^{-1}$. Then the map
from $\left(S^1,pt\right)$ to $GL_3(\wC)$ equal
to $\Aut\left(\wC^2\right)\oplus\lambda_3(t)$ (where $\Aut\left(\wC^2\right)$
has the eigenvalues $\lambda_1(t)$, $\lambda_2(t)$) is homotopic to a map
to $SL_3(\wC)$. Hence the map $S^1\to GL_3(\wC)$ is homotopic to a trivial
map. Let $\rk E=2$ and let the degrees
of $f_i\colon\left(S^1,pt\right)\to S^1$, $j=1,2$, be the opposite numbers
(i.e.,
$\sum\deg f_i=0$). Then $f\colon\left(S^1,pt\right)\to GL_2(\wC)$
is homotopic to a trivial map. Hence for $\rk E\ge 2$ and for $\dim M\ge2$,
all the conditions are satisfied on an open set in the space of principal
elliptic symbols. This open set is nonempty in a connected component
of a trivial symbol (because these conditions can be satisfied over
a smooth closed curve in a coordinate neighborhood in $S^*M$, and over
this curve a map $f$ from $S^1$ to $GL_n(\wC)$ is homotopic to a trivial
map, $n:=\rk_{\wC}E$).
%
\label{RB3020}
\end{rem}

\begin{rem}
Let us generalize the notion of a spectral cut to the case of operators
of complex orders. Let $L\in S_{\log}(M,E)$ be a logarithmic symbol
of a nonzero order $z\in\wC^{\times}$. Let $\left\{U_i\right\}$ be
a finite cover of $M$ by local coordinate charts (with local
trivializations $E|_{U_i}$). Let $\overline{V}_i\subset U_i$ be a cover
of $M$ by (closed) coordinate disks. Let
$$
L=z\log|\xi|+L_0(x,\xi)+L_{-1}(x,\xi)+\ldots
$$
be the components of this logarithmic symbol in $U_i$. Set
\begin{equation}
\alpha(L):=\diam\cup_i\cup_{x\in\overline{V}_i,\;\xi\ne 0}\Im\left(\Spec\left(
L_0(x,\xi)/z\right)\right).
\label{B3925}
\end{equation}
Here, $\Spec L_0$ is the spectrum of a square matrix $L_0$ (its size is
$\rk E$). The symbol $L$ can be represented as $\df_s\exp(sL)|_{s=0}$.
Here, $\exp(sL)=:(\exp L)^s$ is a holomorphic family of classical
PDO-symbols. There is an explicit formula for changing of space coordinates
in PDO-symbols on a manifold, \cite{Sh}, Theorem~4.2. By this formula
we conclude that $\alpha(L)$, (\ref{B3925}), is independent of local
coordinates on $M$ (for given $L$ and a smooth structure on $M$).
We are sure that {\em under the condition}%
\footnote{We suppose that $z\ne 0$ and that $z\notin i\wR$.}
\begin{equation}
|z|^2\alpha(L)/|\Re z|< 2\pi,
\label{B3927}
\end{equation}
{\em any invertible elliptic PDO $A\in\Ell_0^z(M,E)$ with its symbol
$\sigma(A):=\exp L$ has a $\log A\in\fell(M,E)$}. The symbol $\exp L$
is defined as a solution $s_t|_{t=1}$ in $\SEll_0(M,E)$ of the equation
\begin{equation}
\df_ts_t=Ls_t, \quad s_0:=\Id.
\label{B3928}
\end{equation}

{\bf Hypothesis}. {\em Let $A$ be an invertible elliptic PDO of order $z$,
$\Re z\ne 0$. Let $\sigma(A)=\exp L$ for $L\in S_{\log}(M,E)$ (i.e.,
let $\sigma(A)$ be $s_t|_{t=1}$ for the solution of (\ref{B3928})).
Then $\log A\in\fell(M,E)$ with $\sigma(\log A)=L$ exists and is unique
up to a change
of an operator $\log A$ on a finite-dimensional $A$-invariant linear
subspace $K$
in $\Gamma(E)$, $AK=K$. So a family $A^s$ of complex powers for $A$
exists and is unique up to a redefinition of it on a finite-dimensional
$A$-invariant subspace $K$.}

To explain the condition (\ref{B3927}) and the hypothesis
on $L:=\sigma(\log A)$,
let us choose an element $B\in\fell(M,E)$ with the symbol $L$,
$\sigma(B)=L$. Then the element $\exp B\in\Ell_0^z(M,E)$ is defined
as $b_t|_{t=1}$ for the solution of the equation $\df_tb_t=Bb_t$,
$b_0=\Id$, in $\Ell_0(M,E)$. Then $\sigma(\exp B)=\sigma(A)$ and
$b_1:=\exp B$ is invertible. Let $A_t$ be a smooth curve in $\Ell_0^z(M,E)$
such that $A_0=\exp B$, $A_1=A$, and $\sigma\left(A_t\right)=\sigma(A)$
for $t\in[0,1]$. We want to prove that there exists a smooth curve
$B_t$ in $\fell(M,E)$, $\ord B_t=z$, such that $\exp B_t=A_t$, i.e.,
to prove that there exists a smooth family of logarithms%
\footnote{This problem is connected with the problem of using a kind
of the Campbell-Hausdorff formula outside the domain of its convergence.}
$$
B_t:=\log A_t\in\fell(M,E), \quad\sigma\left(B_t\right)=L \quad\text{ for }
t\in[0,1].
$$
To find $B_t$, we have to prove the existence of a solution of an ordinary
differential equation
\begin{equation}
F^{-1}\left(\ad B_t\right)\circ\left(\df_tA_t\cdot A_t^{-1}\right)=\df_tB_t,
\quad B_0:=B.
\label{B3929}
\end{equation}
(Here we use Lemma~\ref{LB616} and Remark~\ref{RB621} below,
$F^{-1}(t):=t/(\exp t-1)$. We use also that $B=\log A_0$ exists.)
Under the condition (\ref{B3927}), we claim that $\ad\left(B_t\right)$
for any $B_t$ with $\sigma\left(B_t\right)=L$ has only a finite number
of eigenvalues from $2\pi i\wZ\setminus 0$, and all these eigenvalues
are of finite (algebraic) multiplicities. So the operator
$F^{-1}\left(\ad B_t\right)$ is defined on an $\ad B_t$-invariant subspace
of a finite codimension. However the equation
(\ref{B3929}) is nonlinear, and it is difficult to prove the existence
of its solution $B_t$.

Suppose we can prove that a smooth family $\log A_t$ exists. Then we can
prove (Proposition~\ref{PB3931}) that the following equality holds
\begin{equation}
\det\left(A_1\right)=\det\left(A_0\right){\det}_{Fr}\left(A_1A_0^{-1}\right).
\label{B3930}
\end{equation}
(Here, $A_j$ are invertible, $\sigma\left(A_0\right)=\sigma\left(A_1\right)$,
$A_1A_0^{-1}\in F$.) This equality is a generalization of (\ref{B628}).
We don't suppose in (\ref{B3930}) that $A_1$ and $A_0$ possess spectral
cuts. We suppose only that a smooth in $t$ family $\left(A_t\right)^s$,
$0\le t\le 1$, of complex powers exists (i.e., that there is a smooth
family of logarithms $\log A_t\in\fell(M,E)$ of order $z$ elliptic PDOs
$A_t$).
\label{RB3924}
\end{rem}

\begin{rem}
A given elliptic symbol $\sigma(A)\in\Ell_0^z(M,E)$, $z\in\wC^{\times}$,
can have different logarithmic symbols $\sigma(\log A)$.
Let $z\notin i\wR$. Then the condition (\ref{B3927}) can be satisfied
for some
$\sigma(\log A)\in S_{\log}(M,E)$ and unsatisfied for another
$\sigma(\log A)$. This condition cannot be formulated as a condition
on $\sigma(A)$.
\label{RB3963}
\end{rem}

\begin{pro}
The equality (\ref{B628}) holds for an invertible
$Q\in\{\Id+\cK\}=:F$ (where $\cK$ is a smoothing operator,
i.e., it has a $\wC^{\infty}$ Schwartz kernel on $M\times M$), and
for an invertible $A\in\Ell_0^d(M,E)$, $d\in\wR^{\times}$, such that
$A$ is sufficiently close to a positive definite self-adjoint PDO.%
\footnote{Under this condition operators $A$ and $QA$ possess a cut
$L_{(\theta)}$ of the spectral plane for almost all $\theta$ close
to $\pi$ (i.e., except a finite number of $\theta$'s).}
\label{PB630}
\end{pro}

\noindent{\bf Proof.} Let $Q=\Id+\cK$ be an operator from $F$
($\cK$ is a compact operator in $L_2(M,E)$. Hence its spectrum is discrete
in $\wC\setminus 0$ with a unique possible accomulation point at zero.)
Let  there be no eigenvalues of $Q$ from $\wR_-$.
Then $\log_{(\pi)}Q$ is defined by the integral analogous to (\ref{B7})
\begin{equation}
{\log}_{(\pi)}Q={i\over 2\pi}\int_{\Gamma_{R,\pi}}{\log}_{(\pi)}\lambda\cdot
\left(Q-\lambda\right)^{-1}d\lambda.
\label{B790}
\end{equation}
Here, $\left(Q-\lambda\right)^{-1}$ is the resolvent of the bounded
linear operator $Q$ in $L_2(M,E)$. (The contour $\Gamma_{R,\pi}$
is the same as in (\ref{B7}) with $\tpi=\pi$.)
The operator $\log_{(\pi)}Q=:C$
is an operator with a $C^\infty$-kernel on $M\times M$.

For any $\eps>0$ all the eigenvalues $\lambda$ of $Q$ except a finite
number of them are in the spectral cone
$\{\lambda\colon-\eps<\arg\lambda<\eps\}$. So, if $\Spec Q$ does not
contain $0$, then in an arbitrary small conical neighborhood of $L_{(\pi)}$
there is a spectral cut $L_{(\theta)}$ such that
$\Spec Q\cap L_{(\theta)}=\emptyset$. For $0\notin\Spec Q$ the logarithm
$\log_{(\theta)}Q=:C$ is defined. It is defined as $\log_{(\tpi)}Q$
by (\ref{B790}) with the integration contour $\Gamma_{R,\tpi}$.

Set $Q_t:=\exp(tC)$, $0\le t\le 1$, $A_t:=Q_tA$. Let $\ord A\in\wR_+$.
We have for $\Re s>\dim M/\ord A$
\begin{equation}
\zeta_{A_t,(\tpi)}(s):=\Tr\left({i\over 2\pi}\int_{\Gamma_{(\tpi)}}\lambda
_{(\tpi)}^{-s}\left(A_t-\lambda\right)^{-1}d\lambda\right).
\label{B791}
\end{equation}
Here, $\Gamma_{(\tpi)}$ is the contour $\Gamma_{(\theta)}$ from (\ref{A5})
with an admissible $\theta$ sufficiently close to $\pi$ and
$\lambda_{(\tpi)}^{-s}$ is defined as in (\ref{B2}).
For such $s$
we have
\begin{multline}
\df_t\zeta_{A_t,(\tpi)}(s)=\Tr\left({i\over 2\pi}\int_{\Gamma_{(\tpi)}}\lambda
_{(\tpi)}^{-s}\left(-\left(A_t-\lambda\right)^{-1}CA_t\left(A_t-\lambda
\right)^{-1}\right)d\lambda\right)=\\
=\Tr\left({i\over 2\pi}\int_{\Gamma_{(\tpi)}}\lambda_{(\tpi)}^{-s}\left(-CA_t
\left(A_t-\lambda\right)^{-2}\right)d\lambda\right)=\\
=\Tr\left({i\over 2\pi}\int_{\Gamma_{(\tpi)}}
\lambda_{(\tpi)}^{-s}\left(-\df_\lambda\left(CA_t\left(A_t-\lambda\right)
^{-1}\right)\right)d\lambda\right)=\\
=-s\Tr\left({i\over 2\pi}\int\lambda_{(\tpi)}^{-(s+1)}CA_t\left(A_t-\lambda
\right)^{-1}d\lambda\right)=-s\Tr\left(CA_{t,(\tpi)}^{-s}\right),
\label{B792}
\end{multline}
since $\left(\!A_t\!-\!\lambda\!\right)^{-1}\!C\!A_t\!\left(\!A_t\!-\!\lambda
\!\right)^{-1}$ and $\left(\!A_t\!-\!\lambda\!\right)^{-2}\!C\!A_t$
are trace class operators in $L_2\!(\!M\!,\!E\!)$
whose trace norms are $O\left(\left|\lambda\right|^{-1}\right)$
for $\lambda\in\Gamma_{(\tpi)}$. So
\begin{equation}
\df_t\zeta_{A_t,(\tpi)}(s)=-s\Tr\left(CA_{t,(\tpi)}^{-s}\right)
\label{B793}
\end{equation}
for $\Re s>\dim M/\ord A$. The term $\Tr\left(CA_{t,(\tpi)}^{-s}\right)$
on the right in (\ref{B793}) is a meromorphic function of $s$
by Proposition~\ref{PB3755} and Remark~\ref{RB4}. It is a trace class operator
for all $s\in\wC$. Hence
$\Tr\left(CA_t^{-s}\right)$ is holomorphic in $s\in\wC$ and it is equal
to $\Tr C$ for $s=0$.

\begin{lem}
Under the conditions of Proposition~\ref{PB630} and for $\ord A\in\wR_+$,
the equality holds
\begin{equation}
\df_t\left(\df_s\zeta_{A,(\tpi)}(s)\big|_{s=0}\right)=-\Tr C.
\label{B795}
\end{equation}
Here, $C:=\log_{(\tpi)}Q$ is a trace class operator defined by (\ref{B790}).
\label{LB794}
\end{lem}

\begin{cor}
Under the conditions of Proposition~\ref{PB630}, we have
\begin{multline}
{\det}_{(\tpi)}(QA)\big/{\det}_{(\tpi)}(A)=\exp\left(
\int_0^1dt\Tr(C)\right)=\exp(\Tr(C))=\\
={\det}_{Fr}(\exp C)={\det}_{Fr}(Q).
\label{B97}
\end{multline}
\label{CB796}
\end{cor}

Proposition~\ref{PB630} is proved.\ \ \ $\Box$

\noindent{\bf Proof of Lemma~\ref{LB794}.}
The factor $\Tr\left(CA_{t,(\tpi)}^{-s}\right)$ on the right in (\ref{B793})
is defined for all $s\in\wC$. (Indeed, $CA_{t,(\tpi)}^{-s}$ is a trace class
operator since $C$ is of trace class and $A_{t,(\tpi)}^{-s}$ is a PDO from
$\Ell_0^{-s\ord A}(M,E)$.)
Note that the value of $\Tr\left(CA_t^{-s}\right)$ at $s=0$ is defined
and is equal to $\Tr(C)$ (since $A_{t,(\tpi)}^{-s}|_{s=0}=\Id$, $A$ is
invertible).
Thus the equality (\ref{B795}) follows from (\ref{B793}).\ \ \ $\Box$

\begin{rem}
The equality (\ref{B628}) may be also obtained from the assertions
as follows.

1. Note that for $A\in\Ell_0^d(M,E)$, $d\in\wR^{\times}$, sufficiently
close to a positive self-adjoint PDO, the ratio
${\det}_{(\tpi)}(QA)\big/{\det}_{(\tpi)}(A)=:f_A(Q)$
is independent of $A\in\Ell_0^d(M,E)$ and of $d\in\wR^{\times}$.
Indeed, let $A\in\Ell_0^{d_1}(M,E)$ and $C\in\Ell_0^{d_2}(M,E)$,
$d_j\in\wR^{\times}$, be two such operators and let $d_1\ne d_2$.
Set $B:=A^{-1}C\in\Ell_0^{d_2-d_1}(M,E)$. Then according to (\ref{B3872})
we have
\begin{multline}
f_{AB}(Q)\big/{\det}_{(\tpi)}(AB)={\det}_{(\tpi)}(QAB)=F(QA,B){\det}_{(\tpi)}
(QA){\det}_{(\tpi)}(B)=\\
=f_A(Q)F(QA,B){\det}_{(\tpi)}(A){\det}_{(\tpi)}(B)=\\
=f_A(Q)F(QA,B)\big/\left(F(A,B)\cdot{\det}_{(\tpi)}(AB)\right).
\label{B541}
\end{multline}
Here, $F(A,B)$ and $F(QA,B)$ are defined by (\ref{B3872}).
%
%
%
%
By (\ref{B3752}) $F(A,B)$ depends on symbols $\sigma(A)$,
$\sigma(B)$ only, $F(A,B)=F(QA,B)$.
Thus $f_A(Q)=f_C(Q)$. (For $d_1=d_2$
it is enough to take $D\in\Ell_0^d(M,E)$
with $d>d_1$ sufficiently close to a positive definite self-adjoint PDO.
We have $f_A(Q)=f_D(Q)=f_C(Q)$.)
Hence $f(Q):=f_A(Q)$ is independent of $A$. Note that
${\det}_{(\tpi)}(AQ)={\det}_{(\tpi)}(QA)=f(Q){\det}_{(\tpi)}(A)$,
since the operator $AQ$ is adjoint to $QA=A^{-1}(AQ)A$.
The value $f(Q)$ is defined for all $Q\in F$ as
${\det}_{(\tpi)}(QA)/{\det}_{(\tpi)}(A)$ and is independent
of an admissible cut $L_{(\tpi)}$ by Remark~\ref{RB3750}.

2. The function $f(Q)$ is multiplicative, i.e.,
$f\left(Q_1Q_2\right)=f\left(Q_1\right)f\left(Q_2\right)$.

Indeed, for PDOs $A\in\Ell_0^{d_1}(M,E)$ and $B\in\Ell_0^{d_2}(M,E)$,
$d_j\in\wR_+$, sufficiently close to positive definite self-adjoint PDOs
we have
\begin{multline}
f\left(Q_1Q_2\right){\det}_{(\tpi)}(AB)={\det}_{(\tpi)}\left(Q_1Q_2AB\right)=
{\det}_{(\tpi)}\left(Q_2ABQ_1\right)=\\
=F(A,B){\det}_{(\tpi)}\left(Q_2A\right){\det}_{(\tpi)}\left(BQ_1\right)=
f\left(Q_1\right)f\left(Q_2\right){\det}_{(\tpi)}(AB).
\label{B545}
\end{multline}

3. Let $A\in\Ell_0^d(M,E)$, $d\in\wR_+$, be a positive self-adjoint PDO.
Let $\{e_i\}$, $i\in\wZ_+$, be an orthonormal basis in the $L_2$-completion
of $\Gamma(M,E)$ consisting of the eigenvectors of $A$. (Such a basis
exists according to \cite{Sh}, Ch.~I, \S~8, Theorem~8.2.)

Let $Q$ be an operator with its matrix elements with respect to the basis
$\{e_i\}$, $Qe_i=\left((\lambda-1)\delta_{1i}+1\right)e_i$,
$\lambda\in\wC^{\times}$. Then $Q\in F$ and we have
\begin{equation}
\log{\det}_{(\tpi)}(QA)=-\df_s\zeta_{QA,(\tpi)}(s)\big|_{s=0}=
\log\lambda+\log\zeta_{A,(\tpi)}(s)\big|_{s=0}.
\label{B3885}
\end{equation}
Hence for this $Q$ we have $f(A)=\lambda$.
Since the $K_1$-functor $K_1(\wC)$ is equal to $\wC^{\times}$ (\cite{Mi})
and since $f(Q)$ is multiplicative in $Q$, we have $f(Q)=\det(Q)$
for $Q$ such that $Q-\Id$ is
a finite size invertible square matrix. (In (\ref{B3885}) $Q-\Id$
is equal to $\lambda-1$.)

4. For an arbitrary $Q\in F$ and for any $s\in\wR$, $N\in\wZ_+$ there
exists a sequence of $Q_i=\Id+K_i\in F$ with finite rank operators
$K_i$ such that $Q_i$ tends to $Q$ as $i\to\infty$ as a sequence
of operators from the Sobolev space $H^s(M,E)$ into $H^{s+N}(M,E)$.

%
Let $N$ be greater than $\ord A+\dim M$, $\ord A\in\wR^{\times}$.
Then ${\det}_{(\tpi)}(Q_iA)$ tends to ${\det}_{(\pi)}(QA)$ as $i$ tends
to infinity.
%
So we have
\begin{gather}
\begin{split}
{\det}_{(\tpi)}\left(Q_iA\right)\big/{\det}_{(\tpi)}(A) & =:f\left(Q_i\right)
={\det}_{Fr}\left(Q_i\right),\\
f(Q):={\det}_{(\tpi)}(QA)\big/{\det}_{(\tpi)}(A) & =\lim_{i\to\infty}{\det}
_{Fr}\left(Q_i\right)={\det}_{Fr}(Q).
\end{split}
\label{B552}
\end{gather}
The convergence $\det_{(\tpi)}\left(Q_iA\right)\to\det_{(\tpi)}(QA)$
as $i\to\infty$ follows from the Cauchy integral formula
for $\df_z\left(\zeta_{QA}(z)-\zeta_{Q_iA}(z)\right)$.
\label{RB3871}
\end{rem}

\begin{pro}
Let a smooth family of logarithms $\log A_t\in\fell(M,E)$, $0\le t\le 1$,
exist for some smooth curve $A_t$ of invertible elliptic operators
in $\Ell_0^z(M,E)$, $\sigma\left(A_t\right)=\sigma(A)$, $0\le t\le 1$.
Then the equality (\ref{B3930}) holds.
\label{PB3931}
\end{pro}

\noindent{\bf Proof.} Set $\zeta_{A_t}(s):=\TR\left(A_t^{-s}\right)$
for $sz\ne 0$. Then by Proposition~\ref{PB3755}
$\Res_{s=0}\zeta_{A_t}(s)=-\res\Id=0$. Hence by this Proposition
$\zeta_{A_t}(s)$ is regular at $s=0$. For $\Re(sz)>\dim M$, the operators
$A_t^{-s}$ are of trace class. In view of Remark~\ref{RB4} we conclude
(analogous to (\ref{B3515}), (\ref{B3516})) that for $\Re(sz)>\dim M$
the equalities hold
\begin{gather*}
\Tr\left(A_t^{-s}\right)=\TR\left(A_t^{-s}\right)=\zeta_{A_t}(s), \\
\df_t\zeta_{A_t}(s)=-s\TR\left(\dot{A}_tA_t^{-1}\cdot A_t^{-s}\right)=-s\Tr
\left(\dot{A}_tA_t^{-1}\cdot A_t^{-s}\right).
\end{gather*}
Here, $\dot{A}_tA_t^{-1}=:C_t$, where $C_t$ is a trace class operator.
Hence
\begin{equation}
\df_t\df_s\left(-\zeta_{A_t}(s)\right)\big|_{s=0}=\Tr\left(C_tA_t^{-s}\right)
\big|_{s=0}.
\label{B3962}
\end{equation}
The expression on the right in (\ref{B3962}) for all $s$ are the traces
of trace class operators. Hence this expression is regular for all $s$,
and we can set $s=0$ on the right in (\ref{B3962}),
$$
\df_t\df_s\left(-\zeta_{A_t}(s)\right)\big|_{s=0}=\Tr\left(C_t\right).
$$
Thus
\begin{equation}
\det\left(A_1\right)/\det\left(A_0\right)=\exp\left(\int_0^1\Tr\left(C_t
\right)dt\right)={\det}_{Fr}\left(A_1A_0^{-1}\right).
\label{B3995}
\end{equation}
The formula (\ref{B3930}) is applicable to this case.\ \ \ $\Box$

{\bf Definition}. Let $A$ be an invertible elliptic PDO of a nonzero
complex order $z\in\wC^{\times}$, $A\in\Ell_0^z(M,E)$, such that
a logarithmic symbol $\sigma(\log A)\in S_{\log}(M,E)$ exists (i.e.,
$\sigma(A)=s_t|_{t=1}$ for a solution $s_t$ of (\ref{B3925})).
Let $B$ be any element of $\fell(M,E)$ with $\sigma(B)=L$. Then
the element $b:=\exp B\in\Ell_0^z(M,E)$ is defined as a solution
$b_t|_{t=1}$ of $\df_tb_t=Bb_t$, $b_0=\Id$. The canonical section
of $G(M,E)$ over $\exp L$ is defined by
\begin{equation}
\tilde{d}_0(B\!)\!:=\!d_1(\exp B\!)\!/\!\det(\exp B\!),\;\det(\exp B\!)\!:=
\!\exp\left(\!-\!\df_s\TR(\exp(-sB\!))\!\big|_{s=0}\!\right),
\label{B3943}
\end{equation}
$\det(\exp B)\in\wC^{\times}$. By Proposition~\ref{PB3952} below
$\tilde{d}_0(B)$
depends on $\sigma(B)=L\in S_{\log}(M,E)$ only. Thus we can define
$\tilde{d}_0(\sigma(\log A))$ by the expression on the right
in (\ref{B3943}) for any $B\in\fell(M,E)$ with $\sigma(B)=\sigma(\log A)$.

\begin{rem}
We don't suppose in the definition of $\tilde{d}_0(\sigma(\log A))$ that
there exist a $\log A\in\fell(M,E)$. We can take any $B\in\fell(M,E)$
with $\sigma(B)=\sigma(\log A)$ in $S_{\log}(M,E)$ and define
$\tilde{d}_0(\sigma(\log A))$ as the expression on the right
in (\ref{B3943}).
\label{RB3945}
\end{rem}

\begin{rem}
The definition of $\tilde{d}_0(\sigma(\log A))$ provides us with
a canonical prolongation of the {\em zeta-regularized determinants}
to a domain where {\em zeta-functions} of elliptic operators
{\em do not exist}.
Namely let $A\in\Ell_0^z(M,E)$, $z\in\wC^{\times}$, be an invertible
elliptic PDO such that $\sigma(\log A)$ is defined. (However we do not
suppose that a $\log A\in\fell(M,E)$ exists. An element $L\in S_{\log}(M,E)$
is a symbol $\sigma(\log A)$, if $\sigma(A)$ is equal to $s_t|_{t=1}$
for a solution $s_t$ of (\ref{B3925}).) Then $\det(A)$ (corresponding
to a given $L=\sigma(\log A)$) is defined by
\begin{equation}
\det(A):=d_1(A)/\tilde{d}_0(\sigma(\log A)).
\label{B3947}
\end{equation}
If $\sigma(\log A)$ exists but $\log A$ does not exist, then $\det(A)$
can be canonically defined by (\ref{B3947}). However zeta-regularized
determinants $\det_\zeta(A)$ are not defined in this case. (Indeed,
for any $\det_\zeta(A)$ to be defined, an appropriate zeta-function
$\zeta_A(s)$ has to be defined. But if a $\log A$ does not exist, then
a family of complex powers $A^{-s}$ does not exist.)

The formula (\ref{B3947}) provides us with a definition
of a {\em canonical determinant} of elliptic PDOs {\em in its natural
domain of definition}. This determinant is a function
of $A\in\Ell_0^d(M,E)$, $d\in\wC^{\times}$, and of $\sigma(\log A)$.
A simple sufficient condition for the existence of $\sigma(\log A)$
is given in Remark~\ref{RB3964} below. For zero order elliptic PDOs
of the odd class on an odd-dimensional closed manifold, a definition
of their canonical determinant is given in Section~\ref{SC},
Corollary~\ref{CB243}.

Some clearing
and explanation of the problem of the existence of a $\log A$ if
$\sigma(\log A)$ exists, is contained in Remark~\ref{RB3924}.
Let $A$ be an invertible elliptic PDO such that $\sigma(\log A)$ exists
and $\ord A\in\wC^{\times}$ but such that the condition (\ref{B3927})
for $\sigma(\log A)$ is not satisfied. Then we are sure that in general
$\log A$ does not exist (though $\sigma(\log A)$ exists
by our supposition).

Note also that if $A$ is an elliptic operator of a real positive order
$d$ and if $\sigma_d(A)$ is sufficiently close to a positive definite
self-adjoint symbol, the function $\zeta_A(s):=\zeta_{A,\tpi}(s)$ and
$\log_{(\tpi)}A$ can be defined by an admissible cut $L_{(\tpi)}$
of the spectral plane
(sufficiently close to $L_{(\pi)}$), Section~\ref{SA}. In this case,
the determinant (\ref{B3947}) coincides with $\det_\zeta(A)$. Namely
in this case,
\begin{equation}
d_0(A)=\tilde{d}_0(\sigma(\log_{(\tpi)}A)), \quad \det(A)=\exp\left(-\df_s
|_{s=0}\zeta_{A,\tpi}(s)|_{s=0}\right)={\det}_\zeta(A).
\label{B3948}
\end{equation}
\label{RB3946}
\end{rem}

\begin{rem}
With respect to the exponential maps in the determinant Lie group $G(M,E)$
and in the group $\Ell_0^{\times}(M,E)$ of invertible elliptic PDOs
of complex orders the situations are different.
Namely these groups are fiber bundles over their quotients,
\begin{align}
p_G & \colon G(M,E)\to\SEll_0^{\times}(M,E),
\label{B3950} \\
p_E & \colon\Ell_0^{\times}(M,E)\to\SEll_0^{\times}(M,E).
\label{B3951}
\end{align}
The fiber of (\ref{B3950}) is $F_0\backslash F=\wC$ and the fiber
of (\ref{B3951}) is $F$ ($F$ and $F_0$ are defined at the beginning
of this section, (\ref{B600})). The image of the exponential map
in $G(M,E)$, (\ref{B3950}), contains the whole fibers of $p_G$.
Indeed, if $\sigma(\log A)\in S_{\log}(M,E)$ exists, then
$\{\exp(\sigma(\log A)+c\cdot 1)\text{ for }c\in\wC\}$
in $\exp\left(\sfrg_{(l)}\right)\to G(M,E)$ is
$\wC^{\times}\cdot d_1(A)=p_G^{-1}(\sigma(A))$. (If a $\sigma(\log A)$
does not exist, then there are no points in $p_G^{-1}(\sigma(A))$
belonging to the image of $\exp$ in $G(M,E)$.)
But in $\Ell_0^{\times}(M,E)$, (\ref{B3951}), the picture is completely
different. Namely let $A\in\Ell_0^{\times}(M,E)$, $\ord A\ne 0$, and
let $\sigma(A)$ have a logarithm $\sigma(\log A)$. However let
the condition (\ref{B3927}) be not satisfied for $\sigma(\log A)$.
Then we are sure that in general there are no $\log A$ with given
$\sigma(\log A)$. But it is clear that there is an element
$B\in\fell(M,E)$ with $\sigma(B)=\sigma(\log A)$. So
$\exp B\in p_E^{-1}(\sigma(A))\subset\Ell_0^{\times}(M,E)$ and
the image of $\exp$ in $\Ell_0^{\times}(M,E)$ contains some points
in $p_E^{-1}(\sigma(A))$. However if the condition (\ref{B3927})
is not satisfied for $\sigma(\log A)$, then the differential of the map
$B\to\exp B$ for $B\in\fell(M,E)$, $\sigma(B)=\sigma(\log A)$, is not
a map onto fibers of $T\left(p_E^{-1}(\sigma(A))\right)$.
\label{RB3949}
\end{rem}

\begin{rem}
There is a rather simple sufficient condition for the existence
of $\sigma(\log\!A\!)\!\in S_{\log}(M,E)$ for a given elliptic symbol
$\sigma(A)\in\Ell_0^{\times}(M,E)$. It is enough that there is a smooth
field of cuts $L_{(\theta)}(x,\xi)$ over points $(x,\xi)$
of a cospherical bundle $S^*M$ such that $L_{(\theta)}(x,\xi)$ is
admissible for the principal symbol $\sigma_d(A)(x,\xi)$ (i.e., that
$\sigma_d(A)(x,\xi)$ has no eigenvalues on $L_{(\theta)}$) and that
there is a smooth function $f\colon S^*M\to\wR$ such that
$\theta(x,\xi)=f(x,\xi)$ modulo $2\pi\wZ$. Under these conditions,
the symbol $\sigma\left(A^z\right)$ is defined by the formulas (\ref{A3}),
(\ref{A5})
of Section~\ref{SA}. Here the existence of $f(x,\xi)$ is used
in the definition of $\lambda^z$ in (\ref{A5}) (since the branch
of $\lambda^z$ over $(x,\xi)$ has to be changed smoothly
in $(x,\xi)\in S^*M$).
The condition of the existence of $f$ is equivalent to a topological
condition that the map $\theta\colon S^*M\to S^1=\wR/2\pi\wZ$
is homotopic to a trivial one.
\label{RB3964}
\end{rem}

\begin{pro}
Let $B_1$,\;$B_2$ be elements of $\fell(M,\!E)$ with the same symbols,
$\sigma\!\left(B_1\!\right)\!=\sigma\left(B_2\right)$, and such that
$\ord B_j\in\wC^{\times}$. Then
$\tilde{d}_0\left(B_1\right)=\tilde{d}_0\left(B_2\right)$ (where
$\tilde{d}_0\left(B\right)$ is defined by (\ref{B3943})), i.e.,
\begin{equation}
d_1\left(\exp B_1\right)/\det\left(\exp B_1\right)=d_1\left(\exp B_2\right)/
\det\left(\exp B_2\right).
\label{B3953}
\end{equation}
\label{PB3952}
\end{pro}

\noindent{\bf Proof.} Let $B_t$, $1\le t\le 2$, be a smooth curve
in $\fell(M,E)$ from $B_1$ to $B_2$ such that
$\sigma\left(B_t\right)=\sigma\left(B_j\right)$ for $t\in[1,2]$. Then
$d_1\left(\exp B_t\right)/\det\left(\exp B_t\right)=:\tilde{d}_0\left(B_t
\right)$ is defined for $t\in[1,2]$. By the definition of $G(M,E)$
we have
\begin{multline}
d_1\left(\exp B_2\right)=d_1\left(\exp B_2\exp(-B_1)\right)d_1\left(\exp B_1
\right)=\\
={\det}_{Fr}\left(\exp B_2\exp(-B_1)\right)d_1\left(\exp B_1\right)
\label{B3954}
\end{multline}
(because $\exp B_2\exp(-B_1)\in F$ and the identification
$F_0\backslash F\rs\wC^{\times}$ is given by the Fredholm determinant).

By Proposition~\ref{PB3931} we have
\begin{equation}
\det\left(\exp B_2\right)={\det}_{Fr}\left(\exp B_2\exp(-B_1)\right)
\det\left(\exp B_1\right).
\label{B3955}
\end{equation}
(Here, $\det\left(\exp B_j\right)$ are defined by (\ref{B3943}).)
So (\ref{B3953}) follows immediately from (\ref{B3954}),
(\ref{B3955}).\ \ \ $\Box$

\begin{cor}
The definition (\ref{B3943}) of $\tilde{d}_0(B)$ is correct.
\label{CB3956}
\end{cor}

\medskip
We have a partially defined section $S\to d_0(S)\in p^{-1}(S)$
of the fibration (\ref{B602}). Since
$$
d_1(A)d_1(B)=d_1(AB)
$$
for $A$ and $B$ from $\Ell_0^{\times}(M,E)$, we have (using
Remark~\ref{RB3750})
\begin{equation}
{\det}_{(\tpi)}(AB)\big/{\det}_{(\tpi)}(A){\det}_{(\tpi)}(B)\cdot d_0(\sigma(A)
\sigma(B))=d_0(\sigma(A))d_0(\sigma(B)),
\label{A310}
\end{equation}
i.e., $F(A,B)d_0(AB)=d_0(A)d_0(B)$, where $F(A,B)=F(\sigma(A),\sigma(B))$
is given by (\ref{B221}). Here we suppose that the principal symbols
of $A$ and $B$ are sufficiently close to positive definite
self-adjoint ones and that $\ord A,\ord B\in\wR^{\times}$.

\begin{thm}
The Lie algebra $\frg(\!M\!,\!E)$ of the Lie group
$G\!(M,E)\!:\!=\!\det_*\!\SEll_0^{\times}\!(M,E)$ is canonically isomorphic
to the Lie algebra $\sfrg$ defined by the central extension
(\ref{B307}) of the Lie algebra $\frg:=S_{\log}(M,E)$.%
\footnote{Note that $\frg(M,E)$ is also canonically isomorphic
to the Lie algebra $\sfrg_{(l)}$ defined by (\ref{B307}).
Here, $l=\sigma(\log A)$ is the symbol of an operator $A\in\Ell_0^1(M,E)$
such that $\log A$ exists (i.e., some root $A^{1/k}$ of $A$, $k\in\wZ_+$,
possesses a cut $L_{(\tpi)}$). The canonical identifications
$W_{l_1l_2}\colon\sfrg_{(l_1)}\to\sfrg_{(l_2)}$ of Lie algebras
$\sfrg_{(l_j)}$ define the Lie algebra $\sfrg$. The associative system
$W_{l_1l_2}$ of isomorphisms
is given by Proposition~\ref{PB403}, (\ref{B404}).}

The identification of the Lie algebras $\frg(M,E)$ and $\sfrg$
is done by the identification of the (local) cocycles for the Lie
groups $\SEll_0^{\times}(M,E)$ and $\exp\left(\sfrg\right)$ defined
by partially defined sections
$S\to d_0(S)$ and by $X\to\sX$ (this section is given by (\ref{B640}) below)
of the $\wC^{\times}$-fiber bundles $G(M,E)\to\SEll_0^{\times}(M,E)$ and
$\exp\left(\sfrg\right)\to\SEll_0^{\times}(M,E)$.
\label{TB570}
\end{thm}

\begin{rem}
On the Lie algebras level we have a central extension
\begin{equation}
0\to\wC @>i>>\frg(M,E) @>p>>\frg\to 0
\label{B3911}
\end{equation}
of $\frg=S_{\log}(M,E)$ and a cocentral extension
$$
0\to CS^0(M,E)\to\frg @>r>>\wC\to 0.
$$
So we have a natural projection
$$
rp\colon\frg(M,E)\to\wC
$$
to a trivial Lie algebra $\wC$. The central Lie subalgebra of $\frg(M,E)$
is $\wC$, (\ref{B3911}).
(However $rpi\colon\wC\to\wC$ is the zero map.) On the Lie groups level
we have the extension
\begin{equation}
1\to\wC^{\times}\to G(M,E)\to\SEll_0^{\times}(M,E)\to 1.
\label{B3917}
\end{equation}
(Here, the central subgroup $\wC^{\times}$ appears from a natural
construction of the determinant Lie group $G(M,E)$ but not
from the exponential map
of the Lie algebras extension (\ref{B3911}).) The Lie group
$\SEll_0^{\times}(M,E)$ is a cocentral extension
\begin{equation}
1\to\SEll_0^0(M,E)\to\SEll_0^{\times}(M,E) @>q>>\wC\to 1,
\label{B3912}
\end{equation}
where $q$ is the order of elliptic symbols.

Note that we have a similar situation in case of the subgroup
$\SEll_0^0(M,E)$ in (\ref{B3912}). Namely there is a central subgroup
$\wC^{\times}:=\wC^{\times}\cdot\Id\underset{i}{\hookrightarrow}\SEll_0^0
(M,E)$.

Set $GS_0^0(M,E):=\SEll_0^0(M,E)/i\wC^{\times}$. Then the multiplicative
residue $\res^{\times}$, (\ref{B12}), defines a homomorphism
\begin{equation}
\res^{\times}\colon GS_0^0(M,E)\to\wC
\label{B3914}
\end{equation}
onto (additive) group $\wC$. We have to note that $\res^{\times}$ was
initially introduced on $\SEll_0^{\times}(M,E)$ (\cite{Wo2}). However
it defines a homomorphism to $\wC$, (\ref{B11}), (\ref{B12}), and
is equal to zero on the normal subgroup
$\wC^{\times}\cdot\Id\hookrightarrow\SEll_0^0(M,E)$. So $\res^{\times}$
induces a homomorphism (\ref{B3914}). We have to underline that
$\res^{\times}(a)$ (for $a\in\SEll_0^0(M,E)$) depends on $a$ only and
not on a smooth curve $a(t)$ from $\Id=a(0)$ to $a=a(1)$ used
in (\ref{B12}). This assertion is equivalent to the equality
\begin{equation}
\res P=0
\label{B3915}
\end{equation}
for any zero order PDO-projector $P\in CL^0(M,E)$, $P^2=P$. (Here,
$\res$ is the noncommutative residue.) The equality (\ref{B3915})
is equivalent to the independence of $\zeta_A(0)$ (for invertible
$A\in\Ell_0^d(M,E)$,
$d\ne 0$, such that complex powers $A^s$ exist) of a holomorphic
family $A^s$, i.e., to the fact that $\zeta_A(0)$ depends on $A$
only. (The equality (\ref{B3915}) is proved in \cite{Wo1}.)

However, ``difficult'' parts in the diagrams \{(\ref{B3917}), (\ref{B3912})\},
and (\ref{B3914}) are different. In (\ref{B3917}) it is not easy to see
from the definition of $\sfrg$ that the central subgroup of $G(M,E)$
is $\wC^{\times}$. (It is proved with the help of the direct definition
of $G(M,E)$ and of Theorem~\ref{TB570}.) The central subgroup
of $\SEll_0^0(M,E)$ is $\wC^{\times}\cdot\Id$ (and it is an easy part).
But the existence of the homomorphism (\ref{B3914}) is equivalent
to the equalities (\ref{B3915}) for all zero order PDO-projectors $P$.
This fact is equivalent to the existence of $\eta$-invariants (and
it is not so clear).
\label{RB3910}
\end{rem}

\noindent{\bf Proof of Theorem~\ref{TB570}.} Let $X\in\SEll_0^d(M,E)$, where
$d=d(X)\in\wR^{\times}$.
Let $\log_{(\pi)}X$ exist. Set $l_X:=\log_{(\pi)}X/d(X)$.
Then $l=l_X$ defines a central extension $\sfrg_{(l)}$ (\ref{B307})
of the Lie algebra $\frg:=S_{\log}(M,E)$. We have the splitting
of the linear space
\begin{equation}
\sfrg_{(l)}=S_{\log}(M,E)\oplus\wC\cdot 1.
\label{B622}
\end{equation}
(Here, $1$ is the generator of the kernel $\wC$ in (\ref{B307}).
This splitting is defined by (\ref{B624}).) Hence the element
$l\in S_{\log}(M,E)$ defines an element $\tilde{l}\in\sfrg_{(l)}$,
$\tilde{l}:=l+0\cdot 1$.

Set $\sX$ be an element
\begin{equation}
\sX:=\exp\left(d(X)\tilde{l}_X\right)=\exp\left({\widetilde{{\log}
_{(\pi)}X}}\right),
\label{B640}
\end{equation}
where $\widetilde{{\log}_{(\pi)}X}$ is the inclusion
of $\log_{(\pi)}X\in S_{\log}(M,E)$
in $\sfrg_{(l_X)}$ with respect to the splitting (\ref{B622}).
{}From now on by $\widetilde{\log X}$ we denote the image
of $\widetilde{\log X}$ in $\sfrg$ under the identification%
\footnote{The identification $W_l\colon\sfrg_{(l)}\rs\sfrg$ is defined
by the identifications $W_{ll_j}$ of $\sfrg_{(l)}$ with $\sfrg_{(l_j)}$.}
$W_{l_X}\colon\sfrg_{(l_X)}\rs\sfrg$.
The element
$\sX$ in (\ref{B640}) is defined as the solution $\sX:=\sX_t|_{t=1}$
of the equation in $G(M,E)$
\begin{equation}
\df_t\sX_t={\widetilde{{\log}_{(\pi)}X}}\cdot\sX_t, \quad \sX_0:=\Id.
\label{B641}
\end{equation}

For an arbitrary $A\in S_{\log}(M,E)$ set $\Pi_{l_X}(A)$ be the inclusion
of $A$ into $\sfrg_{(l_X)}$ with respect to the splitting (\ref{B622}).
Let $Y\in\SEll_0^{d(Y)}(M,E)$, $d(Y)\in\wR^{\times}$, and let
$l_Y:=\log Y\in\frg$ be defined.%
\footnote{Under the latter condition, $Y=Y_t|_{t=1}$ is the solution
of the equation $\df_t Y=\log Y\cdot Y_t$, $Y_0:=\Id$.}

\begin{rem}
For any element $X\in\SEll_0^d(M,E)$ and for any its logarithm
$dl$, $l\in S_{\log}(M,E)$, $r(l)=1$, the element
$\sX_{(l)}:=\exp_{W_l}\left(d\Pi_ll\right)$ in $\exp\left(\sfrg\right)$
is defined ($\Pi_ll$ is considered as an element of $\sfrg$).
\label{RB3878}
\end{rem}

\begin{lem}
We have
\begin{gather}
W_{l_Xl_Y}\Pi_{l_X}(A)=\Pi_{l_Y}(A)+\left(A-(r(A)/2)\left(l_X+l_Y\right),l_X-
l_Y\right)_{\res}\cdot 1,
\label{B580}\\
\left[\Pi_{l_X}(A),\Pi_{l_X}(B)\right]=\Pi_{l_X}\left([A,B]\right)+K_{l_X}
(A,B).
\label{B582}
\end{gather}
Here, $r(\!A\!)$ (defined by (\ref{B301})) is the order of a PDO-symbol
$\exp\!A$
for $A\!\in\!S_{\log}(\!M,E\!)\!=:\frg$ and $K_l(a,b)$ is the $2$-cocycle
of $\frg=CS^0(M,E)$ defined by (\ref{B426}) ($l\in r^{-1}(1)$ and
$a,b\in\frg$).

(The equality (\ref{B580}) means that under the identifications
$W_{l_1l_2}\colon\sfrg_{(l_1)}\rs\sfrg_{(l_2)}$ defined
by Proposition~\ref{PB403} the elements $\Pi_{l_1}(A)\in\sfrg_{(l_1)}$
are mapped to the elements $\Pi_{l_2}(A)+\left(A-(r(A)/2)\left(l_1+l_2\right),
l_2-l_1\right)\cdot 1$ of $\sfrg_{(l_2)}$.)
\label{LB581}
\end{lem}

\begin{cor}
Under notations of Lemma~\ref{LB581}, for $A\in S_{\log}(M,E)$
with $\delta r(A)=0$ we have
\begin{equation}
\delta W_{l_X}\left(\Pi_X(A)\right)=W_{l_X}\left(\Pi_X(\delta A)\right)-
\left(A-r(A)l_X,\delta l_X\right)_{\res}\cdot 1.
\label{B1587}
\end{equation}
\label{CB586}
\end{cor}

\noindent{\bf Proof of Lemma~\ref{LB581}.}
1. For $A:=rl_X+a_0\in\frg$, $r:=r(A)$, and
for $\Pi_{l_X} A:=rl_X+a_0+0\cdot 1\in\sfrg_{(l_X)}$, $l_1:=l_X$, $l_2:=l_Y$,
we have
$$
W_{l_1l_2}\Pi_{l_1}(A)=rl_2+a'_0+\left\{\left(l_1-l_2,a_0\right)_{\res}+
r\left(l_2-l_1,l_2-l_1\right)_{\res}/2\right\}\cdot 1,
$$
where $rl_1+a_0=rl_2+a'_0$, i.e.,
\begin{equation}
W_{l_1l_2}\Pi_{l_1}(A)-\Pi_{l_2}(A)=\left\{\left(l_1-l_2,a_0\right)_{\res}+
r\left(l_2-l_1,l_2-l_1\right)_{\res}/2\right\}\cdot 1.
\label{B583}
\end{equation}
The term on the right in (\ref{B583}) can be transformed as follows
\begin{multline*}
\left(l_1-l_2,a_0\right)_{\res}+
r\left(l_2-l_1,l_2-l_1\right)_{\res}/2=\\
=\left(A-rl_1,l_1-l_2\right)_{\res}+r\left(l_2-l_1,l_2-l_1\right)_{\res}/2=
\left(A-(r/2)\left(l_1+l_2\right),l_1-l_2\right)_{\res}.
\end{multline*}
The formula (\ref{B580}) is proved. \\

2. For $A=r_1l+a_0$, $r_1:=r(A)$, $a_0\in\frg_0$, $l:=l_X$,
$\Pi_{l_X}(A):=r_1l+a_0+0\cdot 1\in\sfrg_{(l)}$, and for $B=r_2l+b_0$,
$r_2:=r(B)$,
$B\in\frg_0$, we have
\begin{multline}
\left[\Pi_{l_X}(A),\Pi_{l_X}(B)\right]:=\left[r_1l+a_0+0\cdot 1,r_2l+b_0+
0\cdot 1\right]_{\sfrg_{(l)}}:=\\
=\left[r_1l+a_0,r_2l+b_0\right]_{\frg}+K_l\left(a_0,b_0\right)\cdot 1=
\Pi_{l_X}\left([A,B]\right)+K_l\left(a_0,b_0\right)\cdot 1\in\sfrg_{(l)}.
\label{B584}
\end{multline}
The formula (\ref{B582}) is proved.\ \ \ $\Box$

\begin{pro}
Let $X\in\Ell_0^{d_1}(M,E)$, $Y\in\Ell_0^{d_2}(M,E)$, and
$XY\in\Ell_0^{d_1+d_2}(M,E)$ possess a cut $L_{(\pi)}$ of the spectral plane.
Let $d_1$, $d_2$, and $d_1+d_2$ be from $\wR^{\times}$. Then the elements
$\sX$, $\sY$, and $\sXY$ are defined and the following equality holds
\begin{equation}
\sX\sY\left(\sXY\right)^{-1}=F(a,b)\in\wC^{\times}.
\label{B586}
\end{equation}
Here, $a:=\sigma(X)$, $b:=\sigma(Y)$, and $F(a,b):=F(X,Y)$ is defined
by (\ref{B221}).
\label{PB585}
\end{pro}

We have the fixed central extension
\begin{equation}
1\to\wC^{\times}\to\exp\left(\sfrg\right) @>>p> \SEll_0^{\times}(M,E)\to 1.
\label{B588}
\end{equation}
Since $\sX\in\exp\left(\sfrg\right)$,
$\sY$ and $\sXY$ are the elements of the same group $\exp\left(\sfrg\right)$,
and since $p\left(\sX\sY\right)=p\left(\sXY\right)=XY$, the expression
on the left in (\ref{B586}) is an element of the kernel $\wC^{\times}$
of (\ref{B588}).

\begin{rem}
Proposition~\ref{PB585} claims that a partially defined cocycle
$$
f(X,Y):=\sX\sY\left(\sXY\right)^{-1}
$$
coincides with the cocycle
$F(\sigma(X),\sigma(Y))$ defined by (\ref{B221}). The cocycle $f(X,Y)$
is defined (in particular) for $X$ and $Y$ sufficiently close to symbols
of positive self-adjoint elliptic PDOs of positive real orders.
\label{RB588}
\end{rem}

\begin{rem}
We use a non-standard and not completely rigorous notion of a ``partially
defined $2$-cocycle'' of a Lie group $G$ (in our setting a subgroup
of $\SEll_0^{\times}(M,E)$ consisting of real order symbols).
We have in mind a function defined on an open set of pairs of elements
of $G$ obeying the cocycle condition on a nonempty open set of triples
of group elements.
The most close known to us notion is the cohomology of semigroups or
monoids (see \cite{McL}, Chapter X.5). Indeed, in formulas
for one of the standard
cochain complexes computing the group cohomology one does not use inversion
of elements of $G$. Namely
$$
(\dd\!c)\left(g_1,\dots,g_{n+1}\right)=\sum_{i+1}^n(-1)^ic\left(g_1,\dots,g_ig
_{i+1},\dots,g_{n+1}\right).
$$
Here $c$ denotes $n$-cochain of $G\ni g_i$ with the values in any trivial
$G$-module, $\dd\!c$ is the coboundary of $c$.

It is known in topology that under some mild conditions, the cohomology
of a semigroup coincides with the cohomology of the universal group
generalized by this semigroup (see \cite{A3}, \S~3.2, pp. 92--93).
Formulas (\ref{A201})--(\ref{A208}) are applicable in a more general
situation than ours. These formulas show explicitly
how to pass from partially defined $2$-cocycles to germs at identity
of cohomologous group cocycles.
Moreover, arguments of Section~\ref{SE1} show that we have
a canonical associative system of isomorphisms between corresponding
central extensions of local Lie groups.
Hence we obtain a canonical central extension of the local Lie group
and of the corresponding Lie algebra.
We will not develop a general formalism of partially defined cocycles
here because in our situation we have made everything explicitly.

\label{Rmax}
\end{rem}

\noindent{\bf Proof of Proposition~\ref{PB585}.}
The following lemmas hold.
\begin{lem}
Let a logarithm $\log X\in\frg$ of $X\in\Ell_0^\alpha(M,E)$,
$\alpha\ne0$, exist. Set $l:=\log X/\alpha$. Then for $\delta X$
with $\delta\ord X=0$ we have
\begin{equation}
\sX_{(l)}^{-1}\delta\sX_{(l)}=\Pi_{(l)}\left(X^{-1}\delta X\right).
\label{B591}
\end{equation}
Here, $X^{-1}\delta X\in\frg:=S_{\log}(M,E)$ and
$\sX_{(l)}^{-1}\delta\sX_{(l)}\in\sfrg$
($\sfrg$ is the Lie algebra obtained by the identifications
$W_{l_1l_2}$ of the Lie algebras $\sfrg_{(l)}$, $l\in r^{-1}(1)$).
\label{LB590}
\end{lem}

\begin{lem}
Set $f(X,Y):=\sX\sY\left(\sXY\right)^{-1}\in\wC^{\times}$. In the domain
of definition for $f(X,Y)$ the equality holds for $\delta X$, $\delta Y$
such that $\delta\ord X=0=\delta\ord Y$
\begin{equation}
\left(\delta_X f\right)\cdot f^{-1}\big|_{(X,Y)}=\left(X^{-1}\delta X,l_X-
l_Y\right)_{\res}+
\left(Y^{-1}X^{-1}\delta X\cdot Y,l_Y-l_{XY}\right)_{\res}.
\label{B592}
\end{equation}
\label{LB591}
\end{lem}

\begin{lem}
The expression on the right in (\ref{B592}) is equal to
\begin{equation}
\left(X^{-1}\delta X,l_X-l_{YX}\right)_{\res}=\delta_X f\cdot f^{-1}\big|
_{(X,Y)}.
\label{B594}
\end{equation}
\label{LB593}
\end{lem}

\begin{rem}
According to (\ref{B3742}), (\ref{B221}) for $\delta X$ with $\delta\ord X=0$
we have
\begin{equation}
\delta_X\log F(X,Y)=\left(\delta X\cdot X^{-1},l_X-l_{XY}\right)_{\res}=
\left(X^{-1}\delta X,l_X-l_{YX}\right)_{\res}.
\label{B596}
\end{equation}
Indeed, by conjugation with $X$ we obtain according to (\ref{B8}) that
$$
\left(\delta X\cdot X^{-1},l_X-l_{XY}\right)_{\res}=\left(X^{-1}\delta X,
l_X-l_{YX}\right)_{\res}.
$$
Thus for such $\delta X$ we have
\begin{equation}
\delta_X\log F(X,Y)=\delta_X f(X,Y).
\label{B597}
\end{equation}
\label{RB595}
\end{rem}

\begin{rem}
For $X=\exp(r_1l)$, $Y=\exp(r_2l)$ (with $l\in r^{-1}(1)\subset\frg$,
$r_j\in\wR^{\times}$) we have
$$
F(X,Y)=f(X,Y)=0.
$$
\label{RB596}
\end{rem}

Hence Proposition~\ref{PB585} follows from Lemmas~\ref{LB590},
\ref{LB591}, \ref{LB593}.\ \ \ $\Box$

Now we can finish the proof of Theorem~\ref{TB570}.

The (local) section $S\to d_0(S)$ of the $\wC^{\times}$-fiber bundle
$G(M,E)\to\SEll_0^{\times}(M,E)$ is also defined by the cocycle
$F(A,B)=f(A,B)$. Thus we conclude that the real subalgebras of real
codimension $1$ of our Lie algebras, consisting of elements of real
orders, are canonically isomorphic (by Remark~\ref{Rmax} we have
an isomorphism of local Lie groups).
Moreover, this isomorphism is complex linear on the subalgebras
consisting of elements of zero order. Hence the complexification
of our isomorphism along one real direction (of $\ord$) gives us
a canonical isomorphism of complex Lie algebras.
Theorem~\ref{TB570} is proved.\ \ \ $\Box$

\begin{rem}
The local section $\sX$ of the $\wC^{\times}$-fiber bundle
$\exp\left(\sfrg\right)\to\SEll_0^{\times}(M,E)$ is the exponential
of the cone $C\subset\sfrg$ of the null vectors for the invariant
quadratic form on $\sfrg$ defined by Proposition~\ref{PB417}.
Indeed, $\log\sX=\ord X\cdot l_X$ (for $\ord X\ne 0$) is an element
of the cone $C_{l_X}\subset\sfrg_{(l_X)}$ of the null vectors
for the invariant quadratic form $A_l$ on $\sfrg_{(l_X)}$ (defined
by (\ref{B428})). These cones $C_l$ are canonically identified
with the cone $C\subset\sfrg$ (under the system of isomorphisms $W_{l_1l_2}$,
Proposition~\ref{PB403}).

In the proof of Theorem~\ref{TB570} we show that elements
$d_0(\sigma(A))\in G(M,E)$ for elliptic symbols $\sigma(A)$ with
$\alpha:=\ord A\in\wR^{\times}$ (and such that the principal symbols
$\sigma_\alpha(A)$ are sufficiently close to positive definite self-adjoint
ones) belong to the exponential of the canonical cone $C\subset\sfrg$.

We define also the elements $\tilde{d}_0(\sigma(\log A))$ for elliptic
$A$ with $\ord A\in\wC^{\times}$ such that $\sigma(\log A)$ exists.
It follows from the definition of $\tilde{d}_0(\sigma(\log A))$,
(\ref{B3948}), that such elements form the exponential image
of a $\wC^{\times}$-cone in the Lie algebra $\frg(M,E)$ of $G(M,E)$,
$\frg(M,E)$ is canonically identified with $\sfrg$ by Theorem~\ref{TB570}.

Thus there are two $\wC^{\times}$-cones in $\sfrg$ whose intersections
with the hyperplane of logarithmic symbols of real orders coincide
(in a neighborhood of $0\in\sfrg$). So these two cones coincide.
\label{RB597}
\end{rem}

\noindent{\bf Proof of Lemma~\ref{LB590}.}
Let $X_t$ be a solution in $\SEll_0^{\times}(M,E)$ of an ordinary
differential equation
\begin{equation}
\df_t X_t=\alpha l_X X_t, \qquad X_0:=\Id.
\label{B670}
\end{equation}
(Under the conditions of Lemma~\ref{LB590}, we have $X=X_1$. The solution
of (\ref{B670}) exists for $0\le t\le 1$.)

Let $\sX_t$ be a solution in $G(M,E)$ (\ref{B623}) of the equation
\begin{equation}
\df_t\sX_t=\Pi_X\left(\alpha l_X\right)\cdot\sX_t, \quad \sX_0:=\Id.
\label{B6711}
\end{equation}
By Lemma~\ref{LB616} below, we have
\begin{gather}
\begin{split}
\delta X\cdot X^{-1}     & =\int_0^1\Ad\left(X_t\right)\cdot\delta_X\left(
\alpha l_X\right)dt, \\
\delta_X\sX\cdot\sX^{-1} & =\int_0^1\Ad\left(\sX_t\right)\cdot\delta_X
\left(\alpha\Pi_X l_X\right)dt.
\end{split}
\label{B671}
\end{gather}
We have also
\begin{equation}
\df_t\left(\Ad\left(X_t\right)\cdot m_0\right)=\ad\left(\alpha l_X\right)
\cdot m_t
\label{B672}
\end{equation}
for $m_0\in\frg:=S_{\log}(M,E)$, $m_t:=\Ad\left(X_t\right)\cdot m_0$.

Under the conditions of this lemma, an element \
$m_0:=\delta_X\left(\alpha l_X\right)$ belongs
to $CS^0(M,E)=:\frg_0\in S_{\log}(M,E)=:\frg$ and we have
\begin{equation}
\Pi_X\df_t m_t=\df_t\Pi_X m_t=\ad\left(\alpha\Pi_X l_X\right)\cdot\Pi_X m_t.
\label{B673}
\end{equation}

To prove (\ref{B673}), note that
\begin{equation}
\Pi_X\df_t m_t:=\Pi_X\left(\ad\left(\alpha l_X\right)\cdot m_t\right)=\ad
\left(\alpha\Pi_X l_X\right)\cdot\Pi_X m_t.
\label{B6731}
\end{equation}
The latter equality follows from (\ref{B624}) and from (\ref{B304})
since for an arbitrary $C\in\frg_0$ we have
\begin{equation}
\left[\Pi_X\alpha l_X,\Pi_X C\right]_{\sfrg_{(l_X)}}=\Pi_X\left[\alpha l_X,
C\right]_{\frg}+K_{l_X}\left(l_X,m_t\right)\cdot 1,
\label{B674}
\end{equation}
and since $K_l(l,C)=0$ for $C\in\frg_0$.
Hence we have two dinamical systems
\begin{gather}
\begin{split}
\df_{t,0}C       & =\ad\left(\alpha l_X\right)\cdot C
\quad\text{ on } \frg_0\ni C, \\
\df_{t,(l_X)}C_1 & =\ad\left(\Pi_X\left(\alpha l_X\right)\right)\cdot C_1
\quad\text{ on } \sfrg_{(l_X)}\ni C_1, \\
\end{split}
\label{B675}
\end{gather}
such that they are in accordance with a linear map
$\Pi_X\colon\frg_0\to\sfrg_{(l_X)}$, i.e., we have
\begin{equation}
\Pi_X\df_{t,0}C=\Pi_X\left(\ad\left(\alpha l_X\right)\cdot C\right)=\ad\left(
\Pi_X\left(\alpha_X l_X\right)\right)\cdot\Pi_X C=\df_{t,(l_x)}\Pi_X C.
\label{B676}
\end{equation}
The equality
\begin{equation}
\Pi_X m_t=\Ad\left(\sX_t\right)\cdot\Pi_X m_0=:\tilde{m}_t
\label{B677}
\end{equation}
follows from (\ref{B676}) and (\ref{B673}) since the equation (\ref{B6711})
has a unique solution. From (\ref{B677}) we have
\begin{equation}
\tilde{m}_t\in\Pi_X\frg_0,
\label{B678}
\end{equation}
since $m_t:=\Ad\left(X_t\right)\cdot m_0\in\frg_0$ for $m_0\in\frg_0$.

It follows from (\ref{B671}), (\ref{B673}), (\ref{B677}) that
\begin{equation}
\Pi_X\left(\delta X\cdot X^{-1}\right)=\Pi_X\int_0^1m_t dt=\int_0^1\Pi_Xm_tdt=
\int_0^1\tilde{m}_tdt=\delta\sX\cdot\sX^{-1}.
\label{B679}
\end{equation}

To prove the equality (\ref{B591}), note that
\begin{gather}
\begin{split}
X^{-1}\delta X    & =\Ad\left(X^{-1}\right)\circ\left(\delta X\cdot X^{-1}
\right), \\
\sX^{-1}\delta\sX & =\Ad\left(\sX^{-1}\right)\circ\left(\delta\sX\cdot\sX^{-1}
\right),
\end{split}
\label{B680}
\end{gather}
$X^{-1}=X_t|_{t=-1}$ for the solution $X_t$ of (\ref{B670}),
$\sX^{-1}=\sX_t|_{t=-1}$ for $\sX_t$ from (\ref{B6711}).

We see from (\ref{B679}) and from (\ref{B6711}) that
\begin{gather}
\begin{split}
\Pi_X\left(\delta X\cdot X^{-1}\right) & =\delta\sX\cdot\sX^{-1}, \quad
\delta X\cdot X^{-1}\in\frg_0,\\
\Pi_X m_t                              & =\Ad\left(\sX_t\right)\cdot\Pi_Xm_0
\end{split}
\label{B681}
\end{gather}
for $m_0\in\frg_0$, $m_t:=\Ad\left(X_t\right)\cdot m_0$. Hence we obtain
\begin{multline}
\sX^{-1}\cdot\delta\sX=\Ad\left(\sX^{-1}\right)\left(\delta\sX\cdot\sX^{-1}
\right)=\Ad\left(\sX_t|_{t=-1}\right)\circ\Pi_X\left(\delta X\cdot X^{-1}
\right)=\\
=\Pi_X\Ad\left(X_t|_{t=-1}\right)\cdot\Pi_X\left(\delta X\cdot X^{-1}\right)=
\Pi_X\left(X^{-1}\delta X\right).
\label{B682}
\end{multline}
The latter equality in (\ref{B682}) follows from (\ref{B674}), (\ref{B6731}),
and from (\ref{B670}) since
$\Ad\left(X_t\right)\cdot\left(X^{-1}\delta X\right)\in\frg_0$.
The lemma is proved.\ \ \ $\Box$

\begin{lem}
Let $A$ be a symbol from $\SEll_0^\alpha(M,E)$ ($\alpha\in\wC$) or let
$A$ be an element of the group $G(M,E)$ (defined by (\ref{B623})).
Let there exist a logarithm $\cL_A$ of $A$, $A:=\exp\left(\cL_A\right)$.%
\footnote{$\cL_A$ is an element of $\frg:=S_{\log}(M,E)$
for $A\in\SEll_0^{\times}(M,E)$ or of $\sfrg$ for $A\in G(M,E)$.
We have $A=A_t|_{t=1}$, where $\df_tA_t=\cL_A\cdot A_t$, $A_0:=\Id$.}
Let $\delta A$ do not change an order of $A$. (For $A\in G(M,E)$ the order
of $p(A)\in\SEll_0^{\times}(M,E)$ is defined,
$p\colon G(M,E)\to\SEll_0^{\times}(M,E)$.) Then we have
\begin{equation}
\delta A\cdot A^{-1}=F\left(\ad\left(\cL_A\right)\right)\circ\delta\cL_A,
\label{B617}
\end{equation}
where
\begin{equation}
F\left(\ad\left(\cL_A\right)\right):=\int_0^1dt\Ad\left(A_t\right),
\quad A_t:=\exp\left(t\cL_A\right).
\label{B620}
\end{equation}
\label{LB616}
\end{lem}

\noindent{\bf Proof.} According to Duhamel principle we have
for $A:=\exp\left(\cL_A\right)$
$$
\delta A=\int_0^1A_t\delta\cL_A A_{1-t}dt.
$$
Hence we have
\begin{equation}
\delta A\cdot A^{-1}=\int_0^1\Ad\left(A_t\right)\cdot\delta\cL_A.
\label{B618}
\end{equation}

\begin{rem}
According to the equality
$$
\int_0^1dt\exp(tz)=(\exp z-1)/z
$$
the expression $F\left(\ad\left(\cL_A\right)\right)$ in (\ref{B620})
has formal properties of $(\exp z-1)/z|_{z=\ad(\cL_A)}$.
\label{RB621}
\end{rem}

\noindent{\bf Proof of Lemma~\ref{LB591}.} We have
$$
\delta_X f\cdot f^{-1}\cdot 1=\sX^{-1}\left(\delta_X f\cdot f^{-1}\cdot 1
\right)\sX
$$
since $f\in\wC^{\times}\cdot 1\in\Ker p$ ($p$ is from (\ref{B588})) and since
$\delta_X f\cdot f^{-1}$ is an element of the kernel $\wC$ in the central
extension (\ref{B307}). Hence
\begin{equation}
\delta_X f\cdot f^{-1}\cdot 1=\sX^{-1}\delta\sX-\sY\left(\sXY\right)^{-1}
\delta_X\left(\sXY\right)\sY^{-1}.
\label{B700}
\end{equation}

According to Lemma~\ref{LB590} we have
\begin{gather}
\begin{split}
\sX^{-1}\delta\sX                               & =\Pi_X\left(X^{-1}\delta X
\right), \\
\left(\sXY\right)^{-1}\delta_X\left(\sXY\right) & =\Pi_{XY}\left((XY)^{-1}
\delta_X(XY)\right)\equiv\\
 & \equiv\Pi_{XY}\left(\Ad\left(Y^{-1}\right)\circ\left(X^{-1}\delta X\right)
\right)
\end{split}
\label{B701}
\end{gather}
because $\delta_X\ord(XY)=0$.

By Lemma~\ref{LB581} we have
\begin{gather}
\begin{split}
\Pi_X\left(X^{-1}\delta X\right)         & =\Pi_Y\left(X^{-1}\delta X\right)+
\left(X^{-1}\delta X,l_X-l_Y\right)_{\res}, \\
\Pi_{XY}\left((XY)^{-1}\delta(XY)\right) & =\Pi_Y\left(\Ad\left(Y^{-1}\right)
\circ\left(X^{-1}\delta X\right)\right)+ \\
 & +\left(\Ad\left(Y^{-1}\right)\circ\left(X^{-1}\delta X\right),l_{XY}-l_Y
\right)_{\res}
\end{split}
\label{B702}
\end{gather}
since $X^{-1}\delta X\in\frg_0$. Hence  we get
\begin{multline}
\delta_X f\cdot f^{-1}\cdot 1=\left\{\Pi_Y\left(X^{-1}\delta X\right)-\sY
\Pi_Y\left(\Ad\left(Y^{-1}\right)\circ\left(X^{-1}\delta X\right)\right)\sY
^{-1}\right\}+\\
+\left(X^{-1}\delta X,l_X-l_Y\right)_{\res}+\left(\Ad\left(Y^{-1}\right)
\circ\left(X^{-1}\delta X\right),l_Y-l_{XY}\right)_{\res}.
\label{B703}
\end{multline}

The assertion of the lemma follows from (\ref{B703}) and
from Lemma~\ref{LB704} below. The latter lemma claims that the first term
on the right in (\ref{B703}) is equal to zero.\ \ \ $\Box$

\begin{lem}
Let $Y$ be an element from $\SEll_0^{\times}(M,E)$ of nonzero order $\alpha$
and such that $\log Y=\alpha l_Y$ is defined. Then the linear operator
$\Pi_Y\colon\frg_0\to\sfrg_{(l_Y)}$ commutes with $\Ad(Y)$ and with
$\Ad\left(\sY\right)$.%
\footnote{To remind, $\sY:=\exp\left(\Pi_Y\left(\alpha l_Y\right)\right)$
lies in $G(M,E)$.}
Namely we have
\begin{equation}
\Ad\left(\sY\right)\circ\Pi_Y Z=\Pi_Y\left(\Ad(Y)\circ Z\right)
\label{B705}
\end{equation}
for $Z\in\frg_0$ ($:=S_{\log}(M,E)$).
\label{LB704}
\end{lem}

\noindent{\bf Proof.} Let $Y_t$ be a solution of an ordinary differential
equation in $\SEll_0^{\times}(M,E)$
$$
\df_t Y_t=\alpha l_Y\cdot Y_t, \qquad Y_0:=\Id.
$$
Let $\sY_t$ be a solution in $G(M,E)$ of
$$
\df_t\sY_t=\Pi_Y\left(\alpha l_Y\right)\cdot\sY_t, \qquad \sY_0:=\Id.
$$
Then we have $\sY=\sY_1$, $Y=Y_1$, and
$$
\Ad\left(\sY_t\right)\Pi_Y Z=\Pi_Y\left(\Ad\left(Y_t\right)\circ Z\right)
$$
according to (\ref{B677}). The lemma is proved.\ \ \ $\Box$

\noindent{\bf Proof of Lemma~\ref{LB593}.} We have from (\ref{B8}) that
\begin{multline}
\left(\Ad\left(Y^{-1}\right)\circ\left(X^{-1}\delta X\right),l_Y-l_{XY}
\right)_{\res}=\left(X^{-1}\delta X,\Ad(Y)\circ\left(l_Y-l_{XY}\right)\right)
_{\res}=\\
=\left(X^{-1}\delta X,l_Y-l_{YX}\right)_{\res}.
\label{B720}
\end{multline}
Hence, from (\ref{B592}) and from (\ref{B720}) we see that
$$
\left(\delta_X f\cdot f^{-1}\right)\big|_{(X,Y)}=\left(X^{-1}\delta X,l_X-
l_{YX}\right).
$$
The lemma is proved.\ \ \ $\Box$

\begin{rem}
The holomorphic structure on the determinant $\wC^{\times}$-bundle
\begin{equation}
p\colon G(M,E)\to\SEll_0^{\times}(M,E)
\label{B1292}
\end{equation}
is defined.
The reason is that all Lie algebras in our situation have natural
complex structures and the isomorphism from Theorem~\ref{TB570} is
defined over $\wC$.
\label{RB1290}
\end{rem}

\begin{pro}
Let $C$ be a positive definite elliptic PDO of order $m>0$,
$C=\exp(mJ)$, $J\in\fell(M,E)$, $J:=\log_{(\tpi)}C$. Then
the splitting (\ref{B622}) of $\sfrg_{(l)}$ with $l:=\sigma(J)$
\begin{equation}
\sfrg_{(\sigma(J))}=\sfrg\oplus\wC\cdot 1
\label{B4028}
\end{equation}
is defined by a homomorphism $f_J\colon CL^0(M,E)\to\wC$,
$$
f_J(L):=\TR(L\exp(-sJ)-\res\sigma(L)/s)|_{s=0}.
$$
Namely for a curve $\exp(tL)\in\Ell_0^0(M,E)$ we have
\begin{equation}
\df_t\log\left(d_1(\exp(tL))/\exp\left(t\Pi_{\sigma(J)}\sigma(L)\right)\right)
\big|_{t=0}=f_J(L).
\label{B4029}
\end{equation}
Here, $d_1(\exp(tL))$ is the image of $\exp(tL)$ in $G(M,E)$ and
$\exp\left(t\Pi_{\sigma(J)}\sigma(L)\right)$ is a solution in $G(M,E)$
of the equation
$$
\df_tu_t=\left(\Pi_{\sigma(J)}\sigma(L)\right)u_t, \quad u_0=\Id,
$$
$\Pi_{\sigma(J)}\sigma(L)$ is the inclusion
of $\sigma(L)\subset\frg_0\subset\frg$ into $\sfrg_{(\sigma(J))}$
with respect to the splitting (\ref{B4028}).
\label{PB4027}
\end{pro}

\noindent{\bf Proof.} The formula (\ref{B4029}) follows
from Proposition~\ref{PB840}, (\ref{B841}), (\ref{B842}) below.\ \ \ $\Box$

\subsection{Topological properties of determinant Lie groups
as $\wC^{\times}$-bundles over elliptic symbols}
\label{SE51}

For the sake of simplicity the following lemma is written in the case
of a trivial $\wC$-vector bundle $E:=1_N$ with $N$ large enough.

\begin{lem}
For a trivial vector bundle $1_N=:E$, where $N$ is large
enough, over an orientable closed manifold $M$, $\dim M>0$,
the $\wC^{\times}$-extension
$G(M,E)$ of $\SEll_0^{\times}(M,E)$ is nontrivial.

Namely the Chern
character of the associated linear bundle over $\SEll_0^{\times}(M,E)$
is nontrivial in $H^*\left(\SEll_0^{\times}(M,E),\wQ\right)$.
\label{LB3006}
\end{lem}

{\bf Proof.} 1. The principal symbols of a family of elliptic operators
from $\Ell_0^{\times}\left(M,1_N\right)$ (parametrized by a map of a smooth
manifold $A$, $\phi\colon A\to\Ell_0^{\times}\left(M,1_N\right)$) define
a smooth map
\begin{equation}
\phi_{symb}\colon A\times S^*M\to U(N).
\label{B3009}
\end{equation}
If $N$ is large enough, then the space of such maps is homotopy
equivalent to the space of maps from $A$ into $U(\infty)$.
The $K$-functor $K^{-1}(A\times S^*M)$ is defined as the set of homotopy
classes $[A\times S^*M;U(\infty)]$ (\cite{AH}, 1.3). The Chern character
$\ch\colon K^{-1}(A\times S^*M)\to H^{odd}(A\times S^*M,\wQ)$ defines
an isomorphism of $K^{-1}\otimes\wQ$ with $H^{odd}$ (\cite{AH}, 2.4). \\

2. The space $\Ell_0\left(M,1_N\right)$ is a bundle over
$\SEll_0^{\times}\left(M,1_N\right)$ with a contractable fiber
$F=\{\Id+\cK\}=\pi^{-1}(\Id)$, where $\cK$ are operators with
$C^{\infty}$-smooth kernels on $M\times M$ (i.e., smoothing operators).
The determinant of the index bundle over $\Ell_0\left(M,1_N\right)$
is isomorphic%
\footnote{This isomorphism of linear bundles is not canonical.
The existence of such an isomorphism is proved in Section~\ref{SE52}.}
to the pull-back $\pi^*L$ to $\Ell_0\left(M,1_N\right)$
of the associated with $G\left(M,1_N\right)$ linear bundle $L$ over
$\SEll_0^{\times}\left(M,1_N\right)$. The Chern character of $\pi^*L$
restricted to a family $A$ of elliptic operators is given
by the Atiyah-Singer index theorem for families%
\footnote{The orientation of $S^*M$ differs from the orientation
in \cite{AS1}, \cite{AS2} and coincides with its orientation in \cite{P}.}
\begin{equation}
\ch\left(\phi_{symb}^*L\right)=\ch(\phi^*\pi^*L)=\int_{S^*M}\cal{T}(S^*M)\ch
(u_A).
\label{B3010}
\end{equation}
Here, $\cal{T}(S^*M)$ is the Todd class for $T(S^*M)\otimes\wC$, $\cal{T}$
corresponds to
$$
\Pi\left({-y_i\over 1-\exp(y_i)}\cdot{y_i\over 1-\exp(-y_i)}\right),
$$
where $y_i$ are basic characters of maximal torus of $O(n)$ and
the Pontrjagin classes $p_j(TX)$ are the elementary symmetric functions
$\sigma_j$ of $\{y_i^2\}$, $\cal{T}=1-p_1/12+\ldots$. The $u_A$
in (\ref{B3010}) is an element of $K^{-1}(A\times S^*M)$ corresponding
to $\phi_{symb}$ (\ref{B3009}). Its Chern character
$\ch\left(u_A\right)\in H^{odd}(A\times S^*M)$ corresponds to an element
$\ch\left(\delta u_A\right)\in H^{ev}(A\times S^*M,A\times S^*M)$
in the exact sequence of the pair $(B^*M,S^*M)$ (\cite{AH}, 1.10).
Here, $B^*M$ is the bundle of unit balls in $T^*M$ and
$\delta\colon K^{-1}(A\times S^*M)\to K^0(A\times B^*M,A\times S^*M)$
is the natural homomorphism. By the Bott periodicity,
$$
K^{-1}(A\times S^*M)=K^1(A\times S^*M).
$$

3. The family $\phi\colon A\to\Ell_0\left(M,1_N\right)$
is a smooth map to the connected component of the operators with their
principal symbols homotopic to a trivial ones. Hence for any $a\in A$
the map $\phi_{symb}(a)\colon a\times S^*M\to U(N)$ is homotopic
to the map to a point in $U(N)$. Up to the multiplication of the element
$u(a):=\left[\phi_{symb}(a)\right]\in K^1(S^*M)$ by a number $n\in\wZ_+$
the latter condition is equivalent to the equality $\ch(u(a))=0$
in $H^{odd}(S^*M,\wQ)$. (Here, we suppose that $N$ is large enough.
The torsion subgroup of $K^1(S^*M)$ is a finite group.)

Let $A$ be an orientable closed even-dimensional manifold. Then there is
a smooth map $\phi_{symb}\colon A\times S^*M\to U(N)$ (where $N$ is
large enough) such that
\begin{gather}
\begin{split}
\ch\left(\phi_{symb}\right)[A\times S^*M] & \ne 0, \\
\ch\left(\phi_{symb}(a)\right)            & =0.
\end{split}
\label{B3012}
\end{gather}

4. Let $A=\Sigma$ be an orientable compact surface. Let a smooth map
$\phi_{symb}$ satisfy (\ref{B3012}). Then the integer multiple
of $\phi_{symb}$
$$
n\cdot\phi_{symb}\colon\Sigma\times S^*M\to U(nN),
$$
$n\in\wZ_+$ is homotopic to a trivial one under the restriction
to $a\times S^*M$ for any $a\in\Sigma$.
So there is a smooth family $\phi_1$ of elliptic PDOs with their principal
symbol map $n\cdot\phi_{symb}$,
$\phi_1\colon\Sigma\to\Ell_0^{\times}\left(M,1_{nN}\right)$.
By (\ref{B3010}) and (\ref{B3012}) we have
\begin{multline*}
\ch\left(\Ind\phi_1\right)[\Sigma]=\ch\left(\phi_1^*\pi^*L\right)[\Sigma]=\\
\int_{S^*M\times\Sigma}\cal{T}(S^*M)\ch\left(u_{\Sigma}\right)=\ch\left(n
\phi_{symb}\right)\left[S^*M\times\Sigma\right]\ne 0.
\end{multline*}
Then $\ch(L)$ is nontrivial in
$H^{odd}\left(\SEll_0^{\times}\left(M,1_{nN}\right),\wQ\right)$
because
$$
\ch(L)\left[\phi_1\Sigma\right]=\ch\left(\Ind\phi_1\right)[\Sigma]\ne 0.
$$
The lemma is proved.\ \ \ $\Box$

\subsection{Determinant bundles over spaces of elliptic operators and
of elliptic symbols.}
\label{SE52}

The line bundle over $\SEll_0^{\times}(M,E)$ associated with the determinant
Lie group $G(M,E)$ can be defined as follows. The determinant line bundle
over the group $\Ell_0^{\times}(M,E)$ of invertible elliptic operators
is canonically trivialized (as $\Ker A=0=\Coker A$
for $A\in\Ell_0^{\times}(M,E)$). Any two operators
$A_1,A_2\in\Ell_0^{\times}(M,E)$ with the same symbols differ
by myltiplying by $B=A_2A_1^{-1}\in\{\Id+\cK\}$, $\cK$ are smoothing.
The identification of fibers $\wC=L_{inv}\left(A_1\right)\ni 1$ and
$L_{inv}\left(A_2\right)\ni 1$ over $A_1$ and $A_2$ is defined as
\begin{equation}
\xi\in L_{inv}\left(A_1\right)\to\xi/{\det}_{Fr}(B)\in L_{inv}\left(A_2
\right).
\label{B3032}
\end{equation}
Here, ${\det}_{Fr}(B)$ is the Fredholm determinant. These identifications
define the line bundle $L$ over $\SEll_0^{\times}(M,E)$ canonically
isomorphic to the linear bundle associated with the principal
$\wC^{\times}$-bundle
$G(M,E)$ over $\SEll_0^{\times}(M,E)$. The holomorphic structure
on the $\wC^{\times}$-bundle $G(M,E)$ over $\SEll_0^{\times}(M,E)$ (defined
in Remark~\ref{RB1290}) gives us the holomorphic structure on the associated
line bundle $L$.

The group $G(M,E)$ is the group of nonzero elements of $L$. The image
$d_1(A)$ of $A\in\Ell_0^{\times}(M,E)$
in $F_0\backslash\Ell_0^{\times}(M,E)=G(M,E)$ (satisfying
the multiplicative property (\ref{B2022})) corresponds to the unit $1_A$
in $\wC=L_{inv}(A)$. The definition (\ref{B3032}) is compatible
with (\ref{B2022}) because for $\xi\in\wC^{\times}$ we have
$$
\xi\cdot 1_{A_1}=\xi\cdot d\left(A_1\right)=\xi\cdot d\left(A_1A_2^{-1}\right)
\cdot d\left(A_2\right)=\xi/{\det}_{Fr}\left(A_2A_1^{-1}\right)\cdot 1_{A_2}.
$$
Here we use for $A_1A_2^{-1}=:B$ the equality
$$
{\det}_{Fr}(B)=d_1(B)\in F_0\backslash F=\wC^{\times}\in F
$$
(where $F$ are invertible operators of the form $\Id+\cK$, $\cK$ is
smoothing).

The determinant bundle $\det_{\Ell}$ over the space of elliptic PDOs
has the determinant line
$\det(\Coker A)\otimes(\det(\Ker A))^{-1}=\det_{\Ell}(A)$
as its fiber over a point $A\in\Ell(M,E)$.
Here, $\det(V):=\Lambda^{\max}V$
for a finite-dimensional vector space $V$ over $\wC$ and $L^{-1}$
is the dual space to a one-dimensional $\wC$-linear space $L$.
(An elliptic PDO $A\in\Ell^q(M,E)$ of any order $q$ defines
the Fredholm operator between Sobolev spaces $H_{(s)}(M,E)$ and
$H_{(s-m)}(M,E)$, where $m:=\Re q$, and $\Ker A\subset C^{\infty}(M,E)$
is independent of $s$, \cite{Ho2}, Theorem~19.2.1 and Theorem~18.1.13
also. The space $\Coker A$ is antidual
to $\Ker A^*\subset C^{\infty}(M,E^*\otimes\Omega)$, where $E^*$
is antidual to $E$ and $\Omega$ is the line bundle of densities on $M$.)

Let $\det_{\Ell}^0$ be the restirction of $\det_{\Ell}$ to the connected
component $\Ell_0(M,E)$ of $\Id$ of the space of elliptic PDOs.
The natural fibration
\begin{equation}
\pi\colon\Ell_0(M,E)\to\SEll_0^{\times}(M,E)
\label{B3051}
\end{equation}
over the space of symbols of invertible elliptic PDOs has contractible
fibers $(\Id+\cK)\cdot A$ (where $A$ is an invertible elliptic PDO
with a given symbol and $\cK$ are smoothing operators,
i.e., their Schwartz kernels are $C^{\infty}$ on $M\times M$).
Hence there are global sections of this fibration.

\begin{pro}
The linear bundle $\det_{\Ell}$ over $\Ell_0(M,E)$ is isomorphic to $\pi^*L$,
where $L$ is the linear bundle over $\SEll_0^{\times}(M,E)$
associated with the determinant Lie group (and $\pi$ is the projection
(\ref{B3051})). This identification is not canonical. Any global
section $s\colon\SEll_0^{\times}(M,E)\to\Ell_0(M,E)$ defines
a canonical identification of line bundles $s^*\det_{\Ell}^0$ and $L$
over $\SEll_0^{\times}(M,E)$.
\label{PB3050}
\end{pro}

This assertion is proved with the help of the following lemma.

\begin{lem}
There is an associative system
$\phi_{A_1,A_2}\colon\det_{\Ell}\left(A_1\right)\rs\det_{\Ell}\left(A_2
\right)$ of canonical linear identifications for $A_j$ from the same fiber
of $\pi$. If $A_1$ and $A_2$ are invertible elliptic PDOs, then
$\det_{\Ell}\left(A_j\right)$ is canonically $\wC$ and $\phi_{A_1,A_2}$
is the multiplication by the Fredholm determinant
$\left({\det}_{Fr}(B)\right)^{-1}$, $B:=A_2A_1^{-1}$.
\label{LB3052}
\end{lem}

\noindent{\bf Proof.} These identifications are defined as follows.
Let $A_0$, $A_1$, $A_2$ be elliptic PDOs from the same fiber of $\pi$
and let $A_0$ be invertible. There are smoothing operators $S_j$,
$j=1,2$, such that
$$
A_j=\left(\Id+S_j\right)A_0.
$$
The determinant line $L_{inv}(A_0)$ is canonically $\wC$. The PDO $A_0$
defines (in a canonical way) the identification of $L_{inv}(A_0)$
with $\left(\det(E_1)\right)\otimes\left(\det(E_0)\right)^{-1}$, where
$E_0\subset\Gamma(M,E)$ is a finite-dimensional space of smooth sections,
$E_1:=A_0E_0$. Let $\tilde{E}_0$, $\tilde{E}_1$ be finite-dimensional
subspaces of $\Gamma(M,E)$ such that $\Ker A_1\subset\tilde{E}_0$ and
the image of the natural map from $\tilde{E}_1$ into $\Coker A_1$
is $\Coker A_1$. Then the determinant line $\det_{\Ell}\left(A_1\right)$
is canonically isomorphic (by the action of the operator $A_1$) with
$\det\left(\tilde{E}_1\right)\otimes\left(\det\left(\tilde{E}_0\right)\right)
^{-1}$. In particular, it is canonically identified (by $A_1$) with
\begin{equation}
\det\left(A_0\tilde{E}_0\right)\otimes\left(\det\left(\tilde{E}_0\right)
\right)^{-1}, \quad\tilde{E}_0:=E_0\left(A_1,A_0\right)=A_0^{-1}K_{-1}\left(S
_1\right),
\label{B3064}
\end{equation}
where $K_{-1}\left(S_1\right)$ is the (algebraic) eigenspace for $S_1$
corresponding to $S_1$-eigenvalue $(-1)$ (i.e., $\dim_{\wC}K_{-1}$ is
the {\em algebraic} multiplicity of $(-1)$ for $S_1$). The operator
$S_1$ is a compact one in $L_2(M,E)$. Hence
$\dim E_0\left(A_1,A_0\right)=\dim K_{-1}\left(S_1\right)<\infty$.
We have the composition of canonical isomorphisms
(for $E_0:=E_0\left(A_1,A_0\right)$) defined by the operators $A_0$ and
$A_1$,
\begin{equation}
L_{inv}\left(A_0\right) @>\psi(A_0)>\lrs> \det\left(A_0\tilde{E}_0\right)
\otimes\left(\det\left(\tilde{E}_0\right)\right)^{-1}@<\psi(A_1)<\lrs<{\det}
_{\Ell}\left(A_1\right).
\label{B3031}
\end{equation}

The truncated Fredholm determinant ${\det}'_{Fr}\left(\Id+S_1\right)$
is defined as the Fredholm determinant of the operator $\left(\Id+S_1\right)$
restricted to the invariant subspace for $\left(\Id+S_1\right)$
complementary to $K_{-1}\left(S_1\right)$ in $L_2(M,E)$.
The identification of the lines
$$
\phi_{A_0,A_1}\colon L_{inv}\left(A_0\right)\rs{\det}_{\Ell}\left(A_1\right)
$$
is the composition of the identifications (\ref{B3031}) multiplied
by $\left({\det}_{Fr}'\left(\Id+S_1\right)\right)^{-1}$. (Note that
if $A_1$ is invertible, then $E_0=0$, $\det\left(E_0\right)$ is canonically
$\wC$, and ${\det}_{Fr}\left(\Id+S_1\right)=\det\left(A_1A_0^{-1}\right)$.
So this definition is compatible with (\ref{B3032}).)
The identification of the lines
\begin{equation}
\phi_{A_1,A_2}\colon{\det}_{\Ell}\left(A_1\right)\rs{\det}_{\Ell}\left(A_2
\right)
\label{B3036}
\end{equation}
is defined as $\phi_{A_0,A_2}\cdot\left(\phi_{A_0,A_1}\right)^{-1}$.

\begin{lem}
The isomorphism (\ref{B3036}) is independent of an invertible PDO $A_0$
from the same fiber.
\label{LB3037}
\end{lem}

\noindent{\bf Proof.} Indeed, let $A'_0$ be another invertible PDO
with the same symbol as $\sigma\left(A_0\right)$. Then we have
\begin{equation}
\phi_{A_0,A_1}=\phi_{A'_0,A_1}\phi_{A_0,A'_0},
\label{B3033}
\end{equation}
where $\phi_{A_0,A'_0}$ is the identification (\ref{B3032}) of the lines
$L_{inv}\left(A_0\right)=\det_{\Ell}\left(A_0\right)$ and
$\det_{\Ell}\left(A'_0\right)$.

To prove (\ref{B3033}), we use the interpretation of the isomorphism
$\phi_{A_0,A_1}$ as follows. We have
\begin{equation}
\phi_{A_0,A_1}\left(1_{A_0}\right)={\det}'_{Fr}\left(A_1A_0^{-1}\right)
\cdot\left(A_0e_0\wedge e_0^{-1}\right),
\label{B3065}
\end{equation}
where $e_0\in\det\left(\tilde{E}_0\right)$, $e_0\ne 0$, and $A_0e_0$
is the image of $e_0$
in $\det\left(\tilde{E}_1\right):=\det\left(A_0\tilde{E}_0\right)$.
(Here, $\tilde{E}_j$ are the same as in (\ref{B3064}). The determinant
line bundle $\det\left(\tilde{E}_1\right)\otimes\left(\det\left(\tilde{E}_0
\right)\right)^{-1}$ is identified with $\det_{\Ell}\left(A_1\right)$
by $\psi\left(A_1\right)$.)
Let $E_1$ be a finite-dimensional invariant subspace corresponding
to algebraic eigenspaces for $A_1A_0^{-1}$ with eigenvalues
$\lambda\in\Spec\left(A_1A_0^{-1}\right)$, $|\lambda|<C$, $C\in\wR_+$.
So $\tilde{E}_1\subset E_1$. Set $E_0:=A_0^{-1}E_1$.
Then we have
\begin{equation}
\phi_{A_0,A_1}=\left({\det}'_{Fr}\left(\left(1-p_{E_1}\right)A_1A_0^{-1}
\right)\right)^{-1}\phi_{A_0,A_1}\left(E_{\bullet}\right),
\label{B3067}
\end{equation}
where $\phi_{A_0,A_1}\left(E_{\bullet}\right)$ is the composition
of identifications (defined by $A_0$ and $A_1$)
\begin{equation}
L_{inv}\left(A_0\right) @>\psi_{E_{\bullet}}(A_0)>\lrs> \det\left(E_1\right)
\otimes\det\left(E_0^{-1}\right) @<\psi_{E_{\bullet}}(A_1)<\lrs< {\det}_{\Ell}
\left(A_1\right),
\label{B3066}
\end{equation}
and $p_{E_1}$ is the spectral projection of $L_2(M,E)$ on the algebraic
eigenspaces for $A_1A_0^{-1}=\Id+S_1$ with eigenvalues $\lambda$,
$|\lambda|<C$. The determinant lines
$\det(E_{\bullet}):=\left(\det\left(E_1\right)\right)\otimes\left(\det\left(
E_0\right)\right)^{-1}$ and
$\det(\tilde{E}_{\bullet}):=\left(\det\left(\tilde{E}_1\right)\right)\otimes
\left(\det\left(\tilde{E}_0\right)\right)^{-1}$ in (\ref{B3066}) and
in (\ref{B3064}) are identified by $A_0$,
\begin{equation}
\psi_{\tilde{E}_{\bullet},E_{\bullet}}\left(A_0\right)\colon\det\left(
\tilde{E}_{\bullet}\right)\rs\det(E_{\bullet}).
\label{B3068}
\end{equation}
The elements in these determinant lines corresponding to the same
element $a\in\det\left(A_1\right)$, $a\ne 0$, are connected
by the identification (\ref{B3068}). (This assertion is compatible
with the ratio of Fredholm determinant factors in the expressions
for $\phi_{A_0,A_1}$ with the help of $\psi_{A_0,A_1}(E_{\bullet})$
and $\psi_{A_0,A_1}\left(\tilde{E}_{\bullet}\right)$.) Hence
$\phi_{A_0,A_1}\left(1_{A_0}\right)$ can be interpreted as an element
of the system of determinant lines $\det(E_{\bullet})$ identified
by $\psi_{\tilde{E}_{\bullet},E_{\bullet}}\left(A_0\right)$
with $\det\left(\tilde{E}_{\bullet}\right)$. This assertion means
that formally $\phi_{A_0,A_1}\left(1_{A_0}\right)$ has the properties
of the expression $A_0e\wedge e^{-1}$, where $e$ is a nonzero ``volume
element'' from ``$\det\left(L_2(M,E)\right)$'' and $A_0e$ is the image
of $e$ in ``$\det\left(H_{(-m)}(M,E)\right)$'',
$m:=\Re\left(\ord A_0\right)$.
Here, $e$ is defined by a basis $\left(e_1,\dots,e_n,\dots\right)$
from a class of admissible basises in $L_2(M,E)$. This class is defined
as an orbit of a given orthonormal basis by the action on it of the group
$F$ of invertible operators of the form $\Id+\cK$, $\cK$ are smoothing.

Let $e$ be the volume element defined by an admissible basis
$\left(e_1,\dots,e_n,\dots\right)$ and let $f$ be the volume element defined
by $\left(f_1,\dots,f_n,\dots\right)=B\left(e_1,\dots,e_n,\dots\right)$,
$B\in F$. Then we have
\begin{equation}
f={\det}_{Fr}(B)\cdot e, \qquad Af={\det}_{Fr}(B)\cdot Ae.
\label{B3069}
\end{equation}
(This interpretation has some analogy with the construction
of the determinant bundle over the Grassmanian of a Hilbert space
in \cite{SW}, \S~3.)

Let $A_0$ be an invertible PDO with the same symbol
as $\sigma\left(A_0\right)$. Hence we have by (\ref{B3065}), (\ref{B3069})
$$
\phi_{A'_0,A_1}\left(1_{A'_0}\right)=A'_0e\wedge e^{-1}=\det\left(A'_0A_0^{-1}
\right)\left(A_0e\wedge e^{-1}\right)=\det\left(A'_0A_0^{-1}\right)
\phi_{A_0,A_1}\left(1_{A_0}\right).
$$
So the equality (\ref{B3036}) is proved since
$$
\phi_{A_0,A'_0}\left(1_{A_0}\right)=\left({\det}_{Fr}\left(A'_0A_0^{-1}\right)
\right)^{-1}\cdot 1_{A'_0}.
$$
The lemma is proved.\ \ \ $\Box$

\noindent{\bf Proof of Proposition~\ref{PB3050}.} Let $s$ be a section
of the fibration (\ref{B3051}), $\pi s=\Id$
on $\SEll_0^{\times}(M,E)$. Then the line bundle $s^*\det_{\Ell}$
over $\SEll_0^{\times}(M,E)$ is isomorphic to the line bundle
$L$ associated with the $\wC^{\times}$-fibration of the determinant Lie group
over $\SEll_0^{\times}(M,E)$. Namely the associative system
$\phi_{A_1,A_2}$ identifies linearly the fibers of $\det_{\Ell}$
for $A_1$, $A_2$ from any fiber of $\pi$ and defines a line bundle $L_1$
over $\SEll_0^{\times}(M,E)$ isomorphic to $L$. The linear bundle
$\det_{\Ell}$ is isomorphic to $\pi^*L_1=\pi^*L$ since $\pi$ is
a fibration with constructible fibers. We have
$$
L=s^*\pi^*L=s^*\pi^*L_1=s^*{\det}_{\Ell}^0.
$$
The canonical identification $L=L_1$ follows immediately from the coincidence
of the identifications (\ref{B3032}) with $\phi_{A_1,A_2}$ for invertible
$A_1$, $A_2$ and from the associativity of $\phi_{A_1,A_2}$ given
by Lemma~\ref{LB3052}.\ \ \ $\Box$

\begin{rem}
(A holomorphic structure on $\det_{\Ell}^0$) A natural holomorphic
structure on $\Ell_0(M,E)$ is defined as follows. We have a natural
projection
\begin{equation}
p\colon\Ell_0(M,E)\to\SEll_0(M,E)
\label{B3919}
\end{equation}
with an affine fiber $\{\Id+\cK\}$, where $\cK$ are smoothing.
(Elements of this fiber may have the zero Fredholm determinant.)
A projection $p_1\colon\Ell_0(M,E)\to ps(M,E)$ on the space
of principal elliptic symbols (of all complex orders) has as its
fiber an affine space $\Id+CL^{-1}(M,E)$. These fibers have a natural
complex structure invariant under the adjoint action of the group
$\Ell_0^{\times}(M,E)$ of invertible elliptic PDOs. The base
$ps(M,E)$ has a natural complex structure (analogous to the one
defined in Remark~\ref{RB3600}. This structure induces complexes
structures on all other connected components of $\Ell(M,E)$
by (left or right) multiplying by representatives of these components.

The line bundle $\det_{\Ell}^0$ over $\Ell_0(M,E)$ has a natural
holomorphic structure. It is the structure induced from a holomorphic
structure on the determinant line bundle $L$ on $\SEll_0^{\times}(M,E)$
(associated with $G(M,E)$%
\footnote{The line bundle $L$ over $\SEll_0^{\times}(M,E)$ is explicitly
defined at the beginning of this subsection. A natural holomorphic
structure on it is defined with the help of Remark~\ref{RB1290}.}%
) under a (local) holomorphic section $r$ of $p$, $r\colon U\to p^{-1}U$.
A holomorphic section of $\det_{\Ell}^0$ over $r(U)$ defines a section
of $\det_{\Ell}^0$ over $p^{-1}(U)$ with the help of the canonical
associative system of identifications (defined in Lemma~\ref{LB3052})
of the fibers of $\det_{\Ell}^0$ over the fibers $p^{-1}(x)$,
$x\in U$. These sections over $p^{-1}(U)$ define a natural
holomorphic structure on $\det_{\Ell}^0$.
\label{RB3918}
\end{rem}

\subsection{Odd class operators and the canonical determinant}
\label{SEOdd}

The odd class PDOs are introduced in Section~\ref{SC}. They are
a generalization of DOs. Let
$\Ell_{(-1)}^{\times}(M,E)\subset\Ell_0^{\times}(M,E)$ be a subgroup
of invertible elliptic PDOs of the odd class.%
\footnote{$\Ell_0^{\times}(M,E)$ is the group of invertible elliptic
PDOs of complex orders.}
Then the subgroup of $\Ell_0^{\times}(M,E)$ generated by elliptic
DOs is contained in $\Ell_{(-1)}^{\times}(M,E)$ and every element
of $\Ell_{(-1)}^{\times}(M,E)$ has an integer order.

The multiplicative anomaly on an odd-dimensional closed manifold
is zero for operators $A,B\in\Ell_{(-1)}^{\times}(M,E)$ such that
$\ord A,\ord B,\ord A+\ord B\in\wZ\setminus 0$. Thus using
the multiplicative property, we can define unambiguously
a determinant $\det(A)$ for zero order $A\in\Ell_{(-1),0}^{\times}(M,E)$
with $\sigma_0(A)$
close to a positive definite self-adjoint one, Corollary~\ref{CB243}.
The canonical determinant $\det_{(-1)}(A)$ for any odd class invertible
zero order elliptic PDO $A$ (on an odd-dimensional $M$) with a given
$\sigma(\log A)\in CS_{(-1)}^0(M,E)$ is defined below, (\ref{B3993}).
These two determinants are equal for odd class elliptic $A$ of zero order
sufficiently close to positive definite self-adjoint ones
and for an appropriate $\sigma(\log A)$. (It is proved below.)

Let $G_{(-1)}(M,E)$ be the determinant Lie group restricted to the odd
class elliptic PDOs, i.e., $G_{(-1)}(M,E)$ be the quotient
$F_0\backslash\Ell_{(-1),0}^{\times}(M,E)$.

Let $\Ell_{(-1),0}^0(M,E)\ni\Id$ be a connected component
of $\Ell_{(-1)}(M,E)$ and let
$$
G_{(-1)}^0(M,E):=F_0\backslash\Ell_{(-1),0}^0(M,E)
$$
be an appropriate determinant Lie group. Then the Lie
algebra $\fell_{(-1)}^0(M,E)$ of $\Ell_{(-1),0}^0(M,E)$ is equal
to $CL_{(-1)}^0(M,E)$ by Proposition~\ref{PB3761}.

Let $l_j:=\sigma\left(\log_{(\theta_j)}A_j\right)/\ord A_j$, where
$A_j\in\Ell_{(-1),0}^{m_j}(M,E)$, $m_j$ are even, $m_j\ne 0$, and
$L_{(\theta_j)}$ are admissible (for $A_j$) cuts of the spectral plane.
Let $\sfrg_{(-1),(l_j)}$ be a one-dimensional central extension
of the Lie algebra $CS_{(-1)}^0(M,E)$ given by the cocycle
$K_{(l_j)}(M,E)$, Lemma~\ref{LB1590}, (\ref{B304}), Remark~\ref{RB2005},
(\ref{B307}), (\ref{B624}).

\begin{rem}
In the definition of logarithmic symbols $\sigma\left(\log A_j\right)$
it is enough to use a smooth field of admissible
for $\left(A_j-\lambda\right)^{-1}$ spectral cuts
$\theta_j\colon P^*M\to S^1=\wR/2\pi\wZ$ as in Remark~\ref{RB3994}.
(This map has to be homotopic to a trivial one.) These fields of cuts
may depend on $A_j$.

For such defined logarithmic symbols
$l_j=\sigma\left(\log A_j\right)/\ord A_j$, Propositions~\ref{PB3975},
\ref{PB3978}, Lemma~\ref{LB3981} (and Corollary~\ref{CB3980}) are valid.
The existence of such fields of spectral cuts is a property of a symbol
$\sigma\left(A_j\right)$ (but not of an even order PDO $A_j$ itself).
If these fields exist, then the Lie algebras (over $\wZ$)
$\sfrg_{(-1),(l_1)}^{\wZ}\rs\sfrg_{(-1),(l_2)}^{\wZ}$ (defined below)
are canonically identified by $W_{l_1l_2}$, Proposition~\ref{PB3978}.
\label{RB3995}
\end{rem}

\begin{pro}
The extensions $\sfrg_{(-1),(l_j)}$ of the Lie algebra $CS_{(-1)}^0(M,E)$
for a closed odd-dimensional $M$ are canonically identified
by an associative system of isomorphisms
$W_{l_1l_2}\colon\sfrg_{(-1),(l_1)}\to\sfrg_{(-1),(l_2)}$ defined
in Proposition~\ref{PB403}, (\ref{B404}). These isomorphisms are $\Id$
with respect to the coordinates in (\ref{B404}).
\label{PB3975}
\end{pro}

\noindent{\bf Proof.} By Corollary~\ref{CB3787} and by Remark~\ref{RB3976}
$l_1-l_2$ belongs to $CS_{(-1)}^0(M,E)$.
So the identification (\ref{B404}), $W_{l_1l_2}(a+c\cdot 1)=a+c'\cdot 1$,
for $a\in CS_{(-1)}^0(M,E)$ is given by
\begin{equation}
c'=c+\left(l_1-l_2,a\right)_{\res}=c
\label{B3977}
\end{equation}
in view of Remark~\ref{RB3785}, i.e., $W_{l_1l_2}=\Id$.\ \ \ $\Box$

However the logarithms of the odd class elliptic PDOs form a Lie
subalgebra (over $\wZ$) $\fell_{(-1)}^{\wZ}(M,E)\subset\fell(M,E)$.
Elements of $\fell_{(-1)}^{\wZ}(M,E)$ take the form $mL+a$, where
$m\in\wZ$, $a\in CS^0(M,E)$, and $2L$ is a logarithm of an element
of $\Ell_{(-1),0}^2(M,E)$ (for example, of $\Delta_E+\Id$, $\Delta_E$
is the Laplacian of a unitary connection $\nabla^E$ on $E$).
Analogous subgroups $\sfrg_{(-1),(l_j)}^{\wZ}$ of $\sfrg_{(l_j)}$ are
defined.

\begin{pro}
The identifications $W_{l_1l_2}$, (\ref{B404}),
$$
W_{l_1l_2}\colon\sfrg_{(-1),(l_1)}^{\wZ}\rs\sfrg_{(-1),(l_2)}^{\wZ}
$$
are $\Id$ (over
$S_{(-1),\log}(M,E)$, the Lie algebra over $\wZ$
of $\SEll_{(-1),0}^{\times}(M,E)$) on an odd-dimensional closed $M$
with respect to the coordinates (\ref{B404}) in the central extensions.
(Here, $l_j$ are under the same conditions as in Proposition~\ref{PB3975}.)
\label{PB3978}
\end{pro}

\noindent{\bf Proof.} An element $ql_1+a+c\cdot 1\in\sfrg_{(-1),(l_j)}^{\wZ}$,
$q\in\wZ$, is identified by $W_{l_1l_2}$ with
$ql_2+a'+c'\cdot 1\in\sfrg_{(-1),(l_2)}^{\wZ}$, where
\begin{gather}
\begin{split}
ql_1 & +a=ql_2+a'\in S_{(-1),\log}(M,E), \\
c'=c & +\left(l_1-l_2,a\right)_{\res}+q\left(l_1-l_2,l_1-l_2\right)_{\res}/2.
\end{split}
\label{B3979}
\end{gather}
By Remark~\ref{RB3785}, $c'=c$ because $l_1-l_2\in CS_{(-1)}^0(M,E)$
by Corollary~\ref{CB3787} and by Remark~\ref{RB3976}.\ \ \ $\Box$

The associative system of identifications
$W_{l_1l_2}\colon\sfrg_{(-1),(l_1)}^{\wZ}\rs\sfrg_{(-1),(l_2)}^{\wZ}$
defines a canonical Lie algebra $\sfrg_{(-1)}^{\wZ}$ over $\wZ$ which is
a central extension of $S_{(-1),\log}(M,E)$ with the help of $\wC$.

\begin{lem}
The cocycle $K_l$, (\ref{B304}), is trivial on $S_{(-1),\log}(M,E)$
for a closed odd-dimensional $M$. Here, $l$ satisfies the same conditions
as $l_j$ in Proposition~\ref{PB3975}.
\label{LB3981}
\end{lem}

\noindent{\bf Proof.} By Remark~\ref{RB3785} it is enough to show that
$[l,a]\in CS_{(-1)}^0(M,E)$ for some even $m$. Then
$\exp(ml)\in\SEll_{(-1),0}^m(M,E)$. There is an invertible
$A\in\Ell_{(-1),0}^m(M,E)$ with $\sigma(A)=\exp(ml)$.
By (\ref{B260}) and by Remark~\ref{RB3976} we have
\begin{equation}
\sigma_{-ms-j}\left(A_{(\theta)}^{-s}\right)(x,\xi)=\left(-1\right)^j\sigma
_{-ms-j}\left(A_{(\theta)}^{-s}\right)(x,-\xi)
\label{B3982}
\end{equation}
for an admissible for $A$ cut $L_{(\theta)}$. So we have
\begin{equation}
\left[\sigma\left(A_{(\theta)}^{-s}\right),a\right]_{-ms-j}(x,\xi)=
\left(-1\right)^j\left[\sigma\left(A_{(\theta)}^{-s}\right),a\right]_{-ms-j}
 (x,-\xi).
\label{B3983}
\end{equation}
Taking $\df_s|_{s=0}$ of (\ref{B3983}), we obtain
$m[l,a]\in CS_{(-1)}^0(M,E)$. The lemma is proved.\ \ \ $\Box$

\begin{cor}
The central extension $\sfrg_{(-1)}^{\wZ}(M,E)$ of $S_{(-1),\log}(M,E)$
is canonically trivial.
\label{CB3980}
\end{cor}

Indeed, by Proposition~\ref{PB3978} the coordinates $c\cdot 1$
in $\sfrg_{(-1),(l_j)}^{\wZ}$ with respect to the splittings
$\sfrg_{(-1),(l_j)}^{\wZ}=S_{(-1),\log}(M,E)\oplus\wC\cdot 1$ do not
change under the identifications $W_{l_1l_2}$. By Lemma~\ref{LB3981}
the $\wC$-extensions $\sfrg_{(-1),(l_j)}^{\wZ}$ of $S_{(-1),\log}(M,E)$
are trivial with respect to these splittings.

\medskip
The canonical splitting $\sfrg_{(-1)}=S_{(-1),\log}(M,E)\oplus\wC$
of this central extension gives us a canonical connection
on the $\wC^{\times}$-bundle $G_{(-1)}(M,E)\to\SEll_{(-1),0}^{\times}(M,E)$.
Hence we can define locally a holomorphic function on the space
of odd class elliptic PDOs on an odd-dimensional closed $M$.
(A natural complex structure on $\Ell_{(-1),0}^{\times}(M,E)$
is defined in Remark~\ref{RB3600}.) Namely, if for $A$ close to $A_0$
we choose as $\hat{d}_0(A)$ a locally flat section over $\sigma(A)$
(with respect to the connection on $G_{(-1)}(M,E)$), then
$$
\widetilde{\det}(A):=d_1(A)/\hat{d}_0(A)\in\wC^{\times}
$$
is holomorphic in $A$. Of course in general we cannot find a global
flat section $\hat{d}_0(A)$. However we can define $\hat{d}_0(A)$
as a multi-valued flat section of $G_{(-1)}(M,E)$ over zero order
symbols of the odd class such that $\hat{d}_0(A)$
is an anlytic continuation of the flat section $\hat{d}_0(A)$ near
$\sigma\left(A_0\right)=\Id$, where $\hat{d}_0(\Id)$ is the identity
of $G_{(-1)}(M,E)$.

Let $\Delta_E$ be the Laplacian on $(M,E)$ for $\left(g_M,\nabla^E\right)$,
where $g_M$ is a Riemannian structure and $\nabla^E$ is a unitary connection.
Then we define $\hat{d}_0(A)$ for $\ord A=2m$, $m\in\zuo$,
as a multi-valued flat
section of $G_{(-1)}(M,E)$ over $\SEll_{(-1),0}^{2m}(M,E)$ such that
$\hat{d}_0\left(\Delta_E^m+\Id\right)=d_1\left(\Delta_E^m+\Id\right)/\det
_{(\pi)}\left(\Delta_E^m+\Id\right)$.

\begin{pro}
For such a section $\hat{d}_0(A)$ the determinant
\begin{equation}
\widetilde{\det}(A):=d_1(A)/\hat{d}_0(A)
\label{B3986}
\end{equation}
gives us a (multi-valued) holomorphic determinant
of $A\in\Ell_{(-1),0}^{\times}(M,E)$ defined in Section~\ref{SC4},
Proposition~\ref{PB1801}.
\label{PB3985}
\end{pro}

\begin{rem}
This holomorphic determinant is a multi-valued function $f(A)$
defined up a constant factor $c\in\wC^{\times}$, $|c|=1$. Here
we define a branch of $f(A)$ equal
to $\det_{(\tpi)}\left(\Delta_E^m+\Id\right)$ at the point
$A_0:=\Delta_E^m+\Id\in\Ell_{(-1),0}^{2m}(M,E)$. We can do this since
\begin{equation}
\left|f\left(A_0\right)\right|^2=\left({\det}_{(\tpi)}\left(\Delta_E^m+
\Id\right)\right)^2=\left|{\det}_{(\tpi)}\left(\Delta_E^m+\Id\right)\right|^2.
\label{B4001}
\end{equation}
Here we use that $\Delta_E^m+\Id$ is self-adjoint and positive definite.
\label{RB4000}
\end{rem}

\begin{cor}
The monodromy of $\widetilde{\det}(A)$ defined by (\ref{B3986})
over closed loops
in $\Ell_{(-1),0}^{\times}(M,E)$ is given by multiplying by roots
of order $2^m$ of $1$, where $m$ depends on $\dim M$ only. (This
assertion follows from Proposition~\ref{PB1802}.)
\label{CB3987}
\end{cor}

Let $A\in\Ell_{(-1),0}^0(M,E)$ be an elliptic PDO of odd class
on an odd-dimensional closed $M$ with a fixed logarithmic symbol
$\sigma(\log A)\in CS_{(-1)}^0(M,E)$. Then $A$ has a canonical
determinant defined with the help of the $\Tr_{(-1)}$-functional,
Proposition~\ref{PB3886}, (\ref{B3887}). This functional is defined
for the odd class PDOs
$CL_{(-1)}^{\bullet}(M,E)$ on an odd-dimensional closed $M$.
Namely let $B\in\fell_{(-1)}(M,E)$ be an operator with
$\sigma(B)=\sigma(\log A)$. Then the element
\begin{equation}
\tilde{d}_{(-1),0}(\sigma(\log A))=d_1(\exp B)/{\det}_{(-1)}(\exp B)
\label{B3988}
\end{equation}
is defined, where $d_1(\exp B)$ is the image
of $\exp B\in\Ell_{(-1),0}^0(M,E)$ in $G_{(-1)}(M,E)$, $F_0\exp B$, and
\begin{equation}
{\det}_{(-1)}(\exp B):=\exp\left(-\df_s\Tr_{(-1)}(\exp(-sB))|_{s=0}\right)\in
\wC^{\times}.
\label{B3989}
\end{equation}
For all $s\!\in\!\wC$ an elliptic operator $\exp(-sB)$ belongs
to $\Ell_{(-1),0}^0(M,\!E)\!\subset\!CL_{(-1)}(M\!,E)$. Hence
$\Tr_{(-1)}(\exp(-sB))$ is defined for all $s\in\wC$ and is regular
in $s$.

\begin{lem}
The element $\tilde{d}_{(-1),0}(\sigma(\log A))\!\in\!G_{(-1)}(M,\!E)$ is
independent of a choice of $B$ with $\sigma(B)=\sigma(\log A)$.
\label{LB3990}
\end{lem}

\noindent{\bf Proof.} Let $B_1\in\fell_{(-1)}(M,E)$ and
$\sigma(B)=\sigma\left(B_1\right)$. Then
$$
d_1\left(\exp B_1\right)={\det}_{Fr}\left(\exp B_1\exp(-B)\right)d_1(\exp B),
$$
\begin{multline}
{\det}_{(-1)}\left(\exp B_1\right)=\exp\left(-\df_s\Tr_{(-1)}\exp\left(-sB_1
\right)\big|_{s=0}\right)=\\
={\det}_{(-1)}(\exp B)\exp\left(-\df_s\Tr_{(-1)}\left(\exp\left(-sB_1\right)-
\exp(-sB)\right)\big|_{s=0}\right).
\label{B3991}
\end{multline}

An operator $\exp\left(-sB_1\right)-\exp(-sB)$ is of trace class
for all $s\in\wC$. It is even a smoothing operator. (An operator
$\exp(-sB)$ is defined in $L_2(M,E)$ by (\ref{K25}), $B$ is bounded
in $L_2(M,E)$, and $\sigma\left(B_1\right)=\sigma(B)$.) Thus
\begin{equation}
\Tr_{(-1)}\left(\exp\left(-sB_1\right)-\exp(-sB)\right)=\Tr\left(\exp\left(
-sB_1\right)-\exp(-sB)\right).
\label{B3992}
\end{equation}
Similarly to (\ref{B3515}), (\ref{B3516}), to Remark~\ref{RB4}, and
to Proposition~\ref{PB3931}, (\ref{B3995}), we conclude that
\begin{multline*}
\exp\left(-\df_s\Tr_{(-1)}\left(\exp\left(-sB_1\right)-\exp(-sB)\right)\right)=
\exp\left(\Tr\left(B_1-B\right)\right)=\\
={\det}_{Fr}\left(\exp B_1\exp(-B)\right).
\end{multline*}
The lemma is proved.\ \ \ $\Box$

\medskip
{\bf Definition}. An (odd class) determinant $A\in\Ell_{(-1),0}^0(M,E)$
with a given logarithmic symbol $\sigma(\log A)\in CS_{(-1)}^0(M,E)$ is
defined by
\begin{equation}
{\det}_{(-1)}(A):=d_1(A)/\tilde{d}_{(-1),0}(\sigma(\log A)),
\label{B3993}
\end{equation}
where $\tilde{d}_{(-1),0}(\sigma(\log A))\in G_{(-1)}(M,E)$ is defined
by the expression on the right in (\ref{B3988}) with any operator
$B\in\fell_{(-1)}(M,E)=CL_{(-1)}^0(M,E)$ such that
$\sigma(B)=\sigma(\log A)$. The expression on the right in (\ref{B3993})
is independent of $B$ with $\sigma(B)=\sigma(\log A)$ by Lemma~\ref{LB3990}.

\begin{pro}
Let $A\in\Ell_{(-1),0}^0(M,E)$ ($M$ is odd-dimensional) be sufficiently
close to positive definite self-adjoint PDOs. Then
\begin{equation}
{\det}_{(-1)}(A)=\det(A),
\label{B4021}
\end{equation}
where $\det(A)$ is defined with the help of the multiplicative property,
Theorem~\ref{TB241}, Corollary~\ref{CB243}. In (\ref{B4021}) we
suppose that an appropriate $\sigma\!(\log A)$, namely
$\sigma\!\left(\log_{(\tpi)}\!A\right)$, is used in the definition
of $\det_{(-1)}(A)$. (This symbol is defined by (\ref{B248}).)
\label{PB4020}
\end{pro}

\noindent{\bf Proof.} 1. Let $L:=\log A\in\fell_{(-1)}(M,E)=CL_{(-1)}^0(M,E)$
exist. The $\det_{(-1)}(A)$ corresponding to $\sigma(L)\in CS^0(M,E)$
is given by
\begin{equation}
{\det}_{(-1)}(A)=\exp\left(\Tr_{(-1)}L\right).
\label{B4022}
\end{equation}
(This formula can be read as
$\det_{(-1)}(A)=\exp\left(\Tr_{(-1)}(\log A)\right)$. The functional
$\Tr_{(-1)}$ is defined by Proposition~\ref{PB3886}, (\ref{B3887}).)

To prove (\ref{B4022}), note that for any $A\in CL_{(-1)}^0(M,E)$
we have $\Tr_{(-1)}(A)=\TR\left(AC_{(\pi)}^{-s}\right)|_{s=0}$
for any positive definite self-adjoint $C\in\Ell_{(-1),0}^{2m}(M,E)$,
$m\in\wZ_+$. In particular,
\begin{equation}
{\det}_{(-1)}(A)=\exp\left(-\df_s\left(\TR\left(\exp(-sL)C^{-s_1}\right)|
_{s_1=0}\right)|_{s=0}\right).
\label{B4023}
\end{equation}
The family $\exp(-sL)C^{-s_1}$ is a holomorphic family of elliptic
PDOs. So the function $\TR\left(\exp(-sL)C^{-s_1}\right)$ is regular
at $s_1=0$
(since $\exp(-sL)$ is a PDO of the odd class and since $M$ is
odd-dimensional) and then at $s=0$. We can rewrite (\ref{B4023}) as
\begin{multline*}
{\det}_{(-1)}(A)=\exp\left(-\left(\TR\left(-L\exp(-sL)C^{-s_1}\right)|_{s_1=0}
\right)|_{s=0}\right)=\\
=\exp\left(-\Tr_{(-1)}(-L\exp(-sL))|_{s=0}\right)=
\exp\left(\Tr_{(-1)}L\right).
\end{multline*}

2. The determinant $\det(A)$ of $A:=\exp L$, $L\in\fell_{(-1)}(M,E)$,
(defined by Corollary~\ref{CB243}) for $L$ sufficiently small
is
$$
\det(A)={\det}_{(\tpi)}((\exp L)C)/{\det}_{(\tpi)}(C).
$$
For all sufficiently small $t$ we have
\begin{gather}
\begin{split}
\det(\exp(tL))=\exp & \left(t\df_s\log\det(\exp(sL)C)|_{s=0}\right), \\
\df_s\log\det(\exp(sL)C)|_{s=0}     & {\overset{(\alpha)}{=}}\TR\left(LC^{-z}
\right)|_{z=0}=\Tr_{(-1)}L.
\end{split}
\label{B4024}
\end{gather}
The equality $\overset{(\alpha)}{=}$ in (\ref{B4024}) follows
from the variation formulas
(\ref{B3517}), (\ref{B3751}), (\ref{B853})
for $\delta\log\det\left(C_s\right)$, $C_s:=\exp(sL)C$,
\begin{multline}
\df_s\log\det\left(C_s\right)\!=\!\left(\left(1\!+\!z\df_z\right)\Tr\left(
LC_s^{-z}\right)\right)\big|_{z=0}\!=\!\Tr\left(LC_s^{-z}\!-\!\res\sigma(L)/z
\ord C_s\right)\big|_{z=0}\!=\\
=\Tr\left(LC_s\right)|_{z=0}.
\label{B4025}
\end{multline}
(By Remark~\ref{RB3785} $\res\sigma(L)=0$ since $L$
is an odd class PDO and $M$ is odd-dimensional). The expression
$\Tr\left(LC_s^{-z}\right)$ is equal to $\TR\left(LC_s^{-z}\right)$
for $\ord C\cdot\Re z>\dim M$. So
$\Tr\left(LC_s^{-z}\right)|_{z=0}=\Tr_{(-1)}L$. Then we conclude that
$$
{\det}_{(-1)}(A)=\exp\left(\Tr_{(-1)}(tL)\right)=\det(A)
$$
for $A=\exp tL$, where $t$ is sufficiently small. The functions
$\det_{(\tpi)}(A)$ and $\det(A)$ are analytic in $A$ (in their domains
of definition). If $\log_{(\tpi)}A$
is defined, then $\det(A)$ is also defined. The domain of definition
of $\det_{(\tpi)}(A)$ is connected. The proposition is proved.\ \ \ $\Box$

\noindent{\bf Proof of Proposition~\ref{PB3985}.} 1. First we prove that
$\widetilde{\det}(A)=\det(A)$ for PDOs $A$ from $\Ell_{(-1),0}^0(M,E)$
sufficiently close to $\Id$. (Here, $\det(A):=f(A)$ is a branch
of a holomorphic determinant, Proposition~\ref{PB1801}, $f(\Id)=1$.)

Let $A:=\exp(tL)$, $L\in CL_{(-1)}(M,E)$, $C\in\Ell_{(-1),0}^{2m}(M,E)$,
$m\in\wZ_+$, be a positive definite self-adjoint PDO, $J:=\log_{(\tpi)}C$.
Then by Proposition~\ref{PB4027}, (\ref{B4029}), we have
\begin{multline}
\df_t\log\left(d_1(\exp(tL))/\exp\left(t\Pi_{\sigma(J)}\sigma(L)\right)\right)
\big|_{t=0}=\TR(L\exp(-sJ)-\res\sigma(L)/s)\big|_{s=0}=\\
=\TR(L\exp(-sJ))|_{s=0}=\Tr_{(-1)}L.
\label{B4030}
\end{multline}
Here we use that $\res\sigma(L)=0$ for $L\in CL_{(-1)}^0(M,E)$
by Remark~\ref{RB3785} and by the definition (\ref{B3887}),
Proposition~\ref{PB3886}, of $\Tr_{(-1)}L$ ($M$ is odd-dimensional).
In view of (\ref{B4022}) we have
$\det_{(-1)}(\exp(tL))=\exp\left(\Tr_{(-1)}(tL)\right)$,
\begin{equation}
\df_t\log{\det}_{(-1)}(\exp(tL))|_{t=0}=\Tr_{(-1)}L.
\label{B4031}
\end{equation}
We conclude from (\ref{B4030}), (\ref{B4031}) and
from Proposition~\ref{PB4020} that
\begin{multline}
\df_t\log\widetilde{\det}(\exp(tL))\big|_{t=0}\equiv\df_t\log\left(d_1(\exp
(tL))/\exp\left(t\Pi_{\sigma(J)}\sigma(L)\right)\right)\big|_{t+0}=\\
=\Tr_{(-1)}L=\df_t\log{\det}_{(-1)}(\exp(tL))|_{t=0}=\df_t\log\det(\exp(tL))
|_{t=0}.
\label{B4032}
\end{multline}
Thus we have two equal characters of the Lie algebra
$\fell_{(-1)}(M,E)=CL_{(-1)}^0(M,E)\ni L$. Hence the corresponding
characters of $\exp\left(\fell_{(-1)}(M,E)\right)$ are also equal.
The exponential map is a map onto a neighborhood of $\Id$
in $\Ell_{(-1),0}^0(M,E)$ (and even in $\Ell_0^0(M,E)$). Indeed,
for any $A\in\Ell_0^0(M,E)$ close to $\Id$ we can take $\log_{(\tpi)}A$,
and it belongs to $\fell_{(-1)}(M,E)$ for $A\in\Ell_{(-1),0}^0(M,E)$
by Proposition~\ref{PB3761}, (\ref{B248}). So the branches of the analytic
functions $\widetilde{\det}(A)$ and $f(A)$ coincide in a neighborhood
of $\Id\in\Ell_{(-1),0}^0(M,E)$.

We know that according to Corollary~\ref{CB4007} and
to Proposition~\ref{PB4002}, $\det(A)=\det_{(\tpi)}(A)$ in a neighborhood
of $\Id$ in $\Ell_{(-1),0}^0(M,E)$ and that $\det(A)$ is a (local) character
of $\Ell_{(-1),0}^0(M,E)$. We have only to prove that $\widetilde{\det}(A)$
defines a (local) character of $\Ell_{(-1),0}^0(M,E)$. It is enough
to prove that for $L_1,L_2\in CL_{(-1)}^0(M,E)$ and for sufficiently
small $t_1,t_2\in\wC$ we have
\begin{multline}
\exp\left(t_1\Pi_{\sigma(J)}\sigma\left(L_1\right)\right)\exp\left(t_2\Pi
_{\sigma(J)}\sigma\left(L_2\right)\right)=\\
=\exp\left(\Pi_{\sigma(J)}\log_{(\tpi)}\left(\exp\left(t_1\sigma\left(L_1
\right)\right)\exp\left(t_2\sigma\left(L_2\right)\right)\right)\right).
\label{B4033}
\end{multline}
(For sufficiently small $t_1$, $t_2$ this logarithm exists
by the Campbell-Hausdorff formula.) The equality (\ref{B4033}) follows
from the equality $K_{\sigma(J)}\left(a_1,a_2\right)=0$
for $A_1,a_2\in CS_{(-1)}^0(M,E)$, Lemma~\ref{LB3981},
Corollary~\ref{CB3980}. \\

2. Let $2m=\ord A$, $m\in\wZ_+$. We prove the equality
$\widetilde{\det}(A)=\det(A)$ in a neighborhood of $A_0:=\Delta_E^m+\Id$.
{}From Theorem~\ref{TB241}, Corollary~\ref{CB4007} we know that
$\det_{(\tpi)}\left(A_1\right)\det_{(\tpi)}\left(A_2\right)=\det_{(\tpi)}
\left(A_1A_2\right)$ for odd class elliptic PDOs close to positive
definite self-adjoint PDOs. Thus by Proposition~\ref{PB4002},
$\det(A)=\det(B)\det\left(A_0\right)$
for $B:=AA_0^{-1}\in\Ell_{(-1),0}^0(M,E)$ in a neighborhood of $A_0$.

Let us prove that
$\widetilde{\det}(A)=\widetilde{\det}(B)\widetilde{\det}\left(A_0\right)$.
Note that $d_1(A)=d_1(B)d_1\left(A_0\right)$ in $G_{(-1)}(M,E)$ and
that the local section $\hat{d}_0(A)$ of $G_{(-1)}(M,E)$ over
$\SEll_{(-1),0}^{\times}(M,E)$ is defined as a solution of the equation
$$
\dot{g}g^{-1}\in\Pi_{\sigma(J)}\sfrg, \quad\hat{d}_0\left(A_0\right):=d_1
\left(A_0\right)/{\det}_{(\tpi)}\left(\Delta_E^m+\Id\right).
$$
So the assertion of Proposition~\ref{PB3985}, (\ref{B3986}),
for $\ord A=2m$ follows from the same assertion for $m=0$.\ \ \ $\Box$

\subsection{Coherent systems of determinant cocycles on the group
of elliptic symbols}
\label{SE1}

Let $a$, $b$ be the symbols of elliptic PDOs $A$, $B$ of positive orders
such that $A$ and $B$ are sufficiently close to positive definite
self-adjoint PDOs (with respect to a smooth positive density on $M$
and to a Hermitian structure on $E$). Then the cocycle
\begin{equation}
f(a,b):=\log F(A,B)
\label{A200}
\end{equation}
is defined on the group $\SEll_0^{\times}(M,E)$ of elliptic symbols
by (\ref{B221}) (and it depends on $a=\sigma(A)$ and on $b=\sigma(B)$
only).

Then the (partially defined) cocycle $f(a,b)$ can be replaced
by a cohomological one
\begin{equation}
f_{x,y}(a,b):=f(xa,by)+f(x,y)-f(xa,y)-f(x,by),
\label{A201}
\end{equation}
where $x$ and $y$ are the symbols of positive definite self-adjoint
elliptic PDOs of positive orders. Note that the terms on the right
in (\ref{A201}) are defined also in the case when the symbols $a$ and
$b$ are rather close to $\Id$. (They are defined also for $\ord a>-\ord x$,
$\ord b>-\ord y$ if the symbols $a$, $xa$, $b$, and $by$ are sufficiently
close to the symbols of positive definite self-adjoint PDOs. Under
these conditions, the formula (\ref{B221}) for the terms on the right
in (\ref{A201}), (\ref{A200}) is derived.) We have
\begin{gather}
\begin{split}
f_{x,y}(a,b)-f(a,b) & =dr_{x,y}(a,b), \\
dr_{x,y}(a,b)       & :=r_{x,y}(ab)-r_{x,y}(a)-r_{x,y}(b), \\
r_{x,y}             & :=f(yx,a).
\end{split}
\label{A202}
\end{gather}
Note that
\begin{gather}
\begin{split}
f_{x,y}(a,b)-f(a,b) & =\log_{(\tpi)}\left({{\det}_{(\tpi)}(XABY){\det}_{(\tpi)}
(XY)\over {\det}_{(\tpi)}(XAY){\det}_{(\tpi)}(XBY)}\right), \\
r_{x,y}(a)          & =\log_{(\tpi)}\left({{\det}_{(\tpi)}(XAY)\over
{\det}_{(\tpi)}(A){\det}_{(\tpi)}(XY)}\right)
\end{split}
\label{A203}
\end{gather}
under the conditions that the determinants on the right in (\ref{A203})
are defined and that $f_{x,y}(a,b)$ and $f(a,b)$ are defined. (Here,
$L_{(\tpi)}=L_{(\theta)}$ is an admissible cut of the spectral plane
with $\theta$ sufficiently close to $\pi$, $A$ and $B$ are elliptic PDOs
with the symbols $a:=\sigma(A)$ and $b:=\sigma(B)$.)

\begin{rem}
Let $x$, $y$, $x'$, and $y'$ be the symbols of positive definite
self-adjoint elliptic PDOs $X$, $Y$, $X'$, and $Y'$ of positive orders.
Then the cochain
\begin{equation}
\rho_{x,y;x',y'}(a):=r_{x,y}(a)-r_{x',y'}(a)
\label{A207}
\end{equation}
is smooth in $a$ in a neighborhood of $\Id\in\SEll_0^{\times}(M,E)$.

Indeed, we have by (\ref{A203})
\begin{equation}
\rho_{x,y;x',y'}(a):=\log_{(\tpi)}\left({{\det}_{(\tpi)}(XAY){\det}
_{(\tpi)}(X'Y')\over {\det}_{(\tpi)}(XY){\det}_{(\tpi)}(X'AY')}\right).
\label{A205}
\end{equation}
We have also by (\ref{C1}) and by (\ref{B221}) for variations $\delta a$
such that $\delta\ord a=0$
\begin{equation*}
\delta\rho_{x,y;x',y'}(a)\!=\!-\!\left(\!\delta a\!\cdot\!a^{-1}\!,\!{\sigma
\left(\!\log_{(\tpi)}\!(AYX)\!\right)\over \ord\!A\!+\!\ord \!X\!+\!\ord \!Y}
\!-
\!{\sigma\left(\!\log_{(\tpi)}(AY'X')\!\right)\over \ord \!A\!+\!\ord \!X'\!+
\!\ord \!Y'}\!\right)\!_{\res}.
\end{equation*}
The term on the right depends on the symbols $\sigma(A)$, $\sigma(X)$,
$\sigma(Y)$, $\sigma(X')$, and $\sigma(Y')$ only. It is equal
to the integral over $M$ of a density locally defined by the homogeneous
components of these symbols.

We have by (\ref{A205})
\begin{equation}
\rho_{x,y;x',y'}(a)+\rho_{x',y';x'',y''}(a)+\rho_{x'',y'';x,y}(a)\equiv 0
\label{A206}
\end{equation}
for the symbols $x$, $x'$, $x''$, $y$, $y'$, and $y''$ of self-adjoint
positive definite elliptic PDOs of positive orders.

By (\ref{A207}), (\ref{A203}) we have
\begin{equation}
f_{x,y}(a,b)-f_{x',y'}(a,b)=\left(d\rho_{x,y;x',y'}\right)(a,b).
\label{A208}
\end{equation}
\label{RA204}
\end{rem}

Hence we have a natural functorial system of (partially defined) cocycles
$f_{x,y}(a,b)$ on the group $\SEll_0^{\times}(M,E)$. All of them are
cohomologous to the cocycle $f(a,b)$ defined by the multiplicative
anomaly (\ref{B221}) of the zeta-regularized determinants. The cocycle
$f(a,b)$ is {\em symmetric}, $f(a,b)=f(b,a)$ (as it is the logarithm
of the multiplicative anomaly). However this cocycle induces
a cohomological to it {\em skew-symmetric} cocycle%
\footnote{The cocycle $K_l$ defines the central extension $\sfrg_{(l)}$
(defined by (\ref{B307}), (\ref{B624})) of the Lie algebra
$\frg:=S_{\log}(M,E)$. By Theorem~\ref{TB570}, the Lie algebra $\sfrg_{(l)}$
is canonically isomorphic to the Lie algebra $\frg(M,E)$ of the determinant
Lie group $G(M,E)$.}
$K_l(\alpha,\beta)$ (defined by (\ref{B304})) on the Lie algebra
$S_{\log}(M,E)$ of the group $\SEll_0^{\times}(M,E)$ as follows.

Note that $f_{x,y}(a,b)$ is defined for $\ord xa$ and $\ord x$ close
to zero if $\ord by>0$ and $\ord y>0$ (and if $xa$, $by$ are sufficiently
close to the symbols of positive definite self-adjoint PDOs).
Indeed,
\begin{equation}
f_{x,y}(a,b):=\log_{(\tpi)}\left({{\det}_{(\tpi)}(XABY){\det}_{(\tpi)}(XY)
\over{\det}_{(\tpi)}(XAY){\det}_{(\tpi)}(XBY)}\right)
\label{A210}
\end{equation}
(where $a=\sigma(A)$ and so on). Hence $f_{1,y}(a,b)$ is defined
for $\ord y>0$ and for $a$, $b$ close to $\Id$.

If $\ord y>0$, $b$ is close to $\Id$, and
$\alpha\in(rp)^{-1}(0)=CS^0(M,E)\subset S_{\log}(M,E)$, we have
by (\ref{B221})
\begin{equation}
\df_tf_{1,y}(\exp(t\alpha),b)|_{t=0}=-\left(\alpha,{\sigma\left(\log_{(\tpi)}
(BY)\right)\over \ord b+\ord y}-{\sigma\left(\log_{(\tpi)}(Y)\right)\over
\ord y}\right)_{\res}.
\label{A211}
\end{equation}

Let $b:=\exp(\gamma\beta)$, $\beta\in CS^0(M,E)$, and let $\gamma$ be
close to $0\in\wR$. Then by (\ref{A211}),
\begin{equation}
\dfg\left(\df_tf_{1,y}(\exp(t\alpha),\exp(\gamma\beta))|_{t=0}\right)\bgo=
-\left(\alpha,\var_\beta\left(\log_{(\tpi)}y\right)\right)_{\res},
\label{A212}
\end{equation}
where $\var_{\beta}\left(\log_{(\tpi)}y\right):=\dfg\sigma\left(\log_{(\tpi)}
(\exp(t\tilde{\beta})Y)\right)\bgo$ for $Y\in\Ell_0^{\times}(M,E)$ with
$\sigma(Y)=y$, $\tilde{\beta}\in CL^0(M,E)$, is an operator with
$\sigma(\tilde\beta)$ equal to $\beta$.

Let $R_y(\alpha,\beta)$ be the bilinear form given by the left side
of (\ref{A212}). Then the antisymmetrization $AR_y(\alpha,\beta)$
of the form $R_y(\alpha,\beta)$ is given by (\ref{A212}) as
\begin{equation}
AR_y(\alpha,\beta)=\left(\beta,\var_\alpha\left(\log_{(\tpi)}y\right)\right)
_{\res}\big/2-\left(\alpha,\var_\beta\left(\log_{(\tpi)}y\right)\right)_{\res}
\big/2.
\label{A214}
\end{equation}
Here, $\alpha$ and $\beta$ are symbols from $CS^0(M,E)$.

Let us compute the right part of (\ref{A214}) for $\ord y=\eps$
up to $o(\eps)$ (for $\eps\to 0$). By Campbell-Hausdorff formula we see
that for $\sigma(\tilde{\alpha})=\alpha$, $\sigma(\tilde{\beta})=\beta$
\begin{gather}
\begin{split}
\var_{\alpha}\left(\log_{(\tpi)}y\right) & =\df_t\sigma\left(\log_{(\tpi)}(\exp
(\tilde{\alpha}t)Y)\right)\big|_{t=0}=\\
 & =\df_t\left(\sigma\left(\log_{(\tpi)}Y\right)+\alpha t+t\left[\alpha,
\log_{(\tpi)}Y\right]\right)\big|_{t=0}+O\left((\ord Y)^2\right)=\\
 & =\alpha+\left[\alpha,\sigma\left(\log_{(\tpi)}Y\right)\right]/2+
O\left((\ord Y)^2\right), \\
\var_\beta\left(\log_{(\tpi)}y\right)    & =\beta+\left[\beta,\sigma\left(\log
_{(\tpi)}Y\right)\right]/2+O\left((\ord Y)^2\right).
\end{split}
\label{A215}
\end{gather}
Here, $O\left((\ord Y)^2\right)$ is considered with respect to a Fr\'echet
structure on $CS^0(M,E)$ defined by natural semi-norms (\ref{B3233})
(with respect to a finite cover $\left\{U_i\right\}$ of $M$).
Hence we have
\begin{equation}
AR_y(\alpha,\beta)=(\beta,[\alpha,\log y])_{\res}\big/2-(\alpha,[\beta,\log y])
_{\res}\big/2,
\label{A216}
\end{equation}
where $\log y:=\sigma\left(\log_{(\tpi)}Y\right)$.
For $\log y=l\in(rp)^{-1}(1)\subset S_{\log}(M,E)$ we conclude that
\begin{equation}
AR_{\exp l}(\alpha,\beta)=K_l(\alpha,\beta)
\label{A217}
\end{equation}
for $\alpha,\beta\in CS^0(M,E)$. The cocycle $K_l$ has a trivial
continuation (\ref{B304}) from $CS^0\!(\!M,E\!)$ to $S_{\log}(M,E)$ under
the splitting (\ref{B301}). Hence the partially defined symmetric cocycle
$f(a,b)$ on $\SEll_0^{\times}(M,E)$ produces a skew-symmetric cocycle
$K_l(\alpha,\beta)$ on its Lie algebra $\frg(M,E)$. (Namely on the Lie
algebra $\sfrg_{(l)}$ canonically isomorphic to $\frg(M,E)$
by Theorem~\ref{TB570}.)

\begin{rem}
Note that $R_y(\alpha,\beta)$ has a singularity of order $1/\ord y$
if $\ord y\sim 0$.
\label{RA218}
\end{rem}

\subsection{Multiplicative anomaly cocycle for Lie algebras}

We want to produce the multiplicative anomaly formula without using
the determinants of elliptic PDOs. This approach is more general
than in Section~\ref{SA}. We begin with the variation formula (\ref{B221})
(or (\ref{A220}) below). This makes sense for central (and cocentral)
extensions of Lie algebras $\frg_0$ with conjugate-invariant scalar
products, Remark~\ref{RB422}, (\ref{B423})--(\ref{B426}).
In computations below it is enough to replace $\left(,\right)_{\res}$
by an invariant scalar product on $\frg_0$ and
$\sigma\left(\log_{(\tpi)}YX\right)$ (and so on) is defined
as logarithms of elements of a formal group corresponding to the Lie
algebra $\frg$, (\ref{B424}). Then Proposition~\ref{PA221} and
Corollary~\ref{CA224} below provide us with a definition (by integrating
of differential forms) of a (partial defined) multiplicative anomaly
cocycle in a general situation of Remark~\ref{RB422}.

The proof of Theorem~\ref{TB570} provides us with a partially defined
cocycle. It is given by the exponential (\ref{B580}) of the quadratic
cone in $\sfrg$, Proposition~\ref{PB417}. This cone is defined
in the situation of Remark~\ref{RB422} also. Here we obtain the results
on the multiplicative anomaly for Lie algebras without using this
quadratic cone. Propositions~\ref{PA221}, \ref{PA222} and
Corollary~\ref{CA224} below, as well as their proofs, are valid
for central extensions of Lie algebras (Remark~\ref{RB422}) after
trivial changing of notations.

For elliptic PDOs $X$ and $Y$ of positive orders sufficiently close
to positive definite self-adjoint elliptic PDOs, let variations
$\delta X$, $\delta Y$ be such that $\delta(\ord Y)=0=\delta(\ord X)$.
Then by (\ref{B221}) we have
\begin{multline}
\delta_{X,Y}\log\left({\det}_{(\tpi)}(XY)/{\det}_{(\tpi)}(X){\det}
_{(\tpi)}(Y)\right)=\\
=-\left(\delta y\cdot y^{-1},{\sigma\left(\log_{(\tpi)}(YX)
\right)\over \ord x+\ord y}-{\sigma\left(\log_{(\tpi)}Y
\right)\over \ord y}\right)_{\res}-\\
-\left(\delta x\cdot x^{-1},
{\sigma\left(\log_{(\tpi)}(XY)\right)\over \ord x+\ord y}-
{\sigma\left(\log_{(\tpi)}X\right)\over \ord x}\right)_{\res},
\label{A220}
\end{multline}
where $x=\sigma(X)$, $y=\sigma(Y)$. The terms on the right depend on $x$
and on $y$ only. Hence we have a differential $1$-form $\omega_{x,y}^1$
on the domain in $\SEll_0^{c_1}(M,E)\times \SEll_0^{c_2}(M,E)$, where
$c_1:=\ord X$, $c_2:=\ord Y$, $c_j\in\wR^{\times}$, $c_1+c_2\in\wR^{\times}$.

\begin{pro}
The form $\omega_{x,y}^1$ is closed in the directions of the components
of the direct product $\SEll_0^{c_1}(M,E)\times\SEll_0^{c_2}(M,E)$.
\label{PA221}
\end{pro}

\begin{cor}
The function $\log F(\!A,B\!)$ in the formula of the multiplicative anomaly
(\ref{B221}) is defined by the integration of the $1$-form $\omega_{x,y}^1$
on $x$ and then on $y$
(since $\log_{(\pi)}F\left(S_{(\pi)}^{c_1},S_{(\pi)}^{c_2}\right)=0$
for powers of a positive definite self-adjoint $S\in\Ell_0^1(M,E)$).
\label{CA224}
\end{cor}

\begin{pro}
The form $\omega_{x,y}^1$ is a (partially defined) $2$-cocycle on
$\SEll_0^{\wR}(M,E)$, i.e., on the group of elliptic symbols of real
orders. This assertion means that
\begin{equation}
\left(d_{cochain}\omega^1\right)(x,y,z)\!:=\!\omega^1(y,z)\!-\!\omega^1(xy,z)
\!+\!\omega^1(x,yz)\!-\!\omega^1(x,y)\!=\!0,
\label{A223}
\end{equation}
if the terms on the right are defined.
\label{PA222}
\end{pro}

\noindent{\bf Proof.} By (\ref{A220}) and by (\ref{A223}) the terms with
$dx\cdot x^{-1}$ in $d_{cochain}\omega^1(x,y,z)$ are
\begin{multline}
\bigl(dx\cdot x^{-1},{\log(xyz)\over \ord x+\ord y+\ord z}-{\log(xy)\over
\ord x+\ord y}-{\log(xyz)\over \ord x+\ord y+\ord z}+\\
+{\log x\over \ord x}+{\log(xy)\over \ord x+\ord y}-{\log x\over \ord x}\bigr)
=0.
\label{A225}
\end{multline}
(Here, $\log(xyz):=\sigma\left(\log_{(\tpi)}(XYZ)\right)$ and
so on.)\ \ \ $\Box$

\noindent{\bf Proof of Proposition~\ref{PA221}.} Set $a:=dx\cdot x^{-1}$.
Then we have
\begin{multline}
d_x\omega_{x,y}^1=-\left([a,a]/2,{\log(xy)\over \ord x+\ord y}-{\log x\over
\ord x}\right)+\\
+\left(a,{d_x\log(xy)\over \ord x+\ord y}-{d_x\log x\over \ord x}\right),
\label{A226}
\end{multline}
where $\log(xy):=\sigma\left(\log_{(\tpi)}(XY)\right)$. We have
by Lemma~\ref{LB616}
\begin{equation}
d_x\log(xy)\!=\!\left(\ad(\log(xy))\left(\exp(\ad(\log(xy)))-1\right)^{-1}
\right)\!\circ\!\left(d_x(xy)\!\cdot\!(xy)^{-1}\right).
\label{A227}
\end{equation}
The term on the right is defined as $(F(\ad(\log(xy))))^{-1}$
for $F\left(\ad\left(\cL_A\right)\right)$ given by (\ref{B620}),
where $A=\exp(\cL_A)$
belongs to $\Ell_0^{\times}(M,E)$. Note that $d_x(xy)\cdot(xy)^{-1}=a$.
The series $z/(\exp z-1)$ on the right in (\ref{A227}) is of the form
\begin{gather}
\begin{split}
z/(\exp z-1)=1-z/2 & +\sum_{k\ge 1}c_{2k}z^{2k}, \\
c_{2k}=-{\zeta(1-2k)/(2k-1)!},\quad & c_{2k+1}={\zeta(-2k)/(2k)!}=0,
\end{split}
\label{A228}
\end{gather}
where $\zeta(s)$ is the zeta-function of Riemann.

Since $a$ is a one-form, we have for $k\in\zuo$
$$\left(a,\ad^{2k}(\log(xy))\circ a\right)_{\res}=\left(-1\right)^k\left(\ad^k
(\log(xy))\circ a,\ad^k(\log(xy))\circ a\right)_{\res}=0.
$$
In the second term on the right in (\ref{A226}) the term $-z/2$
in (\ref{A228}) (for $z=\ad(\log(xy))$ and for $z=\ad(\log(x))$)
correspond to
$$
-\left(a/2,{[\log(xy),a]\over \ord x+\ord y}-{[\log x,a]\over \ord x}\right)
_{\res}=\left([a,a]/2,{\log(xy)\over \ord x+\ord y}-{\log x\over \ord x}
\right)_{\res}.
$$
Hence $d_x\omega_{x,y}^1=0$. The equality $d_y\omega_{x,y}^1=0$ is proved
similarly.\ \ \ $\Box$

\begin{rem}
In (\ref{A220}) and in the proofs of Propositions~\ref{PA221}, \ref{PA222}
we do not use that $\sigma\left(\log_{(\tpi)}XY\right)$,
$\sigma\left(\log_{(\tpi)}X\right),\ldots$ are logarithmic symbols
with respect to the same cut or that they are defined by cuts close
to $L_{(\pi)}$.
We use here only that these expressions are some logarithmic symbols
for $\sigma(XY)$, $\sigma(X),\ldots$\ \ . This assertion makes sense
in the case of a formal Lie group corresponding to a Lie algebra $\frg$
in the situation of Remark~\ref{RB422}.
\label{RB3970}
\end{rem}

\subsection{Canonical trace and determinant Lie algebra}
\label{S6}

It is proved in Theorem~\ref{TB10} that the derivatives at zero
of the zeta-functions for elliptic PDOs of order one are the restriction
of the quadratic form $-T_2(cl+B_0)$ (defined by (\ref{B211}))
on the linear space $\{cl+B\}:=\log\Ell_0^{\times}(M,E)$ to the hyperplane
$c=1$. (Here, $l$ is a logarithm of an elliptic PDO $A$ of order one.)

In this section we deduce the structure of the determinant Lie algebra
$\frg(M,E)$ (corresponding to the Lie group $G(M,E)$ defined
by (\ref{B623})) from Theorem~\ref{TB10} and Proposition~\ref{PB5}.
Their statements
are consequences of the existence of the introduced in Section~\ref{SB}
canonical trace $\TR$ defined on PDOs of noninteger orders.

The text of this subsection can be considered as an alternative proof
of Theorem~\ref{TB570}.

First of all, as a Lie algebra, $\frg(M,E)$ is equal to the quotient
of $\fell(M,E)$ modulo the ideal $\ff_0=\{K|K\text{ is smoothing and }
\Tr K=0\}$. We claim that $\ff_0$ belongs to the kernel of the bilinear
form associate with $T_2$, (\ref{B211}), i.e.,
$$
T_2(x+f)=T_2(x) \quad\text{ for }f\in\ff_0,x\in\fell(M,E).
$$
In fact, recall that $T_2$ gives values of the zeta-regularized
determinants and in the proof of Proposition~\ref{PB3931} we established
the formula relating variations of determinants and traces
for deformations of PDOs by smoothing operators.
Hence $T_2$ induces an invariant bilinear form on $\frg(M,E)$.
It is easy to see that the image under the exponential map of the cone
of null-vectors $\{l|T_2(l)=0\}$ in $G(M,E)$ is exactly the section
$\tilde{d}_0(\sigma(\log A))$ (defined by (\ref{B3943})).

Algebraically, we have a situation studied in Section~\ref{SB}:

1) a Lie algebra $\sfrg':=\frg(M,E)$ endowed with an invariant scalar product
$\left(,\right)$ (obtained by the polarization from $T_2$),

2) a nonzero isotropic central element
$$
1\in\sfrg', \quad (1,1)=0,
$$

3) a nonzero homomorphism (order)
$$
r\colon\sfrg'\to\wC
$$
given by the formula $m(x)=(x,1)$.

\medskip
The quotient algebra $\sfrg'/\wC\cdot 1$ is equal to $S_{\log}(M,E)$.
The scalar product $\left(,\right)$ induces a scalar product
on the codimension one ideal $CS^0(M,E)$ invariant under the adjoint
action.
By Proposition~\ref{PB800} this scalar product coincides
(up to a nonzero constant factor)
with the pairing induced by the noncommutative residue.

Using Remark~\ref{Rmax2} we see that $\sfrg'$ is canonically isomorphic
to $\sfrg$ constructed in Section~\ref{SD}. Thus we proved the coincidence
of $\frg(M,E)$ and the canonical extension without variational formulas.

\section{Generalized spectral asymmetry and a global structure
of determinant Lie groups}
\label{S7}

The global structure of the determinant Lie group $G(M,E)$ (i.e.,
of the central $\wC^{\times}$-extension $F_0\backslash\Ell_0^{\times}(M,E)$
of the group of elliptic symbols $\SEll_0^{\times}(M,E)$) is defined
with the help of a certain kind of global spectral invariants generalizing
spectral asymmetry as follows.

The fundamental group $\pi_1\left(\SEll_0^0(M,E)\right)$ is spanned
by loops $\exp(2\pi itp)$, where $p$ is the symbol of a PDO-projector
of order zero and $0\le t\le 1$. Indeed, the fundamental group
of $\SEll_0^0(M,E)$ is the same as the fundamental group of the principal
symbols $\pi_1(\Aut\pi^*E)$, where $\pi\colon S^*M\to M$ is the natural
projection of the co-spherical bundle. For the vector bundle $1_N$ on $M$
it is proved in the proof of Lemma~\ref{LB1810} that
$\pi_1\left(\Aut\pi^*1_N\right)$ is spanned by the loops $\exp(2\pi ita)$,
$0\le t\le 1$, where $a\in\End\left(\pi^*1_N\right)$ is a projection
$a^2=a$. For any such $a$ there exists a zero order PDO-projector
$A\in CL^0\left(M,1_N\right)$ with the principal symbol $a$ (\cite{Wo3}).
The same assertions are also true for $E$ instead of $1_N$.
A closed one-parameter subgroup $\exp(tq)$, $0\le t\le 1$, of $CS^0(M,E)$
is of the form
\begin{equation}
q=2\pi i\sum m_jp_j,
\label{B832}
\end{equation}
where $m_j\in\wZ$ and $\{p_j\}$ is a finite set of pairwise commuting
zero order PDO-projectors from $CS^0(M,E)$
\begin{equation}
p_j^2=p_j, \qquad p_jp_k=p_kp_j.
\label{B833}
\end{equation}

For $A\in CL^0(M,E)$ the section%
\footnote{The operator $\exp(tA)$ is defined by the integral
$\int_{\Gamma_R}\exp t\lambda\cdot(A-\lambda)^{-1}d\lambda$, where
$\Gamma_R$ is defined as in the integral (\ref{B7}).}
$d_1(\exp(tA))$ gives us a trivialization of the $\wC^{\times}$-bundle
\begin{equation}
p\colon G(M,E)\to\SEll_0^{\times}(M,E)
\label{B837}
\end{equation}
over a curve $\sigma(\exp(tA))\subset\SEll_0^0(M,E)$.

Let $X$ be an elliptic PDO of a real positive order $d:=\ord X$.
Let $X$ be sufficiently close to a positive self-adjoint PDO.
Then its complex powers $X_{(\tpi)}^s$ are defined.
A generalized zeta-function
\begin{equation}
\zeta_{X,(\tpi)}(A;s):=\Tr\left(AX_{(\tpi)}^{-s}\right)
\label{B835}
\end{equation}
for $\Re s>\dim M/d$ has a meromorphic continuation to the whole
complex plane. Its singularities are simple poles at the points
of an arithmetic progression and its residue at zero is equal to
\begin{equation}
\Res_{s=0}\zeta_{X,(\tpi)}(A;s)=\res(\sigma(A))/d,
\label{B836}
\end{equation}
Remark~\ref{RB215}. Here $\res$ is the noncommutative residue \cite{Wo2},
\cite{Kas}.
For a PDO-projector $A=P\in CL^0(M,E)$ of order zero its noncommutative
residue is equal to zero \cite{Wo1}. (Hence $\zeta_{X,(\tpi)}(P;s)$
is nonsingular at zero.)

Such an operator $X\in\Ell_0^d(M,E)$ defines another trivialization
of the bundle (\ref{B837}) over the curve $\sigma(\exp(tA))$
in $\SEll_0^0(M,E)$. Namely
\begin{equation}
\exp\left(t\Pi_X\sigma(A)\right)\in p^{-1}(\sigma(\exp(tA))),
\label{B838}
\end{equation}
where $\Pi_X\sigma(A)$ is the inclusion of $\sigma(A)$ into $\sfrg_{(l_X)}$,
$l_X:=\sigma\left(\log_{(\tpi)}X\right)/d$, with respect to the splitting
(\ref{B622}). The element $\Pi_X\sigma(A)$ depends on the symbols
$\sigma(X)$ and $\sigma(A)$ only.
The Lie algebra $\sfrg_{(l_X)}$ is canonically identified with
the Lie algebra $\frg(M,E)$ of $G(M,E)$ by Theorem~\ref{TB570}.
Under this identification, the quadratic $\wC^{\times}$-cone%
\footnote{The partially defined section $S\to d_0(S):=\tilde{S}$
of the $\wC^{\times}$-fibration (\ref{B837}) is defined by (\ref{B626}).}
$\log\tilde{S}\subset\frg(M,E)$ corresponds to the zero-cone $C_{l_X}$
for the quadratic form $A_{l_X}$ given by (\ref{B418}). Quadratic forms
$A_l$ and cones $C_l$ are invariant under identifications $W_{l_1l_2}$
by Proposition~\ref{PB417} and Corollary~\ref{CB421}. (Note that
$\exp\left(t\Pi_XA\right)$ belongs to a Lie group $\exp\left(\sfrg\right)$,
where $\sfrg$ is the Lie algebra defined by identifications $W_{l_1l_2}$
of $\sfrg_{(l)}$.) The Lie group $\exp\left(\sfrg\right)$ is canonically
local isomorphic to $G(M,E)$. For $t\in\wC$ with $|t|$ small enough,
we denote by $\exp\left(t\Pi_XA\right)$ an element of $G(M,E)$
corresponding to the element of $\exp\left(\sfrg\right)$ defined
by this expression.

Hence we have two trivializations $d_1(\exp(tA))$ and
$\exp\left(t\Pi_X\sigma(A)\right)$ of the $\wC^{\times}$-bundle (\ref{B837})
restricted to $\sigma(\exp(tA))$ for $t\in\wC$ with $|t|$ small enough.

\begin{pro}
The equality holds for such $A$, $X$, and $t$
\begin{gather}
d_1(\exp(tA))\big/\exp\left(t\left(\Pi_X\sigma(A)\right)\right)=\exp(tf(A,X)),
\label{B841}\\
f(A,X):=\left(\zeta_{X,(\tpi)}(A;s)-\res(\sigma(A))/sd\right)\big|_{s=0}.
\label{B842}
\end{gather}
Here, $d:=\ord X\in\wR_+$.
\label{PB840}
\end{pro}

Note that $f(A,X)$ is a spectral invariant of a pair $(A,X)$ of PDOs,
where $A\in CL^0(M,E)$ and where $X\in\Ell_0^d(M,E)$ is sufficiently
close to a self-adjoint positive definite PDO.

\begin{rem}
For a PDO-projector $P$ of zero order, $P\in CL^0(M,E)$, we have
\begin{equation}
f(P,X)=\zeta_{X,(\tpi)}(P;0).
\label{B844}
\end{equation}
\label{RB843}
\end{rem}

\begin{lem}
For a PDO-projector $P$ of zero order, the spectral invariant of the pair
$(P,X)$ with its values in $\wC/\wZ$
\begin{equation}
f_0(P,X):=f(P,X)(\mod\wZ)
\label{B846}
\end{equation}
depends on the symbols $\sigma(P)$ and $\sigma(X)$ only.
\label{LB845}
\end{lem}

\begin{rem}
For a general PDO-projector $P$ of zero order, $f_0(P,X)$ cannot be
universal expressed as an integral over $M$ of a density locally defined
by homogeneous components of symbols $\sigma(P)$ and $\sigma(X)$
in local coordinate charts on $M$.
\label{RB847}
\end{rem}

\noindent{\bf Definition.} A {\em generalized spectral asymmetry} of a pair
$(P,X)$ of PDOs is defined as
\begin{equation}
f_0(\sigma(P),\sigma(X))=\Tr\left(PX_{(\tpi)}^{-s}\right)\big|_{s=0}(\mod\wZ).
\label{B849}
\end{equation}
Here, $P\in CL^0(M,E)$ is a PDO-projector of order zero and
$X\in\Ell_0^d(M,E)$ (with $d\in\wR^{\times}$) is sufficiently close
to a self-adjoint positive definite PDO.

\begin{rem}
Let $X\in\Ell_0^d(M,E)$ with $d\in\wR_+$ be self-adjoint.%
\footnote{For the sake of simplicity we suppose here that $X$ has
no zero eigenvalues.}
Let $P:=(X+|X|)/2|X|$ be the PDO-projector to the subspace spanned
by eigenvectors of $P$ with positive eigenvalues.
(Here, $|X|:=\left(X^2\right)_{(\pi)}^{1/2}$.) The spectral asymmetry
of $X$ (\cite{APS1}--\cite{APS3}) is defined as the value at $s=0$
of the analytic continuation from $\Re s>\dim M/d$ of
$$
\eta_X(s):=\sum\sign\lambda\cdot\left|\lambda\right|^{-s},
$$
where the sum is over the eigenvalues of $X$ including their multiplicities.
The spectral asymmetry of $X$ is connected with $f(P,X)$ as follows
\begin{gather}
\begin{split}
\eta_X(s) & =\Tr\left(P(X^2)_{(\pi)}^{-s/2}\right)-\Tr\left((1-P)(X^2)_{(\pi)}
^{-s/2}\right)\\
          & =2\Tr\left(P(X^2)_{(\pi)}^{-s/2}\right)-\Tr\left((X^2)_{(\pi)}
^{-s/2}\right),\\
\eta_X(0) & =2f(P,X^2)-\zeta_{X^2,(\pi)}(0),\\
f_0\left(\sigma(P),\sigma(X^2)\right) & =\left(\zeta_{X^2,(\pi)}(0)+\eta_X(0)
\right)/2 (\mod\wZ).
\end{split}
\label{B851}
\end{gather}
This example explains the name of the invariant
$f_0\left(\sigma(P),\sigma(X^2)\right)$.
\label{RB850}
\end{rem}

\begin{rem}
For $A=2\pi iP$, where $P\in CL^0(M,E)$ is a PDO-projector of zero order,
we have
\begin{equation}
d_1(\exp A)=\Id\in G(M,E), \quad \exp\left(2\pi i\Pi_X\sigma(P)\right)=
\exp\left(-2\pi if_0(P,X)\right).
\label{B1401}
\end{equation}
Hence the invariant $f_0(P,X)$ ($=f(\sigma(P),\sigma(X))$
by Lemma~\ref{LB845}) defines the element
$\exp\left(2\pi i\Pi_X\sigma(P)\right)\in\wC^{\times}\cdot 1=p^{-1}(\Id)$
(where $1\to\wC^{\times}\to G(M,E) @>>p>\SEll_0^{\times}(M,E)\to 1$ is
the central extension). Hence $f_0(P,X)$ defines the structure
of the subgroup $p^{-1}(\exp(2\pi it\sigma(P)))\subset G(M,E)$ over
a one-parametric closed subgroup $\exp(2\pi it\sigma(P))$ in the base
$\SEll_0^{\times}(M,E)$ of this central extension. Invariants $f_0(P,X)$
define the group structure of this central extension over any one-parametric
closed subgroup in $\SEll_0^{\times}(M,E)$. Suppose that we can compute
invariants $f_0(P,X)$. Then we know the Lie algebra $\frg(M,E)$
of the group $G(M,E)$ (canonically isomorphic to the Lie algebra $\sfrg_{(l)}$
by Theorem~\ref{TB570}), the group structure of $\SEll_0^{\times}(M,E)$,
and the group structure of $G(M,E)$ over closed one-parametric subgroups
in $\SEll_0^{\times}(M,E)$. These data define the global structure
of the determinant Lie group $G(M,E)$. Hence the problem of the algebraic
definition of $G(M,E)$ reduces to the problem of computing the invariants
$f_0(P,X)\in\wC/\wZ$.
\label{RB1400}
\end{rem}

\begin{rem}
The element $\Pi_X\sigma(P)\in\sfrg_{(l_X)}$ is the element
$\sigma(P)+0\cdot 1$ with respect to the splitting (\ref{B624}).
Under the identification $W_{l_Xl_Y}$ of Proposition~\ref{PB403}
(where $l_X:=\sigma\left(\log_{(\tpi)}X\right)/\ord X$ and $l_Y$
is analogous), this element transforms to
\begin{equation}
\Pi_Y\sigma(P)+\left(l_X-l_Y,\sigma(P)\right)_{\res}\cdot 1\in\sfrg_{(l_Y)}
\label{B1403}
\end{equation}
with respect to the splitting (\ref{B624}) for $\sfrg_{(l_Y)}$.
Hence we have
\begin{equation}
\exp\left(2\pi i\Pi_X\sigma(P)\right)/\exp\left(2\pi i\Pi_Y\sigma(P)\right)=
\exp\left(2\pi i\left(l_X-l_Y,\sigma(P)\right)_{\res}\right).
\label{B1404}
\end{equation}
The term on the right $\left(l_X-l_Y,\sigma(P)\right)_{\res}$ is
the integral over $M$ of the density locally defined by symbols
$\sigma(\log X)$, $\sigma(\log Y)$, and $\sigma(P)$.
We have
$$
d\log\left(\exp\left(2\pi i\sigma_X(P)\right)\right)=2\pi i\left(dl_X,
\sigma(P)\right)_{\res},
$$
where $dl_X:=d\left(\sigma(\log X)/\ord X\right)$ is an exact one-form
on the complement to the hyperplane $CS^0(M,E)$ in the Lie algebra
$S_{\log}(M,E)$ of logarithms of elliptic symbols (and $CS^0(M,E)$ is
its Lie subalgebra corresponding to the symbols of order zero,
Section~\ref{SD}).
Hence according to (\ref{B1401})
it is enough to compute $f_0(P,X)$ for an elliptic operator
$X\in\Ell_0^d(M,E)$ with $d\in\wR^{\times}$ such that $\log_{(\tpi)}X$
exists.
\label{RB1402}
\end{rem}

\begin{rem}
A variation $\delta P=:L$ of a zero order PDO-projector $P\in CL^0(M,E)$
(i.e., $P_1=P+\eps L+O(\eps^2)$, $\eps\to 0$, is a family of zero order
PDO-projectors) is connected with $P$ by the equations
\begin{equation}
LP=(1-P)L, \qquad L(1-P)=PL.
\label{B2991}
\end{equation}
Hence $L$ maps $\Im P$ into $\Im(1-P)$ and $\Im(1-P)$ into $\Im P$.
Any $L$ of the form
\begin{equation}
L:=[P,Y]
\label{B2992}
\end{equation}
with $Y\in CL^0(M,E)$ gives us a solution of (\ref{B2991}).
(Note also that $\res[P,Y]=0$.) For a family of PDO-projectors of zero
order we have
\begin{equation}
\delta P=[[\delta P,P],P].
\label{B3151}
\end{equation}
Hence the equality (\ref{B2992}) holds with $Y=[\delta P,P]\in CL^0(M,E)$.
For $L$ of the type (\ref{B2992}) we have
\begin{gather*}
\delta_P\log\left(\exp\left(2\pi i\Pi_X\sigma(P)\right)\right)=2\pi i\delta_P
f_0(P,X), \\
f_0(P,X)=f_0\left(APA^{-1},AXA^{-1}\right)\in\wC/\wZ
\end{gather*}
for any $A\in\Ell^\alpha(M,E)$.
Hence for $A_\eps=\exp(\eps Y)\in\Ell_0^0(M,E)$, $\eps\to 0$, we have
\begin{multline}
\delta_P\log\left(\exp\left(2\pi i\Pi_X\sigma(P)\right)\right)=\delta_X\log
\left(\exp\left(2\pi i\Pi_X\sigma(P)\right)\right)\big|_{\delta X=[Y,X]}= \\
=2\pi i\left(\delta\sigma(\log X)/\ord X,\sigma(P)\right)_{\res}\big|
_{\delta X=[Y,X]}.
\label{B2993}
\end{multline}
(Here, $\delta\ord X=0$.) By Lemma~\ref{LB616} we have
\begin{equation}
\delta\sigma(\log X)\big|_{\delta X=[Y,X]}=\left(\ad(\sigma(\log X))\left(
\exp(\ad (\sigma(\log X)))-1\right)^{-1}\right)\circ\left([Y,X]\cdot X^{-1}
\right).
\label{B2994}
\end{equation}
Hence the variation of $f_0(P,X)$ in a smooth family of zero order
PDO-projectors can be transformed to the variation of an elliptic
operator $X$ given by (\ref{B2993}), (\ref{B2994}).
\label{RB2990}
\end{rem}

\begin{rem}
The generalized spectral asymmetry $f_0(P,X)\in\wC/\wZ$ is independent
of a zero order PDO-projector $P_t$ in a smooth family of such projectors,
if variations $\delta_tP_t$ in this family are PDOs
from $CL^{-\dim M-1}(M,E)$. This assertion follows from (\ref{B3151}),
(\ref{B2993}), and (\ref{B2994}) since
$$
\ord\left(\delta\sigma(\log X)|_{\delta X=[Y,X],Y=[\delta P,P]}\right)\le
\ord Y\le\ord(\delta P)
$$
and since the noncommutative residue $\res(a)$ for a symbol
$a\in CS^{-\dim M-1}(M,E)$ is zero.

The analogous assertion is valid for smooth families of bounded
projectors in a separable Hilbert space. Namely, let $P_t$ be such
a family and let $\delta_t P$ be from trace classes. Then the formula
(\ref{B3151}) for $\delta_tP$ holds. So
\begin{equation}
\Tr\left(\delta_tP_t\right)=\Tr\left(\left[\left[\delta_tP,P_t\right],P_t
\right]\right)=0
\label{B3152}
\end{equation}
because $\left[\delta_tP,P\right]$ is a trace class operator and $P_t$
is bounded. Hence $P_{t_1}-P_{t_2}$ is a trace class operator and
\begin{equation}
\Tr\left(P_{t_1}-P_{t_2}\right)=0.
\label{B3153}
\end{equation}
for any projectors from this family.
\label{RB3150}
\end{rem}

\noindent{\bf Problem.} To compute the generalized spectral asymmetry
invariants $f_0(P,X)\in\wC/\wZ$ in algebraic terms (i.e., without
using the analytic continuation and the Fredholm determinants). \\

\noindent{\bf Proof of Proposition~\ref{PB840}.} We have
\begin{multline}
d_1(\exp(tA))\big/\exp\left(t\Pi_X\sigma(A)\right)=d_1(\exp(tA)\cdot X)d_1(X)
^{-1}\exp\left(-t\Pi_X\sigma(A)\right)=\\
=d_0\!(\exp(\!tA\!)\!\cdot\!X\!)d_0(X)^{-1}\!\exp\!\left(\!-\!t\Pi_X\sigma
(\!A\!)\!\right)\!{\det}_{(\pi)}(\!\exp(\!tA\!)\!\cdot\!X)\!\left(\!{\det}
_{(\pi)}(\!X\!)\!\right)^{-1}.
\label{B852}
\end{multline}
Here, $d_0(S)=:\tilde{S}$ is defined by (\ref{B626}) with $\theta=\pi$
for $S$ sufficiently close to a positive definite self-adjoint PDO of
a nonzero real order. The parameter $t\in\wC$ in (\ref{B852}) is such that
$|t|$ is small enough. In this case, the PDO $\exp(tA)\cdot X$
is sufficiently close to a positive definite self-adjoint PDO.
For the scalar factor on the right in (\ref{B852}), we have the equality
analogous to (\ref{B6})
\begin{multline}
\df_t\log\left({\det}_{(\pi)}(\exp(tA)\cdot X)\big/{\det}_{(\pi)}(X)\right)=\\
=-\df_s\left(-s\Tr\left(A\cdot X_t^{-s}-{\res\sigma(A)\over s\ord X}\right)
\right)\big|_{s=0}=:f(A,X_t)
\label{B853}
\end{multline}
(where $X_t:=\exp(tA)\cdot X$) because $\df_t X_t\cdot X_t^{-1}=A$.
On the right in (\ref{B853}) we take the restriction at $s=0$
of an analytic continuation (for the trace) from $\Re s>\dim M/\ord X$.
The expression on the right in (\ref{B853}) is regular at $s=0$ according
to (\ref{B836}). (Note also that $\df_s(s(\res\sigma(A)/s\ord X))\equiv0$
and it is used in (\ref{B853}).)

The nonscalar factor on the right in (\ref{B852})
\begin{equation}
K_t:=d_0\left(X_t\right)d_0\left(X\right)^{-1}\exp\left(-t\Pi_X\sigma(A)
\right)
\label{B854}
\end{equation}
belongs to the connected component $\wC^{\times}$ of the central subgroup
in the group $\exp\left(\sfrg\right)$. Hence $\log K_t\in\wC$ is defined.
We have to compute $\df_t\log K_t$ for $t\in\wC$ with $|t|$ small enough.
To do this, note first that under the canonical identification of the Lie
algebras $\sfrg$ and $\frg(M,E)$ (given by Theorem~\ref{TB570})
the invariant quadratic $\wC^{\times}$-cone $\log\tilde{S}\subset\frg(M,E)$
corresponds to a $\wC^{\times}$-cone $\log\sX$ in $\sfrg$, where
\begin{equation}
\sX=\exp\left(\Pi_X\sigma\left(\log_{(\tpi)}X\right)\right)
\label{B8541}
\end{equation}
for a PDO $X$ of a nonzero real order sufficiently close to a self-adjoint
positive definite PDO. Hence
\begin{equation}
K_t=\sX_t\left(\sX\right)^{-1}\exp\left(-t\Pi_X\sigma(A)\right),
\label{B855}
\end{equation}
where $\sX_t:=\exp\left(\Pi_X\sigma\left(\log_{(\tpi)}X_t\right)\right)$
and $|t|$ is small enough. We have
\begin{multline}
\df_t\log K_t=\df_t K_t\cdot K_t^{-1}=-K_t\cdot (\Pi_X\sigma(A))\cdot K_t^{-1}+
\df_t\sX_t\cdot\sX_t^{-1}=\\
=-\Pi_X\sigma(A)+\df_t\sX_t\cdot\sX_t^{-1}.
\label{B856}
\end{multline}

According to (\ref{B679}) we have
$$
\df_t\sX_t\cdot\sX_t^{-1}=\Pi_{X_t}\left(\df_t X_t\cdot X_t^{-1}\right)=
\Pi_{X_t}\sigma(A).
$$
By Lemma~\ref{LB581} and by (\ref{B856}), we have
\begin{gather}
\begin{split}
\Pi_{X_t}\sigma(A) & =\Pi_X\sigma(A)+\left(\sigma(A),l_{X_t}-l_X\right)_{\res}
\cdot 1\in\sfrg, \\
\df_t\log K_t      & =\df_t\sX_t\cdot\sX_t^{-1}-\Pi_X\sigma(A)=\left(\sigma
(A),l_{X_t}-l_X\right)_{\res}\cdot 1,
\end{split}
\label{B857}
\end{gather}
where $l_X:=\sigma\left(\log_{(\tpi)}X\right)/\ord X$ (and the same
is true for $l_{X_t}$). Hence
\begin{equation}
\log\left(d_1(\exp(tA))\big/\exp\left(t\Pi_X\sigma(A)\right)\right)=f\left(A,
X_t\right)+\left(\sigma(A),l_{X_t}-l_X\right)_{\res}.
\label{B860}
\end{equation}

By Proposition~\ref{PD3} we have
\begin{equation}
f\left(A,X_t\right)-f(A,X)=-\left(\sigma(A),\sigma\left(\log X_t\right)/\ord X-
\sigma(\log X)/\ord X\right)_{\res}.
\label{B861}
\end{equation}

Proposition~\ref{PB840} follows from (\ref{B860}), (\ref{B861}), and
from (\ref{B857}).\ \ \ $\Box$ \\

\noindent{\bf Proof of Lemma~\ref{LB845}.} Let $X_1$ be a PDO of a real
nonzero order sufficiently close to a positive definite self-adjoint PDO.
Then by Proposition~\ref{PD3} we have
\begin{equation}
f\left(A,X_1\right)-f(A,X)=-\left(\sigma(A),{\sigma\left(\log_{(\tpi)}X_1
\right)\over \ord X_1}-{\sigma\left(\log_{(\tpi)}X\right)\over \ord X}\right).
\label{B866}
\end{equation}

In particular, for $\sigma(X)=\sigma(X_1)$ we have
$f(A,X)=f\left(A,X_1\right)$. It is true even more strong statement.
Namely, if $X_1-X\in CL^{\ord X-\dim M-1}(M,E)$, then the term
on the right in (\ref{B866}) is equal to zero because
$\sigma(A)\in CL^0(M,E)$ and because under this condition,
$$
\sigma\left(\log_{(\tpi)}X_1\right)-\sigma\left(\log_{(\tpi)}X\right)\in CS
^{-\dim M-1}(M,E).
$$
Hence the dependence $f(A,X)$ on $X$ can be expressed with the help
of its dependence on the image $\sigma(X)$ in
$CS^{\ord X}(M,E)/CS^{\ord X-\dim M-1}(M,E)$.

Let $P_1$ and $P$ be PDO-projectors belonging to $CL^0(M,E)$ such that
$$
P_1-P\in CL^{-\dim M-1}(M,E).
$$
Then $(P_1-P)X^{-s}$ for $\Re s>-1/\ord X$ is a trace class operator.
We have
\begin{equation}
f\left(P_1,X\right)-f(P,X)=\Tr\left(P_1-P\right).
\label{B864}
\end{equation}

The assertion $\Tr\left(P_1-P\right)\in\wZ$ immediately follows
from Proposition~\ref{PB871} below.\ \ \ $\Box$

\subsection{PDO-projectors and a relative index}

\begin{pro}
1. Let $P_1$ and $P_2$ be PDO-projectors from $CL^0(M,E)$ such that
$P_1-P_2\in CL^{-\dim M-1}(M,E)$. Consider the operator
$\tilde{P}_2:=P_2|_{\Im P_1}$,
\begin{equation}
\tilde{P}_2\colon\Im P_1\to\Im P_2.
\label{B872}
\end{equation}
Then $\Ker\tilde{P}_2$ and $\Coker\tilde{P}_2$ are finite-dimentional.
For the index of $\tilde{P}_2$ the equality holds
\begin{equation}
\ind\tilde{P}_2=\Tr\left(P_1-P_2\right).
\label{B873}
\end{equation}

2. The same equality holds for a pair $P_1$, $P_2$ of (bounded)
projectors acting in a separable Hilbert space $H$ and such that
$P_1-P_2$ is of trace class. Namely
\begin{equation}
\ind\tilde{P}_2=\Tr\left(P_1-P_2\right)=-\ind\tilde{P}_1.
\label{B3223}
\end{equation}
\label{PB871}
\end{pro}

\begin{cor}
Under the conditions of Proposition~\ref{PB871}, we have
\begin{equation}
\Tr\left(P_1-P_2\right)\in\wZ.
\label{B875}
\end{equation}
\label{CB874}
\end{cor}

\noindent{\bf Proof of Proposition~\ref{PB871}.}
Set $P_1-P_2=:S$.

1. The operator $\tilde{P}_2:=P_2|_{\Im P_1}$ has a finite-dimensional
kernel because
\begin{equation}
P_2=P_1-S,
\label{B890}
\end{equation}
$P_1=\Id$ on $\Im P_1\subset L_2(M,E)$, and
$S\colon L_2(M,E)\to H_{(-\dim M-1)}(M,E)\hookrightarrow L_2(M,E)$
is a compact operator. (Here, $H_{(s)}$ is the Sobolev space.) Hence
the space of solutions for the equation
$Se=e$, $e\in L_2(M,E)$, is finite-dimensional. \\

2. The operator $\tilde{P}_2$ has a finite-dimensional cokernel
because from (\ref{B890}) we have
\begin{equation}
P_2m=P_2P_1m-P_2Sm.
\label{B891}
\end{equation}
The operator $K=P_2S|_{\Im P_2}\colon\Im P_2\to\Im P_2$ is a compact
operator on the Hilbert space $L=\Im P_2$. ($L$  is a closed subspace
of $L_2(M,E)$ because $P_2^2=P_2$ and because $P_2\in CL^0(M,E)$ is
a bounded linear operator on $L_2(M,E)$.) For $m\in L$ we have
$m=m_1+Km$,
where $m_1:=P_1m$. Let the operator $P_2|_{\Im P_1}$ have
an infinite-dimensional cokernel. (The operator
$P_2|_{\Im P_1}\colon\Im P_1\to\Im P_2$ is closed since it is
the restriction of the closed operator $P_2\colon L_2(M,E)\to L_2(M,E)$
to a Hilbert subspace $\Im P_1\subset L_2(M,E)$.)
Then the space of $m\in L$ such that $\|Km\|>\|m\|/2$ (with respect
to the scalar product $\left\|x\right\|^2:=(x,x)$ in $L_2(M,E)$) is
infinite-dimensional. Hence $\codim\tilde{P}_2<\infty$. \\

3. Note that $\Tr\left(P_1-P_2\right)$ depends (if $P_1-P_2$ is a trace
class operator) on the images $\Im P_j\subset L_2(M,E)$ only.
The equivalent assertion is the following.

Let $P$ and $P_1$ be bounded projectors with $\Im P_1=\Im P$. Then
\begin{equation}
\Tr\left(P-P_1\right)=0.
\label{B3155}
\end{equation}

Let $H_1:=\Im P_1$, $H_2:=\Ker P_1$, and let $L_2(M,E):=H=H_1\oplus H_2$
be the direct sum decomposition. Then the projector $P$ is conjugate
to $P_1$, i.e., $P=gP_1g^{-1}$ with
\begin{equation*}
g=\left(
\begin{split}
\Id_{H_1} & \ L \\
0         & \ \Id_{H_2}
\end{split}
\right) \qquad g^{-1}=\left(
\begin{split}
\Id_{H_1} & \ -L \\
0         & \ \Id_{H_2}
\end{split}
\right),
\end{equation*}
where $L$ is of trace class and
$\left(\Id_{H_2}+L\right)\colon H_2\rs\Ker P$. For a family of bounded
projectors $P(t)$,
\begin{equation*}
P(t):=g_tP_1g_t^{-1}, \quad g_t:=\left(
\begin{split}
\Id_{H_1} & \ tL \\
0         & \ \Id_{H_2}
\end{split}
\right),
\end{equation*}
we have
\begin{equation*}
\df_tP(t)=\left[\left(
\begin{split}
0 & \ L \\
0 & \ 0
\end{split}
\right),P(t)\right].
\end{equation*}
Here, $L$ is a trace class operator. So $\df_tP(t)$ is of trace class.
By Remark~\ref{RB3150}, (\ref{B3153}), we conclude that
$\Tr\left(P\left(t_1\right)-P\left(t_2\right)\right)=0$.
The equality (\ref{B3155}) is proved since $P=:P(1)$, $P_1=:P(0)$.

4. The decomposition of $L_2(M,E)$ in the direct sum
of $H_1\left(P_2\right):=\Im P_2$ and $H_2\left(P_2\right):=\Ker P_2$
can be produced (\cite{SW}, \S~3) by the action of an invertible operator
$g$ in $H$ written in a block form with respect to the decomposition
$H=H_1\oplus H_2$ with $H_j=H_j\left(P_1\right)$
\begin{equation}
\left(
\begin{split}
a & \ b \\
c & \ d
\end{split}
\right)\left(
\begin{split}
H_1 \\
H_2
\end{split}
\right)=\left(H_1\left(P_2\right),H_2\left(P_2\right) \right),
\label{B3157}
\end{equation}
where operators $b$ and $c$ are of trace class and $P_2:=gP_1g^{-1}$.
Here, the operators $a$ and $d$ are automatically Fredholm and the index
of $a\colon H_1\to H_1$ is well-defined. We have
\begin{equation}
\Ind\tilde{P}_1=\ind a,
\label{B3158}
\end{equation}
where $\tilde{P}_1\colon\Im P_2\to\Im P_1$ is $P_1|_{\Im P_2}$.
We have the analogous equality for $\ind\tilde{P}_2$,
\begin{equation}
\ind\tilde{P}_2=\ind\alpha,
\label{B3160}
\end{equation}
where
\begin{equation}
g^{-1}:=\left(
\begin{split}
\alpha & \ \beta \\
\gamma & \ \delta
\end{split}
\right), \quad g^{-1}\left(
\begin{split}
H_1\left(P_2\right) \\
H_2\left(P_2\right)
\end{split}
\right)=\left(H_1,H_2\right).
\label{B3163}
\end{equation}
(Here, the operator $g^{-1}\colon H\to H$ is written in the block form
with respect to the decomposition
$H=H_1\left(P_2\right)\oplus H_2\left(P_2\right)$.)

\begin{lem}
Let $g\colon H\to H$ be an invertible operator in a separable Hilbert
space $H$ under the same conditions as in (\ref{B3157}). Then
the equality holds
\begin{equation}
\ind a+\ind\alpha=0.
\label{B3221}
\end{equation}
(Here, $a$ and $\alpha$ are defined by (\ref{B3157}) and by (\ref{B3163}).)
\label{LB3220}
\end{lem}

This lemma is proved in the end of this section.

\begin{rem}
By (\ref{B3160}), (\ref{B3158}), (\ref{B3221}) we have
\begin{equation}
\ind\tilde{P}_2=-\ind\tilde{P}_1.
\label{B3162}
\end{equation}
By (\ref{B861}) the index $\ind\tilde{P}_2$ depends on $P_2$ and
on $\Im P_1$ only. The analogous statement is true for $\ind\tilde{P}_1$.
So by (\ref{B3162}) $\ind\tilde{P}_2$ depends on $\Im P_1$, $\Im P_2$
only. Hence the both sides of (\ref{B866}) depend on $\Im P_1$ and
on $\Im P_2$ only. (Here, we suppose that $P_1-P_2$ is a trace class
operator.)
\label{RB3222}
\end{rem}

Let us continue our proof of Proposition~\ref{PB871}.

5. We can suppose that $\ind\tilde{P}_1=0=\ind\tilde{P}_2$. Indeed,
let $\ind\tilde{P}_2=m\in\wZ_-$ (i.e., $\ind a=-m\in\wZ_+$). Then
there is a (bounded) projector $P_1^n$ in $L_2(M,E)$ such that
$\Im P_1^n\supset\Im P_1$
\begin{gather}
\Tr\left(P_1^n-P_1\right)=-m,
\label{B3210}\\
\ind P_2|_{\Im P_1^n}=-m+\ind\tilde{P}_2.
\label{B3211}
\end{gather}
(In particular, $P_1^n-P_2$ is a trace class operator.) It follows
from (\ref{B3210}) that
$$
\Tr\left(P_1^n-P_2\right)=-m+\Tr\left(P_1-P_2\right), \quad \ind\tilde{P}_1^n
=0.
$$
To produce such a projector $P_1^n$, it is enough to take a finite rank
projector $p$ in $L_2(M,E)$ such that
$$\rk p=-m, \quad \Im p\subset\Ker P_1, \quad \Im P_1\subset\Ker p.
$$
Then $P_1^n:=P_1+p$ is a (bounded) projector in $L_2(M,E)$ satisfying
(\ref{B3210}). The equality (\ref{B3211}) holds for $P_1^n$ since
$\Im P\subset\Im P_1^n$ is a closed subspace in $\Im P_1^n$
of codimension $m$.

6. Let $\ind\tilde{P}_1=0=\ind\tilde{P}_2$. The numbers
$\Tr\left(P_1-P_2\right)$ and $\ind\tilde{P}_j$ depend on $\Im P_1$
and on $\Im P_2$ only. The operators $a$ and $d$ in the transformation
$g$ (\ref{B3157}) are of the form (since $\ind a=0=\ind d$)
$$
a=\left(\Id_{H_1}+l_1\right)q_a,d=\left(\Id_{H_2}+l_2\right)q_\alpha,
$$
where $q_a$, $q_\alpha$ are invertible operators in $H_1$, $H_2$ and
$l_j$ are trace class operators in $H_j$. Transformations
$$
q_a\colon H_1\to H_1, \quad q_d\colon H_2\to H_2, \quad c\to cq_a^{-1},
\quad b\to bq_\alpha^{-1}
$$
do not change $H_j$ and $H_j\left(P_2\right)$. Hence we can suppose
(in the case $\ind\tilde{P}_1=0=\ind\tilde{P}_2$) that the operator $g$
in (\ref{B3157}) has a block form (with respect to $H=H_1\oplus H_2$)
where $a-\Id_{H_1}$, $d-\Id_{H_2}$, $b$, $c$ are trace class operators.
So the following lemma gives us a proof of Proposition~\ref{PB871}.

\begin{lem}
Let $P$ be a (bounded) projector in a separable Hilbert space $H$
with infinite-dimensional $\Ker P:=H_2$ and with $\Im P:=H_1$
($H=H_1\oplus H_2$). Let
$g=\left(
\begin{matrix}
a & b \\
c & d
\end{matrix}
\right)$ be a bounded linear operator in $H$
written in a block form with respect to the decomposition $H_1\oplus H_2$
and such that $a-\Id_{H_1}$, $d-\Id_{H_2}$, $b$, and $c$ are trace class
operators. Then $S:=P-gPg^{-1}$ is a trace class operator in $H$ and
$\Tr S=0$.
\label{LB3212}
\end{lem}

\noindent{\bf Proof.} 1. Let $G$ be the group of invertible operators $g$
in $H$ where $a-\Id_{H_1}$, $d-\Id_{H_2}$, $b$, and $c$ are of trace
class. Then the equality $\Tr S=0$ follows from the assertion that $G$
is connected. Indeed, in this case, for any $g\in G$ there is a smooth
curve $g(t)$ in $G$ from $\Id\in G$ to $g=g(1)$. Then
$\dot{g}(t)\equiv\df_tg(t)$ is of trace class in $H$ and
for $P_t:=g(t)Pg(t)^{-1}$ we see that
$$
\dot{P}_t=\left[\dot{g}(t),P_t\right]
$$
is of trace class. Hence by Remark~\ref{RB3150}, (\ref{B3152}), we have
$$
\Tr\dot{P}+t=0, \qquad \Tr\left(P-gPg^{-1}\right)=0,
$$
because $gPg^{-1}=:P_1$.

2. For any $g\in G$ set $g-\Id=:A=A(g)$. Then $A$ is of trace class.
So $A$ is compact and for any nonzero eigenvalue $\lambda$ of $A$
the corresponding algebraic eigenspace $L_{\lambda}=L_{\lambda}(A)$
is finite-dimensional.

Set $L:=\oplus L_{\lambda}$ over $A$-eigenvalues $\lambda$ with
$|\lambda|\ge\eps$. Then $L$ is a finite-dimensional invariant subspace
with respect to $A$. Let $Q$ be $A$-invariant subspace complementary
to $L$, $H=L\oplus Q$. Then $Q$ is a separable Hilbert space
(with the induced Hilbert norm) and $A=A_L\oplus A_Q$ with respect
to $L\oplus Q$. The group $GL(L)=\Aut_{\wC}L$ is connected.
Let $g_L(t)$ be a smooth curve in $\Aut_{\wC}L$ from $\Id_L$
to $\Id_L+A_L$. The operator norm of $A_Q$ in $Q$ is less than $1/2$
(for $\eps$ small enough). Then
$g(t):=g_L(t)\oplus\left(\Id_Q+tA_Q\right)$ for $0\le t\le 1$ is
a smooth curve in $G$ from $\Id_H$ to $g=\Id_H+A$ (written with respect
to $H=L\oplus Q$). Indeed, $\Id_Q+tA_Q$, $|t|\le 1$, is invertible
in $Q$ since the $L_2$-operator norm of $A_Q$, $\|A_Q\|_2$, is less
than $1/2$.
For the trace norm of $\left(\Id_Q+tA_Q\right)^{-1}-\Id_Q=:B_Q(t)$
the estimate holds (for $|t|\le 1$)
\begin{multline*}
\|B_Q(t)\|_{\tr}\le\|tA_Q\|_{\tr}+\left\|\left(tA_Q\right)^2\right\|_{\tr}+
\ldots+\left\|\left(tA_Q\right)^n\right\|_{\tr}+\ldots\le \\
\le\|tA_Q\|_{\tr}\left(1+\|tA_Q\|_2+\ldots+\left(\|tA_Q\|_2\right)^{n-1}+
\ldots\right)\le 2\|A_Q\|_{\tr}.
\end{multline*}
So $g(t)\in G$. Hence the group $G$ is connected. The lemma
is proved.\ \ \ $\Box$

\noindent{\bf Proof of Lemma~\ref{LB3220}.} By (\ref{B3158}) and
(\ref{B3160}) we have
\begin{equation}
\ind a+\ind\alpha=\ind\tilde{P}_1+\ind\tilde{P}_2,
\label{B3225}
\end{equation}
where $\tilde{P}_1=P_1|_{\Im P_2}\colon\Im P_2\to\Im P_1$ and
$\tilde{P}_2\colon=P_2|_{\Im P_1}$. Operators $\tilde{P}_1$, $\tilde{P}_2$,
and $\tilde{P}_1\tilde{P}_2\colon\Im P_1\to\Im P_1$ are Fredholm. So
\begin{equation}
\ind\tilde{P}_1\tilde{P}_2=\ind\tilde{P}_1+\ind\tilde{P}_2.
\label{B3226}
\end{equation}
However, $\tilde{P}_1\tilde{P}_2=P_1P_2P_1|_{\Im P_1}$ and
$P_1P_2P_1|_{\Im P_1}=\Id_{\Im P_1}+A$, where $A\colon\Im P_1\to\Im P_1$
is of trace class. Hence
\begin{equation}
\ind\tilde{P}_1\tilde{P}_2=0.
\label{B3227}
\end{equation}
The lemma is proved.\ \ \ $\Box$

\section{Determinants of general elliptic operators}
\label{S8}

In (\ref{B3947}) we extended the definition of the zeta-regularized
determinant $\det_\zeta(A)$ to elliptic operators $A$, $\ord A\ne 0$,
with a choice of the logarithm of their symbols $\sigma(\log A)$.
(In (\ref{B3947}) we suppose that some $\sigma(\log A)$ exists but
do not suppose that $\log A$ exists.)

Later on we will call them {\em canonical determinants} and denote
by $\det(A)$
for an operator $A$. We try to generalize these determinants to the case
of general elliptic PDOs (i.e., without of the supposition that their
logarithmic symbols exist).

Let $a_t$, $0\le t\le 1$, be a smooth curve in the Lie algebra $\fell(M,E)$
of logarithms for classical elliptic PDOs such that $a_t\in(rp)^{-1}(c)$,
$c\in\wC^{\times}$. (To remind, $p\colon\fell(M,E)\to S_{\log}(M,E)$ is
the natural projection and $r\colon S_{\log}(M,E)\to\wC$ is the order
homomorphism from the extension (\ref{B303}).) Let $A_t$, $0\le t\le 1$,
be the solution of an ordinary differential equation
\begin{equation}
\df_t A_t=a_tA_t, \qquad A_0:=\Id.
\label{B999}
\end{equation}
Then $A_t$ is the elliptic operator from $\Ell_0^{ct}(M,E)$. The symbol
of the operator $A_{t,t+\eps}:=A_{t+\eps}A_t^{-1}$ has a canonical
logarithm in $S_{\log}(M,E)$ close to zero
(if $\eps>0$ is sufficiently small). Indeed, in this case,
the principal symbol $\sigma_{c\eps}\left(A_{t+\eps}A_t^{-1}\right)$
on $S^*M$ is sufficiently close to $\Id$. Hence the logarithmic symbol
of $\sigma\left(A_{t+\eps}A_t^{-1}\right)$ exists by Remark~\ref{RB3964}%
\footnote{Here it is enough to use a spectral cut $L_{(\pi)}$.}
Thus $\det\left(A_{t,t+\eps}\right)$ is defined.

Let $A\in\Ell_0^c(M,E)$, $c\in\wC^{\times}$, and a smooth curve $a_t$,
$0\le t\le 1$, in $(rp)^{-1}(c)\subset\fell(M,E)$ be such that
$A=A_t|_{t=1}$, where $A_t$ is the solution of (\ref{B999}).
Then the determinant of the pair $(A,a_t)$ is defined by
\begin{equation}
\det\left(A,a_t\right)=\lim_{\sup\{\eps_i\}\to 0}\Pi_i{\det}\left(
A_{t_i,t_{i+1}}\right),
\label{B1000}
\end{equation}
where $\{\eps_i\}$ are finite sets of $\eps_i>0$ such that $\sum\eps_i=1$.
Here, $t_0=0$, $t_i=\eps_0+\ldots+\eps_{i-1}$ for $i\ge 1$,
$t_i+\eps_i=t_{i+1}$.

\begin{rem}
Let $a_t\equiv a$ be independent of $t\in[0,1]$. Then the map
from $a\in S_{\log}(M,E)$ to the value $A_1$ at $t=1$ of the solution
of (\ref{B999}) with $a_t\equiv a$ is the exponential map of $S_{\log}(M,E)$
into $\Ell_0^{\times}(M,E)$ since $A_1=\exp a$.

Let $a\in(rp)^{-1}(c)\subset\fell(M,E)$, $c\in\wC^{\times}$. Then
the determinant (\ref{B1000}) for $A=A_1(a)$ is the zeta-regularized
(and canonical) determinant
\begin{equation}
{\det}(A)=\det(A,a).
\label{B1954}
\end{equation}
In this case, the canonical determinant coincides with the zeta-regularized
determinant
$\det^{\TR}_\zeta(A):=\exp\left(-\df_s\zeta_{A,a}^{\TR}(s)|_{s=0}\right)$.
Here,
the zeta-function of $A$ is defined as $\zeta_{A,a}^{\TR}(s):=\TR(\exp(-sa))$,
$\TR$ is the canonical trace, $\zeta_{A,a}(s)$ is regular at $s=0$
by Proposition~\ref{PB5}.
\label{RB1008}
\end{rem}

Hence the determinant (\ref{B1000}) is an extension of a zeta-regularized
determinant corresponding to the case $a_t\equiv a$ in (\ref{B1000})
(where $a\in(rp)^{-1}\left(\wC^{\times}\right)\subset\fell(M,E)$).%
\footnote{There is an unsolved problem. The determinant of an elliptic
operator $A\in\Ell_0^c(M,E)$, $c\in\wC^{\times}$, has to be defined
as a (multiplicative) functional integral
\begin{equation*}
\Det(A):=\int\det(A,a_t)\cal{D}a_t
\end{equation*}
over the space of curves $a_t$ in $(rp)^{-1}(c)\subset\fell(M,E)$ such
that $A_1=A$ for the solution $A_t$ of (\ref{B999}) (with $a_t$
as the coefficient on the right in (\ref{B999})). The problem is how
to define such an integral and what are the properties of $\Det(A)$.}

\begin{rem}
Let $S$ be a positive definite self-adjoint elliptic operator
from $\Ell_0^1(M,E)$. Then the solution $A_t$ of (\ref{B999}) defines
a solution $B_t=S_{(\pi)}^{-ct}A_t$ of the equation
\begin{equation}
\df_t B_t=b_tB_t, \qquad B_0:=\Id,
\label{B1003}
\end{equation}
and $b_t$ is a curve in $(rp)^{-1}(0)=CL^0(M,E)$,
\begin{equation}
b_t:=S_{(\pi)}^{-ct}a_tS_{(\pi)}^{ct}-c\log_{(\pi)}S.
\label{B1002}
\end{equation}
The operator $A=A_t|_{t=1}$ is defined by $S$ and by a smooth curve
$b_t$ in $CL^0(M,E)$.
\label{RB1001}
\end{rem}

\begin{rem}
A smooth curve $a_t$, $0\le t\le 1$, in $(rp)^{-1}(c)$ is defined
by a smooth curve $\alpha_t$ from the origin in $\fell(M,E)$ such that
$\alpha_t\in(rp)^{-1}(ct)$ and
\begin{equation}
\df_t\alpha_t=a_t, \quad \alpha_t:=\int_0^t a_\tau d\tau.
\label{B1005}
\end{equation}
The class of solutions of the equation (\ref{B999}) for smooth curves
$a_t$ in $(rp)^{-1}(c)\subset\fell(M,E)$ coincides with smooth curves
$A_t$, $0\le t\le 1$, in $\Ell_0^{\times}(M,E)$ such that $A_0=\Id$
and $\ord A_t=ct$.
\label{RB1004}
\end{rem}


\begin{pro}
An invertible elliptic PDO $A\in\Ell_0^c(M,E)$ with $c\in\wC^{\times}$
can be represented as the value at $t=1$ of a solution $A_t$ of (\ref{B999})
with some smooth curve $a_t$ in $(rp)^{-1}(c)\subset\fell(M,E)$.
\label{PB1006}
\end{pro}

\begin{thm}
The determinant $\det(A,a_t)$ (where $A$, $a_t$ are as in (\ref{B1000}))
is defined, i.e., the limit on the right in (\ref{B1000}) exists.
\label{TB1007}
\end{thm}

\begin{cor}
The determinant $\det\left(\!A,a_t\!\right)$ is invariant under smooth
reparametrizations of a curve $\left(a_t\right)$.
\label{CB3002}
\end{cor}

\begin{rem}
Let $A$ be a product $A=A_2A_1$ of elliptic PDOs from $\Ell_0^{\times}(M,E)$.
Let $A_{j,t}$, $j=1,2$, $0\le t\le 1$, be smooth curves
in $\Ell_0^{\times}(M,E)$ such that $\ord A_{j,t}$ are monotonic in $t$
and $A_{j,0}=\Id$, $A_{j,1}=A_j$. Then $\det\left(A,a_t\right)$ is also
defined for a piecewise-smooth curve $a_t$ in $\fell(M,E)$,
\begin{align*}
a_t & :=\left(\df_\tau A_{1,\tau}\cdot A_{1,\tau}^{-1}\right)\big|_{\tau=2t}
=:a_{1,2t} \quad\text{ for }0\le t\le 1/2, \\
a_t & :=\left(\df_\tau A_{2,\tau}\cdot A_{2,\tau}^{-1}\right)\big|_{\tau=2t-1}
=:a_{2,2t-1} \quad\text{ for }1/2\le t\le 1.
\end{align*}
(In general, this curve is disconnected at $t=1/2$.)
We have
$$
\det\left(A_2,a_{2,t}\right)\det\left(A_1,a_{1,t}\right)=\det\left(A_2A_1,a_t
\right).
$$
Here, the orders of PDOs $A_j$ have to be nonzero. However
we {\em don't suppose}
that $A_2A_1$ is an elliptic PDO of a {\em nonzero} order.
\label{RB3003}
\end{rem}

\noindent{\bf Proof of Proposition~\ref{PB1006}.} For an arbitrary
$A\in\Ell_0^c(M,E)$, $c\in\wC^{\times}$, there exists a smooth curve
$A(t)$ in $\Ell_0^{\times}(M,E)$ such that $A(0)=\Id$,
$A(t)\in\Ell_0^{ct}(M,E)$, and $A(1)=A$. Then $\df_t A(t)=a_tA(t)$,
where $a_t\in(rp)^{-1}(c)\subset\fell(M,E)$. Hence for $A\in\Ell_0^c(M,E)$,
$c\in\wC^{\times}$, there exists a curve $a_t$ which satisfies
the same conditions as in (\ref{B1000}).\ \ \ $\Box$

\noindent{\bf Proof of Theorem~\ref{TB1007}.} The product
of determinants on the right in (\ref{B1000}) can be written in the form
\begin{equation}
\Pi_{i=0}^{m-1}{\det}_{(\tpi)}\left(A_{t_i,t_{i+1}}\right)=\Pi_{i=0}^{m-1}
\left(d_1\left(A_{t_i,t_{i+1}}\right)\big/d_0\left(A_{t_i,t_{i+1}}\right)
\right).
\label{B2020}
\end{equation}
Here, $t_0=0<t_1<\ldots<t_m=1$ and $\eps_i:=t_{i+1}-t_i$ are supposed
to be small enough. The element $d_1(A)\in G(M,E)$
for $A\in\Ell_0^{\times}(M,E)$ is defined in Section~\ref{SE}
as the image of $A$ in $F_0\backslash\Ell_0^{\times}(M,E)=:G(M,E)$
(the normal subgroup $F_0$ is defined by (\ref{B600})). The elements
$d_0(A)$ are defined for elliptic PDOs $A$ of real nonzero orders
sufficiently close to positive definite ones as
$d_1(A)/{\det}_{(\tpi)}(A)\in G(M,E)$. By Proposition~\ref{PB627}
the element $d_0(A)\in G(M,E)$ depends on the symbol $\sigma(A)$ of $A$
only. The local section $d_0(A)$ is defined by Theorem~\ref{TB570}
as the exponential of the $\wC^{\times}$-cone of null-vectors
in $\frg(M,E)=\sfrg$ for the invariant quadratic form (\ref{B418})
on $\sfrg$.

The extension of the Lie groups
$$
1\to\wC^{\times}\to G(M,E)\to\SEll_0^{\times}(M,E)\to 1
$$
is central. Hence the product of determinants on the left in (\ref{B2020})
can be represented in the form
\begin{equation}
d_1\left(A_{t_{m-1},t_m}\right)d_1\left(A_{t_{m-2},t_{m-1}}\right)\ldots
d_1\left(A_{t_0,t_1}\right)\big/d_0\left(A_{t_{m-1},t_m}\right)\ldots d_0
\left(A_{t_0,t_1}\right).
\label{B2021}
\end{equation}

By (\ref{B2022}) the numerator of (\ref{B2021}) is equal to $d_1(A)$.
(To remind, $A:=A_t|_{t=1}$.) The denominator in (\ref{B2021}) depends
on symbols $\sigma\left(A_{t_i,t_{i+1}}\right)$, $0\le i\le m-1$, only.
Hence it is enough to prove the assertion as follows.
\begin{pro}
The limit exists
\begin{equation}
\lim_{\sup\{\eps_i\}\to 0}d_0\left(A_{t_{m-1},t_m}\right)\ldots d_0
\left(A_{t_0,t_1}\right).
\label{B2029}
\end{equation}
Here, $\{\eps_i\}$ are finite sets $(\eps_0,\dots,\eps_{m-1})$, $m\in\wZ_+$,
of $\eps_i>0$ such that $\eps_0+\eps_1+\ldots+\eps_{m-1}=1$,
$t_0=0<t_1<\ldots<t_m=1$, $t_{i+1}-t_i=\eps_i$ for $0\le i\le m-1$
and $\eps_i$ are supposed to be small enough
(when $d_0\left(A_{t_i,t_{i+1}}\right)$ are defined).
\label{PB2023}
\end{pro}

\noindent{\bf Proof.} Set $l_t:=\sigma(a_t)/c$ (where
$a_t\in(rp)^{-1}(c)\subset\fell(M,E)$). Let a PDO $B\in\Ell_0^d(M,E)$,
$d\ne 0$, be sufficiently close to a positive
definite self-adjoint PDO.%
\footnote{Here we use only that $\sigma_d(B)|_{S^*M}$ is sufficiently
close to a positive definite and self-adjoint PDO. In this case,
$\sigma\left(\log_{(\tpi)}B\right)$ are defined on $S^*M$. So
$\sigma\left(B^z\right)$ can be defined on $T^*M\setminus M$
by multiplying appropriate terms of this symbol by $t^{z-k}$, $t\in\wR_+$.}
Then under the canonical local identification
$G(M,E)=\exp\left(\sfrg\right)$ of Theorem~\ref{TB570}, $d_0(B)$
corresponds to the element $\exp\left(d\cdot\Pi_l l\right)$
of $\exp\left(\sfrg\right)=\exp\left(\sfrg_{(l)}\right)$, where
$l:=\sigma\left(\log_{(\tpi)}B\right)/d$ and
$\Pi_l\colon\frg\hookrightarrow\sfrg_{(l)}$ is the inclusion
of $\frg:S_{\log}(M,E)$ into $\sfrg_{(l)}$ under the splitting
(\ref{B4050}), (\ref{B622}). (To remind, the Lie algebras $\sfrg_{(l)}$
and $\sfrg_{(l_1)}$ are canonically identified by an associative system
of the Lie algebras isomorphisms $W_{l,l_1}$ given
by Proposition~\ref{PB403}. Hence this system of isomorphisms defines
the canonical Lie algebra $\sfrg$.) Let $W_l\colon\sfrg_{(l)}\rs\sfrg$
be the canonical isomorphism of Lie algebras (defined
by the system $W_{l,l_1}$ of isomorphisms).
Let $F_t\in\exp\left(\sfrg\right)$ be a solution of the equation
\begin{equation}
\df_tF_t=c\cdot W_{l_t}\left(\Pi_{l_t}l_t\right)\cdot F_t, \quad F_0=\Id.
\label{B2025}
\end{equation}
(Here, $c\in\wC^{\times}$ is the constant such that $a_t\in(rp)^{-1}(c)$.)
This equation can be solved by the substitution
\begin{equation}
F_t:=\exp\left(c\cdot t\cdot\tilde{l}\right)\cdot K_t(l),
\label{B3402}
\end{equation}
where $\tilde{l}:=W_l\left(\Pi_l l\right)$, $l\in r^{-1}(1)$ (for instance,
$l:=l_0$), and $K_t(l)\in\exp\left(\sfrg\right)$ is a solution
of the equation
\begin{equation}
\df_tK_t=c\cdot\exp\left(-c\cdot t\cdot\tilde{l}\right)\cdot W_l\left(\Pi_l
f_t+\left(f_t,f_t\right)_{\res}/2\cdot 1\right)\cdot\exp\left(c\cdot t\cdot
\tilde{l}\right)K_t
\label{B2028}
\end{equation}
for $K_0(l):=\Id$, $f_t:=l_t-l\in CS^0(M,E)$. Indeed, let $F_t$ be
the solution of (\ref{B2025}). Then by (\ref{B580}) we have
\begin{gather}
\begin{split}
W_{l_t}\left(\Pi_{l_t}l_t\right) & =W_l\left(\Pi_ll_t+\left(l_t-\left(l_t+
l\right)/2,l_t-l\right)_{\res}\cdot 1\right)=\\
 & =\tilde{l}+W_l\left(\Pi_lf_t+\left(f_t,f_t\right)_{\res}/2\cdot 1\right),\\
\df_tF_t & =c\cdot\tilde{l}F_t+\exp\left(c\cdot t\cdot
\tilde{l}\right)\cdot\df_tK_t(l).
\end{split}
\label{B2027}
\end{gather}
Hence $\df_tK_t(l)$ is given by (\ref{B2028}) (and $K_0(l):=\Id$).

The factor
\begin{equation}
u(t):=\exp\left(-c\cdot t\cdot\tilde{l}\right)W_l\left(\Pi_lf_t+
\left(f_t,f_t\right)_{\res}/2\cdot 1\right)\exp\left(c\cdot t\cdot
\tilde{l}\right)
\label{B3403}
\end{equation}
on the right in (\ref{B2028}) is a smooth curve
$$
u\colon t\in[0,1]\to u(t)\in(rq)^{-1}(0)=q^{-1}\frg_0\subset\sfrg.
$$
(Here, $q\colon\sfrg\to\frg:=S_{\log}(M,E)$ is the natural projection,
$r$ is the order homomorphism from (\ref{B303}), and $\frg_0:=CS^0(M,E)$.)
Hence the equations (\ref{B2028}), (\ref{B2025}) have unique solutions.
(This assertion is proved in Lemma~\ref{LB3400}.)

The approximation similar to Euler polygon line for the ordinary
differential equation (\ref{B2025}) on the determinant Lie group
$G(M,E)$ is defined for any given finite set $\{\eps_i\}$ with
$\eps_i>0$, $\sum_i\eps_i=1$, as the solution of the equation%
\footnote{This solution is a piecewise-smooth continuous curve
$e_0\colon[0,1]\to G(M,E)$, $e_0$ is smooth except points
$e_0\left(t_j,\left\{\eps_i\right\}\right)$. To remind, locally
$G(M,E)$ and $\exp\left(\sfrg\right)$ are canonically isomorphic
by Theorem~\ref{TB570}.}
\begin{gather}
\begin{split}
\df_te_0\left(t,\{\eps_i\}\right) & =c\cdot f_0\left(t,\{\eps_i\}\right)\cdot
e_0\left(t,\{\eps_i\}\right), \quad e_0\left(0,\{\eps_i\}\right)=\Id, \\
f_0\left(t,\{\eps_i\}\right)=\tilde{l}_{t_i} & \quad\text{ for }t\in\left(t_i,
t_{i+1}\right), \quad \tilde{l}_t:=W_{l_t}\left(\Pi_{l_t}l_t\right).
\end{split}
\label{B2030}
\end{gather}

The product of $d_0\left(A_{t_i,t_{i+1}}\right)$ in (\ref{B2029})
is equal to the value at $t=1$ of the solution of the equation
\begin{gather}
\begin{split}
\df_te_1\left(t,\{\eps_i\}\right) & =c\cdot f_1\left(t,\{\eps_i\}\right)\cdot
e_1\left(t,\{\eps_i\}\right), \quad e_1\left(0,\{\eps_i\}\right)=\Id, \\
f_1\left(t,\{\eps_i\}\right)      & =\tilde{l}_i
\quad\text{ for }t\in\left(t_i,t_{i+1}\right),
l_i:=\sigma\left(\log\left(A_{t_i,t_{i+1}}\right)\right)/c\eps_i.
\end{split}
\label{B2031}
\end{gather}
(Here, it is supposed that $\eps_i=t_{i+1}-t_i$ are small enough
for $\log\sigma\left(\left(A_{t_i,t_{i+1}}\right)\right)$ to exist.)

%
The difference $l_i-l_{t_i}\in CS^0(M,E)$
can be estimated as follows.
Set $B(t):=\sigma\left(A_tA_{t_{i+1}}^{-1}\right)$,
$\beta(t):=\sigma\left(\log\left(A_tA_{t_{i+1}}^{-1}\right)\right)$.
Here, $t$ is real and close to $t_{i+1}$.
By (\ref{B999}), (\ref{B2031}), and
by (\ref{B620}), we have
\begin{gather}
\begin{split}
b(t)|_{t=t_i} & =\exp\left(-c\cdot\eps_i\cdot l_i\right), \\
\df_tb(t)|_{t=t_i}b\left(t_i\right)^{-1} & =cl_{t_i}, \\
\df_tb(t)|_{t=t_i}b\left(t_i\right)^{-1} & =
\df_t\sigma\left(A_t\right)\cdot\sigma\left(A_t^{-1}\right)|_{t=t_i}= \\
=F\left(\ad\left(-c\eps_il_i\right)\right)\circ\df_t\beta(t)|_{t=t_i}
 & =F\left(\ad\left(-c\eps_il_i\right)\right)\circ\left(cl_i-c\eps_i\df_t
\gamma(t)|_{t=t_i} \right),
\end{split}
\label{B2040}
\end{gather}
where $F(\ad l)$ is defined by (\ref{B620}) and by Remark~\ref{RB621}.
Here, $\gamma(t):=\beta(t)/c\left(t-t_{i+1}\right)$,
$\gamma\left(t_i\right)=l_i$, $\gamma\left(t_{i+1}\right):=l_{t_{i+1}}$.
So we have
\begin{equation}
\df_tb(t)|_{t=t_i}b\left(t_i\right)^{-1}=cl_i+cF\left(\ad\left(-c\eps_il_i
\right)\right)\circ\left(-\eps_i\df_t\gamma(t)|_{t=t_i}\right).
\label{B3232}
\end{equation}
We conclude that
\begin{equation}
\left(l_{t_i}-l_i\right)=\eps_iF\left(\ad\left(-c\eps_il_i\right)\right)
\circ\left(-\df_t\gamma(t)|_{t=t_i}\right).
\label{B2041}
\end{equation}

The space $CS^0(M,E)$ is a Fr\'echet space with semi-norms defined
as follows. Let $\left\{U_i\right\}$ be a finite cover of $M$
by coordinate charts and let $\left\{V_i\right\}$,
$\overline{V_i}\subset U_i$, be a subordinate finite cover of $M$ such that
$\overline{V_i}$ are compact. The semi-norms are labeled by $k\in\zuo$
and by multi-indexes $\alpha=\left(\alpha_1,\dots,\alpha_n\right)$,
$\omega=\left(\omega_1,\dots,\omega_n\right)$ ($\alpha_j\omega_j\in\zuo$).
For $\alpha\in CS^0(M,E)$ the corresponding semi-norm is
\begin{equation}
\left\|a\right\|_{k,\alpha,\omega}:=\max_i\sup_{x\in V_i}\left(\left|\xi
\right|^k\left\|\df_\xi^\alpha
\df_x^\omega a_{-k}(x,\xi)\right\|\right),
\label{B3233}
\end{equation}
where $a_{-k}$ is a positive homogeneous component of $a$ in coordinates
$U_i\ni x$. This Fr\'echet structure is independent (up to equivalence)
of a finite cover of $M$ by coordinate charts.

The proof of Proposition~\ref{PB2023} uses the following lemmas.

\begin{lem}
The difference $l_{t_i}-l_i$ is $O\left(\eps_i\right)$ (as $\eps_i$
tends to zero) with respect to any finite set of semi-norms (\ref{B3233}).
Namely for any finite set $(k,\alpha,\omega)_j$, $j=1,\dots,N$,
of indexes in (\ref{B2028}) there are constants $C_1>0$, $\eps$,
$1>\eps>0$, such that
$$
\left\|l_{t_i}-l_i\right\|_{(k,\alpha,\omega)_j}<C_1\eps_i
$$
for any $i$, $0<t_i<t_{i+1}:=t_i+\eps_i<1$, $0<\eps_i<\eps$, $1\le j\le N$.
\label{LB3405}
\end{lem}

\begin{cor}
The difference of the coefficients $f_1\left(t,\left\{\eps_i\right\}\right)-
f_0\left(t,\left\{\eps_i\right\}\right)$ in the linear equations
(\ref{B2031}) and (\ref{B2030}) is $O\left(\eps_i\right)$ (as $\eps_i$
tends to zero) with respect to any finite set of semi-norms (\ref{B3233})
uniformly in $t\in\left(t_i,t_{i+1}\right)$ and in $t_i$,
$0\le t_i\le 1-\eps_i$. Namely the logarithmic symbols $l_i$ and $l_{t_i}$
are of order one, $l_i,l_{t_i}\in r^{-1}(1)$. By (\ref{B580}) and
(\ref{B2027}) we have
\begin{equation}
\tilde{l}_{t_i}=\tilde{l}_i+W_{l_i}\left(\Pi_{l_i}\left(l_{t_i}-l_i\right)+
\left(l_{t_i}-l_i,l_{t_i}-l_i\right)_{\res}/2\cdot 1\right).
\label{B3449}
\end{equation}
By Lemma~\ref{LB3405}, $l_{t_i}-l_i$ is $O\left(\eps_i\right)$
(as $\eps_i$ tends to zero) with respect to any semi-norm (\ref{B3233}).
Hence $\left(l_{t_i}-l_i,l_{t_i}-l_i\right)_{\res}=O\left(\eps_i^2\right)$
and $\tilde{l}_{t_i}-\tilde{l}_i$ is $O\left(\eps_i\right)$.
\label{CB3406}
\end{cor}

\begin{lem}
There is a unique solution $F_t$ of the equation (\ref{B2025}). We have
$$
\lim_{\sup\{\eps_i\}\to 0}e_0\left(t,\{\eps_i\}\right)=F_t
$$
uniformly in $t\in[0,1]$, $\sup_i\left\{\eps_i\right\}$.
\label{LB3400}
\end{lem}

\begin{rem}
The convergence in $G(M,E)$ is defined as follows.

1. A sequence $\left\{g_m\right\}\subset G(M,E)$, $m\in\wZ_+$,
is convergent to a point $g\in G(M,E)$, if there is $m_0\in\wZ_+$
such that for $m\ge m_0$
$$
g^{-1}g_m\in\exp\left(W_l\sfrg_{(l)}\right)
$$
(for some fixed $l\in r^{-1}(1)\subset S_{\log}(M,E)$) and if
$W_l^{-1}\log\left(g^{-1}g_m\right)=:u_m\in\sfrg_{(l)}$ are
convergent to zero in $\sfrg_{(l)}$. The points $u_m\in\sfrg_{(l)}$
are written in the form
\begin{equation}
u_m=q_ml+u_m^0+c_m\cdot 1
\label{B3408}
\end{equation}
with respect to the splitting (\ref{B622}) defined by $l$. (Here,
$q_m,c_m\in\wC$, $u_m^0\in CS^0(M,E)$, and $1$ is the central element
in $\sfrg_{(l)}$.) The assertion $u_m\to 0$ in $\sfrg_{(l)}$
(as $m\to\infty$) means that $q_m\to 0$, $c_m\to 0$, and any semi-norm
(\ref{B3233}) of $u_m^0$, $\left\|u_m^0\right\|_{k,\alpha,\omega}$,
tends to zero.

2. Let $f_\eps\colon t\in[0,1]\to f_\eps(t)\in G(M,E)$ be a family
of curves in $G(M,E)$. We say that $f_\eps$ tends to a curve $f_0$
in $G(M,E)$ uniformly in $t$ (as $\eps$ tends to zero), if \\
1) there is $\eps_0>0$ such that for $0<\eps<\eps_0$
$$
f_0(t)^{-1}f_\eps(t)=:\exp\left(W_lu_\eps(t)\right)\in\exp\left(W_l\sfrg_{(l)}
\right),
$$
2) for the components of the elements
$$
u_\eps(t):=q_\eps(t)l+u_\eps^0(t)+c_\eps(t)\cdot 1\in\sfrg_{(l)}
$$
written with respect to the splitting (\ref{B622}) (similarly
to (\ref{B3408})), it holds uniformly in $t$ and with respect to any
semi-norm (\ref{B3233})
$$
q_\eps(t)\to 0, \quad c_\eps(t)\to 0, \quad
\left\|u_\eps^0(t)\right\|_{k,\alpha,\omega}\to 0.
$$
(These conditions are independent of $l\in r^{-1}(1)\subset S_{\log}(M,E)$
by Proposition~\ref{PB403} and by Theorem~\ref{TB570}.)

3. The extension $\sfrg_{(l)}$ of $S_{\log}(M,E)\supset\frg_0:=CS^0(M,E)$
is defined by a cocycle $K_l$, (\ref{B304}). The restriction to $\frg_0$
of this cocycle, $K_l\left(B_0,C_0\right)$, depends only on images
of symbols $B_0$, $C_0$ in $CS^0(M,E)/CS^{-n-1}(M,E)$ ($n:=\dim M$).
By Proposition~\ref{PB403}, the identifications
$W_{l_1,l_2}\colon\sfrg_{(l_1)}\rs\sfrg_{(l_2)}$ with
$l_1-l_2\in CS^{-n-1}(M,E)$ do not change the coordinate $c$
of the central elements $c\cdot 1$ in $\sfrg_{(l_1)}$ and in $\sfrg_{(l_2)}$.

A sequence $\left\{g_m\right\}\subset G(M,E)$ is convergent
to $g\in G(M,E)$, if the following conditions hold.

1) The symbols $s_m:=p\left(g_m\right)$ are convergent
in $\SEll_0^{\times}(M,E)$ to $s:=p(g)$. It means that the orders
$q_m:=\ord s_m\in\wC$
are convergent to $q:=\ord s$ and that the restrictions of $s_m$
to $S^*M$ are covergent to $s|_{S^*M}$. Namely let $\left\{U_i\right\}$
be a finite cover of $M$ by coordinate charts and let $\left\{V_i\right\}$,
$\overline{V_i}\subset U_i$, be a subordinate finite cover of $M$ such that
$\overline{V_i}$ are compact (as in (\ref{B3233})). Then the restrictions
to $S^*M$ of the positive homogeneous components
$\left(s_m\right)_{q_m-k}(x,\xi)$, $k\in\zuo$, (defined by $s_m$ and
by a cover $\left\{U_i\right\}$) are convergent over all $\overline{V_i}$
to $(s)_{q-k}(x,\xi)|_{S^*M}$ together with their partial derivatives
with respect to $(x,\xi)$. (This is a condition of convergence with
respect to semi-norms similar to (\ref{B3233}). Here, the factors with
powers of $|\xi|$ in these semi-norms can be replaced by $1$
since $\xi\in S^*M$.) If such a convergence holds with respect to some
finite cover of $M$ by coordinate charts, then it holds with respect
to any finite cover of $M$ by coordinate charts.

2) The images $c_l\left(u_m\right)$ of elements $u_m\in\sfrg_{(l)}$,
$\exp\left(W_lu_m\right):=g^{-1}g_m$, under the natural projection
\begin{equation}
c_l\colon\sfrg_{(l)}\to\sfrg_{(l)}/\frg=\wC
\label{B3410}
\end{equation}
are convergent (as $m\to\infty$) to zero. Here, $\frg:=S_{\log}(M,E)$
is imbedded (as a linear space) into $\sfrg_{(l)}$ with respect
to the splitting (\ref{B622}) defined by $l$. The projection $c_l$ depends
on the image of $l$ in $\frg/CS^{-n-1}(M,E)$ only ($n:=\dim M$). Namely
for $l_1,l_2\in r^{-1}(1)\subset\frg$ such that $l_1-l_2\in CS^{-n-1}(M,E)$
we have
\begin{equation}
c_{l_2}W_{l_1,l_2}=c_{l_1}.
\label{B3411}
\end{equation}

The group structure of $G(M,E)$ is induced by the group structure
of $\Ell_0^{\times}(M,E)$. This structure is in accordance with
the convergence in $G(M,E)$.
\label{RB3407}
\end{rem}

\begin{lem}
The estimate
\begin{equation}
d_0\left(\sigma\left(A_{t_{i+1}}A_{t_i}^{-1}\right)\right)\exp\left(-c\eps_i
\tilde{l}_{t_i}\right)\!:=\!\exp\left(c\eps_i\tilde{l}_{t_i}\right)
\exp\left(-c\eps_i\tilde{l}_{t_i}\right)\!=\!\Id+O\left(\eps_i^2\right)
\label{B3421}
\end{equation}
holds in $G(M,E)$ uniformly in $i$, $t_i$ (as $\eps_i$ tends to zero).
\label{LB3420}
\end{lem}

\begin{rem}
The estimate (\ref{B3421}) means that its left side has a form
$\exp\left(W_lu\right)$, where $u\in\sfrg_{(l)}$ (for some
$l\in r^{-1}(1)\subset S_{\log}(M,E)$) and that $u$ is
$O\left(\eps_i^2\right)$ in $\sfrg_{(l)}$ (uniformly in $i$, $t_i$).
The latter condition means that with respect to the splitting (\ref{B622})
(defined by $l$) we have
$$
u=0\cdot l+u_0+c\cdot 1\in\sfrg_{(l)}
$$
(because $r\left(l_i\right)=r\left(l_{t_i}\right)=1$ and so $rq(u)=0$),
where $c$ is $O\left(\eps_i^2\right)$ and $u_0$ is $O\left(\eps_i^2\right)$
in $CS^0(M,E)$ with respect to any semi-norm (\ref{B3233}) (as $\eps_i$
tends to zero). This condition is independent of $l\in r^{-1}(1)$
by Proposition~\ref{PB403} and by Theorem~\ref{TB570}.
\label{RB3422}
\end{rem}

Now we return to the proof of Proposition~\ref{PB2023}. We have to prove
the convergence of the product
\begin{equation}
d_0\!\left(\left\{A_{t_i},\eps_i\right\}\right)\!:=\!d_0\!\left(A_{t_{m-1}
t_m}\right)\ldots d_0\!\left(A_{t_0t_1}\right)\!=:\!\exp\!\left(c\eps_{m-1}
\tilde{l}_{m-1}\right)\ldots\exp\!\left(c\eps_0\tilde{l}_0\right)
\label{B3423}
\end{equation}
as $\sup_i\left\{\eps_i\right\}$ tends to zero. By Lemma~\ref{LB3400},
the solution $e_1\left(t,\left\{\eps_i\right\}\right)$ of (\ref{B2030})
tends (in $G(M,E)$) to $F_t$ as $\sup_i\left\{\eps_i\right\}\to 0$.
(Here, $F_t$ is the solution of (\ref{B2025}).) In particular,
\begin{equation}
e_1\left(1,\left\{\eps_i\right\}\right):=\exp\left(c\eps_{m-1}\tilde{l}_{t
_{m-1}}\right)\ldots\exp\left(c\eps_0\tilde{l}_{t_0}\right)
\label{B3424}
\end{equation}
tends to $F_1$. So the product $e_1\left(1,\left\{\eps_i\right\}\right)$
converges (as $\sup_i\left\{\eps_i\right\}\to 0$).

By Lemma~\ref{LB3420} we have
\begin{equation}
\exp\left(c\eps_i\tilde{l}_i\right)=\exp\left(W_lu_i\right)\exp\left(c\eps_i
\tilde{l}_{t_i}\right)
\label{B3425}
\end{equation}
with $u_i=O\left(\eps_i^2\right)$ in $\sfrg_{(l)}$ uniformly in $i$, $t_i$.
By Lemma~\ref{LB3400} we conclude that for any $\left\{\eps_j\right\}$
with $\sup_j\left\{\eps_j\right\}$ small enough, the products
\begin{equation}
P_i\left(\left\{\eps_j\right\}\right):=\exp\left(c\eps_i\tilde{l}_{t_i}\right)
\exp\left(c\eps_{i-1}\tilde{l}_{t_{i-1}}\right)\ldots\exp\left(c\eps_0
\tilde{l}_{t_0}\right)
\label{B3426}
\end{equation}
belong to a bounded set $B$ in $G(M,E)$ for any $\eps_j$, $t_j$.
Indeed, (\ref{B3426}) tends to $F_{t_i}$ uniformly in $i$, $t_i$
as $\sup_j\left\{\eps_j\right\}$ tends to zero. There is an open set
$0\in U\subset\sfrg_{(l)}$ such that the products (\ref{B3426})
belong to $F_{t_i}\exp\left(W_lU\right)$ for all $t_i$
(if $\sup_j\left\{\eps_j\right\}$ is small enough) and $U$ is bounded
in $\sfrg_{(l)}$. (The latter condition means that the direct sum
components of elements of $U$
in $\sfrg_{(l)}=\wC\cdot l\oplus CS^0(M,E)\oplus\wC\cdot 1$
splitted by (\ref{B622}) are bounded. A set $B_0\subset CS^0(M,E)$
is bounded, if it is bounded with respect to all semi-norms (\ref{B3233}).)
So $Q_i:=P_i\left(\left\{\eps_j\right\}\right)^{-1}\exp\left(W_lu_i\right)
P_i\left(\left\{\eps_j\right\}\right)$ is $\Id+O\left(\eps_i^2\right)$
uniformly in $i$, $\left\{t_j\right\}$, if $\sup_j\left\{\eps_j\right\}$
is small enough. (The latter condition means that
$Q_i=\exp\left(W_lk_i\right)$, where $k_i$ is $O\left(\eps_i^2\right)$
in $\sfrg_{(l)}$. Note that $Q_i$ depends not
on $P_i\left(\left\{\eps_j\right\}\right)$ but only on its symbol
$p\left(P_i\left(\left\{\eps_j\right\}\right)\right)\in\SEll_0^{\times}
(M,E)$.) We have
\begin{multline}
d_0\left(\left\{A_{t_i},\eps_j\right\}\right)=\exp\left(W_lu_{m-1}\right)
\exp\left(c\eps_{m-1}\tilde{l}_{t_{m-1}}\right)\ldots\exp\left(W_lu_0\right)
\exp\left(c\eps_0\tilde{l}_{t_0}\right)= \\
=\!\exp\!\left(W_lu_{m-1}\right)\exp\!\left(c\eps_{m-1}\tilde{l}_{t_{m-1}}
\right)
\ldots\exp\left(W_lu_{i+1}\right)\exp\left(c\eps_{i+1}\tilde{l}_{t_{i+1}}
\right)P_i\left(\left\{\eps_j\right\}\right)Q_i\ldots Q_0\!= \\
=P_{m-1}\left(\left\{\eps_j\right\}\right)Q_{m-1}\ldots Q_0=e_1\left(1,
\left\{\eps_j\right\}\right)Q_{m-1}\ldots Q_0.
\label{B3428}
\end{multline}

We see also that the product $Q_{m-1}\ldots Q_0$ tends to $\Id\in G(M,E)$
as $\sup_j\left\{\eps_j\right\}\to 0$. Indeed, $Q_j$ is
$\Id+O\left(\eps_j^2\right)$ uniformly in $j$ and the product
$$
\Pi_j\left(1+C\eps_j^2\right)\le\exp\left(C\sum\eps_j^2\right)\le\exp\left(
C\cdot\sup_j\left\{\eps_j\right\}\right)
$$
tends to zero as $\sup\{\eps_j\}\to 0$. (Here, $C>0$, $\{\eps_j\}$
is a finite set, $\sum\eps_j=1$, $\eps_j>0$.) Proposition~\ref{PB2023}
is proved.\ \ \ $\Box$

%

Theorem~\ref{TB1007} follows from (\ref{B2021}) and
from Proposition~\ref{PB2023}.\ \ \ $\Box$

Hence the definition (\ref{B1000}) of the determinant
$\det\left(A,a_t\right)$ is correct.

\noindent{\bf Proof of Lemma~\ref{LB3405}.} By (\ref{B2041}), it is
enough to prove that for sufficiently small $\eps_i>0$,
\begin{equation}
\cal{L}\left(\eps_i,A_t\right):=F\left(\ad\left(-c\eps_il_i\right)\right)
\circ\df_t\gamma(t)|_{t=t_i}\quad\text{ is }O(1)
\label{B3430}
\end{equation}
uniformly in $t_i$. Here, $\gamma(t)=\sigma\left(\log\left(A_tA_{t_{i+1}}^{-1}
\right)\right)/c\left(t-t_{i+1}\right)\in r^{-1}(1)\subset S_{\log}(M,E)$,
$\gamma_{t_{i+1}}(t):=\gamma(t)$. Let $l\in r^{-1}(1)\subset S_{\log}(M,E)$
be fixed. Then $\gamma_{t_1}^0(t):=\gamma_{t_1}(t)-l\in CS^0(M,E)$
is defined for any point $\left(t_1,t\right)$ of $I^2$, $I=[0,1]$,
sufficiently close to the diagonal in $I^2$. The derivative
$\df_t\gamma(t)|_{t=t_i}$ in (\ref{B3430}) (where
$\gamma(t):=\gamma_{t_{i+1}}(t)$) is equal
to $\df_t\gamma_{t_{i+1}}^0(t)|_{t=t_i}$ and so it is an element
of $CS^0(M,E)$. The assertion (\ref{B3430}) means that for sufficiently
small $\eps_i$ an element $\cal{L}\left(\eps_i,A_t\right)\in CS^0(M,E)$
is defined and that any semi-norm (\ref{B3233})
$\|\cal{L}\|_{k,\alpha,\omega}$ of $\cal{L}$ with respect to a finite
cover $\left\{U_i\right\}$ of $M$ by coordinate charts is bounded
by $C(k,\alpha,\omega)$ uniformly in $t_i$, $\eps_i$. The derivative
$\df_t\gamma_{t_1}^0(t)$ (for $t$ sufficiently close to $t_1$) exists,
if all the homogeneous components $\left(\gamma_{t_1,t}^0\right)_{-k}(x,\xi)$
of the symbol $\gamma_{t_1}^0(t)\in CS^0(M,E)$ written in local coordinates
$U_i$ are smooth in $t$, $x$, $\xi$, $\xi\ne 0$.

Let us prove that $\df_t\gamma_{t_1}^0(t)$ is $O(1)$ in $CS^0(M,E)$
uniformly in $\left(t_1,t\right)$ from some neighborhood of the diagonal
in $I^2$. The symbol $s_{t_1}(t):=\sigma\left(A_tA_{t_1}^{-1}\right)$
is a solution of the equation
\begin{equation}
\df_ts_{t_1}(t)=\sigma\left(a_t\right)s_{t_1}(t), \quad s_{t_1}\left(t_1
\right)=\Id.
\label{B3466}
\end{equation}
Here, $s_{t_1}(t)\in\SEll_0^{c(t-t_1)}(M,E)$ is a smooth curve
in $\SEll_0^{\times}(M,E)$ (i.e., the curve
$\exp\left(-c\left(t-t_1\right)l\right)s_{t_1}(t)$ is smooth
in $\SEll_0^0(M,E)$). This assertion follows from the Peano
differentiability
theorem for ordinary differential equations (\cite{Ha}, V.~3).
For equations equivalent to (\ref{B3466}) its proof is contained in the proof
of Lemma~\ref{LB3400}.
For small $\left|t-t_1\right|$ the symbol $s_{t_1}(t)$ is close
to $\Id$ on $S^*M$ and
$\sigma\left(\log\left(A_tA_{t_1}^{-1}\right)\right)\in S_{\log}(M,E)$
is defined. The curve
$\beta_{t_1}(t):=\sigma\left(\log\left(A_tA_{t_1}^{-1}\right)\right)$
is smooth in $S_{\log}(M,E)$ for small $\left|t-t_1\right|$,
$\beta_{t_1}^0(t)=\beta_{t_1}(t)-c\left(t-t_1\right)l$ is a smooth
curve in $CS^0(M,E)$, i.e., in local coordinates on $M$ all the homogeneous
components of $\beta_{t_1}(t)$ are smooth in $t$, $t_1$, $x$, $\xi$
for small $\left|t-t_1\right|$ and $\xi\ne 0$. We have
$\df_t\beta_{t_1}^0(t)|_{t=t_1}=c\left(l_{t_1}-l\right)\in CS^0(M,E)$
and $c\left(l_{t_1}-l\right)=a_{t_1}-cl$ is a smooth curve in $CS^0(M,E)$
(under the conditions of (\ref{B999})). So
$\gamma_{t_1}(t)-l=\beta_{t_1}(t)/c\left(t-t_1\right)-l$ is bounded
with respect to any semi-norm (\ref{B3233})
$\left\|\gamma_{t_1}(t)\right\|_{k,\alpha,\omega}$ uniformly
in $\left(t_1,t\right)$ from some small neighborhood of the diagonal
in $I^2$. (Here, $\gamma_{t_1}(t)|_{t=t_1}$ is defined as $l_{t_1}$.)

It is enough to prove that $F\left(-\ad\left(\eps_il_i\right)\right)$
transforms a bounded set $B$ in $CS^0(M,E)$ into a bounded set $B_1$
in $CS^0(M,E)$ for all sufficiently small $\eps_i$ uniformly in $i$,
$l_i$. The operator $\ad\left(\eps_il_i\right)$ acts
on $B\subset CS^0(M,E)$ as $\eps_i\left[l_i,b\right]$. (It is proved
above that $l_i-l$ are uniformly bounded in $CS^0(M,E)$.) Let
$\left\{U_j\right\}$ be a finite cover of $M$ by coordinate charts
and let $\left\{V_j\right\}$, $V_j\subset U_j$, be a subordinate cover
with $\overline{V}_j$ compact in $U_j$. Then
$l_i|_{\overline{V}_j}=\log|\xi|\cdot\Id+f_i|_{\overline{V}_j}$,
where $\xi$ corresponds to local coordinates of a chart $U_j$ and $f_i$
belongs to the restriction to $\overline{V}_j$ of a bounded set
$B\subset CS^0\left(U_j,E|_{U_j}\right)$ uniformly in $l_i$.
So we have
\begin{multline}
\left[l_i,b\right]|_{\overline{V}_j}=\sum_{q,k\in\zuo}\bigl\{\sum_{|\alpha|
\ge 1}{1\over \alpha!}\df_\xi^\alpha\log|\xi|\cdot D_x^\alpha b_{-k}
(x,\xi)+ \\
+\sum_{\alpha\ge 0}\left(\df_\xi^\alpha\left(f_i(x,\xi)\right)_{-q}D_x
^\alpha b_{-k}(x,\xi)-\df_\xi^\alpha b_{-k}(x,\xi)D_x^\alpha\left(f_i
(x,\xi)\right)_{-q}\right)\bigr\}\big|_{\overline{V}_j}.
\label{B3445}
\end{multline}
The symbol $\left[l_i,b\right]$ belongs to $CS^0(M,E)$ and
by (\ref{B3445}) its homogeneous components $\left[l_i,b\right]_{-m}$,
$m\in\wZ_+$, can be estimated as follows.
Let the semi-norm $\left\|\cdot\right\|_N$
in $CS^0(M,E)$ be defined as the sum
$\sum\left\|\cdot\right\|_{k,\alpha,\omega}$ over $(k,\alpha,\omega)$
with $0\le k+|\alpha|+|\omega|\le N$. Then the Fr\'echet structure
given by the semi-norms $\left\|\cdot\right\|_N$, $N\in\zuo$,
on $CS^0(M,E)$ is equivalent to the one given by the semi-norms
$\sum\left\|\cdot\right\|_{k,\alpha,\omega}$. We have by (\ref{B3445})
\begin{equation}
\left\|\left[l_i,b\right]|_{\overline{V}_j}\right\|_N\le C\left(N,B_{U_j}
\right)\left\|b|_{\overline{V}_j}\right\|_N.
\label{B3446}
\end{equation}
Indeed, by (\ref{B3445}) and by Leibniz' formula, the estimate holds
\begin{multline}
\sum_{m+|\beta|+|\omega|\le N}\sup_{x\in\overline{V}_j,\xi\ne 0}
\left(\left\|\df_\xi^\beta D_x^\omega\left[l_i,b\right]_{-m}(x,\xi)\right\|
\left|\xi\right|^{m+\beta}\right)\le \\
\le C_N\left(\sup_{|\xi|\ne 0,1\le|\alpha|\le N}\left|\df_\xi^\alpha\log
|\xi|\right|+\sup_{x\in\overline{V}_j,|\xi|\ne 0,
0\le|\gamma|+|\alpha|+q\le N,q\in\zuo}\df_\xi^\alpha D_x^\gamma\left(f_i
(x,\xi)\right)_{-q}\right)\times\\
\times\sup_{\overset{x\in\overline{V}_j,\xi\ne 0}
{0\le|\gamma|+|\alpha|+|k|\le N}}\left(\left\|
D_x^\gamma\df_\xi^\alpha b_{-k}\right\|\cdot\left|\xi\right|^{k+|\alpha|}
\right)\le
C_N\left(1+\left\|f_i|_{\overline{V}_j}\right\|_N\right)\left\|b
|_{\overline{V}_j}\right\|_N
\label{B3447}
\end{multline}
and $\left\|f_i|_{\overline{V}_j}\right\|_N\le C_N\left(B_{U_j}\right)$
for $f_i$ from a bounded set $B_{U_j}$ in $CS^0(M,E)|_{U_j}$.
So the operator norm of $\ad\left(l_i\right)$ in $CS^0(M,E)$ with respect
to the semi-norm $\left\|\cdot\right\|_N$ in $CS^0(M,E)$ is bounded
by $C(N,L)$, where $L\subset CS^0(M,E)$ is a bounded set such that
all the elements $l_i$ belong to $l+L$ for all $\left\{\eps_k\right\}$.

The action of $F\left(\ad\left(-c\eps_il_i\right)\right)$ on an element
$b\in CS^0(M,E)$ is defined by Remark~\ref{RB621} as
\begin{equation}
F(z)\circ b|_{z=-c\eps_i\ad(l_i)}\equiv\sum_{n\ge 1}{z^{n-1}\over n!}\circ b
|_{z=-c\eps_i\ad(l_i)}.
\label{B3448}
\end{equation}
The operator norm of $z:=-c\eps_i\ad\left(l_i\right)$ in $CS^0(M,E)$
with respect
to the semi-norm $\left\|\cdot\right\|_N$ is bounded by $c\eps_iC(N,L)$.
Hence the operator norm of $F(z)$
in $\left(\!CS^0(\!M,E\!),\left\|\!\cdot\!\right\|_N\!\right)$ is bounded
by $\sum\left(c\eps_iC(N,L)\right)^{n-1}/n!$. This series is convergent
uniformly in $\eps_i$, $0\le\eps_i\le 1$. (Note that this convergence
is not uniform with respect to $N\in\zuo$.) So $F(z)$ is a bounded
operator with respect to all semi-norms $\left\|\cdot\right\|_N$.
It is proved above that $\df_t\gamma(t)|_{t=t_i}$ belongs to a bounded
in $CS^0(M,E)$ set uniformly in $i$, $t_i$, $\eps_i$.
So $F(z)\cdot\df_t\gamma(t)|_{t=t_i}$ is bounded in $CS^0(M,E)$
uniformly in $i$, $t_i$, $\eps_i$. Lemma~\ref{LB3405} is proved.\ \ \ $\Box$

\noindent{\bf Proof of Lemma~\ref{LB3420}.} For sufficiently small
$\eps_i$, the product on the left in (\ref{B3421}) belongs
to $\exp\left(\sfrg_{(l)}\right)$ by the Campbell-Hausdorff formula.
In our case this formula takes the form
\begin{multline}
\log\left(\exp\left(c\eps_i\tilde{l}_{t_i}\right)\exp\left(-c\eps_i\tilde{l}_i
\right)\right)=c\eps_i\left(\tilde{l}_{t_i}-\tilde{l}_i\right)+c_i^2\eps_i^2
\left[\tilde{l}_{t_i}-\tilde{l}_i,-\tilde{l}_i\right]/2+ \\
+c^3\eps_i^3\left(\left[-\tilde{l}_{t_i},\left[\tilde{l}_{t_i}-\tilde{l}_i,
-\tilde{l}_i\right]\right]+\left[-\tilde{l}_i,\left[-\tilde{l}_i,
\tilde{l}_{t_i}-\tilde{l}_i\right]\right]\right)/12+\ldots
\label{B3450}
\end{multline}
By Lemma~\ref{LB3405} $l_{t_i}-l_i$ is
$O\left(\eps_i\right)$ in $CS^0(M,E)$ uniformly in $i$, $t_i$, $\eps_i$,
i.e., $\left(l_{t_i}-l_i\right)/\eps_i$ belongs
to a bounded set $B$ in $CS^0(M,E)$. It is proved in Lemma~\ref{LB3405}
that $\ad\left(l_i\right)$ is a bounded operator
in $\left(CS^0(M,E),\left\|\cdot\right\|_N\right)$ for any $N\in\zuo$.

Note that $\tilde{l}_{t_i}-\tilde{l}_i$ belongs
to $\sfrg_0=(rp)^{-1}(0)\subset\sfrg$.
The identifications $W_{l_1,l_2}\colon\sfrg_{(l_1)}\to\sfrg_{(l_2)}$
transform the Lie subalgebra $\sfrg_0\subset\sfrg_{(l_j)}$ into itself
by Proposition~\ref{PB403}, (\ref{B404}). However these identifications
for general $l_1,l_2\in r^{-1}(1)$ {\em do not} act as $\Id$ on $\sfrg_0$.
By (\ref{B3449}), $\tilde{l}_{t_i}-\tilde{l}_i$ is also
$O\left(\eps_i\right)$ in $\sfrg_0$ with respect to the natural extension
of the semi-norm $\left\|\cdot\right\|_N$ to $\sfrg_0$
for any $N\ge n:=\dim M$.
The operator $\ad\left(\tilde{l}_i\right)$ is also bounded
in $\sfrg_0\subset\sfrg_{(l)}$ with respect to $\left\|\cdot\right\|_N$,
$N\ge n$. Note that by Proposition~\ref{PB403}, (\ref{B404}),
the semi-norm $\left\|\cdot\right\|_N$ on $\sfrg_0\subset\sfrg_{(l)}$
is transformed to an equivalent semi-norm $\left\|\cdot\right\|_N$
on $\sfrg_0\subset\sfrg_{(l_1)}$ under the identification
$W_{l,l_1}\colon\sfrg_{(l)}\rs\sfrg_{(l_1)}$ when $N\ge n$.
So it is enough to show that $\ad\left(\tilde{l}_i\right)$ is bounded
in $\left(\sfrg,\left\|\cdot\right\|_N\right)$, $\sfrg_0\subset\sfrg_{(l)}$.
The element
$W_l^{-1}\left(\tilde{l}_i\right)=W_{l_i,l}\left(\Pi_{l_i}l_i\right)$
is given by
\begin{equation}
W_l^{-1}\left(\tilde{l}_i\right)=\Pi_ll_i+\left(l_i-l,l_i-l\right)_{\res}/2
\cdot 1\in\sfrg_{(l)}.
\label{B3480}
\end{equation}

Any element of $\sfrg_0\subset\sfrg_{(l)}$ is of the form $\Pi_la+c\cdot1$,
where $a\in\sfrg_0=CS^0(M,E)$ and $c\in\wC$. So we have by (\ref{B3480}),
(\ref{B582}), (\ref{B624})
\begin{gather}
\begin{split}
\ad\left(W_l^{-1}\left(\tilde{l}_i\right)\right)\left(\Pi_la+c\cdot 1\right)
& =\Pi_l\left(\left[l_i,a\right]\right)+K_l\left(l_i-l,a\right)\cdot 1, \\
K_l\left(l_i-l,a\right) & :=-\left(\left[l,l_i-l\right],a\right)_{\res}=
\left(\left[l_i,l\right],a\right)_{\res}.
\end{split}
\label{B3481}
\end{gather}

We know that $\ad\left(l_i\right)$ is a bounded operator
in $\sfrg_0:=\left(CS^0(M,E),\left\|\cdot\right\|_N\right)$.
The operator $\left(\left[l_i,l\right],a\right)_{\res}$ is a bounded
linear operator from $\left(\frg_0\ni a,\left\|\cdot\right\|_N\right)$
to $\wC$ for $N\ge n$. Hence $\ad\left(\tilde{l}_i\right)$ is bounded
in $\sfrg_0\subset\sfrg_{(l)}$.

So the first term in (\ref{B3450})
is estimated in the semi-norm $\left\|\cdot\right\|_N$ by $C\eps_i^2$,
the second term is estimated by $C_N\cdot C\eps_i^3/2$, the third one
is estimated by $C_N^2\cdot C\eps_i^4/6$. Hence for sufficiently
small $\eps_i>0$ the series (\ref{B3450}) is convergent with respect
to the semi-norm $\left\|\cdot\right\|_N$ on $\sfrg_0\subset\sfrg_{(l)}$
(because this series is convergent in a neighborhood of zero in a normed
Lie algebra). Its $\left\|\cdot\right\|_N$ semi-norm is estimated
by $C_1\eps_i^2$ uniformly in $\eps_i$ for small $\eps_i$.

However it is difficult to prove the simultaneous convergence
of the series (\ref{B3450}) with respect to all semi-norms
$\left\|\cdot\right\|_N$ (for a fixed small $\eps_i$).%
\footnote{It may be so that there are no $\eps_i>0$ such that the series
(\ref{B3450}) is convergent with respect to all semi-norms
$\left\|\cdot\right\|_N$, $N\in\zuo$, simultaneously.}
But the existence of a logarithm for a given element $g\in G(M,E)$
(i.e., the existence of an element $h\in\sfrg$ such that $\exp(h)=g$)
depends on the properties of the principal symbol $(pg)_{\ord g}$
for the image $pg\in\SEll_0^{\times}(M,E)$ of $g$, Remark~\ref{RB3949}.
(The order
of the expression on the left in (\ref{B3411}) is zero.) The convergence
of the Campbell-Hausdorff series (\ref{B3450}) with respect to semi-norms
$\left\|\cdot\right\|_N$, $N\le N_1$, means that in our case
(for sufficiently small $\eps_i>0$) the first homogeneous terms
$\left(\log(pg)\right)_{-k}(x,\xi)$, $k=0,1,\dots,N_1$, exist and that
$D_x^\omega\df_\xi^\alpha\left(\log(pg)\right)_{-k}(x,\xi)$ exist
for $|\alpha|+|\omega|+k\le N_1$. Hence $\log g\in\sfrg$ exists
in our case for sufficiently small $\eps_i>0$. To obtain the estimate
of $\log g$ by $O\left(\eps_i^2\right)$ with respect to all semi-norms
$\left\|\cdot\right\|_N$ (as $\eps_i$ tends to zero), note that
in our case $\log g\in\sfrg_0\subset\sfrg=W_l\sfrg_{(l)}$ is defined
for small $\eps_i$. So the semi-norms $\left\|\log g\right\|_N$ are defined
for all $N\in\zuo$. For a fixed $N\ge n:=\dim M$ the series (\ref{B3450})
is convergent with respect to $\left\|\cdot\right\|_N$ on $\sfrg_0$
for $0\le\eps_i\le\eps(N)$, $\eps(N)>0$.%
\footnote{Note that $\|a\|_{N_1}\ge\|a\|_{N_2}$ for $N_1\ge N_2\ge 0$.
So this series is convergent with respect to $\left\|\cdot\right\|_N$
for all $N\in\zuo$.}
So by the written above estimates of the terms on the right
in (\ref{B3450}) with respect to $\left\|\cdot\right\|_N$, we see that
the $\left\|\cdot\right\|_N$ semi-norm of the series (\ref{B3450}) is
$O\left(\eps_i^2\right)$ for $0\le\eps_i\le\eps(N)$.

The same estimate can be also produced with the help of ordinary
differential equations. Namely set
$v(t):=\exp\left(t\tilde{l}_{t_i}\right)\exp\left(-t\tilde{l}_i\right)$.
Then we have $v(0)=\Id\in G(M,E)$,
\begin{equation}
\df_tv(t)\!=\!\exp\left(t\tilde{l}_{t_i}\right)\!\left(\tilde{l}_{t_i}\!-
\!\tilde{l}_i\right)\exp\left(-t\tilde{l}_i\right)\!=\!v(t)\exp\left(t
\tilde{l}_i\right)\!\left(\tilde{l}_{t_i}\!-\!\tilde{l}_i
\right)\exp\left(-t\tilde{l}_i\right).
\label{B34501}
\end{equation}
We claim that
\begin{equation}
\left\|v\left(c\eps_i\right)-\Id\right\|_N=O\left(\eps_i^2\right)\quad
\text{ in }\sfrg_0
\label{B3451}
\end{equation}
for all $N\in\zuo$ as $\eps_i$ tends to zero. Here, $\left\|\cdot\right\|_N$
is the operator norm in $\left(\sfrg_0,\left\|\cdot\right\|_N\right)$,
i.e., $\|A\|_N\|f\|_N\ge\|Af\|_N$ for any $f\in\sfrg_0$ and $\|A\|_N$
is the infinum of numbers with such a property. If $\|Af\|_N\not\equiv 0$
on $\sfrg_0$, then $\|A\|_N>0$.

Set
\begin{gather}
\begin{split}
q(t) & :=\exp\left(t\tilde{l}_i\right)\left(\tilde{l}_{t_i}-\tilde{l}_i\right)
\exp\left(-t\tilde{l}_i\right)=:\Ad_{\exp(t\tilde{l}_i)}\left(\tilde{l}_{t_i}-
\tilde{l}_i\right)\in\sfrg_0, \\
\df_tq(t) & =\ad\left(\tilde{l}_i\right)\circ q(t), \quad q(0):=
\tilde{l}_{t_i}-\tilde{l}_i.
\end{split}
\label{B3452}
\end{gather}
It is shown in the proof of Lemma~\ref{LB3405} that the operator
$\ad\left(l_i\right)$ in $\left(CS^0(M,E),\left\|\cdot\right\|_N\right)$
is bounded (since $l_i$ belongs to a bounded set
in $\left(CS^0(M,E),\left\|\cdot\right\|_N\right)$ for any $N\in\zuo$
uniformly in $i$, $t_i$, $\eps_i$).
The operator $\ad\left(\tilde{l}_i\right)$  is also bounded
in $\sfrg_0$ with respect to the natural prolongation
of $\left\|\cdot\right\|_N$ from $CS^0(M,E)$
to $\sfrg_0:=(rp)^{-1}(0)\subset\sfrg_{(l_1)}$. (Here, we suppose that
$N\ge n:=\dim M$.) So by Lemma~\ref{LB3405} and by (\ref{B3452}),
(\ref{B3449}) we have for all $N\ge n$, $0<\eps_i\le\eps(N)$
\begin{equation}
\left\|q(t)\right\|_N\le C_N\left\|\tilde{l}_{t_i}-\tilde{l}_i\right\|_N\le
C'_N\eps_i.
\label{B3453}
\end{equation}
By (\ref{B34501}), (\ref{B3453}) we have%
\footnote{We denote the constants in (\ref{B3453}), (\ref{B3454}), and below
depending only on $N$ by the same symbols $C_N$, $C'_N$, etc..}
\begin{equation}
\df_t\left\|v(t)\right\|_N\le\left\|\df_tv(t)\right\|_N\le C_N\|v(t)\|_N\|
q(t)\|_N.
\label{B3454}
\end{equation}
Here, we use that $\|a\cdot b\|_N\le C_N\|a\|_N\cdot\|b\|_N$
for $a,b\in CS^0(M,E)$. We use also that the analogous estimates hold
for $a\in\exp\left(\sfrg_0\right)\subset\exp\left(\sfrg_{(l_1)}\right)$,
$b\in\sfrg_0:=(rp)^{-1}(0)\subset\sfrg_{(l_1)}$. (In that case
$\|a\|_N$ is the operator norm
in $\left(\sfrg_0,\left\|\cdot\right\|_N\right)$. We have
$\|a\|_N\le\exp\left(\|\alpha\|_N\right)$ for $a=\exp\alpha$,
$\alpha\in\sfrg_0$.)
So the following estimates for the operator norms
in $\left(\sfrg_0,\left\|\cdot\right\|_N\right)$ hold by (\ref{B3454}),
(\ref{B3453})
\begin{gather}
\begin{split}
\left\|v(t)\right\|_N      & \le\left\|v(0)\right\|_N\exp\left(C_NC'_N\eps_it
\right), \\
\left\|\df_tv(t)\right\|_N & \le C_NC'_N\eps_i\exp\left(C_NC'_N\eps_it\right)
\left\|v(0)\right\|_N, \\
\left\|v(t)-v(0)\right\|_N & \le\int_0^t\left\|\df_\tau v(\tau)\right\|
d\tau\le\left\|v(0)\right\|_N\left(\exp\left(C_NC'_N \eps_it\right)-1\right),\\
\left\|v\left(c\eps_i\right)-\Id\right\|_N & \le C''_N\eps_i^2
\end{split}
\label{B3455}
\end{gather}
for $0\le\eps_i\le\eps(0)$, i.e., the estimate (\ref{B3451}) is proved.
Thus Lemma~\ref{LB3420} is proved.\ \ \ $\Box$

\noindent{\bf Proof of Lemma~\ref{LB3400}.} The equation (\ref{B2025})
is equivalent to (\ref{B2028}) with $K_0:=\Id$, $f_t:=l_t-l\in CS^0(M,E)$,
i.e.,
\begin{equation}
\df_tK_t=cu(t)K_t, \quad u(t):=\exp\left(-ct\tilde{l}\right)\left(W_l\left(
\Pi_lf_t+\left(f_t,f_t\right)_{\res}/2\cdot 1\right)\right)\exp\left(ct
\tilde{l}\right).
\label{B3460}
\end{equation}
Here, $K_t\in G^0(M,E):=p^{-1}\left(\SEll_0^0(M,E)\right)$,
$u(t)\in\sfrg_0:=W_l\left(p_l^{-1}CS^0(M,E)\right)$, where
$p_l\colon\sfrg_{(l)}\to S_{\log}(M,E)$ and $p\colon G(M,E)\to\SEll_0^0(M,E)$
are the natural projections. (The Lie subalgebra $\sfrg_0\subset\sfrg$
is independent of $l\in r^{-1}(1)\subset S_{\log}(M,E)$.)

The extension
$p\colon G(M,E)\to\SEll_0^{\times}(M,E)$ is central. So for $k_t:=pK_t$
we have the equation in $\SEll_0^0(M,E)$
\begin{equation}
\df_tk_t=u_1(t)k_t, \quad k_0:=\Id, \quad u_1(t):=c\cdot pu(t).
\label{B3461}
\end{equation}
The coefficient $u_1(t)$ belongs to $CS^0(M,E)$,
$$
u_1(t)=q_0(t;x,\xi)+q_{-1}(t;x,\xi)+\ldots+q_{-m}(t;x,\xi)+\ldots
$$
in local coordinates $x$ on $M$. (Here, $q_{-j}$ is positive homogeneous
of degree $(-j)$ in $\xi$.) The symbol $k_t$ belongs to $\SEll_0^0(M,E)$
and
$$
k_t=k_0(t;x,\xi)+k_{-1}(t;x,\xi)+\ldots
$$
in local coordinates. The symbol is a local notion. So (\ref{B3461})
is equivalent to the system of ordinary equations
\begin{gather}
\begin{split}
\df_tk_0    & =q_0k_0, \\
\df_tk_{-1} & =q_0k_{-1}+q_{-1}k_0+\sum_i\df_{\xi_i}q_0D_{x_i}k_0, \\
\ldots\ldots & \ldots\ldots\ldots\ldots\ldots\ldots\ldots\ldots\ldots\ldots
\ldots \\
\df_tk_{-m} & =q_0k_{-m}+\sum{1\over\alpha!}
\df_\xi^\alpha q_{-r}D_{x_i}^\alpha k_{-j}, \\
\ldots\ldots & \ldots\ldots\ldots\ldots\ldots\ldots\ldots\ldots\ldots\ldots
\ldots
\end{split}
\label{B3462}
\end{gather}
(Here, $r,j\in\zuo$ and the sum on the right for $\df_tk_{-m}$ is over
$(r,j,\alpha)$ with $r+j+|\alpha|=m$,
$|\alpha|+r>0$, and $D_x:=i^{-1}\df_x$.) This system has a triangle form.
Its first equation
(written with respect to a smooth local trivialization of $E$) for fixed
$(x,\xi)$ is a linear equation
\begin{equation}
\df_tk_0(t;x,\xi)=q_0(t;x,\xi)k_0(t;x,\xi), \quad k_0(0;x,\xi)=\Id
\label{B3463}
\end{equation}
on $GL_N(\wC)$, $N:=\rk_{\wC}E$. Its coefficient $q_0(t;x,\xi)$ is smooth
in $t$, $x$, $\xi$ (for $\xi\ne 0$), $0\le t\le 1$. So its solution
$k_0$ is unique and smooth in such $t$, $x$, $\xi$. The second equation
is a linear equation on $M_N(\wC)$ with $k_{-1}(0;x,\xi)=0$ and with known
smooth in $(t,x,\xi)$ for $\xi\ne 0$ coefficients $q_0(t;x,\xi)$ and
$\left(q_{-1}k_0+\sum\df_{\xi_i}q_0D_{x_i}k_0\right)(t;x,\xi)$.
So its solution $k_{-1}(t;x,\xi)$ is unique and smooth in $(t,x,\xi)$
for $\xi\ne 0$, $0\le t\le 1$. The equation for $k_{-m}$ (in (\ref{B3462}))
is also linear in $M_N(\wC)$ with $k_{-m}(0;x,\xi)=0$ and with known
smooth in $(t,x,\xi)$ ($\xi\ne 0$) coefficients. So $k_{-m}(t;x,\xi)$
is unique and smooth in such $t$, $x$, $\xi$. Hence the solution $k_t$
of (\ref{B3461}) exists and is unique, and $k_t\in\SEll_0^0(M,E)$.

Therefore we know $k_t:=pK_t$ and have to find $K_t\in G^0(M,E)$.
Let $K_t^0$, $0\le t\le 1$, be a smooth curve in $G^0(M,E)$ with $K_0^0=\Id$
and $pK_t^0=k_t$. Set $K_t:=K_t^0v_t$. Then
$v_t\in p^{-1}(\Id)\simeq\wC^{\times}\subset G(M,E)$, where $\wC^{\times}$
is a central
subgroup of $G(M,E)$. (The Lie algebra $\wC\cdot 1$ of $\wC^{\times}$ is
$W_l\left(p_l^{-1}(0)\right)$. Note that the identifications $W_{l,l_1}$
restricted to $p_l^{-1}(0)=\wC\cdot 1$ act as $\Id$ on $\wC$.)
The equation (\ref{B3460}) is equivalent to
\begin{equation}
\df_tv_t=\left(-\left(K_t^0\right)^{-1}\df_tK_t^0+\left(K_t^0\right)^{-1}
cu(t)K_t^0\right)v_t, \quad v_0=1\in\wC^{\times}.
\label{B3464}
\end{equation}
The coefficient of this linear equation is a smooth function
$\phi\colon[0,1]\to\wC\cdot 1:=W_l\left(p_l^{-1}(0)\right)$.
Indeed, the image of $\phi(t)$ in $S_{\log}(M,E)$ is
$$
-k_t^{-1}\left(\df_tk_t+u_1(t)k_t\right)=0,
$$
and so $\phi(t)\in\wC\cdot 1\subset\sfrg_0\subset\sfrg$. The curve
$K_t^0$ is a smooth curve in $G^0(M,E)$.
So $-\left(K_t^0\right)^{-1}\df_tK_t^0$ and $\Ad_{(K_t^0)^{-1}}cu(t)$
are smooth curves in $\sfrg_0$ (because $u(t)$ is a smooth curve
in $\sfrg_0$). So $\phi(t)\in\wC\cdot 1\subset\sfrg_0$ is smooth.
The solution of (\ref{B3464}) is $v_t:=\exp\left(\int_0^t\phi(t)dt\right)$.
Hence the equation (\ref{B3460}) has a unique solution $K_t$ and
the equation (\ref{B2025}) (equivalent to (\ref{B3460})) has a unique
solution $F_t$.

We have to prove that the solution $e_0\left(t,\left\{\eps_i\right\}\right)$
of (\ref{B2030}) converges to the solution $F_t$ of (\ref{B2025})
uniformly in $t\in[0,1]$, as $\sup_i\left\{\eps_i\right\}$ tends to zero.
Set
$e_0(t,l):=\exp\left(-ct\tilde{l}\right)e_0\left(t,\left\{\eps_i\right\}
\right)$. (Here, $l\in r^{-1}(1)\subset S_{\log}(M,E)$ is the same
as in (\ref{B3402}), (\ref{B3460}).) Then $e_0(t,l):=e_t$ is the solution
of the equation
\begin{gather}
\begin{split}
\df_te_t & =cx(t)e_t, \quad e_0=\Id, \\
x(t)     & :=\exp\left(-ct\tilde{l}\right)\left(W_l\left(\Pi_lf_{t_i}+\left(
f_{t_i},f_{t_i}\right)_{\res}/2\cdot 1\right)\right)\exp\left(ct\tilde{l}
\right)=u\left(t_i\right)
\end{split}
\label{B3467}
\end{gather}
for $t\in\left(t_i,t_{i+1}\right)=\left(t_i,t_i+\eps_i\right)$. (Here,
$u(t)$ is defined by (\ref{B3460}) and $f_t:=l_t-l$. Recall that
the analogous equation for $K_t=\exp\left(-ct\tilde{l}\right)F_t$ is
$\df_tK_t=cu(t)K_t$, $K_0=\Id$.)

In (\ref{B3467}) $e_t$ is a curve in $G^0(M,E)$ and $x(t)\in\sfrg_0$.
The image $e_t^\sigma=pe_t$ in $\SEll_0^{\times}(M,E)$ of the curve
$e_t$ is the solution of the equation
\begin{equation}
\df_te_t^\sigma=x_1(t)e_t^\sigma, \quad e_0^\sigma=\Id, \quad x_1(t)=c\cdot p
x(t)\in CS^0(M,E).
\label{B3468}
\end{equation}
Here, $x_1(t)$ for $t\in\left(t_i,t_{i+1}\right)$ is equal
to $u_1\left(t_i\right)$, where $u_1(t)$ is the coefficient of the equation
(\ref{B3461}). The symbol $e_t^\sigma$ belongs to $\SEll_0^0(M,E)$ and
in local coordinates it takes the form
$$
e_t^\sigma=m_0(t;x,\xi)+m_{-1}(t;x,\xi)+\ldots,
$$
where $m_{-j}$ is positive homogeneous of degree $(-j)$ in $\xi$.
The symbol is a local notion. So the equation (\ref{B3468}) is equivalent
to the system of the form (\ref{B3462}) with $k_{-j}$ changed by $m_{-j}$
and with $q_{-j}(t;x,\xi)$, $t\in\left(t_i,t_{i+1}\right)$, changed
by $q_{-j}^\eps:=q_{-j}\left(t_i;x,\xi\right)$. Here,
$k_{-j}(0;x,\xi)=q_{-j}(0;x,\xi)=\delta_{j,0}\Id$. The first equations
of these systems
\begin{equation}
\df_tk_0=q_0k_0, \quad \df_tm_0=q_0^\eps m_0
\label{B3469}
\end{equation}
for fixed $(x,\xi)$, $\xi\ne 0$, are linear equations on $GL_N(\wC)$,
$N:=\rk_{\wC}E$. So $k_0^{-1}m_0=:r_0\in GL_N(\wC)$ is the solution
of the equation
\begin{equation}
\df_tr_0(t)=\left(k_0^{-1}\left(q_0^\eps-q_0\right)k_0\right)r_0(t)=:s_0(t)
r_0(t), \quad r_0(0)=\Id.
\label{B3470}
\end{equation}
Here, the coefficients $q_0$ and $q_0^\eps$ are
$$
q_0(t)=\Ad_{\exp(-ctl)}\circ f_t, \quad q_0^\eps(t)=\Ad_{\exp(-ctl)}\circ
f_{t_i}
$$
for $t\in\left(t_i,t_{i+1}\right)$. The symbol $\exp(-ctl)$ belongs
to $\SEll_0^{-ct}(M,E)$ and it can be locally expressed by the symbol $l$
(as in Section~\ref{SA}). So for $\sup_i\left\{\eps_i\right\}$ small
enough, the difference $q_0(t)-q_0^\eps(t)$ is small uniformly
in $t\in[0,1]$, $x$, $\xi$ ($\xi\ne 0$). The same assertion is true
for any finite number of partial derivatives
$\df_\xi^\alpha D_x^\omega\left(q_0(t)-q_0^\eps(t)\right)$, i.e.,
$\left\|q_0(t)-q_0^\eps(t)\right\|_N$ for $0\le N\le N_1$ and any fixed
$N_1\in\wZ_+$ is uniformly small in $t$ as $\sup_i\left\{\eps_i\right\}$
tends to zero. (Here, $\left\|\cdot\right\|_N$ is the same semi-norm
over a local coordinate chart $\overline{V}_i$ as in (\ref{B3446}),
(\ref{B3447}).)

The principal symbol $k_0(t):=k_0(t;x,\xi)$ in (\ref{B3470}) is a fixed
smooth curve in $\SEll_0^0(M,E)/CS^{-1}(M,E)$. So
$\Ad_{k_0^{-1}(t)}\left(q_0^\eps-q_0\right)(t)$ is small uniformly in $t$,
$x$, $\xi$ ($\xi\ne 0$) with respect to semi-norms $\left\|\cdot\right\|_N$
over $\overline{V}_i$ for $0\le N\le N_1$ as $\sup_i\left\{\eps_i\right\}$
tends to zero. Hence by (\ref{B3470}) we claim that for any $\eps>0$,
$N_1\in\wZ_+$, there is $\delta>0$ such that for $0\le N\le N_1$
\begin{equation}
\left\|r_0(t)-\Id\right\|_N<\eps
\label{B3471}
\end{equation}
uniformly in $t\in[0,1]$ as $\sup_i\left\{\eps_i\right\}<\delta$.

The second equations of (\ref{B3462}) and of the analogous system
for $m_{-j}$ are
\begin{gather}
\begin{split}
\df_tk_{-1} & =q_0k_1+\left(q_{-1}k_0+\sum_i\df_{\xi_i}q_0D_{x_i}k_0\right),\\
\df_tm_{-1} & =q_0^\eps m_{-1}+\left(q_{-1}^\eps m_0+\sum_i\df_{\xi_i}q_0^\eps
D_{x_i}m_0\right), \\
m_{-1}(0;x,\xi) & =k_{-1}(0;x,\xi)=0.
\end{split}
\label{B3472}
\end{gather}
For fixed $(x,\xi)$, $\xi\ne 0$, these equations are linear with known
coefficients such that the estimates $\left\|q_0-q_0^\eps\right\|_N<\eps$
and
$$
\left\|\left(q_{-1}k_0+\sum_i\df_{\xi_i}q_0D_{x_i}k_0\right)-\left(q_{-1}^\eps
m_0+\sum_i\df_{\xi_i}q_0^\eps D_{x_i}m_0\right)\right\|_N<\eps
$$
hold uniformly in $t\in[0,1]$ for $0\le N\le N_1$
as $\sup_i\left\{\eps_i\right\}<\delta$. (This assertion is true for any
given $\eps>0$, $N_1\in\wZ_+$, if $\delta$ is sufficiently small.)
{}From (\ref{B3472}) we conclude that $\left\|k_{-1}-m_{-1}\right\|_N$
is small uniformly in $t\in[0,1]$ for $0\le N\le N_1$,
if $\sup_i\left\{\eps_i\right\}$ is sufficiently small.

Let this assertion be true for $\left\|k_{-j}-m_{-j}\right\|_N$,
$0\le N\le N_1$ (with any $N_1\in\wZ_+$), if $0\le j\le a-1$ ($a\in\wZ_+$).
Then with the help of the linear equations for $k_{-a}$ from the system
(\ref{B3462}) and with the help of the analogous equations for $m_{-a}$
we conclude that the same assertion is true
for $\left\|k_{-a}-m_{-a}\right\|_N$ uniformly in $t\in[0,1]$
as $\sup_i\left\{\eps_i\right\}$ tends to zero. Therefore, the solutions
$e_0^\sigma(t)\in\SEll_0^0(M,E)$ of (\ref{B3468}) (for different
$\left\{\eps_i\right\}$) tend to the solution $k(t)$ of (\ref{B3461})
uniformly in $t\in[0,1]$ with respect to all semi-norms
$\left\|\cdot\right\|_N$ as $\sup_i\left\{\eps_i\right\}$ tends to zero.

Set $r_t:=K_t^{-1}e_t\in G^0(M,E)$, $0\le t\le 1$. (We know already
that $pr_t\in\SEll_0^0(M,E)$ tends to $\Id$ uniformly in $t\in[0,1]$ with
respect to $\left\|\cdot\right\|_N$ as $\sup_i\left\{\eps_i\right\}\to 0$.)
The curve $r_t$ is the solution of the equation
\begin{equation}
\df_tr_t=\left(c\Ad_{K_t^{-1}}(x(t)-u(t))\right)r_t, \quad r_0=\Id.
\label{B3473}
\end{equation}
(The coefficient in (\ref{B3473}) belongs to $\sfrg_0$. The semi-norms
$\left\|\cdot\right\|_N$ on $\frg_0:=CS^0(M,E)$ have natural continuations
to the semi-norms $\left\|\cdot\right\|_N$ on $\sfrg_0\subset\sfrg_{(l)}$.)
The coefficient in (\ref{B3473}) is small with respect
to $\left\|\cdot\right\|_N$, $0\le N\le N_1$, uniformly in $t\in[0,1]$
as $\sup_i\left\{\eps_i\right\}$ tends to zero. The projection
$pr_t\in\SEll_0^0(M,E)$ is close to $\Id$ with respect
to $\left\|\cdot\right\|_N$ uniformly in $t\in[0,1]$ under the same
conditions. An element $g\in G^0(M,E)$ has a logarithm in $\sfrg_{(l)}$,
if $pg\in\SEll_0^0(M,E)$ has a logarithm in $\sfrg_0:=CS^0(M,E)$.
Note that $\log pr_t$ exists because $pr_t$ is close to $\Id$
in $\left\|\cdot\right\|_N$, $1\le N\le N_1$.
So the equation
(\ref{B3473}) for $r_t$ can be written as the equation
for $\rho_t:=\log r_t\in\sfrg_0\subset\sfrg_{(l)}$ (by Lemma~\ref{LB616}).

Namely, by (\ref{B617}), (\ref{A228}), the equation (\ref{B3473})
is equivalent to the equation for $\rho_t\in\sfrg_0$
\begin{gather}
\begin{split}
\df_t\rho_t & =cF^{-1}\left(\ad\left(\rho_t\right)\right)\Ad_{K_t^{-1}}(x(t)-
u(t)), \quad \rho_0=0, \\
F^{-1}\left(\ad\left(\rho_t\right)\right) & :={z\over \exp z-1}\big|
_{z=\ad(\rho_t)}\!=\!1\!-\!z/2\!-\!\sum_{k\ge 1}{\zeta(1-2k)\over(2k-1)!}z
^{2k}\big|_{z=\ad(\rho_t)}.
\end{split}
\label{B3482}
\end{gather}
(The series $F^{-1}(z)$ is convergent for $|z|<2\pi$.)
Taking into account (\ref{B3482}) we conclude that
\begin{equation}
\df_t\left\|\rho_t\right\|_N\le\left\|\df_t\rho_t\right\|_N\le c\left\|F^{-1}
\left(\ad\left(\rho_t\right)\right)\right\|_N\left\|\Ad_{K_t^{-1}}(x(t)-y(t))
\right\|_N.
\label{B3485}
\end{equation}

One can try to prove that $\left\|\rho_t\right\|_N$ is small
for $t\in[0,1]$, $0\le N\le N_1$ using the estimates analogous
to (\ref{B3485}) and the Picard approximations. However we prefer to use
the structure of (\ref{B3482}) and the information about
$\left\|p\rho_t\right\|_N$.

Note that $\ad\left(a_1\right)=\ad\left(a_2\right)$ in $\sfrg_0$
(for $a_j\in\sfrg_0$), if $a_1-a_2$ belongs to the central Lie subalgebra
$\wC\cdot 1$ of $\sfrg_0$. So $\ad\left(\rho_t\right)$ (as an operator
in $\sfrg_0$) depends on $p\rho_t\in\frg_0$ only.
Set $\ad\left(p\rho_t\right):=\ad\left(\rho'_t\right)$ for any $\rho'_t$
with $p\rho'_t=p\rho_t$. We know the solution $p\rho_t$ of the equation
in $\frg_0$ which is the projection of (\ref{B3482}) to $\frg_0$.
Namely $p\rho_t$ is the solution of the equation
\begin{multline}
\df_t\left(p\rho_t\right)=cF^{-1}\left(\ad\left(p\rho_t\right)\right)p\left(
\Ad_{K_t^{-1}}(x(t)-u(t))\right)\equiv \\
\equiv F^{-1}\left(\ad\left(p\rho_t\right)\right)\Ad_{k_t^{-1}}\left(x_1(t)-
u_1(t)\right),
\label{B3490}
\end{multline}
and we know that $\left\|p\rho_t\right\|_N$ is small uniformly
in $t\in[0,1]$, $N$ for $0\le N\le N_1$ as $\sup_i\left\{\eps_i\right\}$
is small enough. Let $l\in r^{-1}(1)\subset S_{\log}(M,E)$ be fixed.
Then the equation (\ref{B3482}) in $\sfrg_{(l)}$ written with respect
to the splitting (\ref{B622}) is
\begin{equation}
\df_t\rho_t=F^{-1}\left(\ad\left(p\rho_t\right)\right)\left(p\left(\Ad_{K_t
^{-1}}(x(t)-u(t))\oplus f(t)\cdot 1\right)\right),
\label{B3491}
\end{equation}
where $f\colon[0,1]\to\wC$ is a smooth function and $|f(t)|$ is small
uniformly in $t\in[0,1]$ as $\sup_i\left\{\eps_i\right\}$ is small
enough. We know that $\left\|\ad\left(p\rho_t\right)\right\|_N$ is small
in $\frg_0$ for $N\le N_1$. So it is small also in $\sfrg_0$
for $n:=\dim M\le N\le N_1$. Set $\rho_t=p\rho_t\oplus w_t\cdot 1$,
$w_t\in\wC$, with respect to the splitting (\ref{B622}). Then in view
of (\ref{B3490}), (\ref{B3491}) we conclude that $w_t$ is the solution
of an ordinary differential equation
\begin{equation}
\df_tw_t=f(t)+\left(p\rho_t,\left[l,\left(1-z/2-\sum{\zeta(1-2k)\over(2k-1)!}
z^{2k-1}\right)\big|_{z=\ad(p\rho_t)}u_t\right]\right)_{\res}.
\label{B3492}
\end{equation}
Here, $u_t:=\Ad_{k_t^{-1}}\left(x_1(t)-u_1(t)\right)$ and
$\ad\left(p\rho_t\right)$ in (\ref{B3492}) acts on $\frg_0:=CS^0(M,E)$.

To prove that $\rho_t$ is small
in $\left(\sfrg_0,\left\|\cdot\right\|_N\right)$ for $0\le N\le N_1$,
it is enough to prove that $|w_t|$ is small for $t\in[0,1]$,
if $\sup_i\left\{\eps_i\right\}$ is small enough. We know that $p\rho_t$
and $u_t$ are small in $\left(\frg_0,\left\|\cdot\right\|_N\right)$
for $0\le N\le N_1$. It is proved in (\ref{B3446}) that $\ad(l)$ is
a bounded operator in $\frg_0$ with respect to $\left\|\cdot\right\|_N$.
The operator $\ad\left(p\rho_t\right)$
in $\left(\frg_0,\left\|\cdot\right\|_N\right)$ has the norm
$\left\|\ad\left(p\rho_t\right)\right\|_N$ not greater than
$C_N\left\|p\rho_t\right\|_N$ by (\ref{B3447}). So the series on the right
in (\ref{B3492}) is convergent and the estimate
\begin{equation}
\left\|\left[l,\left(1-z/2-\sum{\zeta(1-2k)\over(2k-1)!}z^{2k-1}\right)
\big|_{z=\ad(p\rho_t)}u_t\right]\right\|_N\le C'_N\left\|u_t\right\|_N
\label{B3494}
\end{equation}
is valid for $\sup_i\left\{\eps_i\right\}$ small enough. We have
by (\ref{B3492}),
by the estimate $\left|(a,b)_{\res}\right|\le\|a\|_N\|b\|_N$
for $a,b\in\frg_0$, $N\ge n$, and by (\ref{B3494})
$$
|w(t)|\le\int_0^t|f(\tau)|d\tau+C'_N\int_0^t\left\|p\rho_\tau\right\|_N
\left\|u_\tau\right\|_Nd\tau
$$
for any $N\ge n$. So $|w(t)|$ is small for $t\in[0,1]$,
if $\sup_i\left\{\eps_i\right\}$ is small enough. Hence for such
$\left\{\eps_i\right\}$ the logarithm $\rho_t$ of $r_t$ is small
in $\left(\sfrg_0,\left\|\cdot\right\|_N\right)$ for all $t\in[0,1]$,
$0\le N\le N_1$. Thus
%
%
%
$r_t\in G^0(M,E)$ is uniformly in $t\in[0,1]$
close to $\Id\in G^0(M,E)$ with respect to all $\left\|\cdot\right\|_N$
as $\sup_i\left\{\eps_i\right\}$ tends to zero. Lemma~\ref{LB3400}
is proved.\ \ \ $\Box$

\subsection{Connections on determinant bundles given by logarithmic
symbols. Another determinant for general elliptic PDOs}
\label{S81}

Let $l$ be the symbol of $\log_{(\theta)}A$, where $A\in\Ell_0^1(M,E)$
and $L_{(\theta)}$ is an admissible (for $A$) cut of the spectral plane.
Then the central $\wC^{\times}$-extension $\sfrg_{(l)}$ of the Lie algebra
$S_{\log}(M,E)=:\frg$ is defined and $l$ also defines the splitting
(\ref{B622})
\begin{equation}
\sfrg_{(l)}=\frg\oplus\wC\cdot 1.
\label{B1901}
\end{equation}

Theorem~\ref{TB570} provides us with a canonical isomorphism between
$\sfrg_{(l)}$ and the Lie algebra $\frg(M,E)$ of the determinant Lie
group $G(M,E)$. Hence the splitting (\ref{B1901}) defines a connection
on the $\wC^{\times}$-bundle $p\colon G(M,E)\to\SEll_0^{\times}(M,E)$.
Namely a local smooth curve $g_t\in G(M,E)$, $t\in[-\eps,\eps]$,
is horizontal with respect to this connection, if $\dot{g}_t\cdot g_t^{-1}$
belongs to the subspace $\frg$ of $\frg(M,E)=\sfrg_{(l)}$ with respect
to the splitting (\ref{B1901}).

Let an operator $B\in\Ell_0^{\times}(M,E)$ be fixed.%
\footnote{In this subsection we don't suppose that $B$ has a real or
a nonzero order.}
There exists a smooth curve $b_t\in S_{\log}(M,E)$, $t\in[0,1]$, such
that the symbol $\sigma(B)$ is equal to the value at $t=1$ of the solution
$s_t$ of the equation in $\SEll_0^{\times}(M,E)$
\begin{equation}
\df_t s_t=b_t s_t, \quad s_0=\Id.
\label{B1902}
\end{equation}

Let $b_t$ be such a curve in $\SEll_0^{\times}(M,E)$. Then for a fixed
$l$ ($=\sigma\left(\log_{(\theta)}A\right)$, $A\in\Ell_0^1(M,E)$)
the connection on $G(M,E)$ defined by $l$ gives us a canonical pull-back
of the curve $s_t\subset\SEll_0^{\times}(M,E)$ to the curve
$\tilde{s}_t\subset G(M,E)$. This curve $\tilde{s}_t$ is the solution
of the equation in $G(M,E)$
\begin{equation}
\df_t\tilde{s}_t=\left(\Pi_{(l)}b_t\right)\cdot\tilde{s}_t, \quad \tilde{s}_t=
\Id,
\label{B1904}
\end{equation}
where $\Pi_{(l)}\colon\frg\hookrightarrow\sfrg_{(l)}\!=\!\frg(M,E)$
is the inclusion with respect to the splitting (\ref{B1901}) (and
to the canonical identification given by Theorem~\ref{TB570}). \\

\noindent{\bf Definition.} Let the operator $B\in\Ell_0^{\times}(M,E)$,
the logarithmic symbol $l$ of a first-order elliptic PDO $A\in\Ell_0^1(M,E)$,
and a curve $b_t$ in $S_{\log}(M,E)$ such that the solution $s_t$
of (\ref{B1901}) is equal to $\sigma(B)$ at $t=1$, be fixed. Then
the determinant of $B$ is defined by
\begin{equation}
\det\left(B,\left(l,b_t\right)\right):=d_1(B)/\tilde{s}_1.
\label{B1903}
\end{equation}
Here, $d_1(B)$ is the image of $B\!\in\!\Ell_0^{\times}(M,E)$ in the quotient
$G(M,E)\!:=\!F_0\!\backslash\!\Ell_0^{\times}(M,E)$, where the normal subgroup
$F_0$ of $\Ell_0^{\times}(M,E)$ is defined by (\ref{B600}). The term
$\tilde{s}_1$ in (\ref{B1903}) is the value at $t=1$ of the solution
$\tilde{s}_t$ for (\ref{B1904}) with the coefficient $b_t$.

\begin{rem}
This determinant is invariant under smooth reparametrizations of a curve
$b_t$. This determinant is defined for any smooth curve $s_t$, $0\le t\le1$,
in $\SEll_0^{\times}(M,E)$ such that $s_0=\Id$, $s_1=\sigma(B)$. (Here,
$b_t':=\df_ts_t\cdot s_t^{-1}$.)
\label{RB3001}
\end{rem}

\begin{rem}
We have $p\tilde{s}_t=s_t$, where $p\colon G(M,E)\to\SEll_0^{\times}(M,E)$
is the natural projection. Hence $\tilde{s}_1\in p^{-1}(\sigma(B))$.
We have $\det\left(B,\left(l,b_t\right)\right)\in\wC^{\times}$
since the fibers
of $p$ are principal homogeneous $\wC^{\times}$-spaces because
$F_0\backslash F=\wC^{\times}$ and because $F,F_0$ are normal subgroups
in $\Ell_0^{\times}(M,E)$ (defined in Section~\ref{SE}).
\label{RB1905}
\end{rem}

\begin{rem}
The determinant $\det\left(B,\left(l,b_t\right)\right)$ depends
on a curve $b_t$ in the space of logarithmic {\em symbols}
(in contrast with a curve $a_t$ from the definition (\ref{B1000}),
$a_t\subset\fell(M,E)$, i.e., it is a curve in the space of logarithms
for classical elliptic PDOs). The determinant
$\det\left(B,\left(l,b_t\right)\right)$ is defined for all classical
elliptic PDOs, not only for PDOs of real nonzero orders. In contrast,
the determinant (\ref{B1000}) is defined for PDOs of real nonzero orders.
\label{RB1906}
\end{rem}

\begin{rem}
Let a logarithmic symbol $l_1=\sigma\left(\log_{(\theta_1)}A_1\right)$
be fixed. (Here, $A_1\in\Ell_0^1(M,E)$ and $L_{(\theta_1)}$ is
an admissible for $A_1$ cut of the spectral plane.) Then by (\ref{B580}),
we have
\begin{multline}
\tilde{s}_1(l_1)/\tilde{s}_1(l)=\exp\left(\int_0^1 dt\left(\Pi_{(l_1)}b_t-
\Pi_{(l)}b_t\right)\right)=\\
=\exp\left(\int_0^1 dt\left(b_t-r\left(b_t\right)\left(l_1+l\right)/2,l-l_1
\right)_{\res}\right)=:\exp f\left(b_t;l,l_1\right).
\label{B1908}
\end{multline}
Note that $f\left(b_t;l,l_1\right)$ is the integral over $M\times[0,1]$
of a density locally defined by the symbols of $b_t$, $l$, and of $l_1$.
By (\ref{B1908}) we have
\begin{equation}
\det\left(B,\left(l,b_t\right)\right)\big/\det\left(B,\left(l_1,b_t\right)
\right)=\exp f\left(b_t;l,l_1\right).
\label{B1909}
\end{equation}
\label{RB1907}
\end{rem}

\begin{rem}
By (\ref{B582}) we have the formula for a curvature of the connection
defined by $l=\sigma(\log A)$, $A\in\Ell_0^1(M,E)$,
on the $\wC^{\times}$-bundle
$p\colon G(M,E)\to\SEll_0^{\times}(M,E)$. Namely,
if $\dot{g}_1,\dot{g}_2\in T_g\left(\SEll_0^{\times}(M,E)\right)$ are
two tangent vectors, then the value of the curvature form is given by
\begin{equation}
R_l\left(\dot{g}_1,\dot{g}_2\right)=K_l\left(\dot{g}_1 g^{-1},\dot{g}_2
g^{-1}\right),
\label{B1911}
\end{equation}
where $K_l$ is the $2$-cocycle on $S_{\log}(M,E)$ defined by (\ref{B304})
(and by Lemma~\ref{LB1590}), $\dot{g}_j g^{-1}\in S_{\log}(M,E)=:\frg$.
Let $b_t$ and $b'_t$, $t\in[0,1]$, be two curves in $\frg$ such that
the solutions of (\ref{B1902}) with the coefficients $b_t$ and $b'_t$
have $\sigma(B)$ as their values at $t=1$ and are homotopic curves
in $\SEll_0^{\times}(M,E)$ from $\Id$ to $\sigma(B)$. Then we have
\begin{equation}
\tilde{s}_1\left(b'_t\right)\big/\tilde{s}_1\left(b_t\right)=\exp\left(\int
_{D^2}\phi^*R_l\right),
\label{B1912}
\end{equation}
where $R_l$ is defined by (\ref{B1911}) and
$\phi\colon D^2\to\SEll_0^{\times}(M,E)$ is a smooth homotopy between
$s\left(b'_t\right)$ and $s\left(b_t\right)$ in $\SEll_0^{\times}(M,E)$.
Note that $R_l$ is a $2$-form on $\SEll_0^{\times}(M,E)$ with the values on
$\left(\dot{g}_1,\dot{g}_2\right)\in T_g\left(\SEll_0^{\times}(M,E)\right)$
given by an integral over $M$ of a density locally defined by the symbols
$g$, $g_j$, $l$. We have by (\ref{B1903}), (\ref{B1912})
\begin{equation}
\det\left(B,\left(l,b_t\right)\right)\big/\det\left(B,\left(l,b'_t\right)
\right)=\exp\left(\int_{D^2}\phi^*R_l\right)
\label{B1914}
\end{equation}
with the same meaning of $\phi$ as in (\ref{B1912}). By Remarks~\ref{RB1907},
\ref{RB1910}, we can control the dependence of the integral (\ref{B1903})
on $l$ and on curves $s_t$, $s'_t$ in $\SEll_0^{\times}(M,E)$ from $\Id$
to $\sigma(B)$ from the same homotopy class.
\label{RB1910}
\end{rem}

\begin{rem}
Let $B_1,B_2\in\Ell_0^{\times}(M,E)$ and let $s_1(t)$ and $s_2(t)$
be smooth curves from $\Id$ to $\sigma\left(B_1\right)$ and
to $\sigma\left(B_2\right)$ in $\SEll_0^{\times}(M,E)$.
Set $b_{j,t}:=\df_ts_j(t)$. Let the logarithmic symbol $l=\sigma(\log A)$,
$A\in\Ell_0^1(M,E)$, be fixed. Then we have
\begin{gather}
\begin{split}
d_1\left(B_2B_1\right) & =d_1\left(B_2\right)d_1\left(B_1\right), \\
\widetilde{s_2s_1}     & =\tilde{s}_2\tilde{s}_1.
\end{split}
\label{B1916}
\end{gather}
The latter equality follows from (\ref{B1902}), (\ref{B1904}). Hence
in view of $d_1/\tilde{s}_1\in\wC^{\times}$, we have
\begin{equation}
\det\left(B_2B_1,\left(l,\left(b_{2,t}\cup b_{1,t}\right)\right)\right)=
\det\left(B_1,\left(l,b_{1,t}\right)\right)\det\left(B_2,\left(l,b_{2,t}
\right)\right).
\label{B1917}
\end{equation}
Here, $b_{2,t}\cup b_{1,t}$ corresponds to a piecewise-smooth curve
$s_2\cup s_1$ from $\Id$ to $\sigma\left(B_2B_1\right)$ through
$\sigma\left(B_1\right)$ which coincides with $s_1(2t)$ for $t\in[0,1/2]$
and with $s_2(2t-1)$ for $t\in[1/2,1]$.
\label{RB1915}
\end{rem}

It follows from Remarks~\ref{RB1910}, \ref{RB1915} that to investigate
the dependence of the determinant $\det\left(B,\left(l,b_t\right)\right)$
on the homotopy
class of a smooth curve $s_t$ from $\Id$ to $\sigma(B)$
in $\SEll_0^{\times}(M,E)$, it is enough to compute
\begin{equation}
\det(\Id,(l,2\pi ip))=:k(p,l)
\label{B1918}
\end{equation}
for projectors $p\!\in\! C\!S_0^0(\!M,E\!)$, $p^2\!=\!p$, in the algebra
$C\!S_0^0$
of classical PDO-symbols of order zero. Each of these projectors
corresponds to a cyclic subgroup $\exp\!(\!2\pi itp\!)$,
$0\!\le\! t\!\le\!1$,
in $\SEll_0^0(\!M,E\!)$. Such subgroups span the fundamental group
$\pi_1\!\left(\SEll_0^0(\!M,E\!),\Id\!\right)$.
This statement is proved in the proof of Lemma~\ref{LB1810}
in Section~\ref{SC4}.

\begin{rem}
To compute (\ref{B1918}), we use Proposition~\ref{PB840}). Namely
we have
\begin{multline}
\det(\Id,(l,2\pi ip)):=\Id\cdot\exp\left(-2\pi i\Pi_{(l)}p\right)=
d_1(\exp(2\pi iP))\exp(-f(2\pi iP,A))=\\
=\exp(-2\pi if(P,A)).
\label{B1920}
\end{multline}
Here, $P$ is a PDO-projector $P\in CL^0(M,E)$, $P^2=P$, with $\sigma(P)=p$.
(Such a projector $P$ exists by \cite{Wo3}.) The operator $A$
in (\ref{B1920}) is an invertible elliptic PDO, $A\in\Ell_0^1(M,E)$,
with its symbol $\sigma(A)$ equal to $\exp l$. The spectral $f(P,A)$
of a pair $(P,A)$ is defined by (\ref{B844}). Hence
\begin{gather}
\begin{split}
\det(\Id,(l,2\pi ip))    & =\exp(-2\pi if_0(p,\exp l)),\\
f_0(p,\exp l)\in\wC/\wZ, & \quad f_0(p\exp l)\equiv f(P,A)(\mod\wZ).
\end{split}
\label{B1921}
\end{gather}

By Lemma~\ref{LB845} the generalized spectral asymmetry $f(P,A)(\mod\wZ)$
depends on symbols $\sigma(P)=p$, $\sigma(A)=\exp l$ only.
\label{RB1919}
\end{rem}

\noindent Remarks~\ref{RB1910}, \ref{RB1915}, \ref{RB1919} express
the dependence of the determinant
 $\det\!\left(B,\!\left(l,\!b_t\right)\!\right)$
on $b_t$ and on $l$ through generalized spectral asymmetries
$f_0(p,\exp l))$, $p^2=p$, $p\in CS^0(M,E)$, and through the integrals
(\ref{B1908}), (\ref{B1914}) of densities locally canonically defined
by homogeneous terms of symbols in arbitrary coordinate charts.

\subsection{The determinant defined by a logarithmic symbol as an extension
of the zeta-regularized determinant}

\begin{rem}
For $A\in\Ell_0^1(M,E)$, for $l\in S_{\log}(M,E)$ such that
$\exp l=\sigma(A)$, and for $b_t\equiv l$, we have $s_1=\sigma(A)$
(where $s_t$ is the solution of (\ref{B1902})). Hence
\begin{equation}
\det(A,(l,l)):=d_1(A)/\exp\left(\Pi_{(l)}l\right)=:d_1(A)/\tilde{A},
\label{B1923}
\end{equation}
where $\tilde{A}$ is defined by (\ref{B640}).

We suppose that there exists $l\in S_{\log}(M,E)$ such that
$\exp l=\sigma(A)$. Hence the symbol $\exp(\eps l)$ for $\eps\in\wR_+$
small enough is sufficiently close to a positive definite symbol.
Hence $B:=A^{\eps}$ possesses a spectral cut $L_{(\tpi)}$ close
to $L_{(\pi)}$ and $\zeta_{B,(\tpi)}(s)$ is defined. Set
\begin{equation}
{\det}_{(\tpi)}(A):=\exp\left(-\eps^{-1}\df_s\zeta_{B,(\tpi)}(s)|_{s=0}\right).
\label{B1926}
\end{equation}
By Proposition~\ref{PB627} the element
\begin{equation}
d_0(A):=d_1(A)/{\det}_{(\tpi)}(A)\in p^{-1}(\exp l)
\label{B1924}
\end{equation}
depends on $\sigma(A):=\exp l$ only. Here,
$p\colon G(M,E)\to\SEll_0^{\times}(M,E)$ is the natural projection.
Hence by (\ref{B1923}), (\ref{B1924}) we have
\begin{equation}
\det(A,(l,l))={\det}_{(\tpi)}(A)\cdot d_0(\exp l)/\exp\left(\Pi_{(l)}l\right).
\label{B1925}
\end{equation}
The elements $d_0(\exp l)$ and $\exp\left(\Pi_{(l)}l\right)$ correspond
one to another under the local identification of the Lie groups $G(M,E)$
and $\exp\left(\sfrg\right)\equiv\exp\left(\sfrg_{(l)}\right)$ given
by Theorem~\ref{TB570}. Hence we obtain the assertion as follows.
\label{RB1922}
\end{rem}

\begin{pro}
Let $A\in\Ell_0^1(M,E)$ have a logarithmic symbol $l\in S_{\log}(M,E)$,
i.e., $\sigma(A)=\exp l$, where $\exp l$ is defined as the value
at $\tau=1$ of the solution of the equation in $\SEll_0^{\times}(M,E)$
$$
\df_\tau A_\tau=lA_\tau,\qquad A_0=\Id.
$$
Then the equality holds
\begin{equation}
{\det}_{(\tpi)}(A)=\det(A,(l,l)),
\label{B1928}
\end{equation}
where the zeta-regularized determinant ${\det}_{(\tpi)}(A)$ is defined
by (\ref{B1926}) for $B:=A^\eps$ with $\eps\in\wR_+$ such that $A^\eps$
possesses a spectral cut $L_{(\tpi)}$ close to $L_{(\pi)}$. The determinant
on the right in (\ref{B1928}) is the determinant (\ref{B1903}) with
$b_t\equiv l$ for $t\in[0,1]$, where $A$ is substituted instead of $B$.
\label{PB1927}
\end{pro}

\begin{rem}
Let $A\in\Ell_0^d(M,E)$ be an elliptic operator of a real nonzero order
$d(A)$ such that there exists a logarithmic symbol $d(A)l\in S_{\log}(M,E)$
of $A$, $\exp(d(A)l)=\sigma(A)$. Then $\det_{(\tpi)}(A)$ in the sense
of (\ref{B1926}) is defined (and it is independent of a sufficiently
small $\eps\in\wR_+$. The term $\det(A,(l,d(A)l))$ (i.e., the determinant
(\ref{B1903}) with $b_t\equiv d(A)l$) is also defined. The equalities
hold (analogous to (\ref{B1925}))
\begin{equation}
\det(A,(l,d(A)l))\!=\!{\det}_{(\tpi)}(A) d_0(\exp(d(A)l))\big/\!\exp\!\left(
\Pi_{(l)}d(A)l\right)\!=\!{\det}_{(\tpi)}(A),
\label{B1930}
\end{equation}
since $d_0(\exp(d(A)l))$ corresponds
to $\exp\left(\Pi_{(l)}d(A)l\right)$
under the local identification \\
$G(M,E)=\exp\left(\sfrg\right)$
given by Theorem~\ref{TB570}. Hence the determinant
$\det(A,(l,d(A)l))$ given by (\ref{B1903}) for an elliptic PDO $A$
of a real nonzero order $d(A)$
(and such that a logarithmic symbol $d(A)\cdot l$ of $A$ exists) is equal
to the zeta-regularized determinant $\det_{(\tpi)}(A)$.
\label{RB1929}
\end{rem}

Thus the determinant (\ref{B1903}) gives us {\em an extension
of the zeta-regularized determinant} $\det_\zeta(A)$ to the class
of general elliptic PDOs $\Ell_0^{\times}(M,E)$ of {\em all complex
orders} from the connected component of the operator $\Id\in\Ell_0^0(M,E)$.
Note that the determinant (\ref{B1903}) depends not only on $A$ and on $l$
but also on an appropriate curve $b_t$, $t\in[0,1]$, in the Lie algebra
$S_{\log}(M,E)$ of logarithmic symbols.

\subsection{Determinants near the domain where logarithms of symbols
do not exist}
\label{SS93}

Let $A(z)\in\Ell_0^{\alpha(z)}(M,E)$ be a holomorphic family of elliptic
PDOs of order $\alpha(z)$. We suppose that $\alpha(z)\in\wC^{\times}$.
Here, $z$ belongs to a one-connected neighborhood $U$
of $I:=[0,1]\subset\wC\ni z$. Let for $z\in\left[0,z_0\right)$
a logarithm of $\sigma(A(z))$ exist. We are interested in the asymptotic
behavior as $z\to z_0$ of determinants of $A(z)$
We claim that there is a locally defined by the symbols $\sigma(A(z))$,
$\sigma(\log A(z))$ object which controls $\det(A(z))$ as $z\to z_0$
along $I$.

Namely let $l\in r^{-1}(1)\subset S_{\log}(M,E)$ be a logarithmic
symbol of order one ($r$ is from (\ref{B303})). Then $l$ defines
the splitting (\ref{B1901}) of $\sfrg:=W_l\sfrg_{(l)}$. Hence
a connection on the $\wC^{\times}$-bundle $G(M,E)$ over
$\SEll_0^{\times}(M,E)$ is defined by $l$. A vector
$\dot{g}(t)\in T_{g(t)}G(M,E)$ belongs to a horizontal subspace,
if $\dot{g}(t)g^{-1}(t)\in W_l\frg$. (Here, $\frg:=S_{\log}(M,E)$
is identified with the image of $\frg$ in $\sfrg_{(l)}$ under the splitting
(\ref{B1901}).)

The section $U\to G(M,E)$, $U\ni z\to d_1(A(z))\in G(M,E)$, over
$U\ni z\to\sigma(A(z))\in\SEll_0^{\times}(M,E)$ is defined.%
\footnote{The element $d_1(A)$ is the image of $A\in\Ell_0^{\times}(M,E)$
in $G(M,E):=F_0\backslash\Ell_0^{\times}(M,E)$, Section~\ref{SD}.}
It is holomorphic in $z\in U$. Let $f_0(z)\colon U\ni z\to G(M,E)$
be another section of $p\colon G(M,E)\to\SEll_0^{\times}(M,E)$
which is a holomorphic curve in $G(M,E)$ horizontal with respect
to the connection defined by $l$ and such that $f_0(A(0))=d_1(A(0))$,
$0\in U$. (Note that this connection is holomorphic. Thus such
a holomorphic curve exists and is unique.)

Then $d_1(z)/f_0(z)\in\wC$ is a {\em holomorphic function} of $z\in U$
and $f_0(z)$ is {\em locally defined by the symbols} $\sigma(A(z))$
of our family. (We suppose here that
$d_1(A(0))$ is known. For example, if $A(0)=\Id\in\Ell_0^{\times}(M,E)$,
then $d_1(A(0))=\Id\in G(M,E)$.)

Let a $\log A(z)\in\fell(M,E)$ exist. Then by Remark~\ref{RB4} and
by Propositions~\ref{PB3755}, \ref{PB3822} we have%
\footnote{$\Tr$ is the canonical trace for PDOs of noninteger orders
defined in Section~\ref{SB}. By Proposition~\ref{PB3755} the residue
of the zeta-function on the right in (\ref{B3902}) at $s=0$ is
$-\res(\Id)=0$. Hence the expression on the right in (\ref{B3902})
is defined.}
\begin{equation}
{\det}_\zeta(A(z)):=\exp\left(-\df_s\TR\exp(-s\log A(z))\big|_{s=0}\right).
\label{B3902}
\end{equation}
Let $\sigma(\log A(z))=\alpha(z)l+a_0(z)$, where $l$ is a logarithmic
symbol of an order one elliptic PDO, $a_0(z)\subset CS^0(M,E)$
is holomorphic in $z$ for $z\in[0,z_0)$, and $a_0(z)$ diverges
as $z\to z_0$.

By Proposition~\ref{PB3952}, by Corollary~\ref{CB3956}, and by (\ref{B3943})
we have a section
$$
S\to\tilde{d}_0(\sigma(\log A(z)))
$$
of the $\wC^{\times}$-bundle $G(M,E)$
over $S=S(z):=\sigma(A(z))$, $z\in[0,z_0)$, depending on $\sigma(\log A(z))$
only. If $\log A(z)$ exists, then by the definition
of $\tilde{d}_0(\sigma(\log A(z)))$ the zeta-regularized determinant
(\ref{B3902}) is equal to
\begin{equation}
{\det}_\zeta(A(z))=d_1(A(z))/\tilde{d}_0(\sigma(\log A(z))).
\label{B3903}
\end{equation}
Here, $d_1(A)$ is defined as the class $F_0A$
in the determinant Lie group $G(M,E)=F_0\backslash\Ell_0^{\times}(M,E)$.

However the canonical determinant $\det(A)$ is defined for more wide
class of elliptic PDOs than the class of PDOs $A$ such that $\log A$
exists, Remark~\ref{RB3946}, (\ref{B3947}). Namely if
$\sigma(\log A)\in S_{\log}(M,E)$ is defined, then
\begin{equation}
\det(A):=d_1(A)/\tilde{d}_0(\sigma(\log A)).
\label{B3960}
\end{equation}
This determinant can be defined even if the zeta-regularized determinant
$\det_\zeta(A)$ is not defined, Remark~\ref{RB3946}. (The definition
(\ref{B3960}) does not use $\log A$. However $\log A$ is defined,
if $\zeta_A(s)$ exists.)

\begin{pro}
There is a scalar function
$$
B(z):=\tilde{d}_0(\sigma(\log A(z)))/f_0(\sigma(A(z)))
$$
holomorphic in $z\in[0,z_0)$ and defined by symbols (and by logarithmic
symbols) of our holomorphic family and such that the divergence
of the canonical determinant $\det(A(z))$ as $z\to z_0$ along $I$
is defined by the behavior of $B(z)$ as $z\to z_0$ along $[0,z_0)$.
\label{PB3904}
\end{pro}

\noindent{\bf Proof.} By (\ref{B3903}) and by the definition of $f_0(z)$
we have
$$
\det(A(z))=\left(d_1(A(z))/f_0(z)\right)/B(z).
$$
The factor $d_1(A(z))/f_0(z)$ is holomorphic in $z$ for $z\in U$.\ \ \ $\Box$

%

\vspace{4mm}
{\em Acknowledgements.} These results were produced in June 1993 -
March 1994 when one of us (M.K.) was at the Max-Planck-Institut
f\"ur Mathematik and another (S.V.) was at the Max-Planck-Institut
f\"ur Mathematik (June - September 1993 and from January 1994) and
at the Institute des Hautes \'Etudes Scientifiques (October - December
1993). We are very grateful to the MPIM and to the IHES for their
hospitality, for financial support, and for an excellent intellectual
atmosphere of these mathematical centers. The results and methods
of this paper were reported at the Conference at Rutgers University
in occasion of I.M.~Gelfand 80th birthday in October 1993 and
at the seminars at the IHES and at the Universit\'e de Paris-Sud,
Orsay, in December 1993.

\end{document}